%% file: Doc.tex
\documentclass[12pt]{report}

\usepackage{amssymb,amsmath,amscd,psfig,makeidx}

\usepackage{pifont,tabularx,cite}

\usepackage{floatflt}

\usepackage{multirow}
 
\makeindex

 \makeglossary

\begin{document}                      


 \input{DEFh.i}
 
\renewcommand{\vec}[1]{\mathbf{#1}}

%
%
%
%

\newpage

%
%
%
%




\clearpage
\pagestyle{empty}
{\large\centering\bfseries\upshape\sffamily
%
NORBERT SCHEU\\[1.5cm]
\underline{
ON THE
COMPUTATION OF 
}\\
\underline{
STRUCTURE FUNCTIONS 
 AND
MASS SPECTRA
}\\
\underline{
IN A RELATIVISTIC HAMILTONIAN FORMALISM:
}\\
\underline{
A LATTICE POINT OF VIEW}\\[1.5cm]
 Th\`ese\\
pr\'esent\'ee\\
{\rm
\`a{ }  la{ }  Facult\'e{ }  des{ }  \'Etudes{ }  Sup\'erieures{ }
}\\
de l'Universit\'e Laval\\
pour l'obtention\\
du grade de Philosophi\ae{ } Doctor (Ph.D.)\\[2cm]
D\'epartement de physique\\
FACULT\'E DES SCIENCES ET DE G\'ENIE\\
UNIVERSIT\'E LAVAL\\
QU\'EBEC\\[1cm]
D\'ECEMBRE, 1997\\[1cm]
\vskip 1.5 cm
\copyright  Norbert Scheu, 1997
}
\clearpage

\setcounter{page}{1}  \renewcommand{\thepage}{\roman{page}}


%
%
%
%

 \pagestyle{plain}
\begin{minipage}{0.45\linewidth}
\psfig{figure=libai2.aps,width=\linewidth,angle=180}
\end{minipage}
\begin{minipage}{0.45\linewidth}
{\tiny

\begin{verse}
\vskip 0.5 cm
\underline{Quiet thought at night}
\vskip 0.5 cm
\end{verse}

\begin{verse}
Bright moon light in front of my bed,\\
Maybe... frost on the ground.\\
I raise my head, behold the bright moon,\\
I bow my head--- home-sick.\\  
\end{verse}

\begin{verse}
[L\u{\i}  B\'ai, T\=ang Dynasty poet]
\end{verse}

}
\end{minipage}
%

%
%
%
%

\newpage

\section*{Abstract for HEP-TH}

\subsection*{Modified version of Ph.D. thesis}

  Herein we propose a new numerical technique for solving field theories: the
large momentum frame (LMF). This technique combines several advantages of
lattice gauge theory with the simplicity of front form quantisation. We apply
the LMF on QED(1+1) and on the $\phi^4(3+1)$ theory. We demonstrate both
analytically and in practical examples
(1) that the LMF does neither correspond
to the {\em infinite} momentum frame (IMF) nor to the front-form (FF)
(2) that the LMF is not equivalent to the IMF
(3) that the IMF is unphysical since it violates the lattice scaling window and
(4) that the FF is even more unphysical because FF propagators violate
micro-causality, causality and the finiteness of the speed of light.
We argue that distribution functions measured in deep inelastic scattering
should be interpreted in the LMF (preferably in the Breit frame) rather than
in the FF formalism. In particular, we argue that deep inelastic
scattering probes {\em space}-like distribution functions.

%
%
%
%

\clearpage
\section*{R\'esum\'e I}
\input{CResumeA1.i}

Sign\'e par

 \begin{center}
   (Norbert Scheu)   
\qquad\qquad\qquad\qquad\qquad\qquad
   (Helmut Kr\"oger)
\end{center}


\clearpage
\section*{R\'esum\'e II}
\input{CResumeA2.i}


Sign\'e par

 \begin{center}
   (Norbert Scheu)   
\qquad\qquad\qquad\qquad\qquad\qquad
   (Helmut Kr\"oger)
\end{center}

%
%
%
%

%
%
%
%


\clearpage
\section*{Avant-propos}

\input {remerciements.i}

%
%
%
%

\newpage

 

\tableofcontents

%
%
%
%



\listoftables
\listoffigures

%
%
%
%

\newpage
 \setcounter{page}{1}\renewcommand{\thepage}{\arabic{page}}

\input{Doc.i}



      \bibliographystyle{h-physrev}

\bibliography{Doc}





 \clearpage \setcounter{page}{1} \renewcommand{\thepage}{\Roman{page}}

\printindex      
            

\end{document}

%% file: CResumeA1.i
A non-perturbative computation of \noIndex{hadronic structure functions}
for deep inelastic lepton hadron scattering has not been achieved yet. 
In this thesis we investigate the viability of the Hamiltonian approach
in order to compute hadronic structure functions.  
In the literature, the so-called \noIndex{front form} (FF) approach is favoured over the conventional
the \noIndex{instant form} (IF, i.e. the conventional Hamiltonian
approach) due to claims (a) that
structure functions are related to \noIndex{light-like correlation functions}
and (b) that the front form is much simpler for numerical computations. 
We dispell both claims using general arguments as well
as practical computations (in the case of the \noIndex{scalar model}
and \noIndex{two-dimensional QED}) demonstrating (a) that structure functions are
related to \noIndex{space-like correlations} and that (b) the IF
is better suited for practical computations if
appropriate approximations are introduced. 
Moreover, we show that the FF is \noIndex{unphysical} in general for reasons
as follows:
(1) the FF constitutes an \noIndex{incomplete quantisation} of  
field theories 
(2) the FF "predicts" an \noIndex{infinite speed of light} in one space dimension,
a \noIndex{complete breakdown of microcausality} and the 
\noIndex{ubiquity of time-travel}. 
Additionaly we demonstrate that the FF cannot be approached by so-called
$\varepsilon$ co-ordinates. We demonstrate that these co-ordinates are
but the instant form in disguise. We argue that the FF cannot be considered
to be an \noIndex{effective theory}. Finally, we demonstrate that
the so-called \noIndex{infinite momentum frame} 
is neither physical nor equivalent
to the FF. 

%% file: CResumeA2.i
In this Ph.D. thesis we demonstrate that
\begin{enumerate}
\item  a numerical diagonalisation of a lattice regularised
Hamiltonian may become drastically simpler, for some field theories,
if the lattice moves fast relative to the object which is to be described. 

We propose a new numerical technique based on this simplification.
We apply this technique to the massive Schwinger model and the scalar model  \item $\theta$ vacua and spectral flow arise naturally in our approach

\item structure functions are related to space-like correlation
functions rather than to light-like correlation functions
\item the notion 'closeness to the light-cone' is irreconcilable
with the theory of relativity
\item $\varepsilon$ co-ordinates are completely equivalent to the instant
form for $\varepsilon\neq 0$.
\item the conventional instant-form has to be used in order to 
obtain non-perturbative input for the parton model
\item the front-form is not equivalent to the instant form {\em in general}.
We explain under which circumstances the front-form is able to come
close to the accuracy of the instant form: Prominent examples
are almost non-relativistic theories, scale-invariant theories without
boson-boson couplings and some graphs in perturbation theory 
\item microcausality and the finiteness of the speed of light are 
completely destroyed by light-like boundary conditions --- however large the periodicity may be
\item even if the front-form describes the mass spectrum of a theory
quite accurately, relativistic propagators are badly damaged by 
front form quantisation
\item the front form canNOT be considered to be an effective theory for the 
quantised instant form. The front-form is not even equivalent to the
quantised infinite momentum frame. We remark on are a few, well-defined 
exceptions to this rule. 
\item the UNQUANTISED front form is an effective theory for the UNQUANTISED infinite
momentum frame, however. 
\item beyond the mean field level,
there is, in general, no physical information in zero mode constraints
which arise in the front from. On the contrary, a correct implementation
of zero-mode constraints adds further damage to the already damaged propagators
\item it is not true that computations in the front form are simpler
than in the instant form
\item the FF is not needed in order to treat two-dimensional 
models
in a simple, numerical way. For instance, the front form yields accurate 
mass spectra for
the Schwinger model. So does the
instant form, only with higher accuracy and less effort (only a finite lattice is needed)
\item the instant form effortlessly reproduces chiral perturbation theory 
on small (effective) lattices whereas an infinite lattice or ad hoc counterterms
are necessary to this aim in the front form 
\item the infinite momentum frame is less unphysical than
the front form but unphysical nontheless because it violates
the the lattice scaling window 
\end{enumerate}

%% file: remerciements.i
\begin{verse}
{\scshape\sffamily 
\`a Christina
--- quand m\^eme $\overset{\ddot{ }}{\frown}$ $\overset{\ddot{ }}{\smile}$ }
\end{verse}

Je tiens \`a souligner ma vive reconnaissance 
envers le Prof.~Dr.~Helmut Kr\"oger ---mon directeur de th\`ese---
pour son appui dans ce projet. 
J'ai beaucoup profit\'e de la collaboration avec Helmut, de sa porte
toujours ouverte, des discussions avec lui
et de sa connaissance
 profonde d'un 
grand nombre de domaines en physique. J'ai particuli\`erement 
appr\'eci\'e sa comp\'etence dans le domaine des
ph\'enom\`enes critiques et de la th\'eorie de jauge sur r\'eseau
et son encouragement \`a poursuivre mes id\'ees.

Merci beaucoup aussi aux "habitants" de la tour d'ivoire: \`a
Baabak (et Raamak et le petit Mazdak),
Bertrand, 
Fr\'ed\'eric,
Ghislain, 
Gurgen, 
Gwendoline, 
Hamza (et Laurence), 
Jean-Fran{\c c}ois (Audet), 
Jean-Fran{\c c}ois (Addor), 
Luc, 
Marek, 
Michel, 
Nicolas,
Patrick,
Peter, 
Pierre, 
Robert (et Anne), 
Simon, 
St\'ephane, 
Yorgo,
Yves, 
et Ziad.
J'ai beaucoup aim\'e la bonne et amicale atmosph\`ere qui r\'egnait dans
cette salle et j'en garderai un tr\`es bon souvenir.
Je voudrais aussi mentionner
Alain,
Ali,
D\'enis et Danielle,
Francine Caron et famille,
Frants,
Gilberto et Bartira,
Jean-Fran\c cois,
Lionel,
Marius,
Mike, %
Pierre,
Raymond,
M.Slobodrian,
et
Tim 
. %

Je remercie Messieurs les Professeurs Amiot, Marleau et Potvin
d'avoir accept\'e de corriger ma th\`ese.

Il me fait tr\`es plaisir de remercier 
Claudette, Colette, Diane, Francine et Lise,
les secr\'etaires, car elles ont toujours \'et\'e
gentilles et toujours pr\^etes \`a aider.

$T\tilde{\underset{\iota}{\eta}}$
$\alpha\gamma\nu\tilde\omega \tau\underset{\iota}{\eta}$,
$\tau\tilde{\underset{\iota}{\omega}}$
$\alpha\gamma\nu\tilde\omega \tau \underset{\iota}{\omega}$ 
: 
merci aussi \`a tous ceux que je pourrais avoir oubli\'es dans
l'ardeur de finir la r\'edaction de ma th\`ese.

Je me suis toujours senti bien \`a l'aise 
{\it icitte} $\overset{\ddot{ }}\smile$ au Canada... 
mise \`a 
part ---bien s\^ur!---
*certaines* coupures du budget universitaire $\overset{\ddot{ }}{\frown}$ 
et le temps froid qu'il fait en hiver 
$\overset{\ddot{ }}\smile$ !

\section*{Acknowledgements}
The author wants to express his appreciation for
having been granted the AUFE fellowship from the DAAD (Deutscher Akademischer
Austauschdienst) which has made this Ph.D. project possible.
We are grateful for many discussions with Prof.~Dr.~Dieter Sch\"utte and 
Prof.~Dr.~Xi\`ang-Qi\'an L\`uo.

%% file: Doc.i



 \chapter{Introduction}\label{intro}
 
\input {intro.i}




\chapter{Structure Functions as Short-Distance Physics}\label{CStructure}
\input{CStructure.i}

\chapter{Instant-Form and Front-Form: A Campaign For Real Time}\label{CFF}

\input{CFF.i}

\chapter{ The Massless Schwinger Model}\label{CSchwinger}
\input{CSchwinger.i}

\chapter{ The Massive Schwinger Model}\label{CNum}
\input{CNum.i}

\chapter{ The $\phi^4$ Theory}\label{CPhi4}
\input{CPhi4.i}


\chapter{Discussion}\label{CConclusion}
\input{CConclusion.i}

%% file: intro.i
%
%
%
%
%
%


\reply{In dieser Form erscheinen Kommentare und Fragen fuer die Korrektur. 
Im Index
sind die Seiten mit solchen Kommentaren aufgelistet.}

\section{The Computation of Structure Functions: A Non-Perturbative Problem}

{\ttfamily\small\footnotesize
\begin{citation}

${^{\tiny \subset}}{\acute O}\sigma\omega\nu$ 
$\acute o\psi\iota\varsigma$
$\alpha\kappa o\grave\eta$
$\mu\acute\alpha\vartheta\eta\sigma\iota\varsigma$, 
$\tau\alpha\tilde{\upsilon}\tau\alpha$
$\varepsilon\gamma\grave\omega$
$\pi\rho o\tau\iota\mu\acute\epsilon\omega$ 

\end{citation}
}
{\scriptsize

\begin{citation}

'Tis things visible, audible, perceptible that I prefer. 

[Heracleitos]. Citation found in a paper written by 
G.~Parisi~\Cite{traduttoretraditore}.
\end{citation}

}
One of the most important problems in contemporary particle
physics is the computation of the internal structure of hadrons from first 
principles (i.e. from
\Index{Quantum chromo-dynamics} (QCD)
\index{QCD}).
Information on the internal structure of hadrons
can be obtained through scattering experiments in particle accelerators.
This information  ---the scattering cross section---
can be expressed in terms of 
frame-independent
\Index{structure functions}
 (containing the actual structural information)
and frame-dependent, kinematic factors independent 
of the internal structure of the scattered object.
For a precise definition see~\Cite{RobertsBuch} and Chapter~\secRef{CStructure}.

In the last two decades, a wealth of data on the structure of the proton
has been collected in collider experiments. 
The largest amount of data gathered so far stems from
\Index{deep(ly) inelastic scattering} (DIS) of leptons off 
\index{DIS}
the proton (or off hadrons
in general). This scattering-process 
is particularly important in order to understand how a hadron
is built up in terms of quarks and gluons, its elementary constituents,
since leptons ---being point-like at all experimentally accessible scales---
constitute a clean probe of hadrons.
In the framework of Feynman's parton model, structure functions may be interpreted as linear combinations of quark
\Index{distribution functions}
as long as the resolution $Q$ of the experiment
is sufficiently large when compared to the mass $M_H$ of the hadron. 
A \Index{quark distribution function} 
is the density of quarks with flavour
$i$
carrying a fraction $x_B$ of the total momentum $\vec P$ of the hadron. 
Unfortunately,
the computation of nuclear structure functions
from first principles (i.e. from 
\Index{Quantum Chromo-Dynamics}, QCD)
\index{QCD} 
has not been achieved yet. Perturbative
QCD merely allows to predict 
the dependence of the structure functions on $Q$ whereas 
genuinely non-perturbative methods are called for in order to compute
the $x_B$ dependence of the structure functions. 
\section{Lattice Gauge Theory}
Now, it is generally accepted that 
the most powerful non-perturbative
method in QCD is \Index{Lattice Gauge Theory}(LGT)
(or \Index{Euclidean lattice gauge theory}, ELGT)~\Cite{RotheBuch,MontvayMunsterBuch}. 
\index{LGT}\index{ELGT}
ELGT is so far the only technique capable of computing
the hadronic masses directly from QCD {\it without} phenomenological
assumptions~\Cite{weingarten:masses,lee.weingarten:scalar}. 
ELGT is based on path-integral quantisation with imaginary time. 
In this framework, renormalisable relativistic field theories appear as
theories of statistical mechanics close to a critical point, a fact which 
makes them accessible to powerful Monte-Carlo methods.  
To render numerical computations feasible and free of infinities, 
continuous space-time
{\em must} be replaced by a finite number of space-time points. 
This approximation
---referred to as (lattice) \Index{regularisation} in the literature---
partially 
destroys the Poincar\'e symmetry of the QCD Lagrangian 
(e.g. rotational invariance and boost invariance). The Lagrangian
is now invariant under the {\em discrete} symmetry group of the lattice. 
One can show, however, 
that Poincar\'e 
invariance is restored in the so-called \Index{continuum limit},
i.e. in the limit
where the correlation lengths $\xi$ of Green functions (measured in 
units of the lattice spacing) diverge. In praxis, the Poincar\'e invariance
of the Lagrangian is {\em approximately} restored already
for correlation lengths
that are only slightly larger than the lattice spacing.
While external symmetries are thus automatically restored in the 
continuum limit, this does not hold for
internal symmetries such as gauge invariance. The defining Lagrangian
should therefore, ideally, be exactly invariant under internal symmetries
(unless one is able
to disentangle physical states and spurious states). 
This requires the gauge-group to be compact and gauge-theories to be regularised on a space-time
lattice rather than on a momentum-lattice. This point was first 
realised by Wilson in the case of QCD (\Index{Wilson action}~\Cite{Wilson:1974sk})
and, earlier, by Wegner in the case of a 
discrete gauge theory~\Cite{WegnerZ2}.

In the last years, important progress has been made in ELGT.
The first moments of nucleon structure-functions, for instance, can now be 
computed for the first time
~\Cite{Martinelli:1989rr,Martinelli:1988bh,Martinelli:1987zd,Martinelli:1987si}
~\Cite{Best:1997qp,Gockeler:1997jx,Gockeler:1997jk,Gockeler:1996mu,Gockeler:1996bm,Gockeler:1995de,Gockeler:1996wg}.
These moments, however, are computed using the so-called 
\Index{quenched approximation}. They
represent, roughly, the moments of valence
structure functions rather than the moments of full structure
functions including sea quarks. 
For the latter ones, fermions have to be accounted for dynamically~\Cite{liu.dong:quark,MontvayMunsterBuch}. 
Computations beyond the quenched approximation are 
much more difficult to perform (i.e. they require much more CPU time) but 
there is no reason for why their computation
should not be achieved in the near future. 
A further problem of ELGT is that the direct
computation of structure functions
or distribution functions would require the computation of a four-point
function. A four-point Green function, however, is extremely difficult
to compute with
the present lattice technology~\Cite{wilcox:1993uq}. 
Moments of distribution functions,
in contrast, as well as $1/Q$ corrections thereof,
can be computed via three-point Green functions
which are well under control~\Cite{wilcox:1993uq}.
 
Minkowsky space-time (as opposed to Euclidean space-time
based on imaginary time)
can be replaced by a lattice, too. 
In Minkowsky space-time, the path-integral formalism is less practical
than in Euclidean space-time
and it is more advantageous to work in the framework of the 
\Index{transfer matrix}
formalism~\Cite{RotheBuch,MontvayMunsterBuch} since the knowledge of all eigenvectors and
eigenvalues of the transfer matrix $\Bbb{T}$
\index{T@$\Bbb{T}$} is equivalent to a complete solution
of the theory.

The \Index{transfer matrix} 
is the generator of the discrete group of finite
lattice translations in temporal direction: 
in the
limit of vanishing temporal lattice spacing $a_t\rightarrow 0$ 
the transfer matrix 
\be
\Bbb{T}=\exp(-ia_t H)\approx 1-ia_t H
\ee
can be replaced by a generator of the Poincar\'e group, 
the \Index{Hamiltonian},
and we end up with the familiar Hamiltonian formulation of quantum mechanics.
Yet choosing the temporal lattice spacing $a_t$ smaller than the spatial
lattice spacing $a$ is not as innocent as it may seem. 
It necessitates, in principle, the introduction of additional
relevant operators and coupling constants
\footnote{This is referred to as
\Index{renormalisation of the speed of light}} into the
action which are excluded by the symmetries of a symmetric lattice.
Fortunately, however, the exclusion of these
operators seems to be justified as 
ELGT calculations on anisotropic lattices seem to indicate
~\Cite{Alford:1996nx,morningstar:1997ff,morningstar:1997ix,morningstar:1997ze}.

The Hamiltonian which corresponds to the Wilson Lagrangian of lattice 
gauge theory is the so-called \Index{Kogut-Susskind Hamiltonian}, first
derived in~\Cite{Kogut:1975ag}.
{\bf The aim of this thesis} is to explore the {\bf Hamiltonian approach} towards relativistic 
field theories from the point of view of lattice (gauge) theory. 
The Hamiltonian approach provides us with 
the advantage of the intuitive
particle picture which is somewhat obscured in an imaginary-time formalism such
as Euclidean lattice field theory. Once wave functions are computed, the 
computation of distribution functions  
is straightforward. 
A second advantage is that it is relatively easy to compute
scattering observables such as structure functions 
(or other S-Matrix elements),
once the eigenfunctions of the Hamiltonian are found. 

\section{A Brief Review of Hamiltonian Methods}
Over recent years several researchers have explored Hamiltonian methods.
Prominent examples are the work of L\"uscher~\Cite{Luscher:1984gm} 
and van Baal~\Cite{Koller:1987yk,vandenHeuvel:1994ah}
who 
have discovered that much physics of the low-lying $QCD$-spectrum, at least for small lattices,
can be described by zero-momentum
dynamics plus a suitable treatment of the remaining degrees of freedom.
H.~Kr{\"o}ger et al. used the Hamiltonian formalism in order to compute S-matrix elements~\Cite{Chaara:1994ar,Berube:1991ys,Berube:1990iq,ScattOnLatt,Briere:1989wj}. 
But Hamiltonian methods have not been mainstream in the domain 
of non-perturbative methods. One reason for this is that 
the particle number is not conserved in relativistic QFTs:
%
any interaction in relativistic quantum mechanics is capable of 
producing particle-anti-particle pairs. Ultra-relativistic objects  
such as the proton
are thus complex mixtures of few-body and many-body physics: 
Even the vacuum has a non-vanishing density of gluons, quarks and
anti-quarks. Accordingly, the vacuum contains an infinite number of 
virtual particles in the \Index{thermodynamic limit}, i.e. the limit where 
the lattice size becomes infinite; even the fluctuation of the particle number
diverges.

In the applications of the Kogut-Susskind Hamiltonian to 
QCD~\Cite{Kogut:1975ag}, several groups have developed clever ways to take into account a large number of degrees of freedom, e.g., via the $t$-expansion method by Horn and co-workers~\Cite{vandenDoel:1987xk}.
The $\exp[S]$ method, coming from nuclear physics, 
is a very effective method in order to deal with
the volume divergences in the 
virtual particle number.
A real breakthrough in the application of 
the $\exp[S]$ method
to the Kogut-Susskind Hamiltonian of pure QCD has been achieved recently by~\Cite{Luo:1997za,Schutte:1997du,hamer:1996zj,Luo:1997sa,Luo:1996ha}:  
glueball masses and string tension have been correctly estimated.
Further Hamiltonian approaches are front-form quantisation~\Cite{BrodskyPauli}
(see \secRef{CFF}),
and ---quite recently--- a 
Hamiltonian renormalisation group approach~\Cite{morningstar:1996ig,morningstar:1994fh,morningstar:1994sz}.
For a more thorough review of Hamiltonian lattice gauge theory
we refer the reader to~\Cite{Guo:1997sc,Schutte:1997du}. 
\section{Advantages of the Lorentz-Contraction:\\ Proposal and Test of a New Technique}
%
In this thesis, we present a new method which drastically simplifies the 
numerical diagonalisation of a relativistic lattice Hamiltonian. The 
inspiration to our method comes from Feynman's \Index{parton model}.
The parton model  
necessitates a fast-moving hadron rather than a hadron at rest
in order for the distribution functions to be related to 
the structure functions in a simple way.
We are able to show that, surprisingly, a hadronic state which
moves sufficiently fast 
{\em relative to the lattice} can be
dramatically simpler when compared to a bound state at rest.
Part, but not all, of this simplicity stems from the fact that a fast, 
Lorentz-contracted object fits into smaller lattices as we shall see. 
We shall refer to a frame in which the hadron moves with large
but {\em finite} momentum relative to the lattice as a 
\Index{large momentum frame} (\Index{LMF}). 
The LMF must not be confused with another, similar frame, 
the \Index{infinite momentum
frame} (\Index{IMF}), i.e. a frame wherein all particle 
masses can be neglected compared to the energies of these particles.
While
the IMF is admissible for some elementary perturbative calculations, 
the IMF cannot be used, in general, for the non-perturbative
computation of distribution functions on a {\em finite} lattice:
We demonstrate this in Chapter~\secRef{CFF}. We argue that 
the IMF on a finite 
lattice is unphysical in general since
it is incompatible with the \Index{scaling window}~\Cite{MontvayMunsterBuch} of LGT.
The LMF, in contrast,
allows the limit of infinite momentum (relative to the lattice)
to be approached on a finite lattice without leaving the scaling
window of LGT.
In Chapter~\secRef{CNum} we illustrate this by practical computations.
We apply the LMF technique to two models: quantum electrodynamics in $1+1$ space-time dimensions ---referred to as massive \Index{Schwinger model}
or QED$(1+1)$---
\index{massive Schwinger model}
and the scalar $\phi^4$ model in $3+1$ dimensions. 
It turns out that it is much simpler, in these models, 
to describe a physical particle that moves sufficiently fast relative to the lattice than to describe a particle at rest.
We demonstrate, both theoretically and with practical examples, 
that a physical particle cannot move arbitrarily fast
on a finite lattice, implying, in particular, that the IMF
is {\em unphysical} (with few exceptions
in perturbation 
theory or purely fermionic systems). 
We also demonstrate that the parton distribution functions 
receive significant contributions from the vacuum and that 
there is only one reference frame, the Breit frame, 
in which the vacuum contributions cancel entirely.  

\clearpage

\section{Does the Hamiltonian exist?}
{\ttfamily\small

\begin{citation}

Begriffe und Begriffssysteme erhalten die Berechti-\\
gung nur dadurch, da{\ss}
sie zum {\"U}berschauen von Erleb-\\
niskomplexen dienen; eine andere 
Legitimation gibt\\ 
es f{\"u}r sie nicht. Es ist deshalb nach meiner 
{\"U}berzeugung\\
einer der verderblichsten Taten der Philosophen, da{\ss}\\
sie gewisse begriffliche Grundlagen der Naturwissen-\\
schaft aus dem der Kontrolle 
zug\"anglichen Gebiete des\\ 
Empirsch-Zweckm\"a{\ss}igen
in die 
{\em unangreifbare H{\"o}he des\\
Denknotwendigen} (Apriorischen) versetzt haben.  
\end{citation}

[Albert Einstein: Grundz\"uge der Relativit\"atstheorie.]

}
Historically, during the 60's and 70's,
quantum field theory (\Index{QFT}) in general and Hamiltonian field theory in particular were considered to be amateurish:
they were neglected in favour of the boot-strap programme, the
hope of finding the S-matrix for the forces of nature from principles
such as duality, analyticity, crossing-symmetry and the like. 
Many findings of this time, such as Regge theory and dispersion-relations
remain relevant independently of the underlying theory whereas
the ambitious boot-strap programme itself failed: QFT prevailed. 
The reason for the widespread mistrust of QFT in its earlier stages
was the dominant {\em philosophy} that 
Poincar\'e symmetry had to be treated as an {\em exact} symmetry 
in any sensible computation. Discretisation of space-time ---
the modern LGT approach--- was not yet seriously considered. Taking
{\em exact} Poincar\'e
invariance of the Lagrangian
as an axiom combined with other physically motivated axioms,
it can be shown in the framework of \Index{axiomatic field theory}
~\Cite{HaagsTheorem} that relativistic 
QFT is not well-defined except for non-interacting
theories. 
This problem is solved in the modern approach which interprets renormalisable
QFTs
as systems close to a critical point.
In the framework of LGT, fields
are defined on a finite lattice replacing continuous space-time;
an infinite number of effective field degrees of freedom
is replaced by a finite number of degrees of freedom.
Poincar\'e invariance of the Lagrangian arises {\em dynamically}, i.e. it
is restored to arbitrary accuracy when approaching
the critical point. Even though some quantities, such as bare parameters
or field fluctuations,
diverge in the infinite volume limit, this does not constitute
a problem in the lattice
approach as these quantities are finite on any finite lattice. 
\section{Is Quantisation on Light-Like Quantisation Surfaces Viable?}
Before the modern QFT philosophy was fully developed,
the Hamiltonian formalism had been re-introduced 
in the form of the so-called \Index{front-form}(FF)
\index{FF} 
quantisation~\Cite{dirac:forms}. This approach seemed
to lack the ''problems'' that afflicted the usual or \Index{instant form}
(IF)\index{IF}
Hamiltonian formalism. 
In particular, the vacuum in this formalism seemed to be trivial and
the infinite field fluctuations seemed to be absent. 
Some researchers went even so far as to claim that the FF approach
was well defined due to the absence of infrared-singular field fluctuations
whereas the IF was not. 
The FF approach is partially
successful when applied to some theories in
$1+1$ dimensions: It describes observables such as
distribution functions and mass spectra of 
e.g. \Index{quantum electro-dynamics} in
two dimensions (QED(1+1))~\Cite{Heyssler:1995yt}\Fehlt{preHeyslerQCD}
\index{QED@QED$(1+1)$}
or QCD(1+1) with little numerical effort compared to
LGT~\Cite{Eller:1987nt,Mo:1992sv}. 
Our approach can be seen as a generalisation of the 
FF approach in the sense that the approximations on which the
IMF approach is based are considerably
less severe than the (implicit!)
approximations
the FF is built upon:
whenever the FF convincingly describes physics, so does the LMF-- usually with more ease and
more accuracy. The contrary does not hold true (with the possible
exception of coincidences). 
There is, nonetheless, still a discussion going on whether the FF is 
an exact method 
equivalent to path-integral quantisation the same
way the instant-form quantisation is equivalent to path-integral 
quantisation. This question is discussed (in a
{\em perturbative} framework) in Ref.~\Cite{Ligterink:1995tm,Schoonderwoerd:1997tn}.
It is known that the
FF retains only half of the field degrees of freedom that are necessary
in the IF quantisation, since half of the equations of motion
are constraints (i.e. they do not involve the FF-time)
~\Cite{BrodskyPauli}.
However, it is often argued that the "missing" degrees
of freedom are only necessary in the IF. The reduction of the number
of degrees of freedom is then considered to be a major advantage of the FF 
since this simplifies the FF Hamiltonian enormously when compared
to an IF Hamiltonian.
{\bf 
It is therefore important to demonstrate ---once and for all---
that the FF is indeed an approximation rather than a rigorous
way of quantising relativistic field theories.} This is done
in Sec.\secRef{CFF}. 
The fact that the FF and the IF are not equivalent 
has important consequences for the interpretation of DIS experiments
since Feynman's parton model is often interpreted in terms
of FF distribution functions (in addition to other possibilities
such as the IMF and the Breit frame). 
Sometimes it is even claimed that the FF is the {\em only} way to
interpret the parton model; in particular, it is often
claimed that distribution functions must be interpreted in terms
of light-like correlation functions, which --- if it were true ---
would necessitate the FF approach in order to properly interpret DIS. 
We demonstrate, however, that structure functions
are related to distribution functions obtained by conventional IF quantisation
rather than distribution functions obtained by FF quantisation. 
In particular, we demonstrate that distribution functions can
be related to light-like correlation functions only if an unphysical frame is
introduced.

We have also provided some intuitive examples illustrating the nature of the approximations that go with a FF quantisation. 
In order to show that the FF and the IF approaches are not
equivalent {\it in general} it suffices to demonstrate their inequivalence
for one special field theory. We have therefore chosen one of the
simplest field theories, the $\phi^4$ theory to make our point. 
In particular, we were able to show for this theory that the
so-called left-movers, i.e. the degrees of freedom 
missing in the FF are in a subtle way responsible for the crucial
property of microcausality. 
We show that ---contrary to the LMF approximation proposed in this thesis---, 
{\bf {\em micro}causality does not hold in the FF} even in a
non-interacting scalar field theory.
It is well-known that {\em causality} of time-ordered propagators 
is hampered if they are derived via FF quantisation, i.e. waves
with positive energies are not necessarily moved forward in (real) time. 
One usually argues that (a) this is no serious reason for abandoning
the FF approach and (b) 
this defect can be 
repaired~\Cite{McCartor:1994mu,Mandelstam:1983cb,Leibbrandt:1984pj}. 
Violation of micro-causality, however, cannot be discarded so easily. 
Either one refrains from using light-like periodic boundary conditions ---in which
case FF quantisation is not defined--- or else one is faced
with {\em observable} unphysical "predictions" such as time-travel
and an infinite speed of light in one spatial direction. 
We also show that the left-movers make a substantial
self-energy contribution to the mass-spectrum of interacting bosons
whereas
interacting fermions do not receive this contribution. 
In Sec.\secRef{CSchwinger} we take another simple example related
to a non-interacting field theory where the FF is unable to 
reproduce the results of the IF: The (massless) Schwinger-model. 
We show that calculation of the mass-spectrum of the
Schwinger-model is as simple 
in the FF as in the IF if the unphysical axial gauge is chosen. 
In the limit of small fermion masses, the LMF method reproduces
results from chiral perturbation theory with ease whereas the FF needs
an infinite lattice in order to do so. 

\section{Organisation}

This thesis is organised as follows:

In Chapter~\secRef{CStructure} we compute some DIS structure
functions in the \Index{impulse approximation} (\Index{IA}). 
We argue that this approximation
cannot be frame-independent if the vacuum is non-trivial
and that the Breit frame is the most advantageous choice of frame in the
sense that the impulse approximation is most accurate in this frame.  
We demonstrate that distribution functions must be interpreted
in terms of space-like correlation functions: these
correlation functions become light-like if and only if an infinite, unphysical
boost is performed which collapses all space-like and light-like
quantities onto a light-front. 
 
In Chapter~\secRef{CFF} we demonstrate the non-equivalence (in general)
of the FF and the IF formalism
and the strongly unphysical character of light-like boundary conditions.
We trace the problems of the FF to the need to introduce boundary 
conditions which --- as we are able to show --- 
provoke an unacceptable, complete breakdown of microcausality. 
We introduce the notion of \Index{kinematical equivalence
of quantisation-frames} and show
that the so-called $\varepsilon$ co-ordinates do not legitimate
the notion of ''closeness to the light-cone''. Quantisation
in these co-ordinates is a mere re-parametrisation of the 
IF for $\varepsilon\neq 0$ and equivalent to the FF for $\varepsilon=0$.
Expressed differently: quantisation in $\varepsilon\neq 0$ co-ordinates is IF quantisation
in a more clumsy form and the limit $\varepsilon\rightarrow 0$ is 
not continuous in general. 

We demonstrate that the IMF and the FF violate elementary 
requirements of lattice (gauge) theory. The IMF is in general 
unphysical. The FF is even more unphysical
in that it is an effective theory to the IMF {\em only}
for the classical, unquantised theory. 
In quantised form, the FF cannot be considered to be an effective
theory to the IMF. There are exceptions to this rule as
QED($1+1$), for instance.
We explain under which circumstances the FF can be almost equivalent
to the IF. 

In Chapter~\secRef{CSchwinger} we use the exactly known solution
to massless QED in two space-time dimensions in order to study 
the impact of various approximations. In particular we justify
the assumptions that went into our numerical computations in 
Chapter~\secRef{CNum}. 

In Chapter~\secRef{CNum} we apply our LMF method to QED in two 
space-time dimensions.
We demonstrate that the masses of the vector boson
and of the scalar boson are accurately described
for the whole range of the fermionic mass. In particular
our method is able to reproduce the linear 
fermion-mass-dependence of the mass spectrum
on a {\em finite} lattice whereas an infinite lattice
is necessary in the FF or in the IMF.

In Chapter~\secRef{CPhi4} we include already published papers
on the four-dimensional
$\varphi^4$ model wherein we compare our method
with 
Ref.~\Cite{luscher:1987ay,Brezin} showing
that we are able to describe reasonably well
the critical properties and the critical line 
of the scalar model. 

In Chapter~\secRef{CConclusion} we draw a few conclusions.

\section{Notation}
 We are using natural units $\hbar=c=1$ throughout this thesis ($\hbar$
is Planck's constant, $c$ is the velocity of light). 
\index{h@$\hbar$}
\index{c@$c$}
Please note that we properly distinguish co-variant four-vectors $x_\mu$ and
contra-variant four-vectors $x^\mu$ since we are using non-orthogonal 
frames in Chapter~\secRef{CFF}. Our notation will therefore be close
to~\Cite{ItzyksonBuch}.
\index{x@$x^\mu$}
\index{x@$x_\mu$}
Accordingly, if $x$ denotes a four-vector,
the expression $x^2$ does in general {\em not} mean $x\cdot x$
unless it is unambiguously clear from the context. 
We shall use the abbreviations $x^\pm\dn x^0\pm x^3$, 
$x_\pm={1\over 2}x^\mp$ (the so-called \Index{light-front co-ordinates} or
\Index{light-cone co-ordinates}. 

In every chapter except for Chapter~\secRef{CFF} we use a diagonal
metric $g$
\index{g@$g$} 
with $g_{11}=g_{22}=g_{33}=-g_{00}=-1$. In these chapters, we can
replace $(x^\mu)^2$
by $x^2_\mu$ since then $x_\mu=\pm x^\mu$.
Four-vectors are represented in terms of their co-ordinates in the form 
$x=(x^0,x^1,x^2,x^3)=(x^0,\vec x)$. Sometimes we label a four-vector
$x^.=(x^0,\vec x)$ or a tensor with a point 
in order to underline that $\vec x$ consists of contra-variant
components.

We also
have to distinguish the distribution $\delta^{(3)}(\vec k-\vec k')$ from
the Kronecker symbol $\delta_{x,y}=1\text{ or } 0$. 
When working in a finite box with length $2L$
in three-direction and $2L_\perp$ 
in one- and two-direction (where the momenta $\vec k$
become discrete) it is explicitly assumed that
the $\delta$ distribution becomes
a function defined as 
\be
\delta^{(3)}(\vec k-\vec k')
=\delta_{\vec k,\vec k'} ({L\over \pi})({L_\perp\over \pi})^2
\ee
and $\int d^3x$ abridges 
\be
\int_{-L\perp} ^{L_\perp}
\int_{-L\perp} ^{L_\perp}
\int_{-L     } ^{L      }
dx^1dx^2dx^3
\qquad.
\ee
Only if the momentum lattice spacing $\triangle k\dn {\pi\over L}$ is one
do $\delta_{\vec k,\vec k'}$ and $\delta^{(3)}(\vec k-\vec k')$ coincide. 
It is convenient to define 
other quantities such as annihilation operators similarly such that
$b^{\dag} _{\vec k}\cket{0}$ \footnote{$\cket{0}$ is the perturbative vacuum}
is normalised to
\be
\brac{0} b_{\vec k} b^{\dag} _{\vec k'} \cket{0}=\delta_{\vec k,\vec k'}
\ee
whereas $b(\vec k)$ is defined such as 
\be
\brac{0} b(\vec k) b^{\dag} (\vec k') \cket{0}=\delta^{(3)}(\vec k-\vec k')
\ee
In the literature, the \Index{helicity} 
${\cal H}=\pm {1\over 2}$ of a proton is often
defined to take on the values $\pm 1$.  
In order to facilitate comparisons with both notations, we shall
write helicities with a bracket $({2\cal H})=\pm 1$.

As this thesis uses the formalism of several, disparate branches of physics
such as quantum field theory, scattering theory, lattice gauge theory, 
numerics, many-body theory, solid state physics, 
constraint quantisation, group theory
and some notation from general relativity we are faced with the problem
that one symbol may mean different things in these domains. In order to avoid
changing familiar symbols 
we distinguish these symbols through four 
different fonts: $italic$, $\sf sans$ $\sf serif$, 
$\cal CALIGRAPHIC$ and $\Bbb{BLACKBOARD}$
rather than through the introduction of entirely new symbols, 
a procedure which conserves the familiarity of
symbols and reduces the ambiguity at a time. 
If it is not clear from the context that a quantity $\cal{O}$ 
is an \Index{operator} rather than a number then this 
operator $\hat{\cal O}$ is identified as such with a hat
.
\index{$\hat{ }$}
The transposition of a matrix $T$ is denoted $T^\top$, its hermitian
conjugate $T^{\dag}$.
\index{${ }^\top$}
\index{${ }^{\dag}$}
\Index{Normal ordering} of an operator $\hat{\cal O}$ is
written as $:\hat{\cal O}:$.\index{:}
We shall be using the \Index{Heisenberg picture} of quantum mechanics
throughout this thesis. 
For the reader's convenience, most symbols that have been used
are listed in a separate index at the
end of this thesis. The index is alphabetically ordered except for operators
(e.g. $\conv{}$) which appear before the letter $a$. Greek letters are ordered
according to the first letter of their Latin transcription, 
e.g. $\alpha=alpha$
is treated as $a$, $\omega=omega$ is treated as $o$. 

\section{Methodology}

The numerical part was done using a combination of C++, 
UNIX and \Index{mathematica}.
A programme for the algebraical manipulation of quantised
fields and their creation operators was designed and 
matrix elements of the Hamiltonian were algebraically
computed using this programme and automatically translated into C++ code in the
case of the three-dimensional scalar model or into \Index{mathematica} code in the case
of one-dimensional QED. 
The energy spectra were obtained through a numerical diagonalisation
of the thus computed matrices. Since small matrices already sufficed
in order to reproduce very accurate results, the numerical diagonalisations
could be effortlessly performed with \Index{mathematica} routines.

%% file: CStructure.i
\begin{center}
 \begin{table}[t]

\begin{tabular}{|c|c|c|c|c|}
\hline
&Polarisation&Pauli-Lubansky & Helicity & Spin/Transversity \\\hline
Hadronic&${\polar{S}}=(\polar{S^0},\vec {\polar{S}})$ &
$S=(S^0,\vec S)$ & ${\mathcal H}$ & ${\mathcal T}$ \\\hline 
Partonic&${\polar{s}}=(\polar{s^0},\vec {\polar{s}})$ & 
$\pauLub{s}=(\pauLub{s^0},\vec{\pauLub{s} })$ &  $s$ & $-$\\\hline 
\end{tabular}

\caption{Overview of polarisation vectors}\label{OverviewPolarisation}
\end{table}
\end{center}

\section*{Introduction}
\Index{Deep inelastic scattering} of leptons off a hadron 
provides us with information on the internal structure
of hadrons.
The incoming lepton with four-momentum $k$ scatters off the hadron
with four-momentum $P=(E,\vec P)$,\index{P@$P$}\index{E@$E$}\index{M@$M$}
where
$E=P^0=\sqrt{M^2+\vec P^2}$ denotes the energy and $M$ the
mass of the hadron. After the scattering process, the four-momentum
$k'$ of the lepton is measured.\index{k@$k$}\index{k@$k'$}\index{q@$q$}\index{P@$P$}
\index{P@$P'$}
We shall only consider \Index{inclusive scattering}\footnote{Inclusive
scattering means that the final state of the proton is not measured.}.
We are not interested in the exact state
$\cket{P',X}$
\index{PX@$\cket{P',X} $}
of the debris of the hadron {\em after} the scattering process. 
The inclusive differential scattering cross-section $d\sigma$ 
is proportional to the contraction $W^{\mu\nu}l_{\mu\nu}$ of the
hadronic tensor $W^{\mu\nu}$ and the leptonic tensor $l^{\mu\nu}$ if
first order perturbation theory is valid, $q\dn k-k'$ is the (space-like) momentum
of the exchanged virtual photon.
While the leptonic tensor $l^{\mu\nu}$ associated with the
incoming lepton may be calculated perturbatively,
this is not the case for the hadronic tensor
\index{Wmunu@$W^{\mu\nu}$}
\begin{equation}   \label{DefWmunu}
W^{\mu\nu}
=
{(2\pi)^6 E}
\int {d^4 y\over (2\pi)^4 } \; 
e^{+iq\cdot y} 
{\left< PS \right|}
:[{\hat j}^\mu(y),{\hat j}^\nu(0)]:
{\left| PS \right>}
\end{equation}
which contains the information on the internal structure of the hadron. 
In order to compute this tensor theoretically, knowledge of the
hadronic wave function 
$
\cket{PS}
$
\index{PS@$\cket{PS} $}
with normalisation
\be
\bra PS|P'S \ket = \delta^{(3)}(\vec P-\vec P')
\ee
is required. 
Here, ${\hat j}^\mu(x)$ 
\index{J@${\hat j}^\mu$}
stands for the hadronic \Index{current operator} the definition 
of which will be detailed below. The four-vector $S$
\index{S@$S$}
\index{M@$M$}
is the 
\Index{Pauli-Lubansky four-vector}
~\Cite{ItzyksonBuch,DiplomArbeit}
---a relativistic generalisation of the spin three-vector---
with the properties
\be
 S\cdot P=S^0P^0-\vec S\cdot\vec P=0 \text{  and  }
 S\cdot S=-M^2
\ee
which characterises the spin of the hadron as follows:
One can always find a Lorentz
frame in which $P=(M,0,0,0)$ 
and $S=(0,\vec S)$. In {\em this} frame, $M^{-1}\vec S={{\vec J}}$
coincides with the \Index{total angular momentum three-vector} ${{\vec J}}$
\index{J@${{\vec J}}$}
which, in turn, coincides with the spin.
Without loss of generality,
we shall henceforth assume that the hadron moves \Index{right},
\be
P=(P^0,P^1,P^2,P^3)=(E,\vec P)=(E,0,0,P^3)\qquad M^2\dn P\cdot P
\ee
i.e. 
the hadron moves in the positive 3-direction $P^3>0$.
\index{E@$E$}
\index{P@$\vec P $}
The direction of $\vec P$ defines the \Index{longitudinal} direction throughout
this thesis. 
If the hadron is \Index{longitudinally polarised} with 
\Index{helicity} 
\be
{\mathcal H}
\dn
{{{\vec J}}\cdot \vec P  \over |\vec P|}
=
{1\over 2}{\vec S\cdot \vec P  \over E|\vec P|}
=
{1\over 2}{S^0 \over |\vec P| } 
=
\pm  {1\over 2}
\ee
\index{H@${\mathcal H}$}
then $\vec S$ is collinear with 
the spin vector and with $\vec P$ 
\index{SH@$\polar{S}_H$}
\be
S
=(S^0,0,0,S^3)
=({2\mathcal H})(P^3,0,0, E)
=
({2\mathcal H})M{\polar{S}}_H  
\ee
where ${\polar{S}}_H={1\over M}(P^3,0,0, E)$ is the \Index{helicity polarisation axis}. 
If the polarisation direction is perpendicular to the momentum of the hadron,
one says that the hadron is \Index{transversely polarised.} 
In this case, the Pauli-Lubansky four-vector reads
\be
S=(0,S^1,0,0)
=({2\mathcal T})(0,M,0,0)
=({2\mathcal T})M{\polar{S}}_T
\ee
if one (arbitrarily) chooses the 1-direction as polarisation 
axis. Here,
\index{ST@$\polar{S}_T$}
${\polar{S}}_T=(0,M,0,0)$ is the \Index{transverse polarisation axis} and
${\mathcal T}=\pm {1\over 2}$ is the \Index{transverse spin}
\index{T@${\mathcal T}$}
i.e. the quantised spin component in this direction.  
The reader might want to notice that some authors normalise $-S\cdot S$ to one. 
Furthermore, a covariant normalisation
\be
\bra PS,\text{cov} | P'S,\text{cov} \ket = 
2E (2\pi)^3 \delta^{(3)}(\vec P-\vec P')
\ee
of the hadron state 
$\cket{PS}=\cket{PS,\text{cov}}/\sqrt{2E (2\pi)^3 }$ is often used in the literature
in which case the hadronic tensor
reads
\begin{equation}
W^{\mu\nu}
=
{1\over 4\pi}
\int d^4 y \; e^{+iq\cdot y} 
\brac{PS,\text{cov}} 
  :[{\hat j}^\mu(y),{\hat j}^\nu(0)]: 
\cket{PS,\text{cov}} \qquad.
\end{equation}
The hadronic tensor $W$ may be decomposed into a symmetric(S) part
\be
W^{\mu\nu} _S
\dn
{1\over 2} (W^{\mu\nu}+W^{\nu\mu})
\ee
\index{WmunuS@$W^{\mu\nu} _S$}
independent of polarisation effects
and an anti-symmetric(A) part 
\be
W^{\mu\nu} _A
\dn
{1\over 2} (W^{\mu\nu}-W^{\nu\mu})
\ee
\index{WmunuA@$W^{\mu\nu} _A$}containing the polarisation effects. 
Both parts may, in turn, be written as a linear combination
of Lorentz-scalar, dimensionless
\Index{structure functions}~\Cite{RobertsBuch,Jaffe:1996zw}
$F_1(x_B,Q)$, $F_2(x_B,Q)$, $g_1(x_B,Q)$, $g_2(x_B,Q)$ which contain the
structural information proper
\begin{align}
\begin{split}
W^{\mu\nu}_S 
&= (-g^{\mu\nu} - {q^\mu q^\nu \over Q^2})\;F_1(x_B,Q)  \\
&+(P^\mu + {P\cdot q\over Q^2}q^\mu)
 (P^\nu + {P\cdot q\over Q^2}q^\nu)\;{F_2(x_B,Q^2)\over P\cdot q}  \\
&  = 
(-g^{\mu\nu} - {q^\mu q^\nu \over Q^2}) 
\;F_1(x_B,Q)  \\
&+(P^\mu + {1\over 2x_B}q^\mu)
 (P^\nu + {1\over 2x_B}q^\nu) 
\;{2x_B F_2(x_B,Q^2)\over Q^2}
\end{split} \\
\begin{split}
W^{\mu\nu}_A&=
-i\epsilon^{\mu\nu\alpha\beta} {q_\alpha\over (P\cdot q)^2}\times\\
&\left[ 
  \left(g_1(x_B,Q^2)+g_2(x_B,Q^2)\right) (P\cdot q)\; S_\beta      
-  g_2(x_B,Q^2) (q\cdot S)\;               P_\beta
\right]
\end{split}
\end{align}
and Lorentz-covariant kinematic tensors which
do not depend on the structure of the hadron. 
The structure functions depend on the 
two invariants \Index{momentum transfer}
(or scattering \Index{resolution})
\index{Q@$Q$}
\be
Q\dn \sqrt{-q\cdot q}
\ee
and on the \Index{Bjorken scaling variable}
\be
x_B\dn
{Q^2\over 2P\cdot q}
\ee
\index{xB@$x_B$}
where $q\dn (P'-P)$. This is so because there are only two scalar 
quantities that can be formed from the kinematic
four-vectors $P$ and $P'$ characterising the hadron.  
Hadronic tensors computed on a lattice, however, can be expected to depend 
additionaly (weakly) on $\vec P$ because a lattice breaks Poincar\'e symmetry.
Frame-dependent approximations also introduce $\vec P$ into the hadronic
tensor. Of course, this dependence has to disappear in the continuum
limit--- as long as the momentum $\vec P$
lies in a domain where the approximations are accurate. 
We have tacitly assumed here, that parity is an exact symmetry. In
other words, we do not consider \Index{weak interactions} (which
are not parity invariant). If we had taken weak interactions into account,
more structure functions would
have appeared~\Cite{RobertsBuch}. Here, we shall only consider electromagnetic interactions
between the hadron and the probing lepton. 


\section{The Hadronic Tensor and Structure Functions}


In the \Index{Breit frame}, the hadronic tensor takes on a particularly 
simple form. In this frame, there is a particularly simple relation between 
structure functions and the hadronic tensor. 
The \Index{Breit frame} is defined as the frame where $\vec q$ is collinear
with $\vec P$ and $q$ is \Index{at rest} $q^0\equiv 0$. Frames 
with $q^0=0$ are the {\em only} frames where $Q$ corresponds to the
resolution ability of the experiment. For in frames where $q^0 \neq 0$, the 
wave-length of the exchanged virtual photon is not ${2\pi\over Q}$
but rather ${2\pi\over |q^3|}={2\pi\over \sqrt{(q_0)^2+Q^2}}$. 
In the Breit frame, the hadronic tensor reduces to
\begin{equation}
W(x_B,Q;{\mathcal H})=
\left(
\begin{array}{cccc}
W^{00}&0&0&0\\
0&W^{11}&W^{12}&0\\
0&W^{21}&W^{22}&0\\
0&0&0&0\\
\end{array}
\right)
\end{equation}
with
\begin{equation}
W^{12}(x_B,Q^2;{\mathcal H})=
i{2\mathcal H}( g_1 - g_2 {M^2\over (P^3)^2})
\approx
i{2\mathcal H}   g_1  
\end{equation}
if the hadron is longitudinally polarised
and to
\begin{equation}
W(x_B,Q;{\mathcal T})=
\left(
\begin{array}{cccc}
W^{00}&0&W^{02}&0\\
0&W^{11}&0&0\\
W^{20}&0&W^{22}&0\\
0&0&0&0\\
\end{array}
\right)
\end{equation}
with
\begin{equation}
W^{02} = -W^{20} =
-i{2\mathcal T}{M \over P^3} (g_1+g_2)
\approx 0
\end{equation}
if the hadron is transversely polarised. 
The diagonal components read
\begin{align} \label{W00}
W^{00} 
&= {1\over 2}F_L =
 - F_1 + {2x_B E^2 \over Q^2} F_2
=
 - F_1 + {E^2 \over P^3 Q} F_2 \\ \nonumber 
&= 
 - F_1 + (1+ ({2x_B M \over Q})^2) {1\over 2x_B} F_2 
\approx
- F_1 + {1\over 2x_B} F_2 \\ \label{WTT}
W^{11} & = W^{22} = F_1
\end{align}
independently of the polarisation. Equation \Ref{WTT} holds in any frame where
both  $\vec P$ and $\vec q$ point in 3-direction whereas  $W^{33}=0$ and Eq. \Ref{W00} only hold in the Breit frame ($q^0=0$). 
For later purposes it is useful to define the total momentum $P^3$ and
energy $P^0$ in 
the Breit frame as 
\be
P_B={Q\over 2x_B} \text{ and } E_B=\sqrt{M^2+P_B^2}\qquad .
\ee
\index{PB@$P_B$}
\index{EB@$E_B$}

\section{The  Hadronic Tensor: Formal Definition}

The current operators 
\be
\hat j^\mu(x^0,\vec x)=U(x^0) \hat j^\mu(0,\vec x) U^{\dag} (x^0)
\ee
appearing in the definition \Ref{DefWmunu} of the hadronic tensor
are in general very complicated, interaction-dependent objects for
$x^0\neq 0$ as the \Index{time-evolution operator} 
\index{U@$U$}
\be\label{UMatrix}
U(x^0)
\dn e^{-iH x^0}
\dn e^{-iH_0 x^0 -iH_I x^0}
\ee
depends on the interaction-part $H_I$ of the
lattice-regularised Hamiltonian of QCD (or
another quantum field theory). $H_0$ stands for the kinetic energy. 
In what follows, we are not interested in the exact form of $H$ and
the problems involved in its definition. It suffices to know that 
$H_{\text{QCD}}$ is well defined, 
when constructed on a lattice in 
configuration space~\Cite{Kogut:1975ag,MontvayMunsterBuch}
using Wilson's compact lattice variables.
\reply{Was..betrifft}
As to a 
Hamiltonian on a {\em momentum} lattice, we remark that such 
a Hamiltonian would require the use of non-compact gauge fields $A^\mu$
which necessitate, in turn, complete gauge-fixing with all its complications
such as the Gribov horizon etc\reply{Gribov zitieren?}. 
It may or may not be possible to write down such
an object.
In what follows, 
we shall use the momentum-space formalism usually
employed in DIS for the practical reason
that the corresponding 
expressions in configuration
space would be by far more complicated and less intuitive.

Equation \Ref{UMatrix} may be written as a perturbative series.
The conditions under which 
this is a good approximation will be discussed later.
Feynman's parton model is based on the \Index{impulse approximation}
(\Index{IA})
i.e. on zeroth order perturbation theory. In this case $U(x^0)\dn e^{-iH x^0}$
and the currents can be represented in terms of free fermion fields
\begin{equation}
\psi(x)=
e^{-iH_0 x^0} \psi(0,\vec x) e^{+iH_0 x^0}
=
\sum_s \int d^R k\; 
  (u_s(\vec k) e^{-ik\cdot x} b_s        (\vec k) +
   v_s(\vec k) e^{+ik\cdot x} d^{\dag} _s(\vec k)
  )
\end{equation}
obeying the anti-commutation relations 
\begin{equation}
\{ \psi(t,\vec x),\psi^{\dag}(t,\vec y) \}
=
\int {d^{[R]} k}
\sum_s ( u_s\otimes u^{\dag} _s + v^{\dag} _s\otimes v _s )
=
{\eins}\; \delta^{(3)}(\vec x-\vec y) \qquad.
\end{equation}
The symbol ${\eins}$
\index{1@${\eins}$}
is the $(4\times 4)$ unity matrix in spinor space. 
For convenience we have introduced relativistic (R) integration measures
\index{dR@$d^Rk$}\index{dR@$d^{[R]} k$}
\begin{align}\label{dR}
d^R k \stackrel{def}{=} {d^3 k \over \sqrt{ (2\pi)^3 2\omega(\vec k) }}  \\
d^{[R]} k \stackrel{def}{=} {d^3 k \over { (2\pi)^3 2\omega(\vec k) }}
\end{align}
similar to Ref.~\Cite{ItzyksonBuch}.
The fermion fields are expressed
in terms of the spinors
$u$ and $v$ which are normalised to
\begin{align}
\bar u_s u_{s'} = +2m \delta_{ss'} \\
\bar v_s v_{s'} = -2m \delta_{ss'}.
\end{align}
and in terms of the fermionic creation and annihilation operators obeying
the standard anti-commutation relations 
\begin{align}
\{b _{s}(\vec k),b^{\dag} _{s'}(\vec k') \} 
&= 
\delta^{(3)}(\vec k-\vec k') \delta_{ss'}   \\
\{d _{s}(\vec k),d^{\dag} _{s'}(\vec k') \} 
&= 
\delta^{(3)}(\vec k-\vec k') \delta_{ss'} 
\end{align}
where $s=\pm {1\over 2}$ designates the \Index{parton helicity}. 
The spinors obey the completeness relations
\begin{equation}
u_s(\vec k)\otimes\bar u_s(\vec k) 
= 
(\not \! k   + m) 
{1+(2s) \gamma_5 \not \!{\polar{s}}\over 2}
\;
\stackrel{m\rightarrow 0}{\rightarrow}
\;
\not \! k 
{1+(2s) \gamma_5\over 2}
\end{equation}
and
\begin{equation}
v_s(\vec k)\otimes\bar v_s(\vec k) 
=
(\not \! k   - m) 
{1+(2s) \gamma_5 \not \!{\polar{s}}\over 2}
\;
\stackrel{m\rightarrow 0}{\rightarrow}
\;
\not \! k  
{1-(2s) \gamma_5  \over 2}
\end{equation}
where 
\index{s@${\polar{s}}$}
\be
{\polar{s}}(\vec k) = {\omega(\vec k)\over m} 
({|\vec k|\over\omega(\vec k)},{\vec k\over|\vec k|})
\ee
defines, analogously to ${\polar{S}}_H$,
the \Index{helicity-direction} (or \Index{spin quantisation axis})
associated with the fermion. 
The Pauli-Lubansky
four-vector of the fermion 
\be
\pauLub{s} \dn (2s)m{\polar{s}}
\ee
is defined such that it is normalised to $-m^2$. 
\swallow{hier war die Erlaerung}
If transverse momenta can be neglected ---which is assumed in the framework
of the parton model--- it would be equally convenient
to choose a spin polarisation 
\index{s@${\polar{s}}_{\text{spin}}$}
\be
{\polar{s}}_{\text{spin}}={1\over m_\perp}(k^3,0,0,k^0)
\qquad\text{with}\qquad
m_\perp\dn\sqrt{m^2+k_\perp ^2}
\ee
instead of a helicity polarisation. 
There are two ways to compute the hadronic tensor in the 
impulse approximation. One may expand the field commutator in 
terms of bilinears --- which is usually done~\Cite{Jaffe:1996zw}--- 
or one may do everything on the level
of the creation and annihilation operators, which we shall do
herein since we consider it to be closer to intuition. 
We note ---at this place--- that the fermionic field $\psi(x)$ is not
gauge invariant because a gauge-transformation 
$\psi\rightarrow \psi e^{i\alpha}$ changes its phase.  
If the fermions move in a gluonic background, we should 
replace the local fermion field $\psi(x)$ with the non-local field 
$\psi_A(y)\dn U_A(y)\psi(y)$
\index{psiA@$\psi_A$}
\index{UA@$U_A$} 
where $U_A(y)$ (with $U_A(0)={\eins}$) 
is a non-local string of gauge fields~\Cite{Jaffe:1996zw} connecting the points $0$ and $y$.

\section{Structure Functions and Distribution Functions}

The current commutator $[j^\mu(y),j^\nu(0)]$ appearing in the hadronic
tensor involves a product of four fermionic fields $\psi$, each of which is
a sum of quark operators $b,b^{\dag}$ and anti-quark operators $d^{\dag},d$.
Consequently, the current-commutator consists of $2^4=16$ terms
with all possible combinations of particles and anti-particles. The hadronic
tensor may thus be written in the form
\be \label{hadronicTTT}
W=
{(2\pi)^6 E\over (2\pi)^4}\int d^4y\;
d^R k\;d^R k'\;
d^R l\;d^R l'\;\sum_{r=1} ^{16}
T^A _r T^B_r T^C_r
\ee
where $T^{A/B/C}_r$ are constants 
\begin{align}
T^{A} _r
=
T^{A\mu\nu} _r
(s,s',\sigma,\sigma';\vec k,\vec k ',\vec l,\vec l')
\end{align}
\index{T@$T^A$}\index{T@$T^B$}\index{T@$T^C$}
whose dependence on Lorentz indices, momenta and spin we suppress
in order to avoid awkward expressions. We shall write 
the $T$-symbols as 16-dimensional vectors (since the index $r$ runs
over 16 values)
\be
T^B\dn <\hat T^B> \dn \brac{PS}\hat T^B\cket{PS} 
\qquad,
\ee
\begin{equation}
T^A=
\left(
\begin{array}{l}
  \bar   u_s(\vec k) \gamma^\mu u_{s'}(\vec k')  \cdot     \bar  u_\sigma(\vec l) \gamma^\nu u_{\sigma'}(\vec l')  \\
  \bar   u_s(\vec k) \gamma^\mu u_{s'}(\vec k')  \cdot     \bar  u_\sigma(\vec l) \gamma^\nu v_{\sigma'}(\vec l')   \\
  \bar   u_s(\vec k) \gamma^\mu u_{s'}(\vec k')  \cdot     \bar  v_\sigma(\vec l) \gamma^\nu u_{\sigma'}(\vec l')  \\
  \bar   u_s(\vec k) \gamma^\mu u_{s'}(\vec k')  \cdot     \bar  v_\sigma(\vec l) \gamma^\nu v_{\sigma'}(\vec l')  \\
  \bar   u_s(\vec k) \gamma^\mu v_{s'}(\vec k')  \cdot     \bar  u_\sigma(\vec l) \gamma^\nu u_{\sigma'}(\vec l')   \\
  \bar   u_s(\vec k) \gamma^\mu v_{s'}(\vec k')  \cdot     \bar  u_\sigma(\vec l) \gamma^\nu v_{\sigma'}(\vec l')   \\
  \bar   u_s(\vec k) \gamma^\mu v_{s'}(\vec k')  \cdot     \bar  v_\sigma(\vec l) \gamma^\nu u_{\sigma'}(\vec l')  \\
  \bar   u_s(\vec k) \gamma^\mu v_{s'}(\vec k')  \cdot     \bar  v_\sigma(\vec l) \gamma^\nu v_{\sigma'}(\vec l')     \\
  \bar   v_s(\vec k) \gamma^\mu u_{s'}(\vec k')  \cdot     \bar  u_\sigma(\vec l) \gamma^\nu u_{\sigma'}(\vec l')  \\
  \bar   v_s(\vec k) \gamma^\mu u_{s'}(\vec k')  \cdot     \bar  u_\sigma(\vec l) \gamma^\nu v_{\sigma'}(\vec l')    \\
  \bar   v_s(\vec k) \gamma^\mu u_{s'}(\vec k')  \cdot     \bar  v_\sigma(\vec l) \gamma^\nu u_{\sigma'}(\vec l')     \\
  \bar   v_s(\vec k) \gamma^\mu u_{s'}(\vec k')  \cdot     \bar  v_\sigma(\vec l) \gamma^\nu v_{\sigma'}(\vec l')  \\
  \bar   v_s(\vec k) \gamma^\mu v_{s'}(\vec k')  \cdot     \bar  u_\sigma(\vec l) \gamma^\nu u_{\sigma'}(\vec l')  \\
  \bar   v_s(\vec k) \gamma^\mu v_{s'}(\vec k')  \cdot     \bar  u_\sigma(\vec l) \gamma^\nu v_{\sigma'}(\vec l')   \\
  \bar   v_s(\vec k) \gamma^\mu v_{s'}(\vec k')  \cdot     \bar  v_\sigma(\vec l) \gamma^\nu u_{\sigma'}(\vec l')   \\
  \bar   v_s(\vec k) \gamma^\mu v_{s'}(\vec k')  \cdot     \bar  v_\sigma(\vec l) \gamma^\nu v_{\sigma'}(\vec l')    \\
\end{array}
\right)\qquad,
\end{equation}
\begin{equation}
\hat T^B=
\left(
\begin{array}{cl}
 { }[ b_s ^{\dag}(\vec k)   b         _{s'}(\vec k') ,       b^ {\dag} _\sigma(\vec l)   b         _{\sigma'}(\vec l')  ]   \\  
 { }[ b_s ^{\dag}(\vec k)   b         _{s'}(\vec k') ,       b^ {\dag} _\sigma(\vec l)   d^{\dag}  _{\sigma'}(\vec l')  ]   \\  
 { }[ b_s ^{\dag}(\vec k)   b         _{s'}(\vec k') ,       d         _\sigma(\vec l)   b         _{\sigma'}(\vec l')  ]   \\  
 { }[ b_s ^{\dag}(\vec k)   b         _{s'}(\vec k') ,       d         _\sigma(\vec l)   d^{\dag}  _{\sigma'}(\vec l')  ]    \\ 
{ }\\
 { }[ b_s ^{\dag}(\vec k)   d^{\dag}  _{s'}(\vec k') ,       b^ {\dag} _\sigma(\vec l)   b         _{\sigma'}(\vec l')  ]    \\ 
 { }[ b_s ^{\dag}(\vec k)   d^{\dag}  _{s'}(\vec k') ,       b^ {\dag} _\sigma(\vec l)   d^{\dag}  _{\sigma'}(\vec l')  ]   \\ 
 { }[ b_s ^{\dag}(\vec k)   d^{\dag}  _{s'}(\vec k') ,       d         _\sigma(\vec l)   b         _{\sigma'}(\vec l')  ]  \\ 
 { }[ b_s ^{\dag}(\vec k)   d^{\dag}  _{s'}(\vec k') ,       d         _\sigma(\vec l)   d^{\dag}  _{\sigma'}(\vec l')  ]   \\
{ }\\
 { }[ d_s        (\vec k)   b         _{s'}(\vec k') ,       b^ {\dag} _\sigma(\vec l)   b         _{\sigma'}(\vec l')  ]  \\ 
 { }[ d_s        (\vec k)   b         _{s'}(\vec k') ,       b^ {\dag} _\sigma(\vec l)   d^{\dag}  _{\sigma'}(\vec l')  ]  \\ 
 { }[ d_s        (\vec k)   b         _{s'}(\vec k') ,       d         _\sigma(\vec l)   b         _{\sigma'}(\vec l')  ]   \\ 
 { }[ d_s        (\vec k)   b         _{s'}(\vec k') ,       d         _\sigma(\vec l)   d^{\dag}  _{\sigma'}(\vec l')  ]  \\ 
{ }\\
 { }[ d_s        (\vec k)   d^{\dag}  _{s'}(\vec k') ,       b^ {\dag} _\sigma(\vec l)   b         _{\sigma'}(\vec l')  ]  \\ 
 { }[ d_s        (\vec k)   d^{\dag}  _{s'}(\vec k') ,       b^ {\dag} _\sigma(\vec l)   d^{\dag}  _{\sigma'}(\vec l')  ]   \\ 
 { }[ d_s        (\vec k)   d^{\dag}  _{s'}(\vec k') ,       d         _\sigma(\vec l)   b         _{\sigma'}(\vec l')  ]   \\ 
 { }[ d_s        (\vec k)   d^{\dag}  _{s'}(\vec k') ,       d         _\sigma(\vec l)   d^{\dag}  _{\sigma'}(\vec l')  ]   
\end{array}
\right)
\quad,\quad
T^C
=
\left(
\begin{array}{l}
  e^{i(              k\cdot y                - k'\cdot y    +qy )} \\
  e^{i(              k\cdot y                - k'\cdot y     +qy  )} \\
  e^{i(              k\cdot y                - k'\cdot y     +qy  )} \\
  e^{i(              k\cdot y                - k'\cdot y    +qy   )} \\
{ }\\
 e^{i(              k\cdot y              +   k'\cdot y    +qy   )} \\
 e^{i(              k\cdot y              +  k'\cdot y     +qy  )} \\
 e^{i(              k\cdot y               +  k'\cdot y    +qy   )} \\
 e^{i(              k\cdot y               +  k'\cdot y    +qy   )} \\
{ }\\
 e^{i(            - k\cdot y                - k'\cdot y    +qy   )} \\
 e^{i(            - k\cdot y                - k'\cdot y    +qy   )} \\
 e^{i(            - k\cdot y                - k'\cdot y     +qy  )} \\
 e^{i(            - k\cdot y                - k'\cdot y     +qy  )} \\
{ }\\
 e^{i(            - k\cdot y                + k'\cdot y     +qy  )} \\
 e^{i(            - k\cdot y                + k'\cdot y     +qy  )} \\
 e^{i(            - k\cdot y                + k'\cdot y     +qy  )} \\
 e^{i(            - k\cdot y                + k'\cdot y     +qy  )} \\
\end{array}
\right)\qquad,
\end{equation}
Performing the integral $\int d^4y$ in \Ref{hadronicTTT}
leaves us with a lower-dimensional integral
$$
(2\pi)^6 E\;
\int d^R k\;d^R k'\;
d^R l\;d^R l'\;
\sum_{s,s',\sigma,\sigma'}
$$
over momentum conserving $\delta$ distributions
\begin{equation}
\tilde T^C
\dn
\int {d^4 y\over (2\pi)^4}\;  T^C
=
\left(
\begin{array}{l}
\delta^{(4)}(   k                - k'      +q   ) \\
\delta^{(4)}(   k                - k'      +q   ) \\
\delta^{(4)}(   k                - k'      +q   ) \\
\delta^{(4)}(   k                - k'      +q   ) \\
\delta^{(4)}(   k                +  k'     +q   ) \\
\delta^{(4)}(   k                +  k'     +q   ) \\
\delta^{(4)}(   k                +  k'     +q   ) \\ 
\delta^{(4)}(   k                +  k'     +q   ) \\
\delta^{(4)}(  -k                 - k'      +q   ) \\  
\delta^{(4)}(  -k                 - k'      +q   ) \\ 
\delta^{(4)}(  -k                 - k'      +q   ) \\  
\delta^{(4)}(  -k                 - k'      +q   ) \\
\delta^{(4)}(  -k                 + k'      +q  ) \\ 
\delta^{(4)}(  -k                 + k'      +q  ) \\ 
\delta^{(4)}(  -k                 + k'      +q  ) \\ 
\delta^{(4)}(  -k                 + k'      +q  ) \\
\end{array}
\right)
\qquad .
\end{equation}
Expanding the commutators and sandwiching them with the vector ${\left| PS \right>}$
yields 
\begin{equation*}
T^B
=
\left(
\begin{array}{lr}
 <b_s ^{\dag}(\vec k)     b         _{\sigma'}(\vec l') >         \delta^{(3)}(\vec k'-\vec l)\delta_{s'\sigma} 
&+
<  :b         _{s'}(\vec k')        b^ {\dag} _\sigma(\vec l):   >\delta^{(3)}(\vec k-\vec l')\delta_{s\sigma'} \\
  <b_s ^{\dag}(\vec k)    d^{\dag}  _{\sigma'}(\vec l') >         \delta^{(3)}(\vec k'-\vec l)\delta_{s'\sigma} \\
 <  b         _{s'}(\vec k')        d         _\sigma(\vec l) >  \delta^{(3)}(\vec k-\vec l')\delta_{s\sigma'}  \\
0  \\
<d^{\dag}  _{s'}(\vec k') ,       b^ {\dag} _\sigma(\vec l) >     \delta^{(3)}(\vec k-\vec l)\delta_{s\sigma'}  \\
0  \\
  <b_s ^{\dag}(\vec k)      b         _{\sigma'}(\vec l') >       \delta^{(3)}(\vec k'-\vec l) \delta_{s'\sigma} 
&+
<d_{s'} ^{\dag}(\vec k')      d         _{\sigma}(\vec l) > \delta^{(3)}(\vec k-\vec l') \delta_{s\sigma'}  \\
<:b_s ^{\dag}(\vec k)     d^{\dag}  _{\sigma'}(\vec l'): >          \delta^{(3)}(\vec k'-\vec l) \delta_{s'\sigma} \\
<d_s        (\vec k)     b         _{\sigma'}(\vec l') > 
\delta^{(3)}(\vec k'-\vec l)\delta_{s'\sigma}  \\
<  :d_s        (\vec k)     d^{\dag}  _{\sigma'}(\vec l'): > 
\delta^{(3)}(\vec k'-\vec l) \delta_{s'\sigma}
&+
<  :b         _{s'}(\vec k') ,       b^ {\dag} _\sigma(\vec l): > \delta^{(3)}(\vec k-\vec l')\delta_{s\sigma'}  \\
0  \\
 <  b         _{s'}(\vec k') ,       d         _\sigma(\vec l)> \delta^{(3)}(\vec k-\vec l')\delta_{s\sigma'}   \\
0  \\
< d^{\dag}  _{s'}(\vec k')        b^ {\dag} _\sigma(\vec l)  > 
\delta^{(3)}(\vec k-\vec l')\delta_{s\sigma'}  \\
<d_s        (\vec k)      b         _{\sigma'}(\vec l') > 
\delta^{(3)}(\vec k'- \vec l)\delta_{s'\sigma}  \\
<:d_s        (\vec k)     d^{\dag}   _{\sigma'}(\vec l'):>         \delta^{(3)}(\vec k'-\vec l) \delta_{s'\sigma}
&+
<  d^{\dag}  _{s'}(\vec k')        d         _\sigma(\vec l)  > \delta^{(3)}(\vec k-\vec l') \delta_{s\sigma'} \\
\end{array}
\right)
\end{equation*}
which, in turn, simplifies to 
\begin{equation}
\left(
\begin{array}{cl}
\left[f_{s,\sigma'}(\vec k)-f_{\sigma,s'}(\vec l)\right]     
  &\delta^{(3)}(\vec k'-\vec l) \delta^{(3)}(\vec k-\vec l')\delta_{s'\sigma} \\ 
 -\Xi^*  _{\sigma',s}(\vec l')
&\delta^{(3)}(\vec k+\vec l') \delta^{(3)}(\vec k'-\vec l)\delta_{s'\sigma}   \\ 
     \Xi        _{\sigma,s'}(\vec l) 
 &\delta^{(3)}(\vec k'+\vec l) \delta^{(3)}(\vec k-\vec l')\delta_{s\sigma'}   \\
0 
\\
  \Xi^{*}  _{s',\sigma}(\vec k')      
    &\delta^{(3)}(\vec k'+\vec l) \delta^{(3)}(\vec k-\vec l)\delta_{s\sigma'}  \\
0   \\
\left[f_{s,\sigma'} (\vec k) + \bar f_{s',\sigma}(\vec k')\right] 
 &\delta^{(3)}(\vec k'-\vec l) \delta^{(3)}(\vec k-\vec l') \delta_{s'\sigma}    \\
 -\Xi^{*}  _{\sigma',s}(\vec l') 
  &\delta^{(3)}(\vec k+\vec l) \delta^{(3)}(\vec k'-\vec l) \delta_{s'\sigma} \\
-\Xi_{s,\sigma'}        (\vec k)       
 &\delta^{(3)}(\vec k+\vec l') \delta^{(3)}(\vec k'-\vec l)\delta_{s'\sigma}\\
\left[- \bar f_s        (\vec k)  - f_{s'}(\vec k') \right] 
  &\delta^{(3)}(\vec k'-\vec l)  \delta^{(3)}(\vec k-\vec l')\delta_{s'\sigma} \\
0  \\
     +\Xi         _{\sigma,s'}(\vec l)
  &\delta^{(3)}(\vec k'+\vec l) \delta^{(3)}(\vec k-\vec l')\delta_{s\sigma'} \\
0 \\
 \Xi^{*}  _{s',\sigma}(\vec k') 
   &\delta^{(3)}(\vec k'+\vec l) \delta^{(3)}(\vec k-\vec l')\delta_{s\sigma'} \\ 
-\Xi_{s,\sigma'}        (\vec k)     
   &\delta^{(3)}(\vec k+\vec l') \delta^{(3)}(\vec k'- \vec l)\delta_{s'\sigma} \\ 
\left[\bar f_{\sigma's}(\vec k')-\bar f_{s'\sigma}(\vec l')\right]        
   &\delta^{(3)}(\vec k'-\vec l)  \delta^{(3)}(\vec k-\vec l')\delta_{s'\sigma}  \\ 
\end{array}
\right)
\end{equation}
if the expectation values are replaced by 
what we shall call \Index{raw distribution functions} $f_s$
\index{distribution function!raw}
\begin{align}
{\left< PS \right|}b^{\dag} _s(\vec k)b_{\sigma'}(\vec l'){\left| PS \right>}
&=
f_{s,\sigma'}(\vec k,\vec P,S) 
\delta^{(3)}(\vec k  - \vec l') \\
%
%
%
{\left< PS \right|}d^{\dag} _s(\vec k)d_{\sigma'}(\vec l'){\left| PS \right>}
&=
\bar f_{s,\sigma'}(\vec k,\vec P,S) 
\delta^{(3)}(\vec k  - \vec l) \\
%
     f_{s}(\vec k,\vec P,S) &\dn      f_{s,s}(\vec k,\vec P,S)\\
\bar f_{s}(\vec k,\vec P,S) &\dn \bar f_{s,s}(\vec k,\vec P,S)
\end{align}
and \Index{raw pairing functions} $\Xi_{s,\sigma}$
\index{fs@$f_s$}
\index{Xis@$\Xi_s$}
\begin{align}
{\left< PS \right|}d _s(\vec k)b_{\sigma'}(\vec l'){\left| PS \right>}    &=
-\Xi  _{s,\sigma'}(\vec k,\vec P,S) 
\delta^{(3)}(\vec k  + \vec l') \\
%
%
%
{\left< PS \right|}b^{\dag} _s(\vec k)d^{\dag} _{\sigma'}(\vec l'){\left| PS \right>}   &=
-\Xi ^* _{\sigma',s}(\vec l',\vec P,S) 
\delta^{(3)}(\vec k  + \vec l') \\
%
%
%
{\left< PS \right|}b _s(\vec k)d_{\sigma'}(\vec l'){\left| PS \right>}   &=
+\Xi  _{\sigma',s}(\vec l',\vec P,S) 
\delta^{(3)}(\vec k  + \vec l') \\
%
%
%
{\left< PS \right|}d^{\dag} _s(\vec k)b^{\dag} _{\sigma'}(\vec l'){\left| PS \right>}   &=
+\Xi ^* _{s,\sigma'}(\vec k,\vec P,S) 
\delta^{(3)}(\vec k  + \vec l') 
\end{align}
in order to distinguish them from what is called \Index{parton distribution}
functions in the literature.
 
The hadronic spin does not coincide with the total angular momentum
${{\vec J}}$ if
the spin quantisation axis $\vec {\polar{S}}_H$ is
not collinear to the hadron momentum $\vec P$ --- in the case of transverse
polarisation ${\polar{S}}_T$ for instance. In such cases, the spin is no 
longer kinematical.
Up to now we did not use any particular photon momentum.
Now we assume that we are in the Breit-frame, i.e. $q=(0,0,0,-Q)$.
In this case, the delta-distribution (first four terms)
\begin{equation}
\begin{split}
\delta( k'- k -q)
&=
\delta( \omega(\vec k')-\omega(\vec k)) \;
\delta( k^{3'} - k ^3 + Q) \;
\delta^{(2)}(\vec k^{\perp'} - \vec k^\perp)         \\
&=
{\omega(\vec k _Q)\over Q}\;
\delta(k^3-Q/2) \;
\delta(k^{3'}+Q/2) \;
\delta^{(2)}(\vec k^{\perp'} - \vec k^\perp)
\end{split}
\end{equation}
can be expressed in terms of the vectors
\begin{align}
\vec k _Q 
&\stackrel{def}{=}
(k^1,k^2,Q/2)   \\
\vec k _{-Q} 
&\stackrel{def}{=}
(k^1,k^2,-Q/2)
\end{align}
since $k$ and $k'$ are on the energy shell due to the impulse approximation.
In frames with $\vec P\parallel \vec q$ but $q^0\neq 0$ we would have
to replace $k^3_Q$ by
\begin{equation}
k_r ^3
\dn
-{q^3 \over 2}
-
{q^0\over Q} \sqrt{ m^2 + (Q/2)^2 }
\approx
-{q^0+q^3 \over 2}
=
-q_-
\end{equation}
and $k^3_{-Q}$ by
\begin{equation}
k_l ^3
\dn
{q^3 \over 2}
-
{q^0\over Q} \sqrt{ m^2 + (Q/2)^2} 
\approx
-{q^0-q^3 \over 2}
=
-q_+
=
-{(Q/2)^2 \over k_r^3}
\end{equation}
and modify the weights $\omega(k_Q)/Q$ as well. 
Using these expressions, a six-fold integral of the form
\begin{equation}
\begin{split}
&(2\pi)^6
\int d^R k
 d^R k'
 d^R l
 d^R l'
\;
\delta^4(k'-k+q)
\delta^{(3)}(\vec k - \vec l')
\delta^{(3)}(\vec k'- \vec l )
T^{AB}(\vec k,\vec k',\vec l,\vec l')   \\
&=
\int
{d^2 k^\perp \over 4Q\omega(\vec k _Q)}\;
T^{AB}(\vec k _Q,\vec k _{-Q},\vec k _{-Q},\vec k _Q)
\end{split}
\end{equation}
is reduced to an integration over the 
transverse momenta $\vec k^\perp$ weighted with
${ Q \over \omega(\vec k _Q)}$. Here, $T^{AB}$ abridges $T^AT^B$. 
\index{TAB@$T^{AB}$}
Analogously, the delta-distribution (appearing in the last four terms)
\begin{equation}
\begin{split}
\delta( k'- k +q )
&=
\delta( \omega(\vec k')-\omega(\vec k)) \;
\delta( k^{3'} - k ^3 - Q) \;
\delta^{(2)}(\vec k^{\perp'} - \vec k^\perp)    \\
&=
{\omega(\vec k _Q)\over Q}\;
\delta(k^3+Q/2) \;
\delta(k^{3'}-Q/2) \;
\delta^{(2)}(\vec k^{\perp'} - \vec k^\perp)
\end{split}
\end{equation}
leads to
\begin{equation}
\begin{split}
&(2\pi)^6
\int 
 d^R k
 d^R k'
 d^R l
 d^R l'
\;
\delta^4(k'-k+q)
\delta^{(3)}(\vec k - \vec l')
\delta^{(3)}(\vec k'- \vec l )
T^{AB}(\vec k,\vec k',\vec l,\vec l')   \\
&=
\int
{d^2 k^\perp \over 4Q\omega(\vec k _Q)}\;
T^{AB}(\vec k _{-Q},\vec k _{Q},\vec k _{Q},\vec k _{-Q})
\end{split}
\end{equation}
which means that the r\^oles of $k$ and $k'$ are interchanged.
Here we have used the fact that
$
\omega(k^3+Q)=\omega(k^3)
$
implies that $k^3=-{Q\over 2}$.
Delta distributions of the form (appearing in the middle)
\begin{equation}
\delta^{(4)}( k'+ k +q )
=
\delta( \omega(\vec k')+\omega(\vec k)) \;
\delta( k^{3'} - k ^3 - Q) \;
\delta^{(2)}(\vec k^{\perp'} - \vec k^\perp)
=
0
\end{equation}
vanish with a space-like four-vector $q$ since 
$\omega(\vec k)=-\omega(\vec k')$ cannot be fulfilled. 
The same holds for the terms of the form 
 \begin{equation}
\delta^{(4)}( k'+ k -q ).
\end{equation}
We end up with  
\begin{equation}
 \begin{split}
&W^{\mu\nu}=\sum_{s,s',\sigma,\sigma'}\int d^2 k^\perp  {E \over 4Q\omega(\vec k _Q)}\\
 (&
T^{A\mu\nu} _1(s,s',\sigma,\sigma';\vec k_{Q},\vec k_{-Q},\vec k_{-Q},\vec k_{Q}) 
\left[
 f_{s,\sigma'}(\vec k _Q)-f_{\sigma,s'}(\vec k _{-Q})   
\right]
   \\
-& 
T^{A\mu\nu} _2(s,s',\sigma,\sigma';\vec k_{Q},\vec k_{-Q},\vec k_{-Q},-\vec k_{Q})
\;\Xi^*  _{\sigma',s}(-\vec k_Q)  
   \\
+&
T^{A\mu\nu} _3(s,s',-s',s,\sigma,\sigma';\vec k_{Q},\vec k_{-Q},-\vec k_{-Q},\vec k_{Q})
    \; \Xi         _{\sigma,s'}(-\vec k_{-Q}) 
   \\
+ &
T^{A\mu\nu}_{14}(s,s',\sigma,\sigma';\vec k_{-Q},\vec k_{Q},-\vec k_{Q},\vec k_{-Q}) 
  \; \Xi^{*}  _{s',\sigma}(\vec k _Q)
   \\
 -&
T^{A\mu\nu}_{15}(s,s',s',-s,\sigma,\sigma';\vec k_{-Q},\vec k_{Q},\vec k_{Q},-\vec k_{-Q}) 
 \; \Xi_{s,\sigma'}      (\vec k _{-Q})   
   \\
 +&
T^{A\mu\nu}_{16}(s,s',\sigma,\sigma';\vec k_{-Q},\vec k_{Q},\vec k_{Q},\vec k_{-Q})
\left[
\bar f_{\sigma',s}(\vec k _Q)-\bar f_{s',\sigma}(\vec k _{-Q})
\right]  
)
\qquad.
\end{split}
\end{equation}
The fluctuation functions $\Xi$ do not appear if we repeat the
same calculation in the FF (for a definition and references
see Chapter\secRef{CFF}). They correspond to a reflection
of particles ''backward in time''. The equivalent
of these terms in ELGT has been described 
in~\Cite{KFLiu}. At first sight it would seem that the presence
of these functions would spoil the interpretation of structure functions
in terms of distribution functions alone. It would also seem to mean
that a relation between distribution functions and structure function
can only be established in the FF. Fortunately, however, 
the leptonic tensors associated with scattering backward in
time are order ${\cal O}(m)$ 
in the limit $m\rightarrow 0$. The same holds for helicity-flip processes. 
If fermion masses and transverse momenta
can be neglected, only the helicity-nonflip distribution
functions $f_{\sigma\sigma}$ contribute to the hadronic tensor $W^{\mu\nu}$. 
In this limit, the hadronic tensor reads:
\begin{equation}\label{StruFuDiFu}
\begin{split}
W^{\mu\nu} 
&\approx
\sum_{s}
 \int d^2 k^\perp  {E \over 4Q\omega(\vec k _Q)} 
\left(
     l_s ^{\mu\nu}f_s(\vec k _Q)            - l_s ^{\mu\nu}f_s(\vec k _{-Q})    +
\bar l_s ^{\mu\nu} \bar f_{s}(\vec k _Q)-\bar l_s^{\mu\nu}\bar f_{s}(\vec k _{-Q})
\right) \\
&
\approx\sum_{s}
 \int d^2 k^\perp  {E \over 2Q^2} 
\left(
     {\mathsf f}_{s}(\vec k _Q)      l_s^{\mu\nu}   
+
\bar {\mathsf f}_{s}(\vec k _Q)\bar l_s^{\mu\nu}
\right) 
+
{\cal O}(m/Q)
\end{split}
\end{equation}
where we have defined
the \Index{leptonic tensor}
\begin{equation}
\begin{split}
l_s^{\mu\nu}(\vec k,\vec k')
\dn
l_\oplus^{\mu\nu}(\vec k,\vec k')
+
(2s)
l_\ominus^{\mu\nu}(\vec k,\vec k')
\end{split}
\end{equation}
\begin{equation}
\begin{split}
\bar 
l_s^{\mu\nu}(\vec k',\vec k)
\dn
l_\oplus^{\mu\nu}(\vec k',\vec k)
-
(2s)
l_\ominus^{\mu\nu}(\vec k',\vec k)
\end{split}
\end{equation}
\begin{equation}
l_\oplus^{\mu\nu}(\vec k,\vec k')
\dn
4 k^\mu k^{\nu} 
+
2k^\mu q^{\nu} + 2k^\nu q ^{\mu} - {Q^2 } g^{\mu\nu}
\end{equation}
\begin{equation}
l_\ominus^{\mu\nu}(\vec k,\vec k')
\dn
2im \epsilon^{\mu\nu\alpha\beta} q_\alpha {\polar{s}}_\beta
\end{equation}
and the \Index{parton distribution functions} 
\index{distribution function!parton}
(as opposed to the {\em raw} distribution
functions $f$ or distribution functions proper) 
\begin{align}                          \label{PartonDiFu}
{\mathsf f}_s(\vec k^\perp,+k^3;\vec P)
&\dn
f_s(\vec k^\perp,+k^3;\vec P)-\bar f_{s}(\vec k^\perp,-k^3;\vec P)   \\
\bar {\mathsf f}_{s}(\vec k^\perp,+k^3;\vec P)
&\dn
\bar f_s(\vec k^\perp,+k^3;\vec P)- f_{s}(\vec k^\perp,-k^3;\vec P)
=
-{\mathsf f}_s(\vec k^\perp,-k^3;\vec P)
\end{align}
The particular form of \Ref{StruFuDiFu} has an intuitive interpretation.
In the impulse approximation, partons are considered to be
free. A fermion
with momentum $\vec k_Q$ is scattered to a different place $\vec k_{-Q}$
on the momentum lattice. Should the place $\vec k_{-Q}$ be already 
occupied by another parton of the same type, however, then the Pauli 
exclusion principle prevents it from being deposited there. The process
of 'chasing' the occupant of $\vec k_{-Q}$ away would be a scattering event
of higher order. 
 
%
\section{Parton Distribution Functions: Two Definitions}
\begin{center}
\begin{table}[t]

\begin{tabular}{|c|c|c|c|}
\hline
&distribution& unpolarised  & polarised         \\\hline 
parton distribution& ${\mathsf f}_s(\vec k;\vec P)$ & ${\mathsf f}(\vec k;\vec P)$ &$-$       \\\hline 
raw distribution& ${f}_s(\vec k;\vec P)$ & $f(\vec k;\vec P)$ & $-$                  \\\hline 
integrated parton distribution& ${\mathsf q}_s(x_B,Q)$ & ${\mathsf q}(x_B,Q)$ & ${\mathsf g}(x_B,Q)$ \\\hline
integrated raw distribution& ${\mathtt q}_s(x_B,Q)$ & ${\mathtt q}(x_B,Q)$ & $g(x_B,Q)$   \\\hline 
\end{tabular}

\caption{Overview of distribution functions}\label{tabDistributions}
\end{table}
\end{center}
We may define integrated \Index{parton distribution functions} ${\mathsf q}$ 
\index{distribution function!parton}
\index{q@${\mathtt q}_s$}
\index{q@${\mathsf q}_s$}
\begin{align}
{\mathtt q}_s(x_B,Q)
&\dn
E_B\int d^2k^\perp {f}(\vec k^\perp,{Q\over 2},{Q\over 2x_B})
\approx 
{Q\over 2x_B}\int d^2k^\perp {f}_s(\vec k^\perp,{Q\over 2},{Q\over 2x_B})\\
{\mathsf q}_s(x_B,Q)
&\dn
E_B\int d^2k^\perp {\mathsf f}(\vec k^\perp,{Q\over 2},{Q\over 2x_B})
\approx 
{Q\over 2x_B}\int d^2k^\perp {\mathsf f}_s(\vec k^\perp,{Q\over 2},{Q\over 2x_B})
\end{align}
which correspond to what is usually referred to as 'distribution functions'
in the terminology of DIS.
From these we finally form the \Index{unpolarised parton distribution function}
\index{q@${\mathsf q}$}
\be
{\mathsf q}(x_B,Q)\dn {\mathsf q}_{+{1\over 2}}(x_B,Q)+{\mathsf q}_{-{1\over 2}}(x_B,Q)
\ee
and the \Index{polarised parton distribution function}
\index{g@${\mathsf g}$}
\be
{\mathsf g}(x_B,Q)\dn ({2\mathcal H})
\left[{\mathsf q}_{+{1\over 2}}(x_B,Q)-{\mathsf q}_{-{1\over 2}}(x_B,Q)
\right]
\ee
(analogous definitions for the anti-quark distributions $\bar {\mathsf q}$).
This definition implies the crossing relations~\Cite{Jaffe:1996zw}
$
\bar {\mathsf q}(x_B)=-{\mathsf q}(-x_B)
$
.
An overview of the distribution functions introduced so far 
is given in~\tabRef{tabDistributions}. 
The parton distribution function ${\mathsf q}$ may be written as 
${\mathsf q}(x_B,Q)={\mathsf q}(x_B,Q|P_B S)$ 
where 
\index{l@$\ell$}
\index{q@$\tilde{\mathsf q}^\mu$}
\begin{align}\label{IntegrationsKontur}
{\mathsf q}(x_B,Q|PS)
&\dn
\int_\ell
{dy^\mu\over 2\pi} 
\tilde{\mathsf q}_\mu(y|PS)  
e^{iy\cdot q/2}                 \qquad;\quad
\ell
\dn \{q s\text{ ; s}\in\Bbb{R} \}        \\
\tilde{\mathsf q}^\mu(y|PS) 
&\dn
(2\pi)^3E
\brac{PS}
\bar \psi_A(Y)\gamma^\mu\psi_A(Y+y)
\cket{PS}   \\
&=
(2\pi)^3E
\brac{PS}
\bar \psi(Y)\gamma^\mu U_A(Y+y)\psi(Y+y)
\cket{PS}    
\qquad.
\end{align}
is a {\em space-like} line-integral. $Y$ is an arbitrary four-vector
which shall be chosen as $Y=0$ from now on. This can be verified by 
inserting free fields into \Ref{IntegrationsKontur}.
The gauge string $U_A(y)\dn \exp(i\int_0 ^y dy^{'\mu} g A_\mu(y'))$
ensures the gauge-invariance of ${\mathsf q}$. It contains a {\em space-like}
contour-integral over $A$ linking the points $0$ and $y$. 
In the Breit frame, the integration contour $\ell$ lies on the
quantisation hyper-surface $x^0$
and $U_A$ contains an integral over the $A_3$ component of the gauge-field.
The integration contour appears to be light-like only in a frame
that moves with the speed of light relative to the Breit frame--- but this
perception is wrong. The momentum transfer $q$ is always space-like, even in the
IMF. The properties ''space-like'' and ''time-like'' are Lorentz-invariant
properties:
they cannot be changed by boosts. 
In the axial gauge $A_3=0$, the gauge-string $U_G={\eins}$ is the unit-operator.
For any space-like $q$, it is a space like \Index{axial gauge} $A\cdot q=0$ 
which eliminates $U_A$---{\em not} the light-like gauge $A_-=0$
sometimes called \Index{light-cone gauge}. This gauge coincides
with the gauge
$q\cdot A=0$ only if the exchanged photon with momentum $q$ is {\em real},
i.e. $q\cdot q=-Q^2=0$. 
Only in this case does the light-cone gauge constitute an advantage. 
The claim in the literature~\Cite{Jaffe:1996zw,RobertsBuch} 
that structure functions have to
be expressed in terms of light-like correlation functions is
based on a {\em frame dependent} argument.  
This argument~\Cite{Jaffe:1996zw} which allegedly proves that DIS
is dominated by light-like correlation functions
is based on the assumptions that (a) the 
four-momentum $P$ of the hadron is fixed and (b) the momentum transfer $Q$
becomes {\em infinite} (as opposed to large but finite) 
while $x_B$ is zero. 
Expression \Ref{IntegrationsKontur} allows to trace the implications
of such assumptions. 
The Lorentz-invariant \Index{relative velocity}\footnote{
The reason for calling $v_B$ the relative velocity is that 
the proton with 4-momentum $P$ moves with the velocity $v_B$ in the
rest-frame of $q$ defined as one of the frames where $q^0=0$. 
$P$ and $q$ define two Lorentz-invariants. One possible choice
is $x_B$ and $Q$; an other choice is $v_B$ and $x_B$. 
The last choice is obviously problematic in the Bjorken limit.
}
\be
v_B
=1/\sqrt{1+\left({MQ\over P\cdot q}\right)^2}
=1/\sqrt{1+\left({2Mx_B\over Q    }\right)^2}
\ee
between $P$ and $q$ becomes $1$ in the limit where the experimental
resolution $Q$ diverges. In this limit, an unphysical boost with
boost-velocity $v=1$ would be needed in order to relate the 
Breit frame to a frame with finite $P$. 
Hence, the limit $Q\rightarrow \infty$ renders the choice of frame
irreversible. In this limit it appears as if 
$q$ where light-like. Consequently, the integration contour $\ell$ along 
the $q$ direction appears to be light-like, too. 
But $q$ never becomes light-like: 
$q$ remains space-like even for the somewhat grotesque choice of an
infinite experimental resolution $Q=\infty$, $q$ because
$-q\cdot q=Q^2=\infty$ is not zero, as required for a 
space-like four-vector. 
The integration contour $\ell$ becomes light-like for $Q=\infty$ but 
remains space-like for any finite
experimental resolution $Q$. 
Instead of choosing a frame with fixed $P$ we might as well 
choose a frame with fixed $q/Q$. In this case, it is the hadron which
approaches
the speed of light in the limit $Q\rightarrow \infty$.
Physics only depends on the relative 
velocity $v_B$ between $P$ and $q$.
The absolute velocity of $P$ or of $q$ is completely irrelevant. 
An argument which crucially depends on keeping $P$ fixed instead of
$q$, can not be trusted since it is a frame-dependent argument: 
Indead, repeating the argument given in Ref.~\Cite{Jaffe:1996zw}
in a frame where the orientation 
of $q$ is kept fixed, leads to completely different conclusions.

Care must always be taken when a boost with boost-parameter $v=1$ 
goes into an argument. These boosts
are singular (and should therefore not really be called boosts) as they contract
a four-dimensional universe onto a three-dimensional sub-space: the light-front.
After the action of such a "boost", all four-vectors appear to be
light-like, or "light-cone" dominated. 
Please note that the Fourier transform $\tilde W^{\mu\nu}(y)$ is
indeed "light-cone" dominated in the sense that $|\tilde W^{\mu\nu}(y)|$
is largest close to the light-cone $y\cdot y=0$. We do not deny this. 
What we claim is this: distribution functions are related 
to space-like line-integrals. There is no need to choose
a frame where the integration contours become light-like
and there is no need to choose $Q=\infty$ (instead of merely
large $Q\gg M$) as there is no point in considering experiments
with infinite experimental resolution. 
For finite resolution $Q$, however, it is not justified to treat the integration
contour as light-like. An infinite boost with $v=1$ would be needed
in order to justify such a step. 
 
\proPub{
We close this section with a few words on terminology. 
If a function $\tilde W(x)$ is almost zero outside the region 
$R_\epsilon=\{x| 0\le x^2<\epsilon\}$, we would prefer to characterise
$\tilde W(x-y)$
as being
dominated by small, {\em time-like} distances $\tau$ 
since the relativistic measure of distance
is $\tau\dn \sqrt{(x-y)\cdot (x-y)}$ --- not $r=|\vec x-\vec y|$.
Applied to DIS this means: the {\em interval of time} $2\tau$ during which the 
scattering process takes place is {\em short}.

Unfortunately, another  
process is called short-distance dominated in present terminology. We shall
argue that 
"small-(hyper-)volume-dominated
process", would be more a more accurate expression --- albeit more lengthly. 
This process, a process with time-like momentum transfer\footnote{
for instance inclusive $e^+e^-$ annihilation, see~\Cite{Jaffe:1996zw}
}
$q=(Q,0,0,0)$ 
probes values of $\tilde W(x)$
in a small \Index{space-time volume}, i.e. the major contribution
to the Fourier transformed function $W(q)$ stems from a finite space-time
region with $x\cdot x\approx 0$ and $|x^0|<1/Q$.
We recall that a hyper-volume is a Lorentz-invariant whereas
a difference of two spatial components $|\vec x-\vec y|$ is not. 
The difference of the notions "small hyper-volume" 
and "small distance" is subtle but crucial.

We are aware of the fact that these points may
contradict some researchers 
we hold in deep respect and whose publications have taught us much (maybe
not enough).  
If we are wrong or if we have misrepresented ideas, we would like to apologize
in advance. The same applies to Chapter~\secRef{CFF}.
} 

\section{Special Case: Breit Frame}

 The leptonic tensors in the Breit frame take on the form
\begin{equation}\label{lmunuBjBreit}
l_s(k_Q,k_{-Q})
\approx
Q^2
\left(
\begin{array}{cccc}
0&0&0&0\\
0&1&+i(2s)&0\\
0&-i(2s)&1&0\\
0&0&0&0\\
\end{array}
\right)
\end{equation}
\begin{equation}\label{minuslmunuBjBreit}
\bar
l_s(k_{-Q},k_Q)
\approx
Q^2
\left(
\begin{array}{cccc}
0&0&0&0\\
0&1&-i(2s)&0\\
0&+i(2s)&1&0\\
0&0&0&0\\
\end{array}
\right)
\end{equation}
if both transverse momenta $\vec k^\perp$ and fermion masses $m$ 
are small when compared to the longitudinal momenta $k^3$. 
\begin{equation}
l^{00} 
=
4[m^2+\vec k_\perp ^2]
=
4\omega^2(\vec k_Q)-Q^2
\approx 0
\end{equation}
\begin{equation}
l^{11} 
=
4 (k^1)^2+Q^2
\approx
Q^2
\end{equation}
\begin{equation}
l^{22} 
=
4 (k^2)^2+Q^2
\approx
Q^2
\end{equation}
Inserting the expression\Ref{lmunuBjBreit} into the hadronic
tensor~\Ref{StruFuDiFu} allows us to compute the structure functions
in terms of the parton distribution functions ${\mathsf q}$ and ${\mathsf g}$
\be \label{qStru}
F_2(x_B,Q) \approx 2x_B F_1(x_B,Q) 
\approx x_B {\mathsf q}(x_B,Q)+x_B \bar{\mathsf q}(x_B,Q)
\ee 
\be \label{g1Stru}
g_1(x_B,Q)\approx {\mathsf g}(x_B,Q)+\bar{\mathsf g}(x_B,Q)
\ee 
and 
\be \label{g2Stru}
 {1\over 2}g_2(x_B,Q)\approx 0
\ee
If we had taken several quark flavours $i$ into account we would have obtained
\be
F_2(x_B,Q) \approx 2x_B  F_1(x_B,Q) 
\approx x_B \sum_i e_i ^2    
\left[
{\mathsf q}(x_B,Q;i)+\bar{\mathsf q}(x_B,Q;i)
\right]
\ee 
\be
g_1(x_B,Q)\approx 
{1\over 2}\sum_i e_i^2     
\left[
{\mathsf g}(x_B,Q;i)+\bar{\mathsf g}(x_B,Q;i) 
\right]
\ee 
where $e_i$ is the charge of the quark with flavour $i$,
${\mathsf q}(x_B,Q;i)$ and ${\mathsf g}(x_B,Q;i)$ are its respective parton distribution functions.

\section{Distribution Functions in Other Frames}

If we use a frame with $\vec P\parallel \vec q$ but with $q^0\neq 0$
then the relations between raw distribution functions and structure
functions become slightly more complicated. In the limit where 
fermion masses and transverse momenta can be neglected, the
structure function $F_1$ reads
\begin{equation}\label{F1IA}
\begin{split}
F_1^{\text{IA}}(x_B,Q;P^3)
&=  
     {\mathtt q}(x_B , P^3)
-
\bar {\mathtt q}(-{Q^2 \over (2P^3)^2 x_B},P^3)   \\
&+
\bar {\mathtt q}(x_B , P^3)
-
     {\mathtt q}(-{Q^2 \over (2P^3)^2 x_B},P^3)
\end{split}
\end{equation}
which for $P^3=P_B={Q\over 2x_B}$ coincides with the expression
we gave for the Breit frame as it should.
We have written $F_1^{\text{IA}}(x_B,Q;P^3)$ 
instead of 
$F_1(x_B,Q)$ since this formula is only accurate to the extent
that the impulse approximation can be trusted. The full structure function
\be
F_1(x_B,Q)=F_1^{\text{IA}}(x_B,Q;P^3)+\triangle F_1^{\text{IA}}(x_B,Q;P^3) 
\ee
should depend on the invariants $x_B$ and $Q$ only; it
should not depend on the momentum of the hadron. The expression 
$F_1^{\text{IA}}(x_B,Q;P^3)$ obtained in the impulse approximation, however,
does depend on $P^3$. Yet it does so weakly. The same holds for 
higher-order corrections $\triangle F_1^{\text{IA}}(x_B,Q;P^3)$
to the impulse approximation. 
\proDoc{
We recall a feature of non-perturbative physics which is fundamentally
different to what the practitioner of perturbation theory is used to: 
creation operators $b^{\dag}_s(\vec k)$ are not irreducible
representations of the interaction-dependent Lorentz group. They are 
irreducible representations of the Euclidean group $E(3)$\footnote{
I.e. the stability group of the quantisation surface} only.
The generators of the Euclidean group are \Index{kinematical}, i.e. they
do not depend on the interaction. A rotation transforms a creation
operator into another creation operator. A boost, however, is
\Index{dynamical}: it contains interactions
and, therefore, a boost transforms a quark creation operator into a complex
mixture of quark operators and gluon operators. 
This is why ---in the presence of {\em non-perturbative} interactions---
{\em creation operators do depend on the quantisation surface}. 
A distribution function $f(\vec k;\vec P)$ of virtual particles 
defined on a given quantisation surface and distribution functions of virtual 
particles defined on a different quantisation surface are essentially 
different. Distribution functions are not Lorentz-covariant.
Contrary to the intuition gained in perturbation theory,
the description of a hadron in terms of {\em free} quarks and gluons
{\em does} depend on the quantisation surface whereas physical observables 
must not depend on the quantisation surface.  
}

If the vacuum is not trivial, then only the full structure function $F_1$
can be independent of the hadron momentum $\vec P$ as we are going
to argue now. An approximation may work better in one frame than it does
in another frame especially if it is defined in terms of frame-dependent
creation operators. Indeed, 
it will turn out that $F_1^{\text{IA}}(x_B,Q;P^3)$ is 
almost boost invariant for momenta $P^3$ comparable to $P_B$ whereas
the dependence on $P^3$ becomes substantial for $P^3\approx 0$ and
for $P^3\rightarrow \infty$. The break-down of the impulse approximation
at $P^3=0$ is standard knowledge; 
the break-down of this approximation at infinite momenta, has not
yet been realised in the literature.  
 
The equation\Ref{F1IA} has some salient consequences. 
In any interacting QFT there is pair-production yielding a non-trivial
vacuum. This implies that the raw distribution function
$f_s^{\text{vac}}(\vec k)$ of the physical vacuum cannot vanish. 
Raw distribution functions $f_s(\vec k,\vec P)$ associated 
with a physical particle can be decomposed into the parity-invariant vacuum part
\be
f_s^{\text{vac}}(\vec k)
\dn 
\brac{ \text{vac} } 
  b^{\dag}_s(\vec k)b_s(\vec k) 
\cket{\text{vac}}
\ee
and the residual part 
\be
\triangle f_s(\vec k,\vec P)
\dn
f_s(\vec k,\vec P)
-
f_s^{\text{vac}}(\vec k)
\ee
associated with the moving particle.
Our point is that while there may well be a sufficiently large 
momentum $|\vec P |$  
such that $\triangle f_s$ vanishes for left-movers, this can no
longer be true for the raw
distribution $f_s$ itself. The raw distribution function $f_s$ 
cannot vanish for left-movers---
whatever the total momentum $\vec P$ may
be. The presence of a physical particle cannot seriously modify
the vacuum state since a particle may be localised inside
a finite microscopic region whereas the the vacuum arises through 
the spontaneous creation and annihilation of virtual particles
all over the universe (or all over the lattice). Parity symmetry
of QCD implies that the vacuum state must contain the same 
number of left-movers and right-movers. The state of a fast physical 
particle which moves right, contains more fast right-movers than 
left-movers. The wave-function of a fast hadron
consists of a long-distance part
and a small-distance part. 
The {\em long-distance part} consists of
virtual particles which arise from the spontaneous
creation of virtual particles all over the universe. 
The {\em long-distance part} of the hadronic wave function can be expected
to be almost identical to the vacuum wave function. Therefore 
it contains the same number of left-movers and right-movers. 
The small distance part, in contrast,
consists of virtual particles whose presence is due to the presence of
the hadron itself. It is the partons associated with the short-distance
part which are accessible in a DIS experiment. 
Expressed briefly: the hadron is embedded in the vacuum and
a non-trivial vacuum always contains left-movers.
We shall present an example of this phenomenon in Chapter\secRef{CFF}.

In the Breit frame, this does not constitute any problem 
as the {\em parity invariant}
vacuum distribution is subtracted away in Eq. \Ref{StruFuDiFu}
and only $\triangle f_s$ survives. 
This mechanism need not even be invoked though, as the vacuum distribution
can be expected to be concentrated in the region of long wave-lengths
$|\vec k|\lessapprox M$ around the origin $\vec k=0$ 
momentum space. The impulse approximation
is only valid for large $Q\gg M$ anyway and, therefore, 
the momentum 
$|\vec k_{-Q}|\approx Q/2$ of the scattered parton is not small. The
scattered parton is therefore deposited outside the region 
$|\vec k|\lessapprox M$ where the vacuum distribution is concentrated. 
In other words: $\vec k_{-Q}$ cannot
probe the vacuum distribution for large $Q$.
In Chapter~\secRef{CSchwinger} we shall compute the vacuum distribution
function of QED($1+1$) and demonstrate that it is indeed concentrated
inside the region $\vec k<M$. 
Consequently, in reference frames that are close to the Breit frame, the
vacuum does not influence the structure functions either. 
The IA becomes less 
reliable, however, if either the rest-frame or the IMF are approached. 
{\em Close to the rest-frame}, the momentum $P^3 x_B$ of the parton
before scattering
becomes so small that the vacuum is probed. 
{\em Close to the IMF}, 
the momentum $-{Q^2\over 4P^3 x_B}$ of the left-moving scattered 
parton starts to probe
the vacuum if $P^3$ becomes larger than ${Q^2\over 4M}$ or if $x_B$ becomes
too small. 
{\em Close to the Breit frame}, in contrast,
the vacuum is only probed if $Q$ (or $x_B$) is
too small. 
Therefore, the structure function $F_1^{\text{IA}}$
as calculated in the impulse approximation
is approximately boost-invariant only for frames close to the Breit-frame.
Close to the rest-frame and to the IMF, the structure functions depend
severely on the frame (which they should not),
a fact which signals the breakdown of the 
impulse approximation in these frames. 
Of course, this is no problem for the parton model as such: 
Observables must not depend on the frame, but an approximation
is not required to be boost invariant: the quality of an approximation
may depend on the velocity of the physical particle {\em relative} 
to the lattice. 
This simply means that 
the hadronic tensor has to be computed non-perturbatively in the IMF
or the rest-frame where corrections to the parton model become more important. 

One comment on the IMF is in order. 
In the IMF, the vacuum distribution can be eliminated
by choosing a very small volume but the IMF is unphysical 
as discussed in Chapters\secRef{CFF}
and\secRef{CSchwinger}. 

Finally, we remark that the domain in momentum space
\be \label{Plateau}
{M\over x_B}\ll P^3\ll {Q^2\over 4x_B M} 
\ee
where $F_1^{\text{IA}}$ is almost independent of $P^3$ increases
in extension
if $Q$ is increased or if $1/x_B$ is decreased\footnote{There is a problem for 
$x_B\approx 1$ which only becomes apparent when calculating higher-order
corrections to the impulse approximation. For more details the reader is
referred to~\Cite{CloseBuch,RobertsBuch,Jaffe:1996zw}).}.

\section{Beyond the Impulse Approximation}
So far we have been working in the framework of the 
\Index{na{\"\i}ve parton model}:
we did not take higher order $QCD$ corrections into account. 
In this framework, the domain\Ref{Plateau} becomes infinitely large in
the so-called \Index{Bjorken limit}
\begin{Definition}[Bjorken limit: strong form]
The Bjorken limit is the 
the limit where 
$Q\rightarrow \infty$ diverges and $x_B$ remains constant.
\end{Definition}
As in the last section, we argue that this limit
is not useful at all in its {\em strong} form, $Q=\infty$,
as the strong Bjorken limit corresponds to the limit of infinite
experimental resolution. In praxis, it suffices to use the Bjorken
limit in a weaker form
\begin{Definition}[Bjorken limit: weak form]\label{WeakBjorkenLimit}
The (weak) Bjorken "limit" is attained if $Q$ is sufficiently
large and $x_B$ sufficiently small such that both $M/Q<1$ and
$1/x_B<1$ are small. 
\end{Definition}
Cf.~\Cite{RobertsBuch,CloseBuch} and references therein. 
It would be a serious mistake to claim that the impulse approximation
becomes better ---or even exact--- in the Bjorken limit. 
Even though the effective QCD coupling $\alpha_s(Q)$ becomes smaller\footnote{
One should realise that the effective QCD coupling is always larger
than the effective QED coupling for {\em every} experimental resolution
smaller than the great unification scale(!)}
with increasing $Q$, higher order corrections to
the impulse approximation become {\em more} important
(not less) in this limit. In order to justify the parton model,
we have but two choices:
(1) choosing a resolution $Q$ which is {\em large enough} compared to $M$ 
such that $\alpha_s(Q)$ is sufficiently small to validate
perturbation theory and yet {\em small enough}
such that higher order QCD corrections are sufficiently suppressed. 
(2) using the \Index{renormalisation group equations} (RGE)
which allows us to describe QCD corrections up to arbitrary orders. 

The second choice leads to the RGE improved parton model. 
Let us discuss the first choice first. 
The \Index{impulse approximation} ---the zeroth order limit
of perturbative QCD--- can only be accurate {\em if} first order QCD is accurate
as well. 
First order QCD is accurate if 
\begin{enumerate}
\item
$Q$ is larger than the typical scale $\Lambda_{\text{QCD}}$ of QCD. 
\item
the parton masses $m_i$ have approximately the same order of magnitude 
as $Q$ (in order to avoid logarithmic collinear singularities) $m_i\sim Q$.
\item
the momentum of the struck parton is smaller than the cut-off $\Lambda$
before and after the scattering process
\item
the cut-off $\Lambda$ has the same order of magnitude as $Q$ (in order
to  perturbatively relate the effective coupling $g_{\text{eff}}(Q)$ to the bare coupling
$g(\Lambda)$)
\end{enumerate}
If $Q$ is too large when compared to the parton masses, higher order QCD
corrections become too important to be neglected: the na{\"\i}ve impulse
approximation breaks down. Fortunately, an arbitrary number
of the most important perturbative higher-order corrections (ladder graphs) 
can be analytically resummed for $M\gg Q$ as explained in~\Cite{RobertsBuch}.
This resummation, which may also be seen as an iteration
of first order perturbation theory, is described by the celebrated 
Dokshitzer-Gribov-Lipatov-Altarelli-Parisi (\Index{DGLAP})
equations~\Cite{RobertsBuch,CloseBuch}. 
Improving the impulse approximation (i.e. zeroth order perturbation theory)
by taking ladder diagrams into account, the structure function
\begin{equation}\label{Convolution}
\begin{split}
F_1&(x_B,Q |P^3)
=
\hat F_1(Q|y_B P^3)\conv 
\left[{\mathsf q}(P^3)+\bar{\mathsf q}(P^3)
\right]  \\
&\dn
\int {dy_B\over y_B}\; 
\hat F_1(x_B/y_B,Q|y_B P^3) \, 
\left[ {\mathsf q}(y_B|P^3)+\bar {\mathsf q}(y_B|P^3))
\right] 
\end{split}
\end{equation}
\index{conv@$\conv$}
can be written as a \Index{convolution} $F_1\conv {\mathsf q}$ of 
the quark distribution functions and a function $\hat F_1(z_B)$ which
may be interpreted as the structure function of a single quark\footnote{
Beyond zeroth order perturbation theory, the gluon
participates in the scattering process and
acquires a structure-function $F_1^{\text{Gluon}}(z_B)$ of its own. 
Due to confinement, the expressions quark- and gluon- structure functions
have to be taken {\em cum grano salis} [with a grain of salt], of course.
}.
It can be demonstrated via the \Index{operator product 
expansion}\Cite{RobertsBuch} that the corrections 
to \Ref{Convolution} are repressed by powers of $M/Q$  
(this ---as an aside--- 
is the rationale for Definition~\Ref{WeakBjorkenLimit}). These 
so-called \Index{higher twist corrections} are due to 
non-zero parton masses,
more general perturbative diagrams, 
non-perturbative final state/confining interactions etc.
To the extent that \Ref{Convolution} is valid,
scattering to arbitrarily large order in ladder diagrams
off {\em point-like quarks}
may be replaced by zeroth order scattering off {\em quarks 
with internal structure} described by their structure function $\hat F_1$. 
This justifies our using the impulse approximation:
{\em Apart from this modification, our results based on
the impulse approximation remain intact}. 
This modification constitutes the highly successful \Index{renormalisation
group improved parton model}. 
The \Index{na{\"\i}ve parton model} based on the impulse approximation,
in contrast,
would correspond to $\hat F_1(z_B)=\delta(1-z_B)$.
If the parton distribution functions ${\mathsf q}(x_B|P^3)$
are approximately boost-invariant, i.e. if the dependence on $P^3$
is only weak, then the \Index{na{\"\i}ve parton model}
predicts \Index{Bjorken scaling}: The structure function $F_1(x_B,Q)$ becomes
independent of $Q$. This is {\em not} what one observes
experimentally. It is not true that the structure function
$F_1$ looses its dependence on $Q$ in the 
limit $Q\rightarrow \infty$
---contrary to what is often stated in the literature. The dependence on 
$Q$ does not disappear; it is logarithmic. This means that $F_1$ may have
to be measured over several orders of magnitude
in $Q$ in order to observe an appreciable $Q$ dependence. Yet this
$Q$ dependence is extremely strong so as to render $F_2(x_B,Q)$
proportional to the delta function $\delta(x_B)$~\Cite{CloseBuch}
in the limit of infinite
experimental resolution $Q\rightarrow \infty$.
In other words: experiments at infinite resolution
---if they were possible--- would yield trivial, divergent results\footnote{
Divergences in field-theories are not unphysical as such: 
QCD predicts finite results for finite experiments. For infinite 
experiments, e.g. $Q=\infty$, it predicts infinite results--- a fact
which should not surprise us too much. Moreover, QCD must be seen
as an effective theory of general theory which includes
gravitation and other forces: the standard model
no longer describes reality if $Q$ is large
enough to see {\em new physics}. 
}.
 We are always interested in finite experimental
resolution and therefore, the Bjorken limit should be defined
as $Q\gg M$ rather than $Q=\infty$. 
For the sake of completeness we briefly mention that other problems
arise for the IA if the parton masses $m$ are too small when compared to 
the hadron mass $M$. In this case, $\hat F_1(z_B)$ never equals $\delta(1-z_B)$
even if $Q$ is very small. The basic picture, however, is
salvaged by the fact that $\hat F_1$ can be written
as a convolution of a $m/M$ dependent function and a $Q/M$ dependent
function~\Cite{RobertsBuch}. 
This justifies the use of perturbation theory (i.e. the DGLAP equations) 
in order to describe the $Q$ dependence 
of the structure functions even in the limit $m/M\rightarrow 0$
where the $m/M$ dependent convolution-factor of $\hat F_1$
is not perturbatively accessible.  
The reader interested in this subject who wants to 
go beyond the brief (and thus necessarily over-simplified) explanations 
in this section is referred to~\Cite{traduttoretraditore,Altarelli:1982ax,RobertsBuch,CloseBuch}.

\subsection{The Breit Frame and the Continuum Limit}
In the \Index{Breit frame}, the \Index{Bjorken limit} corresponds to the \Index{continuum limit} as follows. 
For a sensible computation, the cut-off $\Lambda$ has to be 
chosen larger than the total momentum $P^3$. At the same time,
$\Lambda$ must not be much larger than $Q$ in order for perturbation
theory to be applicable. 
Therefore $P^3={Q\over 2x_B}\sim \Lambda$. Consequently, the (strong)
Bjorken limit 
$x_B=\text{const}$, $Q\rightarrow \infty$ coincides with the continuum
limit $\Lambda/M\rightarrow \infty$. \proDoc{This has
an intuitive interpretation: $Q$ corresponds to the experimental resolution and
$\Lambda$ is the lattice resolution. High
experimental resolution can only be achieved if
the lattice resolution is high, too. 
}

%% file: CFF.i
{\ttfamily\small\footnotesize
\begin{quote}
"It was during the course of my researches at the Campaign
for Real Time that I..." \\
"The what?" said Arthur again. 
\end{quote}
[Douglas Adams: Life, the Universe and Everything]
}

\section{The Problem of Front-Form Quantisation}
In non-relativistic quantum-mechanics, there is only one possible 
quantisation surface: the \Index{instant of time} $t=x^0=0$.
\index{t@$t$}
In relativistic quantum-mechanics, however, there is a larger choice 
of quantisation surfaces since time is no absolute concept but
depends on the observer.  
How can we tell whether 
a three-dimensional (hyper-) surface of four-dimensional 
space-time may serve as quantisation surface?
This question must be discussed on
the classical level first since 
the \Index{initial surface} of the classical equations
of motion corresponds to the \Index{quantisation surface} of the quantised 
system. 
The Euler-Lagrange 
equations of classical, Newtonian mechanics are deterministic:
For instance, the non-relativistic, {\em classical} Euler-Lagrange 
equations of motion are deterministic in the sense that
the knowledge of every observable at one instant of
time suffices in order to predict the entire future or the past.
Expressed in the terminology of partial differential equations this reads:
the instant of time $t=0$ is an \Index{initial surface} of the equations
of motion because initial data specified on this hyper-plane determine
a unique solution to the equations of motion. 

Classical relativistic physics is deterministic as well: 
the relativistic equations of motion allow the state of the 
universe to be calculated at any time from the information specified
at one \Index{instant of time}.
Contrary to non-relativistic physics, however,
there is a larger class of three-dimensional hyper-surfaces 
of the four-dimensional space-time which may be regarded as an instant of time.
The non-uniqueness of the initial surface is a direct consequence 
of 
the finiteness of the speed of light. Every space-like hyper-plane may be regarded as 
an \Index{instant of time} whose points are not causally connected 
and consequently, any {\em space-like} hyper-surface 
may serve as initialisation-surface (or as 
quantisation-surface in the quantum case). 

{\em Non-relativistically}, in contrast,
information may propagate arbitrarily fast.
Consequently, only space-time events occurring at the same
({\em absolute}) time are causally independent; rephrased in geometrical
terms this reads: an instant
of time has only one possible orientation in space-time.  
In brief, the mathematical term \Index{initial surface}
is intimately related to the more intuitive terms \Index{instant of time}
and \Index{set of causally independent points}. This observation will
enable us to understand the problems treated in this chapter more easily. 

Naturally, the question arises whether quantisation surfaces 
other than space-like hyper-planes would serve as well in the relativistic 
case. 
The first to ask this question was P.A.M. Dirac~\Cite{dirac:forms}. In 1950 he proposed
to use light-like hyper-planes $x^+=0$ 
as quantisation-surfaces. 
Note that a light-like hyper surface $x^+=0$ is not defined as 
an instant of \Index{real time} $x^0$: it is defined in terms of an
abstract time $x^+$ --- called \Index{light-cone} time
which does not coincide with the proper time of {\em any} real observer. 
For this form
of quantisation, Dirac coined the term \Index{front form} (FF) quantisation
in order to distinguish it from the more familiar quantisation
on a space-like hyper-plane which he referred to as
\Index{instant form} (IF).
\index{FF}\index{IF}
Dirac's idea, soon forgotten, re-emerged again for several times under different names
such as \Index{light-cone}, \Index{null-plane} or 
\Index{light-front} quantisation. 
For a more thorough historical overview the reader is referred to 
~\Cite{BrodskyPauli}.
\begin{floatingfigure}{0.4\linewidth}

\hbox{
\mbox{
\psfig{clip=,figure=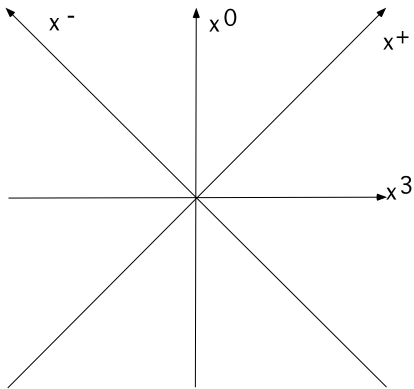,width=0.4\linewidth,angle=0}  }
      }   
\caption{The IF co-ordinates $x^0,x^3$ and their corresponding
FF co-ordinates $x^\pm\dn x^0\pm x^3$}
\end{floatingfigure}
Sometimes, the FF is referred to as \Index{infinite momentum frame} (IMF)
\index{IMF}
~\Cite{Weinberg:1966jm}.
The IMF is basically a frame where every particle moves with
the speed of light. For more precise a definition see section \secRef{secTheIMF}.  
The confusion of the IMF and the FF arises because the two approaches are very similar (when expressed in light-cone co-ordinates $x^\pm$). This
similarity is deceptive, though, as as we shall show in \secRef{secTheIMF}. 
Dirac also found another possible quantisation
surface --- a space-time parabola --- which he called \Index{point-form}. 
Later on, in 1977, H. Leutwyler and J. Stern~\Cite{Leutwyler:1978vy} gave a group-theoretical
characterisation of all possible quantisation-surfaces adding two additional
forms to Dirac's catalogue.  
They pointed out, furthermore, that a light-like hyper-plane has
the largest \Index{stability group} of all hyper-planes, 
i.e. the largest number of generators
of the Poincar\'e group which leave the respective hyper-plane invariant. 
\begin{table}[t]\begin{center}

\begin{tabular}{|c|c|c|c|}
\hline
Relativistic form & Quantisation-surface 
& 
\begin{tabular}{c} Algebra of the \\Stability group \end{tabular}\\
 \hline
Front form   & $x^+=e^{(+)}\cdot x=0$ 
& $\vec P_\perp,{\mathsf J}_{12},{\mathsf J}_{03};{\mathsf J}_{\perp-},P_-$\\  \hline
Instant form & $x^0=e^{(0)}\cdot x=0$ 
& $\vec P,{\mathsf J}_{mn}$ \\  \hline 
Point form   & $x\cdot x=\kappa^2$, $x^0>0$ 
& ${\mathsf J}_{\mu\nu}$\\  \hline
Line form   & 
$\begin{matrix}
 (x^0)^2-(\vec x^\perp)^2=\kappa^2,\\
  x^0>0
 \end{matrix}$ 
& $P_3,{\mathsf J}_{12},{\mathsf J}_{03}$\\  \hline
\begin{tabular}{c} Extended front form  \end{tabular}
& $x^+ x^-=\kappa^2$, $x^0>0$ 
& $\vec P_\perp,{\mathsf J}_{12},{\mathsf J}_{03}$\\  \hline
\end{tabular}

\caption[Relativistic forms]{Relativistic forms. 
The parameter $\kappa$ is a real number.}\label{StabilityGroups}
\end{center}\end{table}
The stability group of a light-like plane is spanned by seven generators, 
one generator more than the IF or the point-form respectively. The
complete catalogue of relativistic forms 
is represented in~\tabRef{StabilityGroups}
A large stability group is usually advantageous because the generators of this
group are \Index{kinematical}, i.e. they do not depend on the interactions
~\Cite{Leutwyler:1978vy}. But although this may be true, a larger stability-group is 
{\em not necessarily} advantageous, as we shall show below
and in Chapter\secRef{CNum}. 
There are useful generators of the Poincar\'e group 
which facilitate the diagonalisation of a Hamiltonian and 
useless generators that do not. For instance, momenta in the IF are
useful, boosts in the IF form are not useful:
the
fact that 
the boost
in the direction of the quantisation-plane
na{\"\i}vely appears to be 
kinematical in the FF is not as useful as are rotations and translations
in the IF since this boost does not commute with the FF Hamiltonian. 

In 1971 it was pointed out by Neville and Rohrlich~\Cite{Neville.Rohrlich} 
and later by Steinhardt~\Cite{steinhardt:problems} that
initial-conditions specified on only one light-like
hyper-surface do not suffice in order to 
unambiguously integrate the equations of motion of a scalar field theory. 
{\em Two} light-like initialisation-planes
are necessary in order to properly determine the space-time evolution
of a relativistic field. 
Quantisation on only {\em one} light-like quantisation
plane is mathematically ill-defined because relativistic equations
of motion cannot be un-ambiguously solved if the fields are known
on one light-like hyper-surface only. 
This becomes plausible already from the observation that 
one point on a light-like hyper-plane may influence another point 
on this hyper-plane via light-like
signals. Consequently, a light-like surface is no instant of time in the
sense that points on a light-like surface are not causally independent.
One may argue, following Dirac, that signals always propagate slower than the speed of light if
massive particles are described only.
Dirac, however, considered many-body physics in ref.~\Cite{dirac:forms}
and did {\em not} have QFTs
in mind. 
In fact, he was trying to construct alternatives to QFTs.
In this section we shall be concerned with field-theories where
Diracs argument is not applicable since in a relativistic
{\em field} theory, causal propagators
always connect light-like
directions regardless if massless particle fields are present
or not. 

This discovery has some uncomfortable consequences. Firstly, we note that it is difficult
to define a FF Hamiltonian approach since two quantisation surfaces 
require two Hamiltonians generating temporal evolution with respect
to {\it two} ''time''-directions of space.  
Secondly, two quantisation-planes have a smaller stability-group as the
na{\"\i}ve quantisation on one light-like plane would suggest, since
the translation which leaves one of the two planes invariant necessarily
moves the other plane, hence becomes dynamical. This additional
dynamical translation $P_-$
is the second Hamiltonian. The generators ${\mathsf J}_{\perp-}$ become
dynamical as well (thus we venture the opinion that
the FF on two quantisation surfaces should be identified
with the last form of the catalogue~\tabRef{StabilityGroups} for
$\kappa=0$). In brief, correct quantisation on two light-like quantisation
surfaces shrinks the stability group to a group with {\em four generators}
---three less than the FF. 
\Index{Transitivity}\footnote{Transitivity means that the stability group is able to connect any points on the quantisation surface. See
\Cite{Leutwyler:1978vy}.
}
of the stability group is lost as well. 

In 1994, Heinzl and Werner~\Cite{heinzl.werner:light}
were able to find a loop-hole avoiding
the necessity of two light-like hyper-surfaces.  
They demonstrated in the case of the {\em free} scalar theory
that one light-like hyper-plane suffices as an initial surface
{\em if} periodic or anti-periodic light-like 
\Index{boundary conditions}(\Index{BC}) are chosen 
on the initialisation-plane. 
This seemed to provide a justification for FF quantisation on 
one hyper-plane as long as  
(anti-) periodic boundary-conditions are specified (as e.g. in  
\Index{discretised light-cone 
quantisation}~\Cite{BrodskyPauli}).
However, periodic {\em light-like} boundary conditions  
are unphysical as we shall demonstrate in this chapter. 

{\it Firstly}, periodic BCs
eliminate boosts as members of the FF stability-group since
boosts change these boundary-conditions. We shall show that boosts are therefore
no longer automatically kinematical transformations. 
In Chapter\secRef{CNum} we demonstrate 
the mass-spectrum of QED($(1+1)$ is boost-invariant 
to a much higher degree in the IF when compared to the FF.
Again, as in the
case of two initialisation planes,  the 
number of the really kinematical operators becomes the same as in the IF.  

{\it Secondly}, light-like 
periodic BCs destroy important symmetries such as parity and 
rotational invariance {\it even} in the infinite (light-like) volume limit. 
While space-like BCs are invariant under parity and time-reversal,
light-like BCs are not.

{\it Thirdly}, and most importantly, we shall show
that light-like boundary conditions completely destroy
microcausality even for arbitrarily
large light-like volumes. Consequently they
are {\em unphysical}, contrary to space-like boundary conditions. 
An implication of this is that the speed of light becomes
infinite. 

It is important to note that the term ''unphysical''
does {\it not} mean that {\em no} physical observable {\em at all} can
be correctly described. Some observables such as mass spectra are sometimes less
sensitive to the damage than propagators, especially for computations
performed close to the non-relativistic limit
where all relativistic forms must coincide. The damage due to light-like
boundary conditions  also depends on the theory under consideration. Some
specific examples will be given. 

\section{Planes, Vectors and Frames}

In the ensuing sections, we are concerned with general co-ordinate
frames and their interpretation. We shall introduce
three frames which have to be distinguished
for a non-perturbative, theoretical description of a 
relativistic particle with 4-momentum $P$:

(F-1)  An arbitrary \Index{frame of reference} 
related to the relativistic particle 
which may be taken to be the rest frame of the particle under study. 

(F-2) The \Index{Inertial Frame} or \Index{Quantisation Frame},
i.e. the frame where the observer is at rest. 
This instant of time with respect to this frame
defines the quantisation surface upon which
initial data and charges are defined.

(F-3) The \Index{lattice rest frame}.

Our intention is to study how different quantisation surfaces
are related to each other. This has to be done with respect
to the fixed but otherwise arbitrary \Index{frame of reference} (F-1). 
In particular,
we are interested in the extreme case of co-ordinate transformations
which transform
a space-like lattice into a light-like lattice. Unfortunately,
there is no non-singular
Lorentz transformation which does this. Therefore, 
the lattice rest frame has to be transported by a 
\Index{general co-ordinate transformation}
which does not leave the metric invariant. 
One of these co-ordinate transformations, the $\varepsilon$ co-ordinates~\Cite{Prokhvatilov:1989eq,Lenz:1991sa,Naus:1997zg},
will be studied in section~\secRef{secTheFFandEpsilon}. 
For the ensuing discussion, 
it is helpful to briefly review some 
salient aspects of \Index{general co-ordinate transformations}. 
The reader not familiar with
the concepts presented in this section is referred to
standard text-books on
mathematical methods or
general relativity as e.g. Ref.~\Cite{WeinbergGravityBuch}. 
A 4-vector 
\begin{equation}
u=u^\mu e_{(\mu)}
\end{equation}
is independent of the \Index{co-ordinate frame} $e_{(\mu)}$ 
\index{e@$e_{(\mu)}$}
whereas
its \Index{contra-variant} components $u^\mu$ or
its \Index{co-variant} components $u_\mu$ do depend on the frame.
For any frame $e_{(\mu)}$ of co-variant vectors there is a \Index{dual frame} of 
contra-variant vectors $e^{(\mu)}$ such that
\begin{equation}
e_{(\mu)}\cdot e^{(\nu)} = \delta^\nu _\mu
\end{equation}
holds. 
An orthonormal co-ordinate system $e_{(\mu)}$ is a co-ordinate system 
with diagonal Minkowski metric $g_{\mu\nu}\dn e_{(\mu)}\cdot e_{(\nu)}$
where $g=\eta$, $\eta^{00}=1,\eta^{11}=\eta^{22}=\eta^{33}=-1$.
\Index{General co-ordinates} $\bar x^\mu=T^\mu_{\;\nu}x^\nu$
\index{xmu@$\bar x^\mu$} are obtained from
the ortho-normal co-ordinates $x^\mu$ by means of the 
\Index{co-ordinate transformation}
$T^\mu_{\;\nu}$. In non-orthogonal co-ordinates, the contra-variant 4-vector 
$\bar e^{(0)}=\bar g^{\mu\nu}\bar e_{(\nu)}$ need not
coincide with the direction $\bar e_{(0)}$ of the time axis. 
What is the intuitive meaning of $\bar e_{(\mu)}$ and $\bar e^{(\mu)}$?
Let us assume that $x^\mu$ is the \Index{rest frame of the observer}.
In an \Index{ortho-normal frame}, 
the rest frame of the observer may be characterised
as one of the
frames in which the observer's world-line $x(\tau)$ with $\tau\in \Bbb{R}$
assumes the simple form $x(\tau)=\tau e_{(0)}=\tau e^{(0)}$. In this frame 
$x^0=x_0$ coincides with the observer's \Index{proper time} 
$\tau=\sqrt{x\cdot x}$\index{tau@$\tau$} 
and the equation $x^0=x_0=0$ defines an instant of time. 
In general co-ordinates, several characterisations
are possible: We define the general \Index{rest frame of the observer}
as the frame wherein the observer's world line reads
$x(\tau)=\tau e^{(0)}$. In this frame, $\bar x_0$ coincides with
the observer's \Index{proper time} and $\bar x^0=0$ defines an
\Index{instant of time} .We could have defined this frame
equally well via $x(\tau)=\tau \bar e_{(0)}$. In that case, the respective
r{\^o}les of $\bar x_0$ and $\bar x^0$ would be interchanged and 
$\bar x^0=\bar y^0$ would
no longer mean that the events $x$ and $y$ are simultaneous. 
The choice of $\bar x_0$ as proper time implies that
the direction $\bar e_{(0)}$ of the time-axis
is of minor importance whereas the dual
time direction  $e^{(0)}$ characterises the observer's world-line
and 
represents, furthermore, the normal vector of the instant of time $x^0=0$ where
a simultaneous measurement of initial conditions takes place\footnote{
In a quantised system, this will be the quantisation surface where
the state-of-the-universe is prepared.}.
The co-variant vector $e_{(0)}$, however, is merely 
the direction of an abstract time-axis, 
a quantity which is physically irrelevant in a general frame.
We shall see that $\bar e_{(0)}$ may even be space-like
whereas $\bar e^{(0)}$ {\em must} be time-like in order to define 
a space-like hyper-surface. 
Henceforth, we shall use $\bar e^{(0)}$ synonymously with the instant of
time $\bar x^0=0$ it defines. 
We summarise: time
imbodies two aspects: the \Index{metric aspect} of the observer's proper time $x_0=\tau$
and the \Index{aspect of simultaneity} $\bar x^0=\bar y^0$. Only in orthogonal frames
can these two aspects be described by the same time variable $x^0=x_0$. 

Another intricacy of general co-ordinate systems 
---which
is easily glimpsed over by the practitioner used to
the ortho-normal
Minkowski metric--- is the definition of a space-time volume
element: while the volume element 
$d^4x=dx^0dx^1dx^2dx^3$ is invariant
under orthogonal co-ordinate transformations
only (i.e. under Lorentz-transformations), its generalisation 
\be
d^4x/\sqrt{\det{g}}
=
d^4\bar x/\sqrt{\det{\bar g}}
\ee 
is invariant under general, non-singular co-ordinate transformations. 
The metric of the light-cone co-ordinates $\bar x^0=x^+$,
$\bar x^3=x^3$, for instance, 
is 
\be
\bar g=
\begin{pmatrix}
0 & 2\\
2 & 0\\
\end{pmatrix}
\ee
(transverse co-ordinates suppressed) and therefore 
\be
k\cdot k=k_+ g^{+-}k_-   + k_- g^{-+} k_+=
4k_+k_-
\qquad.
\ee

\section{The Quantisation Hyper-Surface}

If a charge $Q$ corresponding to a charge density $J^0(x)$ is measured
at the instant of time $x^0=0$,  it may be represented as 
\be\label{ChargeRest}
Q(e^{\text{quant}})=Q(x^0=0)=
\int d^3x J^0(0,\vec x)=\int d^4 x J^0(x) \delta(x^0)
\ee
in an {\em orthonormal} co-ordinate system $e_{(\mu)}$. 
Note that this charge depends on the hyper-surface $e^{\text{quant}}=e^{(0)}$ 
upon which it is defined, a detail that will
turn out to be crucial.
The definition of $Q$ is not linked to a specific co-ordinate system. 
We may equally well define the same charge 
\be\label{Charge}
Q(e^{\text{quant}}) \dn
\int d^4 \bar x  {1\over \sqrt{|\det\bar g^{}|}} J^0(x)
 \delta(e^{\text{quant}}\cdot x)
\ee
in any other general co-ordinate frame $\bar e_{(\mu)}$ even if 
it is not a \Index{quantisation frame}, i.e. even if
$e^{\text{quant}} \neq \bar e^{(0)}$: what we have gained
is that we have disentangled
the hyper plane $e^{\text{quant}}$ on which the charge is defined
from a particular co-ordinate system.

The relativistic 4-momentum $P^\mu$ may be considered as a charge.  
For definiteness, let us now consider a scalar field-theory with Lagrangian
\begin{equation}\label{LPhi4}
{\cal L}
=
{1\over 2} \partial^\mu \varphi(x) \partial_\mu \varphi(x)
-
{m_0^2\over 2} \varphi^2(x)
-V\left( \varphi(x) \right)
\end{equation}
where $V(\varphi)$ is a polynomial of the fields, e.g. 
$V(\varphi)={g_0\over 4!} \varphi^4.$ 
Invariance of $\cal L$ under spatio-temporal translations
leads to the conservation of the \Index{energy-momentum tensor} 
\index{Theta@$\Theta^{\mu\nu}$}
\begin{equation}\label{ThetaMuNu}
\Theta^{\mu\nu}
=
\partial^\mu\varphi\partial^\nu\partial\varphi
-
g^{\mu\nu} {\cal L}  \qquad.
\end{equation}
The charges associated with this current --- the 4-momentum that is ---
\begin{equation}\label{QuadriTextBooks}
P_\mu ^{(0)}
\dn
P_\mu(e^{(0)})
=
\int dx^1dx^2dx^3 \;
\Theta^0 _{\;\mu} 
\end{equation}
are conserved {\em if} the boundary conditions  are properly chosen on the quantisation
surface. Fixed BCs, for instance, must be excluded here since they
would break translational invariance. 
The spatial momenta $\hat P_i(e^{(0)})$ 
which leave the quantisation surface $e^{(0)}$
invariant are \Index{kinematical}, i.e. they 
do not depend on the interaction. The formal reason for this is
that the components 
$\Theta^0 _{\;i}=
\partial^0\varphi\partial_i\varphi$ of the energy-momentum tensor
\Ref{ThetaMuNu} do not depend on the Lagrangian (because $g^0_{\;i}=0$).

The conserved charge $P^0(e^{\text{quant}})$ associated
\index{P@$P$}
\index{equant@$e^{\text{quant}}$}
with temporal translations, however, is dynamical. It is
identical to the Hamiltonian $H(e^{\text{quant}})$ 
obtained by canonical quantisation
if the velocities $\partial_0\varphi(x^0=0,\vec x)$ appearing
in $\Theta^0 _0$ are expressed in terms
of their conjugate momenta 
\begin{equation}
\pi(0,\vec x)
= {\partial {\cal L} \over \partial \partial_0\varphi(0,\vec x)} 
=
\partial^0 \varphi(0,{\vec x})
\end{equation}
subjected to the usual commutation-relations 
\begin{equation}\label{UsualCommutationRelations}
[\pi(0,\vec x),\varphi(0,\vec y)]=-i\delta(\vec x-\vec y)).
\end{equation}
\index{pi@$\pi(0,{\vec x})$}
\index{phi@$\varphi(0,{\vec x})$}
In the same way, the charges $P_i(e^{\text{quant}})$ with $i=1,2,3$
\index{P@$\vec P(e^{\text{quant}})$}
can be identified with the generators translations in $e_{(i)}$-direction.
In summary, the charges $\hat P_i(e^{\text{quant}})$ 
are generators of space-time translations of the
fields $\hat \varphi(x)$ in $e^{(\mu)}$-direction:
they fulfill the equations  
\begin{equation}\label{Generateurs}
[\hat P_\mu,\hat \varphi(x)]=i\partial_\mu \hat \varphi(x)
\end{equation}
as is easily verified (cf.~\Cite{ItzyksonBuch}).

On a light-like hyper-surface, \Index{constraints} are present, i.e.
field velocities $\partial_0\varphi$ cannot be expressed 
in terms of field momenta $\hat \pi$. We shall treat this case later on in this chapter. 
Later on we would like to study what happens if the quantisation surface 
$e^{\text{quant}}$ changes relative to some fixed reference frame
$e_{(\mu)}$. To this end we have to render Eq.\Ref{QuadriTextBooks}
\begin{equation}\label{QuadriImpulsions}
\hat P_\mu(e^{\text{quant}})
=
\int {d^4 x \over \sqrt{|\det g^{}|}} \;
\delta(x\cdot e^{\text{quant}}) \;
 e^{\text{quant}} _\nu
\Theta^{\;\nu} _{\mu}    
\end{equation}
independent of the specific reference-frame used. 
Any non-singular co-ordinate frame with 
\be
e^{(0)}\parallel e^{\text{quant}}
\ee
may be called a \Index{quantisation frame} associated with 
the quantisation surface $e^{\text{quant}}$.
Note that $\hat P_0(e^{(0)})$ is the generator of translations
in the direction of the \Index{time axis} $e^{(0)}$;
if $e^{(0)}$ is not parallel to $e_{(0)}$, then  
the \Index{flow of artificial time} does not coincide with the observer's world line:
$\hat P_0(e^{(0)})$ must then be interpreted  
as the sum of a translation inside the instant of time $\bar x^0=0$
and $\hat P^0(e^{(0)})$, the generator which moves the observer's
proper time $\bar x_0$. In the FF, proper time $\bar x_0=x_+$
coincides with the {\em spatial} co-ordinate $\bar x^3=2\bar x_0$.
Setting $e^{\text{quant}}=e^{(0)}$ we obtain the IF Hamiltonian
\be
P_0(e^{(0)})=
\int d^3 x
\left[
{1\over 2}(\hat\pi)^2
+
{1\over 2}(\nabla \varphi)^2
+
{1\over 2}m_0^2\varphi^2+V(\varphi)
\right]_{x^0=0}
\ee
with the commutation relations
\begin{align}
[\hat\pi(0,\vec x),\varphi(0,\vec y)]
&=-i\delta(\vec x-\vec y)    \label{aCommutIF}      \\
\intertext{and}
[\varphi(0,\vec x),\varphi(0,\vec y)]       \label{bCommutIF} 
&=0
\end{align}
where $\nabla\dn-(\partial_1,\partial_2,\partial_3)$.
Setting $e^{\text{quant}}=e^{(+)}=e^{(0)}+e^{(3)}$ we obtain the FF Hamiltonian
\be\label{FFHamiltonian}
P_+(e^{(+)})=
{1\over 2}\int dx^-d^2x^\perp
\left[
{1\over 2}(\nabla_\perp\varphi)^2+
{1\over 2}m_0^2\varphi^2+V(\varphi)
\right]_{x^+=0}
\ee
where $\nabla_\perp\dn-(\partial_1,\partial_2)$.
Using the Dirac-Bergmann constraint-quantisation procedure one can 
deduce~\Cite{heinzl.werner:light,SundermeyerBuch}
the commutation relations
\be\label{CommutFF}
[\varphi(x^+=0,x^-,\vec x^\perp),\varphi(y^+=0,y^-,\vec y^\perp)]
=
{1\over 2i\partial_-}\delta(x^--y^-)\delta(\vec x^\perp-\vec y^\perp)
\ee
which quantise the theory
{\em if} the field is periodic in $x^-$ direction.
Zero modes
$\int dx^- \varphi(x^+,x^-)$
of $\varphi(x)$ 
are subjected to the constraint
\be \label{ZeroModesFF.1}
(-m_0^2+\nabla_\perp^2)\int dx^- \varphi
=
\int                     
dx^-{\partial\over \partial \varphi} V(\varphi)
\ee
which is an artifact of periodic BCs.

\clearpage

\section{The Boundary Vector}
 {\ttfamily\small\scriptsize
\begin{quote}
Time travel is increasingly regarded as a menace. History 
is being polluted.
The {\em Encyclopedia Galactica} has much to say on the theory and practice of time travel, most of which is incomprehensible to anyone who hasn't
spent at least four lifetimes studying advanced 
hyper-mathematics, and since it was impossible to do this before time
travel was invented,
there is a certain amount of confusion as to how the idea 
was arrived at in the first place. 
One rationalization of this problem states that 
time travel was, by its very nature, discovered simultaneously at all periods of history, but this is clearly bunk.  \\
The trouble is 
that a lot of history
is now quite clearly bunk as well. \\

[Douglas Adams: Life, the Universe and Everything]

\end{quote}
}

We shall be concerned with systems on a finite lattice and, therefore,
it is convenient to introduce another hyper-plane with
normal vector 
$e^{\text{BC}}$ on which the boundary conditions of the box are specified. 
Periodic boundary conditions  may be written in the form
\begin{equation}
\varphi(x+{\frak{L}\,})
=
\varphi(x-{\frak{L}\,})
\end{equation}
where ${\frak{L}\,}$ 
\index{L@${\frak{L}\,}$}
is a 4-vector with components
\begin{equation}
{\frak{L}\,}=
({\frak{L}\,}^0,{\frak{L}\,}^1,{\frak{L}\,}^2,{\frak{L}\,}^3)
\end{equation}
In order to obtain a boundary box with finite volume, three boundary vectors ${\frak{L}\,}_{(i)}$ ($i\in[1,2,3]$) 
\begin{equation}
{\frak{L}\,}_{(i)}\cdot e^{\text{BC}} =0
\end{equation}
with invariant lengths $L_{(i)}\dn \sqrt{-{\frak{L}\,}_{(i)}\cdot{\frak{L}\,}_{(i)}}$
are necessary.
We shall call $2L_{(i)}$
the (invariant) \Index{size of the Lattice}.
As we are primarily interested in boundary conditions  in 3-direction
we shall skip, henceforth, the index $(3)$ which indicates the 3-direction
\begin{equation}
{\frak{L}\,}\dn {\frak{L}\,}_{(3)}
\qquad
{L}\dn {L}_{(3)}
\;.
\end{equation}
Similarly, a cut-off $\Lambda$
\index{Lambda@$\Lambda$}
in momentum space corresponding to a lattice spacing $a={\pi\over \Lambda}$
\index{a@$a$ (lattice spacing) }
may be characterised by the four vector 
${\sf a}\dn a{\frak{L}\,}/L$.
\index{a@${\sf a}$}

%
%
%

\begin{table}\begin{center}

\begin{tabular}{||l|c||}
\hline\hline
Three-dimensional hyper-plane &
normal vector     \\ \hline\hline
Instant of time with respect to the co-ordinate frame $e_{(\mu)}$ &
$e^{(0)}$ \\
Instant of time where charges are measured &
$e^{\text{quant}}$ \\
Boundary plane containing the boundary 4-vectors ${\frak{L}\,}_{(i)}$ &
$e^{\text{BC}}$ \\ \hline\hline
\end{tabular}

\caption{Three-dimensional hyper-planes and their corresponding normal
4-vectors}\label{NormalVectors}

\end{center}\end{table}

The three hyper-planes
$e^{\text{quant}}$, $e^{(0)}$ and $e^{\text{BC}}$ 
---as summarised in \tabRef{NormalVectors}---
may in general be different without leading to inconsistencies.  
For practical purposes, however, not identifying these planes would not
be a  
good idea. It is clearly so exceedingly convenient 
to use a frame $e^{\text{quant}}=e^{(0)}$ whose instant of time
$x\cdot e^{(0)}=0$ coincides with the quantisation-surface
$x\cdot e^{\text{quant}} = 0$  
that in QFT text-books, this choice is always implicitly made. 
Even more expedient, from the practical point of view, is the identification 
\begin{equation}\label{eUndBC}
e^{\text{quant}}\equiv e^{\text{BC}}
\end{equation}
of the quantisation hyper-plane and the hyper-plane on which boundary-conditions are chosen unless one is willing to unnecessarily deal with 
dynamical boundary terms and charges which are not conserved in time. 
This fact may be formulated as:
\begin{Theorem}[Charge non-conservation]\label{NonConservatioBC}
Let $Q(e^{(0)})$ be the charge associated with the conserved current
$J^\mu(x)$, i.e. $\partial_\mu J^\mu(x)=0$. All fields are periodic
with the periodicity ${\frak{L}\,}=(0,0,0,{\frak{L}\,}^3)$.
Then only $Q(e^{\text{BC}})$ is conserved whereas $Q(e^{\text{quant}})$ with $e^{\text{BC}}\neq e^{(0)}$ 
is not conserved in general. 
\end{Theorem}
Proof: It suffices to give {\em one} example of a non-conserved charge. 
Take 
\be
J^{}(x)=
\begin{pmatrix}
\rho(x) \\
v\rho(x)
\end{pmatrix}
\ee
with
\be
\rho(x^0,x^3)
=
\begin{cases}
\delta(x^3-v x^0+{\frak{L}\,}^3) & \text{if } 0\le x^0<+{\frak{L}\,}^3/v\\
\delta(x^3-v x^0-{\frak{L}\,}^3) & \text{if } -{\frak{L}\,}^3/v<x^0<0\\
\end{cases}
\ee
and $0<v<1$. 
The reader will easily convince himself (cf. ~\figRef{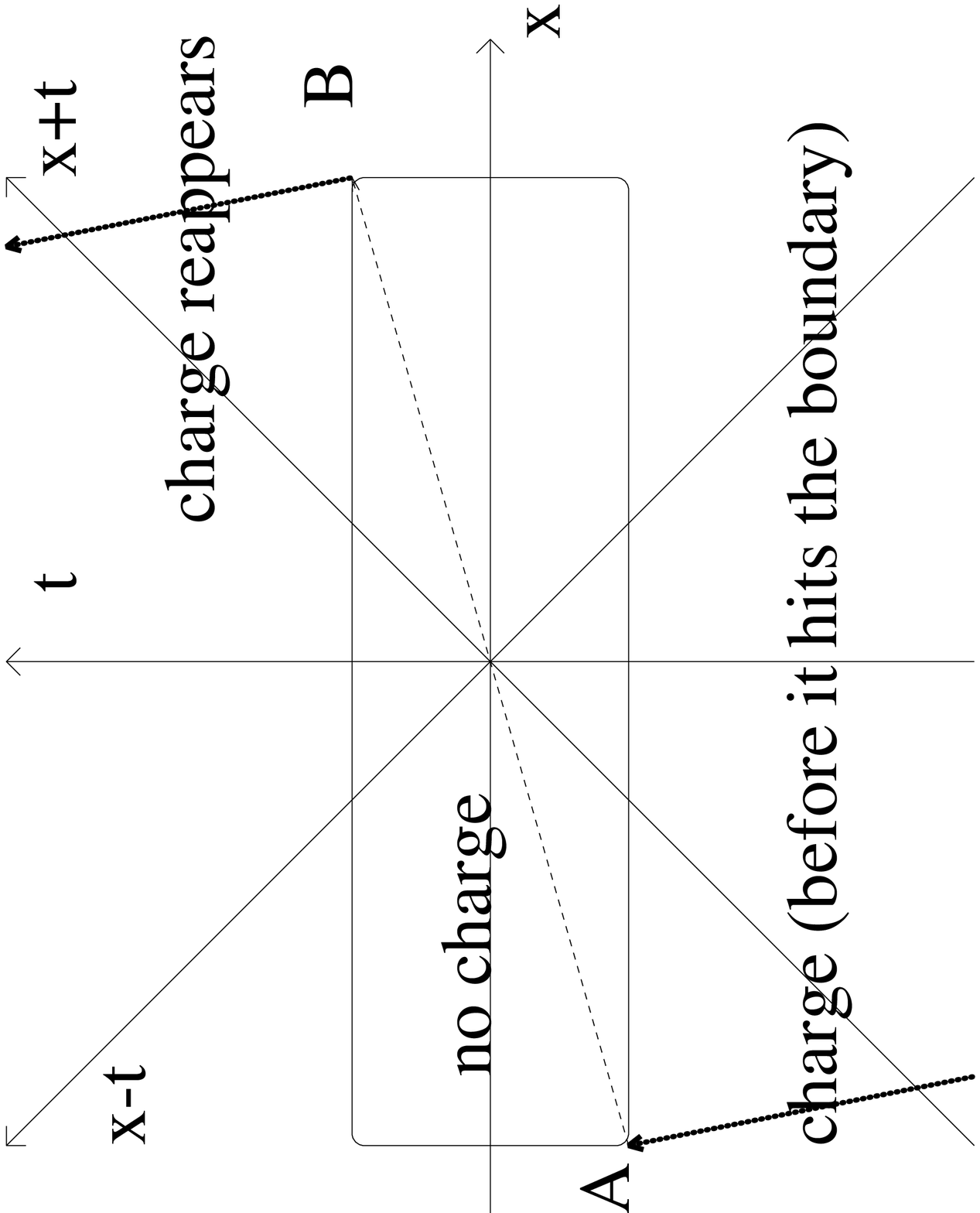}
and~\figRef{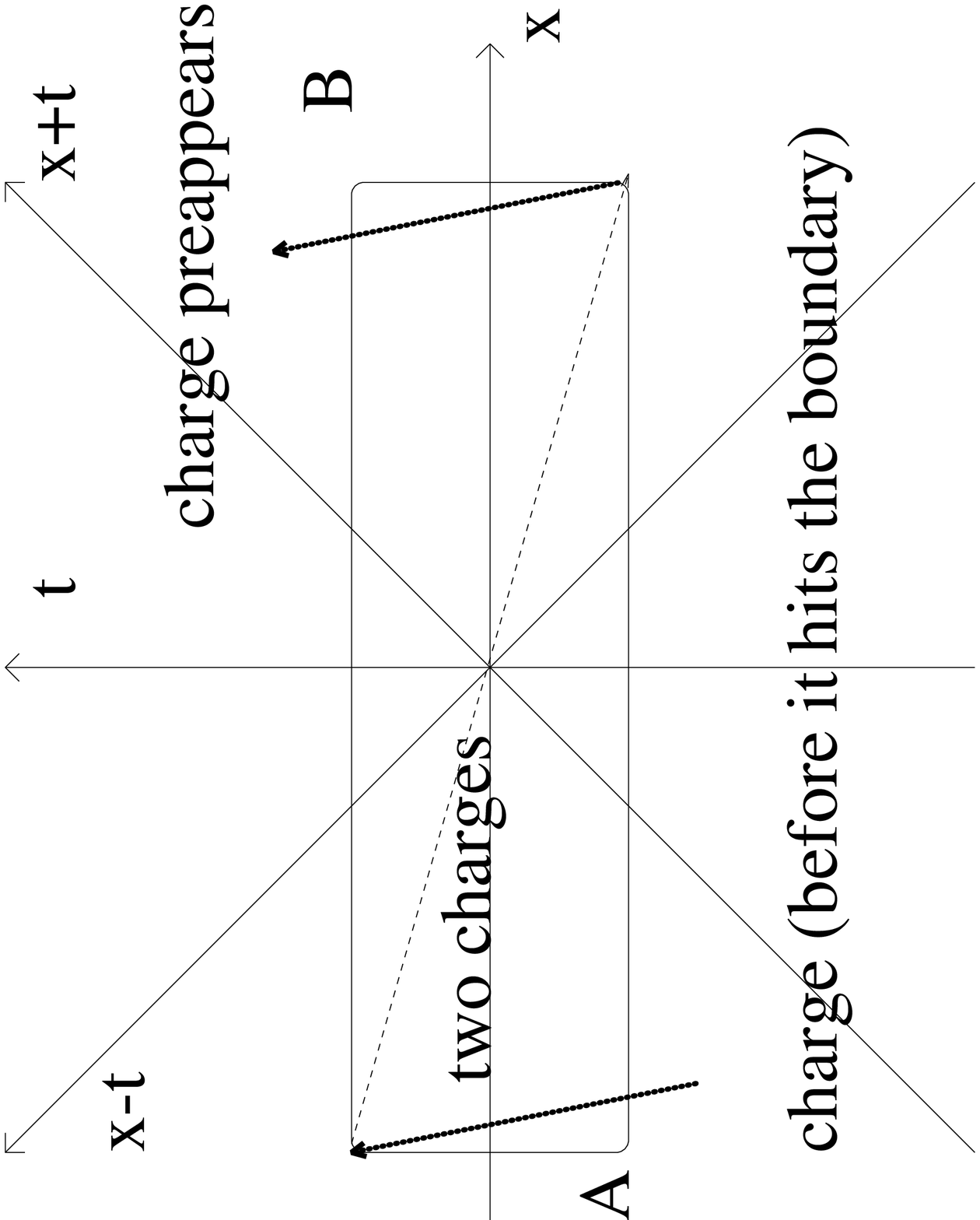}
were $\frak{L}=\overrightarrow{AO}=\overrightarrow{OB}$)
that the charge jumps between 
$Q(\bar e^{(0)})=0$
and
$Q(\bar e^{(0)})=1$
in the frame 
\be
\bar x^0={x^0-v x^3\over \sqrt{1-v^2}}\qquad
\bar x^3={x^3-v x^0\over \sqrt{1-v^2}}
\ee
where $\bar J(\bar x)$ at rest.
\begin{figure}
\begin{minipage}[tl]{0.47\linewidth}
\centering
 \psfig{clip=,figure=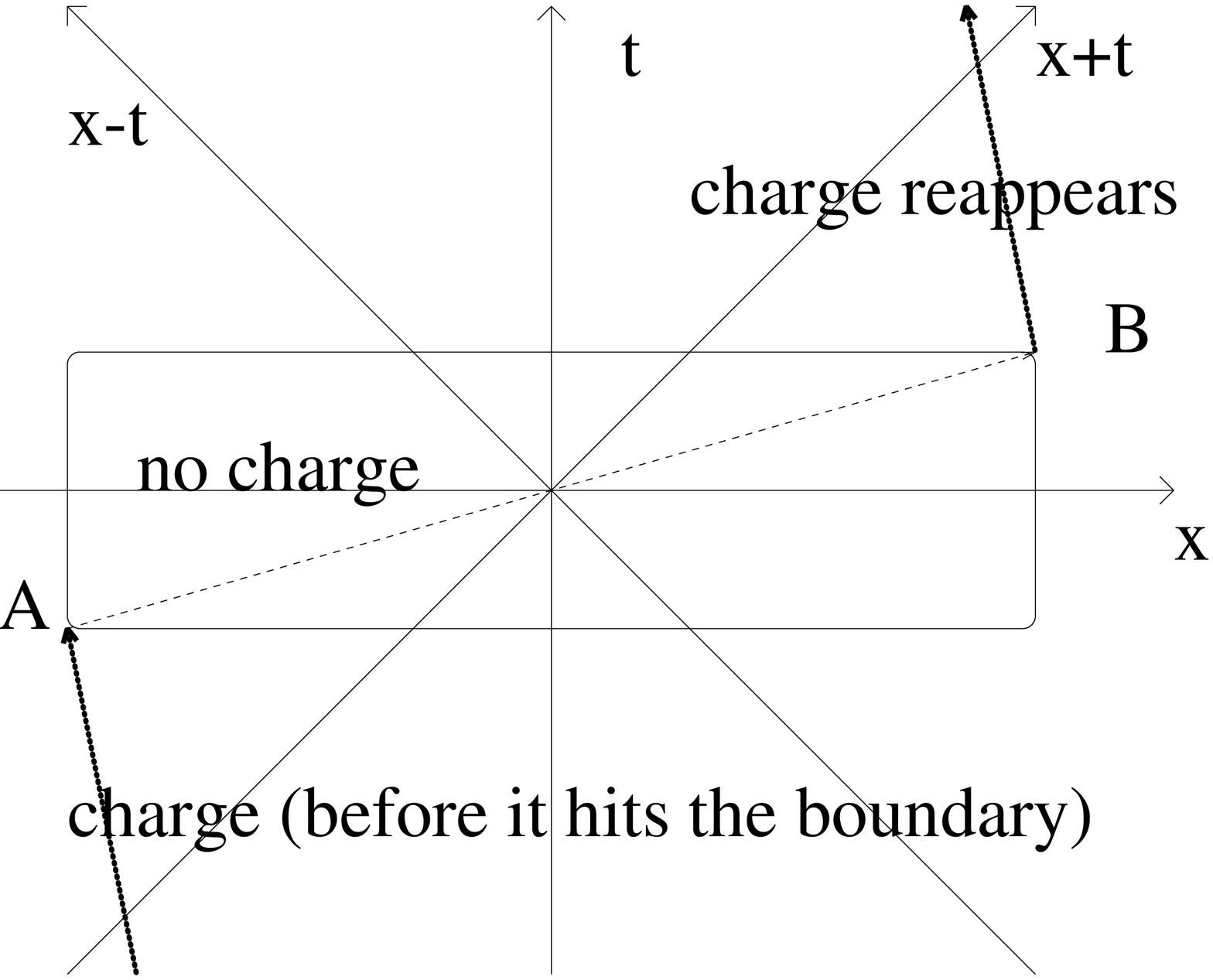,width=\linewidth,angle=-90}  
  \caption{Illustration of charge non-conservation. The
boundary conditions join A and B
}\label{BCIllust.eps}
\end{minipage}\qquad  
%
%
%
\begin{minipage}[tr]{0.47\linewidth}
\centering
 \psfig{clip=,figure=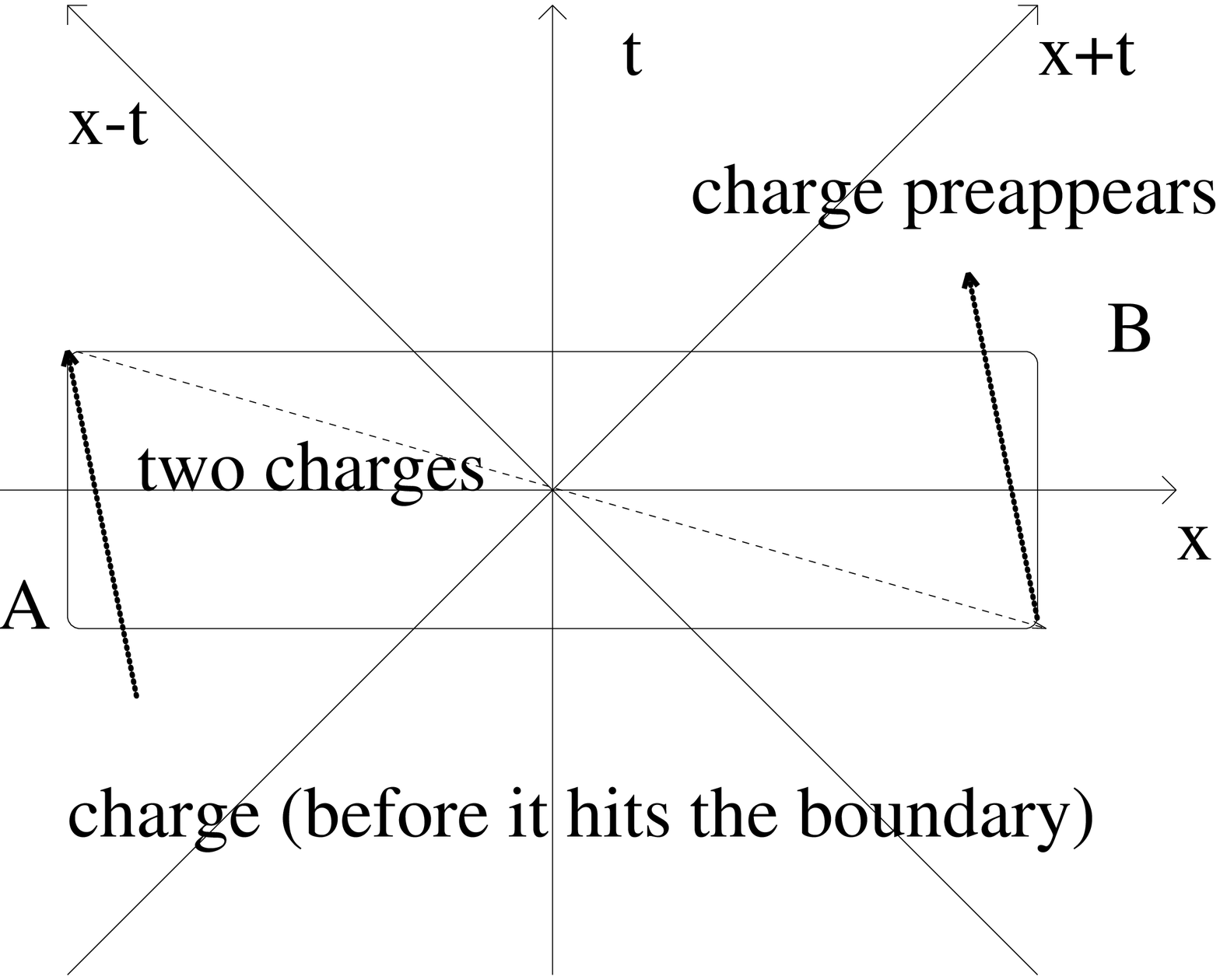,width=\linewidth,angle=-90}  
  \caption{Illustration of microcausality violation. The
boundary conditions join A and B}\label{BCIllust2.eps}
\end{minipage}   
\end{figure}
The deeper mathematical reason behind this
is the fact that the conservation of charges 
follows from Gau{ss}' theorem under the condition that the integration
contour over the boundaries vanishes. The charge~\Ref{Charge} is not
properly defined if the boundary terms do not vanish. 
In order for a charge to be
conserved one would have to include the boundary terms into its definition. 
In particular, the momenta $\hat P^\mu(e^{\text{quant}})$ constructed in eq. 
\Ref{QuadriImpulsions} would cease to be conserved charges if 
\Ref{eUndBC} did not hold and, consequently, they would no longer be
identical to the generators of translations.
Periodicity breaks the Poincar\'e invariance 
of the Lagrangian density because 
it singles out the rest frame of ${\frak{L}}$. 
This does not merely imply non-conservation of charge, it
implies, furthermore, that microcausality is violated in every frame except for the 
lattice rest-frame with $e^{(0)}=e^{\text{quant}}$.
This is demonstrated
in \figRef{BCIllust2.eps} which represents a charge that 
disappears at the boundary (A) and "re"appears in the past (B).
The inhabitants of the $2L$-periodic universe, however, would
need (at least) the time $L$
in order to find out that they are in 
a frame where charge is not conserved\footnote{To be more precise, an observer moving
with velocity $v$ relative to the lattice would need the time 
${\frak{L}}^3=L/\sqrt{1-v^2}$ in order to 
gather the information collected by 
detectors which measure the total charge at a given instant of time 
$x^0-vx^3=$const.
} 
. 
This is the same
time it would take in order to find out that the universe is $2L$-periodic.
The finiteness of the speed of light 
justifies the replacement of a potentially infinite universe by
a small model universe with {\em space-like} periodic BCs.
Effects of periodic BCs are {\em unobservable} if the periodicity $L$ is
large enough. 
An experimentalist has access to a finite space-time region of the universe
only. 

The inhabitants of a universe with light-like BCs (i.e. $L=0$), in contrast, 
are always in the "wrong" frame, i.e. in a frame where charge is not conserved
since no observer can move with (almost) the speed of light. 
Worse still, they 
would find out immediately by sending a light-signal in 
three-direction: this signal would return almost instantly in the past.
Another signal travelling with less than the speed of light could be used
in order
to transport information instantly to any place in the universe.
This is the first of a serious of {\em observable} unphysical
predictions resulting from light-like BCs.
  
We conclude that there is no preferred 
time direction $e_{(0)}$ 
when using periodic boundary conditions\footnote{except for the orthogonality
condition $e^{(0)}=e_{(0)}$ we have excluded}.  
The contra-variant 4-vector $e^{(0)}$, in contrast, should be parallel
to $e^{\text{BC}}$ in order for the na{\"\i}ve
charges to be conserved.  
Thence we shall always assume that $e^{\text{quant}}\equiv e^{\text{BC}}$ unless the contrary
is explicitly stated. 
\begin{table}\begin{center}

\begin{tabular}{||l||c|c||}
\hline\hline
Type of BC&
Periodic BCs &
Fixed BCs   \\ \hline\hline
Conserved quantity      &
$\underset{0\le\mu\le 3}{\forall} P_\mu(e^{\text{BC}})$  &
$\underset{e:e^2<0}{\forall} P_0(e)$    \\ \hline\hline
\end{tabular}

\caption{Conserved quantities in the presence of boundary conditions}\label{PeriodicOrFixedBCs}

\end{center}\end{table}

Thes conclusions, however, 
are a particularity of periodic BCs.
For fixed boundary conditions, in contrast, the importance of the co-variant 
time-direction and the contra-variant orientation of the quantisation surface
is somewhat reversed. 
Only the generator
$P_{(0)}$ of translations in $e_{(0)}$-direction is
then a conserved quantity whereas spatial momenta cannot be conserved
because fixed BCs break translation symmetry. 
In this case, the contra-variant co-ordinate axis or the quantisation surface
have no preferred orientation. This is illustrated in \tabRef{PeriodicOrFixedBCs}.

\section{Kinematical Equivalence of Relativistic Frames}

In this subsection, we introduce the expression 
\Index{kinematical equivalence} of co-ordinate systems. 
In order to avoid lengthy expressions, we shall suppress the distracting
transverse co-ordinates $x^1$ and $x^2$ whenever
they are not important. 
\begin{Definition}
We call the co-ordinate systems $x^\mu$ and $\bar x^\mu$ 
\Index{kinematically equivalent} (abridged $x^\mu\triangleq\bar x^\mu$)
\index{$\triangleq$}
if their respective instants of time
$x^0=0$ and $\bar x^0=0$ coincide. 
\end{Definition}
 This implies, in particular, that two events $x_A$ and $x_B$ which are
simultaneous in one frame $x_A ^0=x_B ^0$ are simultaneous also with
respect to the kinematically equivalent frame $\bar x_A ^0=\bar x_B ^0$.
Let us write the co-ordinate transformation 
$T:x^\mu\rightarrow\bar x^\mu=T^\mu _{\;\nu} x^\nu$
in a way 
\begin{equation}
T=
\left(
\begin{array}{cc}
\digamma & -v\digamma \\
v'\digamma' & -\digamma' \\
\end{array}
\right)
\end{equation}
similar to a co-ordinate boost (transverse co-ordinates are suppressed, $\digamma\digamma' \neq 0$). 
Setting $v=v'$,
$\digamma=\digamma'={1\over \sqrt{1-v^2}}$ renders $T$ a combined boost
and parity transformation.
Setting $\digamma=\digamma'=1$, 
$v=-1$ and $v'=1$ yields the FF co-ordinates $\bar x^0=x^+=x^0+x^+$, 
$\bar x^3=x^-=x^0-x^3$.
The rationale for writing $T$ in this form is that we are interested in the
relation between the FF and the IF.
The  \Index{contra-gredient}
\index{$\contgred{}$} 
of $T$ 
\begin{equation}
\contgred{T}
\dn
(T^{-1})^\top=
{1\over (1-vv')\digamma\digamma'}
\left(
\begin{array}{cc}
\digamma' & \digamma' v'\\
-\digamma v & -\digamma \\
\end{array}
\right)
\end{equation}
is the co-ordinate transformation 
$\contgred{T}:x_\mu\rightarrow\bar x_\mu=\contgred{T}_\mu ^{\;\nu} x_\nu$
which transforms co-variant co-ordinates.  
Kinematical equivalence between $x^\mu$ and $\bar x^\mu$ implies that the canonical quantisation procedure
\footnote{Or, if constraints are present, the Dirac-Bergmann algorithm}
yields equivalent results in the sense of the following theorem
\begin{Theorem}\label{KinematicP}
The momenta $P_\mu(e^{(0)})$ obtained through quantisation 
with respect to the frame $ x^\mu$ and the momenta $\bar P_\mu(\bar e^{0})$
obtained through quantisation with respect to the 
kinematically equivalent frame $\bar x^\mu$
\begin{equation} \label{PPconnection}
\bar P_\mu(\bar e^{(0)})
=
\contgred{T}^{\;\nu} _{\mu} P_\nu(e^{(0)})
\end{equation}
are related in an {\em interaction-independent} way 
by a mere 
co-ordinate transformation
\index{T@$\contgred{T}$}
\end{Theorem}
Expressed differently, let 
$P(\bar e^{(0)},\bar e_{(0)})$ and $P(e^{(0)},e_{(0)})$ 
be the 4-momentum 
obtained in the kinematically equivalent frames $\bar e_{(\mu)}$
and $e_{(\mu)}$ respectively. Then both momenta define the same object,
i.e. 
\be
P(\bar e^{(0)},\bar e_{(0)})=P(e^{(0)},e_{(0)})=P(e^{(0)})
\ee
and one can skip the label $e_{(0)}$. The only difference between
the frames
$\bar e_{(\mu)}$
and $e_{(\mu)}$
is that they project out different components $P_\mu=e_{(\mu)}\cdot P$
from the same object. 

The fact that the transformation between
$P(e^{(0)})$ and $\bar P(\bar e^{(0)})=\bar P(e^{(0)})$
is kinematical follows already from the fact that transformations mapping the
initialisation hyper-surface onto itself do not depend on interactions.
This is, in fact, the very reason for why in~\Cite{Leutwyler:1978vy} kinematical
transformations are {\it defined} as transformations leaving the
quantisation surface invariant.
It is instructive, though, to show eq. \Ref{PPconnection} by construction. 
The most general non-singular linear
transformation leaving the equation $x^0=0$ invariant
\begin{equation}
T=
\left(
\begin{array}{cc}
\digamma & 0 \\
v'\digamma' & -\digamma' \\
\end{array}
\right)
\end{equation}
is obtained by setting $v=0$. This transformation can be decomposed into a product of $T$ with $\digamma=1$
and a mere rescaling of the temporal co-ordinate $x^0\rightarrow
\digamma x^0$. Since two conserved Hamiltonians quantised with respect 
to the time co-ordinates $x^0$ and $\digamma x^0$ respectively are 
equivalent ---albeit rescaled---, it suffices to study $T(\varepsilon)$ with $\digamma\equiv 1$. 
The canonical momenta 
\begin{equation}
\pi(0,{\vec x})
=\partial^0\varphi(0,{\vec x})
=
\bar\partial^0\varphi(0,{\vec x})
=
\bar\pi(0,{\vec x})
\end{equation}
appearing in $\Theta$ then 
remain the same under the $T(\varepsilon)$ transformation
---by construction. 
This fact is not restricted to the scalar model. Canonical momenta $\pi(x)$ 
are always contra-variant zero-components $\pi(x)=\pi^0(x)$ of the
4-vector $\pi^\mu(x)=\partial {\cal L}/\partial(\partial_0\varphi)$
even if $\varphi$ is not a scalar field. 
The quantisation surface $e^{(0)}$ remains the same and therefore
it follows from \Ref{QuadriImpulsions}
\be
\bar P_\mu(\bar e^{(0)})
=
\bar P_\mu(e^{\text{quant}})
=
\bar P_\mu(e^{(0)})
=
\contgred{T}^{\;\nu} _{\mu} P_\nu(e^{(0)})
\ee
that we may construct $\bar P_\mu(\bar e^{(0)})$ as a linear combination
of operators $P_\mu(e^{(0)})$ obtained through a quantisation 
in kinematically equivalent co-ordinates. The 4-vectors 
$P(e^{(0)})$ and $P(\bar e^{(0)})$ are the same objects;
only their co-ordinates $\bar P_\mu$ or $P_\mu$ differ. 
Observables such as mass spectra do not depend on which co-ordinates
one quantises in ----if these co-ordinates are kinematically equivalent. If, however,
$e^{(0)}\neq \bar e^{(0)}$, then quantisation with respect to inequivalent
co-ordinate systems yields different 4-momenta. Worse still. If 
$P_\mu(e^{(0)})$ is a conserved quantity, $P_\mu(\bar e^{(0)})$
cannot be conserved due to Theorem~\Ref{NonConservatioBC}.
We conclude that the arbitrary\footnote{except that the time
axis must not lie in the quantisation surface.}
general time direction $\bar e_{(0)}$ and the generator
$\bar P_{(0)}$ of translations
in this direction are both bereft of physical information. 
In general (i.e. non-orthogonal) co-ordinates, $P^{(0)}(\bar e^{(0)})$ 
is the \Index{physical Hamiltonian} which advances
the proper-time $\bar x_0$ of the observer. This 
Hamiltonian is identical in 
all kinematically equivalent frames
---except for rescaling. 
\section{Defining Boost Invariance}
Now we are able to dispel the claim that, contrary to the IF,
the mass spectrum of the
FF is automatically boost-invariant (under boosts in the three-direction 
to be more precise). This claim is usually based 
on the observation that boosts in three directions do not change 
the FF quantisation surface, i.e. they do not change a light-like
hyper-surface $e^{(+)}$ and consequently, they are kinematical. 
But we have to ask first what boost-invariance
means. We give the following {\em necessary} criterion
\begin{Definition}[Boost invariance]
For a theory characterised by the 4-momentum operators 
$\hat P(e^{\text{quant}})$ to
be boost-invariant in three direction, the expectation value 
\be
M^2(P)
\dn
\bra P,\iota |\hat M^2|P,\iota \ket 
\ee
of the mass-squared operator $\hat M^2:=\hat P\cdot\hat P$
must not depend on the kinematical component $P_3(e^{\text{quant}})$
(or on $ P_-(e^{\text{quant}})$).
Here, $\cket{P,\iota}$ denotes an eigenstate of $\hat P$, $\iota$ stands
for conserved quantum numbers other than the momentum. 
\end{Definition} 
In the IF, this simply means
that computed physical masses must not depend on the three-momentum $P_3$
for $|\vec P|\ll \Lambda$.  
In the FF, this means
that physical masses must not depend on the minus-momentum $P_-$. 
In praxis, however, QFTs need to be regularised, e.g. by introducing
a lattice. A lattice, characterised by its boundary
vector ${\frak{L}\,}$ or by its cut-off 4-vector ${\sf a}$, 
breaks the Poincar\'e invariance of the Lagrangian density 
and therefore,
the computed physical masses can at best be {\em approximately} independent
of the momenta. For renormalisable theories, the degree to which $M(\vec P)$
is independent of $\vec P$ should increase towards total independence
in the continuum limit. 
It is evident that neither the FF nor the IF can be exactly boost-invariant
{\em on a lattice}
according to the criterion given above. 
In Chapter\secRef{CNum}, we shall even see that, ironically, the IF is 
boost-invariant to a much higher accuracy than is the FF
in a practical computation. 
Why, then, is this phenomenon nowhere described?
In the FF literature, no distinction is made,
between boost-invariance 
under \Index{co-ordinate} boosts and invariance under  
the \Index{physical boosts} discussed above. 
In the FF, a \Index{co-ordinate boost} in three direction
transforms a frame with $\bar e^{(0)}=e^{(+)}$ 
into a kinematically equivalent frame, i.e. it leaves 
the quantisation surface $e^{(+)}$
intact. In the IF, a \Index{co-ordinate boost} renders the form of the 4-momenta
as well as the expression for the commutation relations more cumbersome. 
But: physics is not changed by choosing co-ordinates 
which
are not adapted to the physical problem in the same way that physics is not
changed if a spherically symmetric system is expressed in elliptical co-ordinates. 
The only penalty for using the ''wrong'' co-ordinates
is a practical one: cumbersome expressions, as we have 
have discussed above. In particular the 
commutation relations\Ref{UsualCommutationRelations} look ugly
when expressed in a frame whose instant of time does not coincide
with the quantisation surface. 
The reason for why \Index{co-ordinate boosts} are kinematical in the FF 
is the same reason for why these boosts are kinematical in the IF, too:
{\em Co-ordinate boosts are always kinematical}. They do not
involve interactions--- by definition.

The invariance of the FF under {\em kinematical} boosts has
some unpleasant side-effects. The cut-off $\Lambda=\pi/a^-$ and the lattice size ${\frak{L}\,}^-$ are 
irrelevant quantities in the FF. In two space-time dimensions
there is no continuum limit
or infinite volume limit as it does not matter if 
${\frak{L}\,}^-\rightarrow \infty$
or if ${\frak{L}\,}^-\rightarrow 0$: the mass spectrum is always the same;
only the number of lattice sites $N_L\dn {2{\frak{L}\,}^-\over a^-}$ is relevant. 
The deeper reason behind this is that 
a space-like boundary vector ${\frak{L}\,}$ defines an invariant
renormalisation scale
$L\dn \sqrt{-{\frak{L}\, }\cdot{\frak{L}\,}}$ whereas a light-like 
boundary vector ${\frak{L}\, }\cdot{\frak{L}\,}=0$ does not: the FF
is scale invariant in two space-time dimensions. We shall make
use of this peculiar property on several occasions in this thesis. 
 

\section{The Front Form and $\varepsilon$ Co-ordinates}\label{secTheFFandEpsilon}
 \subsection{A Sketch of the Problem}

%
Often, the FF is defined as the limit $\varepsilon\rightarrow 0$
in so-called $\varepsilon$-co-ordinates~\Cite{Prokhvatilov:1989eq,Lenz:1991sa,Naus:1997zg},
a co-ordinate system othogonal with respect to 
the Euclidean metric $\delta^{\mu\nu}$ but non-orthogonal
with respect to the Minkowski metric $g^{\mu\nu}$.
A quantisation in these
co-ordinates\footnote{
There are several conventions in order to define
$varepsilon$ co-ordinates. Here we use the definition 
in~\Cite{Lenz:1991sa,Naus:1997zg} replacing
---without loss of generality--- $\varepsilon/L$ by $\varepsilon$.}
$\bar x^\mu(\varepsilon)\dn T^\mu _{\;\nu} x^\nu$ with 
\index{x@$\bar x^\mu(\varepsilon)$}
\index{Tepsilon@$T(\varepsilon)$}
\begin{equation}
T(\varepsilon)
=
\left(
\begin{array}{cc}
1+{\varepsilon} & 1-\varepsilon \\
1  & -1 \\
\end{array}
\right)
\end{equation}
is canonical (as is the IF quantisation) as long as $\varepsilon\neq 0$. 
No constraints as in the FF appear. 
Only for $\varepsilon=0$ do the FF constraints arise; then the procedure
of canonical quantisation has to be replaced by the more sophisticated
Dirac-Bergmann quantisation algorithm~\Cite{dirac:generalized,SundermeyerBuch}.
At first sight, it seems that these co-ordinates
provide a smooth transition between both forms
since $\bar x^0(0)=x^+$ and $\bar x^3(0)=x^-$. 
The $\varepsilon$ co-ordinates seem to achieve what boosted co-ordinates
cannot: a smooth transition between usual co-ordinates and co-ordinates
on the light-front.
We argue, however, that any construction based on the notion \Index{closeness to the
light-front} is doomed\footnote{Cf. Chapter~\secRef{CStructure}}; 
it cannot be reconciled with 
the special theory of relativity. 
We direct attention to the fact that a particular co-ordinate
system does not carry physical information. The actual
carrier of information is the boundary 4-vector defining the BCs. 
For it is this 4-vector which actively breaks Poincar\'e invariance. 
The boundary 4-vector ${\frak{L}\,}=(0,{\frak{L}\,}^3)$ has to 
be {\em actively} boosted in terms of $\varepsilon$ in order for ${\frak{L}\,}$ 
to lie inside the hyper-plane $\bar x(\varepsilon)=0$ lest non-conserved
charges arise.  Therefore, the components ${\bar\frak{L}}^\mu(\varepsilon)$
diverge for $\varepsilon\rightarrow 0$
if the lattice size $L\dn \sqrt{-{\frak{L}\,}\cdot {\frak{L}\,}}$ is
kept constant. This means that 
physical information is destroyed for $\varepsilon=0$
since lattices with different invariant size $2L$ are all mapped onto the same
co-ordinates $\bar{\frak{L}\,}^-(0)=\infty$. Consequently, \Index{ambiguities} must be expected
and, indeed, we shall encouter them later on in this chapter. Similarly, lattices with
different \Index{lattice spacing} $a$ are all mapped onto the same lattice
spacing $\bar a^-=\infty$.  
One might want to avoid these infinities\footnote{
Solely the infinities --- not the ambiguities --- can be avoided!} by 
an explicit reduction in the {\em physical} size of the lattice $L\rightarrow L(\varepsilon)$ such
that $L$ goes 
to zero in the limit $\varepsilon\rightarrow 0$.
But even then is the transition $\varepsilon\rightarrow 0$ not smooth. 
For we are now prepared to show that 
if the physical size of the system becomes infinitesimally small
in the limit $\varepsilon\rightarrow 0$,  then the resulting Hamiltonian
is kinematically equivalent to the IMF rather than to the FF. 
We shall see that the commutative diagram
\index{$\triangleq$}
\begin{equation} \label{CDEpsilon1}
\begin{CD}
{\cal L}  
           @>\text{quantise}> \varepsilon=0 >  
\hat P(\varepsilon=0) =  \hat P(e^{(+)}) \\
           @V{\text{quantise}}V \varepsilon\neq 0 V   
           @A{\text{subtract left-movers}}A
             {\text{from self-energy $\triangle\omega$ }}A \\
\hat P(\varepsilon)\triangleq \hat P(e^{(0)})   
           @>{L\rightarrow 0}>{\varepsilon\rightarrow 0}>
\hat P(\varepsilon\rightarrow 0)\triangleq \hat P(e^{(0)})(\text{IMF})    \\ 
\end{CD}
\end{equation}
only closes if the contributions of left-movers to the self-energy
of a right-mover are {\em explicitly} removed (including
the self-energy terms they create). Without explicit destruction
of these self-energy terms, the limit $\varepsilon\rightarrow 0$ leads
to the IMF and not to the FF. 
The diagram
\index{$\triangleq$}
\begin{equation} \label{CDEpsilon2}
\begin{CD}
{\cal L} 
           @>\text{quantise}>\varepsilon=0>  
\hat P(\varepsilon=0) =  \hat P(e^{(+)}) \\
           @V\text{remove left-movers, constrain}V\text{zero modes}V   
           @A{\varepsilon\rightarrow 0}A
             {L\rightarrow 0 }A \\
{\cal L}_{\text{R}}    
           @>\text{quantise}>\varepsilon=0>
\hat P_R(\varepsilon)\triangleq \hat P_R(e^{(0)})  \\ 
\end{CD}
\end{equation}
obtained by {\em first} removing left-movers and
then quantising the theory, commutes however. This means
that the limit $\varepsilon\rightarrow 0$ is continuous on the
classical level but not on the quantum level. We shall demonstrate
this in the next section.

One comment is in order: A connection between the FF and the zero volume IF has been
inferred already in the article~\Cite{Lenz:1991sa}.
{\bf Herein}, we go some steps further, completely separating the small volume
effects from a particular co-ordinate system
and demonstrating that the $\varepsilon\rightarrow 0$
limit does {\em not} yield the FF but rather the IMF:
the transition from $\varepsilon \rightarrow 0$ to $\varepsilon=0$
is {\em discontinuous} ---even if the lattice size shrinks to zero
in this limit.  
Quantisation and the $\varepsilon\rightarrow 0$ limit
do not commute
\footnote{The only exception to this rule are certain fermionic models such
as the massive Schwinger model in the unphysical 
axial gauge treated in Chapter\secRef{CSchwinger}
where the IMF is almost equivalent to the FF.}.
{\bf Additionally}, we are able to clarify
the nature of the constraints arising in this picture. 
{\small
We demonstrate that light-like and space-like BCs are allowed to co-exist
in the IMF and that the selection of the former are responsible
--even in an IF quantisation-- for the elimination of left-movers. }
{\bf Finally}, there are two points where we disagree with the authors 
of~\Cite{Lenz:1991sa}: Firstly, we show that the size of the lattice cannot be
chosen to be arbitrarily small even if only Lorentz-contracted
objects which fit into the lattice are described. 
Secondly, we show that the FF can, in general, {\em not} be 
regarded as an effective theory of the
IF on small lattices ---let alone on large lattices. 
\subsection{Kinematical Equivalence}

The kinematical equivalence of $\varepsilon$ quantisation and conventional
IF quantisation for $\varepsilon\neq 0$ is swiftly demonstrated using the formalism developed in the last section.
The $\varepsilon$ co-ordinates defined via $T(\varepsilon)$ are
kinematically equivalent to a boosted frame 
\index{x@$\breve x^\mu$}
$\breve x^\mu \dn B^\mu _{\;\nu} x^\nu$ 
defined via the co-ordinate boost
\begin{equation}
B(\varepsilon)
=
{1\over \sqrt{ 1-v^2 }}
\left(
\begin{array}{cc}
1 & -v \\
v & -1 \\
\end{array}
\right)
=
{1\over 2\sqrt{\varepsilon}}
\left(
\begin{array}{cc}
1+\varepsilon & 1-\varepsilon \\
\varepsilon-1 & -1-\varepsilon \\
\end{array}
\right)
\end{equation}
with boost-velocity
\begin{equation}\label{VitesseAbsolue}
v(\varepsilon)=-{1-\varepsilon \over 1+\varepsilon }
\qquad.
\end{equation}
They are also equivalent to the boosted and rescaled frame
\index{x@$\tilde x^\mu$}
$\tilde x^\mu=b B^\mu _{\;\nu} x^\nu$
where
 \be
b(\varepsilon)
=
2\sqrt{\varepsilon}
=
2\sqrt{1+v \over 1-v }
\ee
is the scale transformation. We have sloppily referred to $B(\varepsilon)$ as a ''boost'' even though $B$ is 
actually a combination of a boost and a parity transformation. 
The parity transformation in $B$ originates from 
our sticking to the wide-spread
convention of using $x^+$ as time co-ordinate and $x^-$ as space co-ordinate. A pure boost would have
sufficed if we had used $x^-$ as the time co-ordinate.  

The kinematical equivalence of $\bar x^\mu$ and $\tilde x^\mu$ 
follows from the fact that the hyper-surfaces
$\tilde x^0=0$
and
$\bar x^0 = T^0 _{\;\nu} x^\nu=0$
are equivalent.
The co-ordinates associated
with $T$ or $bB$ 
define the same hyper-surface $x^0=v x^3$ moving with velocity $v$
relative to the fixed reference frame $x^\mu$.
Other co-ordinates such as $\breve x^\mu \dn B^\mu _{\;\nu} x^\nu$
would be kinematically equivalent, too, but we have chosen the frame
which does not rescale time $x^0$ (i.e. $\bar x^0=x^0$ or $\digamma=1$). 
Note that both $B$ and $b$ are singular in the $\varepsilon=0$ limit. 
The combined transformation $bB$ is singular, too, in the sense that 
it is not invertible. Only $\bar x^\mu(\varepsilon)$ is non-singular
for $\varepsilon=0$. 
For every finite $\varepsilon$, however, each of these transformations is well-defined and non-singular.

The metric of the $\tilde x^\mu$ co-ordinates
\be
\tilde g^{\mu\nu}
=b^2(\varepsilon) g^{\mu\nu}
=4\varepsilon g^{\mu\nu}
\qquad
\ee
is diagonal but rescaled. 
The 4-momenta $\bar P_\mu(\bar e^{(0)})$ of the $\varepsilon$ quantisation %
\begin{align}
\bar P_0(\bar e^{(0)})&=\tilde P_0(\tilde e^{(0)})-\tilde P_3(\tilde e^{(0)}) \\
\bar P_3(\bar e^{(0)})&=2\varepsilon\tilde P_3(\tilde e^{(0)})
\end{align}
are ---according to Theorem~\Ref{KinematicP}---trivially related
to the 4-momenta $\tilde P_\mu(\tilde e^{(0)})$ from $\tilde x^\mu$ 
quantisation, i.e. quantisation in the non-orthogonal $\varepsilon$ 
co-ordinates can be reduced to a much simpler quantisation
in orthogonal, rescaled co-ordinates. Here, we have used the fact that
the matrix 
\be
T'\dn T\cdot (Bb(\varepsilon))^{-1}=
\begin{pmatrix}
1 & 0 \\
{1\over2\varepsilon} & {1\over 2\varepsilon} 
\end{pmatrix}
\ee
which relates the two frames $\bar x^\mu=T^{'\mu}_{\;\nu} \tilde x^\nu$,
has the contra-gredient 
\be
\contgred{T}'=
\begin{pmatrix}
1 & -1 \\
0 & 2\varepsilon 
\end{pmatrix}
\qquad.
\ee
\subsection{Boosting the Lattice}
We require the boundary vector ${\frak{L}\,}$ to lie inside the quantisation
hyper-plane, 
i.e. $\bar {\frak{L}\,}\cdot \bar e^{(0)}(\varepsilon)=0$,
in order to avoid non-conservation of $P^\mu$.
Cf. Theorem\Ref{NonConservatioBC}. 
Let $L\dn \sqrt{-{\frak{L}\,}\cdot{\frak{L}\,} }$ 
be the invariant length 
\index{L@$L$}
of the lattice then 
and let the boundary 4-vector lie inside the quantisation
surface $\bar x^0=0$, i.e. ${\bar\frak{L}\,}^0=\tilde {\frak{L}\,}^0=0$.
For $L$ fixed, the periodicity 
$\bar  {\frak{L}\,}^3=L/\sqrt{\varepsilon} $, 
$\tilde{\frak{L}\,}^3=2\sqrt{\varepsilon}L$ 
diverges in any of the quantisation frames which we have introduced. 
If we want to obtain a {\em finite} periodicity 
${\frak{L}\,}^-=L_c$ in the FF limit
\index{Lc@$L_c$}
$\varepsilon\rightarrow 1$,
we have to reduce the physical length 
$L$ to zero in the FF limit
by setting $L(\varepsilon)=L_c\sqrt{\varepsilon}$. 
Only the explicit reduction of the invariant length in terms of $\varepsilon$
is able to convert the space-like ${\frak{L}\,}\cdot {\frak{L}\,}<0$ 4-vector
$\frak L$ into a time like 4-vector ${\frak{L}\,}(0)\cdot {\frak{L}\,}(0)=0$.
If $\tilde P^3$ lies on the mass-shell
\be
\tilde P_0^2-\tilde P_3 ^2=M^2/b^2=(M/2)^2/\varepsilon
\ee
then the dispersion relation reads
\be
\tilde P_0 = \sqrt{\tilde P_3 ^2 - M^2/b^2}
=
 \sqrt{\tilde P_3 ^2 - M^2 \varepsilon}
\ee
in these co-ordinates
and consequently 
it reads
\be\label{EpsDisperse}
\bar P_0
=
{1\over 2\varepsilon} ( \sqrt{\bar P_3^2+M^2\varepsilon} - \bar P_3)
\ee
in $\varepsilon$ co-ordinates. 
Eq.\Ref{EpsDisperse} seems to become gradually
identical to the dispersion relation 
\be
\bar P_0\approx 
{M^2\over 4\bar P_3}
\ee
in $\varepsilon$ co-ordinates in the
$\varepsilon\rightarrow 0$ limit where the square-root 
may be developed in powers
of $\varepsilon$. This argument is deceptive
since $P^3$ can only assume multiple($n$) values of 
\be
\triangle\bar k_3
={\pi\over \bar {\frak{L}\,}^3}
={\pi\over  L}\sqrt{\varepsilon}
 \ee
and therefore 
\be
\bar P_0
=
{\varepsilon\over 2} (\sqrt{ n^2 (\pi/L)^2+M^2}) 
\ee
can only be brought into the FF form if $L$ is small-- irrespective
of $\varepsilon$. 
The same holds for the $\tilde x^\mu$ co-ordinates with
momentum lattice spacing
and
\be
\triangle\tilde k_3
={\pi\over \tilde {\frak{L}\,}^3}
={\pi\over 2 L\sqrt{\varepsilon}} \;,
\ee
of course. 

\subsection{Conclusion}

We conclude that, in spite of contrary claims in the literature~\Cite{Burkardt:1997bd,Burkardt:1996ct},
$\varepsilon$ quantisation is completely equivalent to IF quantisation in boosted co-ordinates as long as $\varepsilon\neq 0$.
Both quantisations define the same object $P(\tilde e^{(0)})=P(\bar e^{(0)})$.
The Hamiltonian of $\varepsilon$-quantisation can be seen as
the projection $\bar e^{(0)}\cdot P$ in $\bar e^{(0)}(\varepsilon)$ direction; the Hamiltonian
in $\tilde x^\mu$ co-ordinates is the projection $\tilde e^{(0)}\cdot P$
in $\tilde e^{(0)}$-direction. 
By no means do $\varepsilon$ co-ordinates justify the unfortunate notion 
of ''\Index{closeness to the light-cone}'', 
a notion completely alien to the theory
of relativity. In fact, the special theory of relativity is built upon the
very absence of such a notion: 
every observer measures the same speed of light, i.e.
the relative velocity between any instant of time and any light-front
is the speed of light. 
The \Index{relative velocity}\footnote{or functions of it such as the
rapidity}
\be
\triangle v_{kl}\dn {k\cdot l\over \sqrt{(k\cdot l)^2- l\cdot l\;k\cdot k}}
\ee
is the proper, Lorentz-invariant, measure of closeness between 
two hyper-surfaces with normal
vectors $k$ and $l$. 
If one hyper-surface, say $l$, is light-like $l\cdot l=0$ then 
the relative velocity $\triangle v_{kl}=1$ is {\em always} the speed
of light, for any space-like hyper-surface $k$. Cf.~\secRef{CStructure}.

Only {\em relative} velocities have physical significance, absolute velocities
have not. 
The \Index{absolute velocity}
$v^3=k^3/k^0$ of $k$ of the lattice
depends on the co-ordinate frame: consequently, it
is irrelevant.
Now, the parameter $\varepsilon$ is but a complicated way of expressing
the {\em absolute} velocity $v$ (cf. Eq. \Ref{VitesseAbsolue}) of the quantisation
surface $\bar e^{(0)}(\varepsilon)$ and therefore, any $\varepsilon\neq 0$
is equally close or far to the light-front. A 4-vector lies either {\em on}
the light-front or {\em off} the light-front. There is no in-between. 

\section{The Infinite Momentum Frame}\label{secTheIMF}
\subsection{Operational definition of the IMF}

As the co-ordinates $\bar x^\mu(\varepsilon)$ and $\tilde x^\mu(\varepsilon)$
are kinematically equivalent, we should be able to reproduce
all properties of $\varepsilon$ co-ordinates
in ordinary co-ordinates. This is what we shall do in this section. 
The limit $\varepsilon\rightarrow 0$ will lead us to the IMF 
if the invariant lattice size $L=\sqrt{1+v\over 1-v}L_c=\sqrt{\varepsilon}L_c$ is explicitly reduced
to zero.  
We have to distinguish two inequivalent definitions of the IMF to start with. 
\begin{Definition}
Let $\vec P$ and $M$ be the momentum and the mass
of a particle defined on a momentum lattice
with momentum lattice spacing $\triangle k^3$ in three-direction.
We say that we are in the \Index{weak infinite momentum frame}
if $|\vec P|/M\rightarrow \infty$. 
We say that we are in the \Index{strong infinite momentum frame}
if $|\vec P|/M\rightarrow \infty$ and $\triangle k^3/M\rightarrow \infty$. 
\end{Definition}
The following theorems hold
\begin{Theorem}
The strong infinite momentum frame is equivalent to choosing 
an arbitrarily small lattice size $L<1/M$. Therefore it is unphysical. 
\end{Theorem}
Proof: $\triangle k^3={\pi\over L}\rightarrow \infty$ is equivalent
to $L\rightarrow 0$. 
It is well known that the correlation length $a\xi\dn {1\over M}$ 
has to be larger
than the lattice size $2L$ in order for the theory 
to be physical~\Cite{MontvayMunsterBuch}(except for the non-relativistic
limit or QFTs such as the Schwinger model which
not require a scaling window).  
\begin{Theorem}
If the \Index{effective lattice size} defined as $N\dn {P^3\over \triangle k^3}$
\index{N@$N$}
is finite, then the weak IMF is equivalent to the strong IMF
\end{Theorem}
Proof: $P^3/M=N\triangle k^3/M\rightarrow \infty$ 
is equivalent to $\triangle k^3/M\rightarrow \infty$.

In other words: Only if $N=\infty$, has the IMF a chance of being physical
(it may or may not, depending on the theory and on the way the
limit $N\rightarrow \infty$ is defined).
\begin{Theorem}
An infinite effective lattice corresponds to a theory
which is continuous 
in the variable $x_B\dn k^3/P^3$ 
if $k^3$ is replaced by  $x_B P^3$.
\end{Theorem}
Proof: $\triangle x_B={\triangle k^3\over P^3}=1/N=0$. 
\begin{Theorem}
In the IMF, the continuum limit $\Lambda/M\rightarrow \infty$ is 
meaningless.  
\end{Theorem}
Proof: $P^3$ has to be smaller than $\Lambda$. 
Thus $P^3/M\rightarrow \infty$ implies $\Lambda/M\rightarrow\infty$.
The weak IMF on a finite lattice is therefore ill-defined in general due
to the infinities which arise without cut-off. The strong infinite momentum
frame is well defined if $N<\infty$ or else if the ultra-violet divergences
can be removed by vacuum-subtraction alone. The reader is referred to
Chapters~\secRef{CSchwinger} and~\secRef{CNum} for examples.  

\subsection{Constraints and Boundary Conditions in the IMF}\label{ConstraintsAndBoudary}
We have seen that a all reference frames with
space-like instances of time are equivalent and
that only an explicit reduction of the lattice size $2L$ brings
the dispersion relation of $\varepsilon$ co-ordinates into FF shape. 
Therefore, it is natural to ask the question: does the {\em simultaneous}
limit
\be
\lim_{\varepsilon,L\rightarrow 0} \hat{\bar P}_\mu(\varepsilon)
\ee
reproduce
the FF operators $P(e^{(+)})$?
Again, this question has to be answered negatively; again, the specific
form of the co-ordinate system does not matter. 
We are going to show now, 
that it is possible to construct the FF as a {\em classical} effective
theory of the {\em strong} IMF without ever moving the instant of 
time $x^0$ towards the light-cone. The {\em quantised FF}, however, is not
equivalent to the {\em quantised} IMF.
In what follows we shall be 
using an orthogonal co-ordinate system $x^\mu$ in which the
boundary 4-vector ${\frak{L}\,}$ takes on the form 
${\frak{L}\,}_{(3)}=(0,0,0,{\frak{L}\,}^3)$ with ${\frak{L}\,}^3=L$. 
We are imposing boundary-conditions in 3-direction and in the transverse
direction but we shall not explicitly mention the latter ones. 
A free scalar field $\varphi(x)$ subjected to space-like periodic boundary conditions  in 3-direction
\be
\varphi(x+{\frak{L}\,}_{(3)})
-
\varphi(x-{\frak{L}\,}_{(3)})
=
\varphi(x+{\frak{L}\,})
-
\varphi(x-{\frak{L}\,})
=
0
\ee
may be expanded
\begin{equation}  \label{VarphiExpansion}
\varphi(x)
=
\sum_{{\vec k} }
{1\over \sqrt{(2\pi^3)^3 2\omega({\vec k} )}}
( 
a_{{\vec k} } e^{-i\omega({\vec k} ) x^0 + i{\vec k}\cdot{\vec x} }
+
a^{\dag} _{{\vec k} } e^{+i\omega({\vec k} ) x^0 - i{\vec k}\cdot{\vec x} }
)
\end{equation}
in terms of
creation $a^{\dag}_{{\vec k} }$ and annihilation $a_{{\vec k} }$ 
operators which obey the usual
\be
[a^{\dag}_{\vec k},a_{{\vec k}'}]
=
\delta_{{\vec k} ,{\vec k}'}
\ee
commutation relations. 
The three-component $k^3$ takes on the discrete values
\be
k^3=\triangle k^3 n = {\pi\over {\frak{L}\,}^3} n
\ee 
where $n$ denotes an integer. The free fields $\varphi$ and 
$\partial_0\varphi=\hat \pi$ at
$x^0=0$ are ---by construction--- 
a realisation of the commutation algebra~\Ref{aCommutIF},\Ref{bCommutIF}. 
Inserting the fields $\varphi(0,\vec x)$ into the Hamiltonian 
\begin{equation}\label{Hamiltonian}
H[\varphi]=P_0(e^{(0)})
= \int d^3 x \; \left( \frac{1}{2} ({\partial_0 \varphi})^2 +
  \frac{1}{2}(\vec \nabla \varphi)^2 + {m_0^2\over 2} \varphi^2 + {g_0\over 4!}
\varphi^4 \right),
\end{equation}
yields the expression
\begin{eqnarray}\label{HamiltonFourier}
H &=&
\sum_{\vec k} 
\left[\omega(\vec k)+\triangle\omega(\vec k)\right]
 a^\dagger_{\vec k}a_{\vec k} +
\triangle k^3
\sum_{\vec k\vec l\vec m} 
{(\triangle k_\perp)^2 g_0\over 4(2\pi)^3 4!}\times \\ \nonumber
& &\left[
   4\cdot {a^\dagger_{\vec k}a_{\vec l}a_{\vec m}a_{\vec k+\vec l+\vec m}\over
    \sqrt{ \omega(\vec k)\omega(\vec l)\omega(\vec m)\omega(\vec k+\vec l+\vec m)}  }+ 
6\cdot {a^\dagger_{\vec k}a^\dagger_{\vec l}a_{\vec m}a_{\vec k+\vec l-\vec m}\over
   \sqrt{ \omega(\vec k)\omega(\vec l)\omega(\vec m)\omega(\vec k+\vec l-\vec m)}  }+ \right.\\
 \nonumber   
& & \left. 4\cdot {a^\dagger_{\vec k+\vec l+\vec m}a^\dagger_{\vec l}a^\dagger_{\vec m}a_{\vec k}\over
   \sqrt{\omega(\vec k+\vec l+\vec m)\omega(\vec l)\omega(\vec m)\omega(\vec k)}    }
\right] + {\cal R}
\end{eqnarray}
where 
\be \label{Tadpoles}
\triangle\omega(\vec k)
\dn
12 \triangle k^3
g_0
{(\triangle k_\perp)^2 \over 4(2\pi)^3\omega(\vec k)4!} 
\sum_{\vec l} 
{1\over \omega(\vec l) }.
\ee
is the \Index{self-energy}
\index{do@$\triangle\omega$} of the virtual particle $a^{\dag}(\vec k)\cket{0}$
and ${\cal R}$
\index{R@${\cal R}$} represents
\Index{pairing} terms of the form $\Re a_{\vec k}a_{-\vec k}$ and 
$
\Re a_{\vec k}a_{\vec l}a_{\vec m}a_{\vec k+\vec l-\vec m} \qquad.
$
Due to momentum-conservation, ${\cal R}$ contains at least one left-moving 
annihilation
operator $a(\vec k)$ if a right-moving operator appears in a product
of two or four pairing operators. 
The momenta components $k^i$ run over the domain 
$-\Lambda^{(i)} \le k^i\le \Lambda^{(i)}$ in multiples
of $\triangle k^i={\pi\over \frak{L}^i}$. These terms are strongly suppressed
on small lattices because pairs of left-movers
and right-movers are suppressed in this case, as we shall 
show in this section.

Taking the lattice size $L$ much smaller than the particle masses
$m_0$ 
and the transverse momenta $\vec k^\perp$
 (i.e. $\Lambda^\perp\ll \Lambda^3=\Lambda$
)\footnote{Note that this implies a strongly {\em anisotropic} lattice since
the lattice-spacing in three-direction has to be much smaller 
than the lattice-spacing in the perpendicular directions}
renders the momentum lattice spacing $\triangle k^3={\pi\over L}$ divergent. 
This, in turn, allows us to approximate 
the kinetic energy 
\be\label{omega(L)}
\omega(k)=
\sqrt{m_0^2+\vec k^2} \approx |k^3|+{m_\perp^2\over |k^3|}+{\cal O}({\frak{L}\,}^3)
\ee
(with the \Index{transverse mass} $m_\perp^2\dn m_0^2+k_\perp ^2$)
\index{m@$m_\perp$}
and its inverse
\be\label{Iomega(L)}
1/\omega(k)\approx 1/|k^3|+{\cal O}({\frak{L}\,}^3)
\ee
in a way that mimics the FF. The deeper reason for why the IMF and the FF have a similar appearance is
that any FF lattice necessarily has zero invariant size which implies 
that all momenta are infinite $n\triangle k\rightarrow \infty$. 
Claims that the FF ---contrary to the IMF--- is 
frame independent are unfounded. 

\subsection{How To Construct an Effective Hamiltonian}
The divergence of the scale $\triangle k^3= \pi L^{-1}$ allows us to replace the
Hamiltonian $H$ by an \Index{effective} Hamiltonian $H_{\text{eff}}$
which describes
bound-states with masses much smaller than this scale . 
To this end we divide the Hamiltonian into the kinetic energy 
$T=\sum_{\vec k} \omega(\vec k)$ and the interaction term $H_I\dn H-T$.
\index{T@$T$}
\index{HI@$H_I$}
We are only interested in bound states with momenta $\vec P\neq 0$
in three-direction. These
momenta $\vec P$ are multiples of $\triangle k^3$. Thus they diverge. 
For this reason, we may calculate the mass squared operator
\be
M^2=(H+P^3)(H-P^3)\approx 2P^3(H-P^3)=2P^3(T+H_I-P^3)
\ee
in a way that resembles the FF ($P^3\approx P^0$ for $v^3\approx 1$).
The kinetic energy minus $P^3$
\be
T-P^3
=
\sum_{\vec k;k^3= 0}
a^{\dag}_{\vec k}a_{\vec k}
\sqrt{m_0^2 + k_\perp ^2}
+
\sum_{\vec k;k^3\neq 0}
a^{\dag}_{\vec k}a_{\vec k}
\left[
{m_0^2 + k_\perp ^2\over 2|k^3|}+|k^3|-k^3
\right]
\ee
diverges linearly with $\triangle k^3\propto L^{-1}$ 
if at least one particle is present
which moves in a direction opposite to $\vec P$. This holds 
because the energy $2|k^3|$ of a left mover is order ${\cal O}( L^{-1})$
whereas a right-mover's energy is order ${\cal O}(L)$.  
Hence the contribution of \Index{left-movers} $k^3<0$ 
to the mass squared is {\sf order} ${\cal O}(L^{-2})$, the
contribution of \Index{right-movers} {\sf order} unity ${\cal O}(L^{0})$
and  
the 
contribution of particles with $k^3=0$ (\Index{zero modes})
is {\sf order} ${\cal O}(L^{-1})$. 
Consequently, only right-movers contribute to the {\em low}-energy spectrum:
the kinetic masses of left-movers and zero modes diverge. 
The effective Hamiltonian $H_{\text{eff}}$ may therefore 
be constructed by excluding modes with divergent energy. 
There are two ways of excluding divergent modes: 

(1) \Index{weak exclusion} or \Index{exclusion on the quantum level}, i.e. quantising all modes and then 
removing the divergent modes from the normal-ordered 
Hamiltonian~\Ref{HamiltonFourier}.

(2) \Index{strong exclusion} or \Index{exclusion on the classical level},
i.e. exclusion of the divergent fields from the Lagrangian 
density ${\cal L}$ {\em before}
quantisation. The resulting Lagrangian density ${\cal L}_R$ and Hamiltonian
$H_R$ become non-local quantities except for $m_0=0$.

We are going to show now, (1) that only the first approach yields
the correct effective Hamiltonian and (2) that the second approach is
completely equivalent to the FF approach with periodic, light-like BCs.
\subsection{The Classically Effective Hamiltonian}

Let us describe the second approach first. 
The (free) 
field $\varphi(x)=\varphi_r(x)+\varphi_l(x)+\varphi_0(\vec x^\perp)$ may be written as a sum of a right-moving field
\begin{equation}\label{phi:r}  
\begin{split} 
\varphi_r(x)
\dn 
\sum_{{\vec k};k^3>0 }
{1\over \sqrt{(2\pi^3)^3 2k^3}}
( 
a_{{\vec k} } 
e^{-ik^3 x^- -i{m^2\over 2k^3} x^0 +i{\vec k}^\perp\cdot{\vec x}^\perp}
+
a^{\dag} _{{\vec k} } 
e^{+ik^3 x^- + i{m^2\over 2k^3} x^0 -i{\vec k}^\perp\cdot{\vec x}^\perp}
)
\end{split}
\end{equation}
a left-moving field
\begin{equation}   
\varphi_l(x)
\dn 
\sum_{{\vec k};k^3<0 }
{1\over \sqrt{(2\pi^3)^3 2|k^3|}}
( 
a_{{\vec k} } 
e^{-ik^3 x^+ + i{m^2\over 2k^3} x^0 +i{\vec k}^\perp\cdot{\vec x}^\perp}
+
a^{\dag} _{{\vec k} } 
e^{+ik^3 x^+ -i{m^2\over 2k^3} x^0 -i{\vec k}^\perp\cdot{\vec x}^\perp}
)
\end{equation}
and the zero mode $\varphi_0(\vec k^\perp)$. 
Formally we may define the 
linear \Index{projection operators} $\Pi_r$, $\Pi_l$
and $\Pi_0$ 
\index{Pir@$\Pi_r$}\index{Pil@$\Pi_l$}\index{Pi0@$\Pi_0$}
which project onto right-movers $\varphi_r$, left-movers
$\varphi_l$ or zero modes $\varphi_0$
respectively. E.g. $\varphi_r(x)=\Pi_r\varphi(x)$.
Now we define the \Index{classically effective Hamiltonian}
\be
H_R\dn 
H\left[\Pi_r\hat\pi,(\Pi_r+\Pi_0)\varphi\right]
\ee
as a functional of right-moving fields $\varphi(x)$ and
field-momenta $\hat\pi(x)$ only. 
The reader will have noticed that we have excluded the canonical
momentum $\hat\pi_0=\partial_0\varphi$
of the zero modes but not the zero mode $\varphi_0$ itself. 
%
%
The rationale for this is the fact that $\varphi_0=0$ would lead
to constraints on righ-movers via the equations of motion. 
Imposing the condition $\varphi_0=0$
on the classical level is too strong if $g_0\neq 0$
since it is not compatible with the equations of motion
\be
\left[-\partial_0 ^2 +\partial_3^2 \right]\varphi
=
\left[
-\nabla_\perp^2+m_0^2
\right]
\varphi
+{g_0\over 3!} \varphi^3
\ee
as they stand. We can see this by applying $\int {dx^3\over 2L}$
on both sides. This  unveils the constraint
\be\label{IMFconstraint}
(\nabla_\perp^2-m_0^2)\varphi_0
=
{g_0\over 3!} \int dx^3
\varphi^3
\ee
where we have used that (a) the integral over the partial derivative 
\be
\int_{-L}^L dx^3\partial_3 \partial_3\varphi(x^3)=\partial_3\varphi(L)-
\partial_3\varphi(-L)=0
\ee
vanishes with periodic BCs and (b) the temporal derivative 
$\partial_0\varphi_0$ vanishes by assumption. 
This consistency constraint has the same form as the constraint which arises
in the Dirac-Bergmann quantised FF, cf. Eq. \Ref{ZeroModesFF.1}. 
If we had set $\varphi_0=0$ then the constraint~\Ref{IMFconstraint}
would have been a constraint on $\varphi_r$. Retaining $\varphi_0\neq 0$, however,
makes this a constraint on the zero mode. 
This constraint can in principle be solved for the zero mode.
The kinetic operator $T-P^3$ reads 
\begin{equation}
T-P^3
=
\int d^3x\;
{1\over 2}(\partial_+\varphi)^2
+
{1\over 2}\int d^3x\;\varphi\Bigg[
-\nabla_\perp^2
+
m_0^2
\Bigg]\varphi
\end{equation}
in terms of the fermionic fields. Projecting out the left-moving
field $\varphi_l$ and the zero-mode velocity $\partial_0\varphi_0$ 
and using the constraint~\Ref{IMFconstraint},
we obtain
\begin{equation}
\begin{split}
T-P^3
&=
\int {d^3x\over 2}
\Bigg[
(\partial_+\varphi_r)^2
+
(\nabla_\perp\varphi_r)^2
+
m_0^2\varphi_r^2
\Bigg]
+
\int {d^2x\over 2}
\varphi_0
\Bigg[
\partial_0^2
-\nabla_\perp^2
+
m_0^2
\Bigg]
\varphi_0 \\
&=
\int {d^3x\over 2}
\Bigg[
(\partial_+\varphi_r)^2
+
(\nabla_\perp\varphi_r)^2
+
m_0^2\varphi_r^2
\Bigg]
+
\int {d^3x\over 2}
{g_0\over 3!} 
\varphi^3
\qquad.
\end{split}
\end{equation}
This demonstrates that the zero modes no longer contribute
to the kinetic mass except for an indirect contribution via interactions. 
The right-movers obey the commutation relations
\be\label{CommutIFFF}
[\varphi_r(0,\vec x),\varphi_r(0,\vec y)]
= 
{1\over 2i\partial^3_x}\delta(\vec x-\vec y)
= 
{1\over 2i\partial_3^y}\delta(\vec x-\vec y)
\ee
which means that $2\partial_3\varphi_r|_{x^0=0}=2\partial_-\varphi_r|_{x^0=0}$ plays the r\^ole of the 
canonical momentum of $\varphi_r$.  
Now we neglect the order ${\cal O}(L^{-1})$ contribution $\partial_+\varphi_r$
and end up with
the \Index{classically effective Hamiltonian}
\be\label{FFHamiltonian2}
H_R=
P^3+
\int d^4x\;\delta(x^0)
\left[
{1\over 2}(\nabla_\perp\varphi)^2+
{1\over 2}m_0^2\varphi^2+V(\varphi)
\right]
\ee
with
\be\label{CommutFF2}
[\varphi(x^0=0,\vec x),\varphi(y^0=0,\vec y)]
=
{1\over 2i\partial_3}\delta(\vec x-\vec y)
\ee
which has now exactly the same form as the FF Hamiltonian 
$P^-(e^{(+)})=2P_+(e^{(+)})$~\Ref{FFHamiltonian}
except that $x^+$ and $x^0$ are interchanged. 
Alternatively, the reader will easily convince her/himself that the Dirac-Bergmann quantisation of the Lagrangian density
\index{LR@${\cal L}_R$}
\be
{\cal L}_R\dn
-2\partial_3\varphi\partial_0\varphi-(\partial_3\varphi)^2-
\left[
{1\over 2}(\nabla_\perp\varphi)^2+
{1\over 2}m_0^2\varphi^2+V(\varphi)
\right]
\ee
leads to the Hamiltonian $H_R$ with the commutation relations~\Ref{CommutFF2} and the zero-mode constraint~\Ref{IMFconstraint}. 
Please note that this Lagrangian density---which results from the elimination of high-energy modes on the classical level---
is not equivalent to the physical Lagrangian density ${\cal L}$ we started with. 
This completes our construction. 

If we had imposed \Index{anti-periodic BCs} on the field $\varphi$ the discussion would have become
much easier because anti-periodic BCs eliminate the zero mode. 
The principal result, however, would have been the same: 
{\em the elimination of high-energy modes on the classical level reproduces the FF}. 

\subsection{The Quantum Effective Hamiltonian}
%
We define the \Index{quantum effective Hamiltonian}
\index{HR@$H_{\text{eff}}$}
\be
H_{\text{eff}}\dn (\Pi_r+\Pi_0) H[\hat\pi,\varphi]
\qquad
\ee
by removing left-movers through a weak constraint, i.e. instead of the strong
constraint $\varphi_r=0$ we merely impose the condition $\varphi_r\cket{\text{eff}}\equiv 0$ 
in order to restrict the total Fock-space to states $\cket{\text{eff}}$ 
which do not contain left-movers. 
This Hamiltonian is guaranteed to be the low energy effective Hamiltonian
for the strong IMF since eigenstates of the IMF Hamiltonian
which mix left-movers and right-movers have infinite energy in the 
limit $\triangle k^3\rightarrow \infty$.
The same does not hold true for the classically effective Hamiltonian
$H_R$ which is formally equivalent to the FF Hamiltonian. 
Even though these two constructions, $H_R$ and $H_{\text{eff}}$,  
are equivalent on the classical level, they are no
longer equivalent on the quantum level. The second 
construction ---which is equivalent to the FF--- 
does not have the same low-energy mass spectrum as the IMF. It can
therefore not be called effective Hamiltonian. 
This is so because $H_R$ fails to reproduce the correct
self-energy term $\triangle\omega(\vec k)$.
The self-energy receives equally important contributions from normal-ordering of left-moving particles and right-moving particles.
\begin{equation} \label{TadpolesIMF}
\begin{split}
\triangle \omega(\vec k)
&=
12 \triangle k^3
g_0
{(\triangle k_\perp)^2 \over 4(2\pi)^3|k^3|4!} 
\left[
\sum_{\vec l;l^3= 0}
{1\over \omega(0,\vec l^\perp)}
+
\sum_{\vec l;l^3\neq 0} 
{1\over |l^3| }
\right] \\
&= 
\triangle \omega_r(\vec k)+\triangle \omega_l(\vec k)
+\triangle \omega_0(\vec k)
\qquad,
\end{split}
\end{equation}
where $\omega_r(\vec k)$,  $\omega_r(\vec k)$ and
 $\omega_r(\vec k)$  are 
defined as the contribution to $\omega(\vec k)$
stemming from left-movers,
right-movers and the zero mode respectively.
 
 Removing left-movers
from the fields (as opposed to removing them from the Hamiltonian
after normal ordering) fails to reproduce the correct self-energy 
term and, consequently, the FF Hamiltonian can {\em not} be
considered as an effective Hamiltonian to the IMF. It describes
a completely different theory. 
With these results we can now answer the question, whether 
quantisation in $\varepsilon$ co-ordinates smoothly becomes
identical to FF quantisation in the limit $\varepsilon\rightarrow 0$,
as long as the invariant lattice size $L$ becomes zero in this limt:
The answer is no. 
As the results of $\varepsilon$ quantisation and IF quantisation
are completely equivalent, the only relevant quantity is 
the invariant lattice size $L$. 
{\bf The limit $L\rightarrow 0$ 
leads us to the IMF, not to the FF unless the FF constraints
are imposed by hand.}
We have seen, however, that the FF mass operator
cannot be considered as an effective IMF mass operator. 
{\em Therefore,
the limit $\varepsilon\rightarrow 0$, $L\rightarrow 0$ 
is discontinuous.}  
This negative result holds for bosons only. The fermionic self-energy of
a right-moving particle in QED$(1+1)$ receives contribution from
right-moving particles only. We shall see this
in Chapter \secRef{CNum} (the FF Hamiltonian for Fermions is derived 
in~\Cite{Mustaki:1990uq,Kalloniatis:1994fk}).
The self-energy contributions of left-movers 
in the bosonic case, in contrast, are far from
being negligible. In fact, the mass-spectrum of the scalar $\varphi^4$
theory  
is dominated by the self-energy contribution because physical
states are almost purely one-particle states\footnote{In fact,
the self-energy $\triangle\omega$ is responsible for the 
mean-field part of the critical exponents whereas the tiny
logarithmic corrections stem from many-body effects.}.
(See Chapter~\secRef{CPhi4}).

Another reason for why one should remove left-movers 
from the Hamiltonian and not from the fields is that
two fields $\varphi_r(\vec x)$ at different positions 
do not commute due to~\Ref{CommutIFFF}
which means that they are causally connected. 
This is part of a violation of microcausality which
shall be treated in the next section: {\bf the classical effective theory
strongly violates microcausality.}  
  
The explicit removal of left-movers from the fields $\varphi(x)$ 
is intimately connected with the implementation of light-like
BCs. If only right-moving modes are allowed, a field which
is periodic in a space like direction with periodicity $L$ is periodic also 
in the light-like direction\footnote{This statement holds
for $|x^0|<L$ with arbitrary accuracy in the limit $L\rightarrow 0$. The {\em physical} reason
for the restriction $|x^0|<L$ will be discussed in the next subsections.
}
with boundary 4-vector ${\frak{L}}'=(L/2,0,0,-L/2)$. 
If both left-moving modes and right-moving
modes are present in $\varphi(x)$, then this field is only periodic
in spatial direction. 

In conclusion: {\em the FF is an unphysical approximation of yet another
unphphysical approximation: the IMF}.
The FF is neither an effective theory of the IF nor is it an
effective theory of the IMF, except for the non-relativistic limit where all relativistic
forms coincide.

\subsection{Some Comments on Zero Modes in the FF}

Light-like periodic BCs are incompatible with relativistic
equations of motion as they stand 
\be
4\partial_+\partial_-\varphi(x)
=
(m_0^2-\nabla_\perp^2)\varphi(x)
+{\partial\over \partial \varphi} V(\varphi(x))
\qquad;
\ee
they have to be supplemented by a constraint on the zero-modes
$\tilde \varphi(x^+)=\int {dx^-\over 2{\frak{L}\,}^-} \varphi(x^+,x^-)$
which follows from integrating these equations
\be\label{ZeroModesFF.2}
0
=
(m_0^2-\nabla_\perp^2)\tilde \varphi(x^+)+
\int_{-{\frak{L}\,}^-} ^{+{\frak{L}\,}^-} 
    {dx^-\over 2{\frak{L}\,}^-}{\partial\over \partial \varphi} V(\varphi(x))
\ee
with respect to $x^-$. 
This constraint corresponds to the IF constraint~\Ref{IMFconstraint} 
which arises if one tries to construct an effective Hamiltonian
on the level of fields.
The constraint~\Ref{ZeroModesFF.2} does not arise from the FF
as such nor is it connected with
the quantisation surface $e^{\text{quant}}$; being another by-product 
of light-like BCs it is independent of the quantisation surface. 
If the field $\varphi(x)$ is subjected to anti-periodic BCs, the constraint~\Ref{ZeroModesFF.2} 
does not arise. 
There is no justification whatsoever for using a constraint which results
from the unphysical character of periodic BCs(see Chapter~\secRef{CPhi4})
in order tho extract physics such as spontaneous symmetry breaking.

Our FF construction on a space-like quantisation plane sheds some light on why 
this constraint appears to describe spontaneous symmetry breaking. In the framework of our construction, this constraint
arises out of the requirement that the kinetic energy of zero-modes vanish.  
This, in turn, requires that the velocity $\partial_0\varphi_0$ of the zero-mode vanish as well. 
These are, perchance, the ingredients of Landau's celebrated {\em mean field} treatment of the scalar $\varphi^4$ theory.
The Landau mean field
theory uses a field which is constant in terms of $x^0$ and $x^3$
(the same condition as in \Ref{IMFconstraint}). {\em Coincidentally}, 
this is exactly the property of the zero mode subjected
to light-like BCs. 
This finding explains a paradox which would otherwise be quite puzzling: 
a QFT in an infinitesimal volume\footnote{A vanishing physical volume is always implied by light-like BCs} 
would seem to describe, at least
approximately so, spontaneous symmetry breaking which {\em solely} occurs
in the infinite volume limit. 
The real danger of the FF is not that the FF is consistently wrong in all circumstances. Worse than that. Sometimes, as in this case,
the FF comes close to reality for the wrong reasons or because the FF is extremly ambiguous (due
to the omission of half of the degrees of freedom). A mere change of BCs, for instance, may drastically change the outcome even
in situations where this should not happen: Anti-periodic BCs do not induce the
constraint~\Ref{ZeroModesFF.2} after all.


\section{The Breakdown of Causality in the Front Form}

\subsection{The Causality Region}

\begin{figure}
\begin{minipage}[tl]{0.47\linewidth}
\centering
 \psfig{clip=,figure=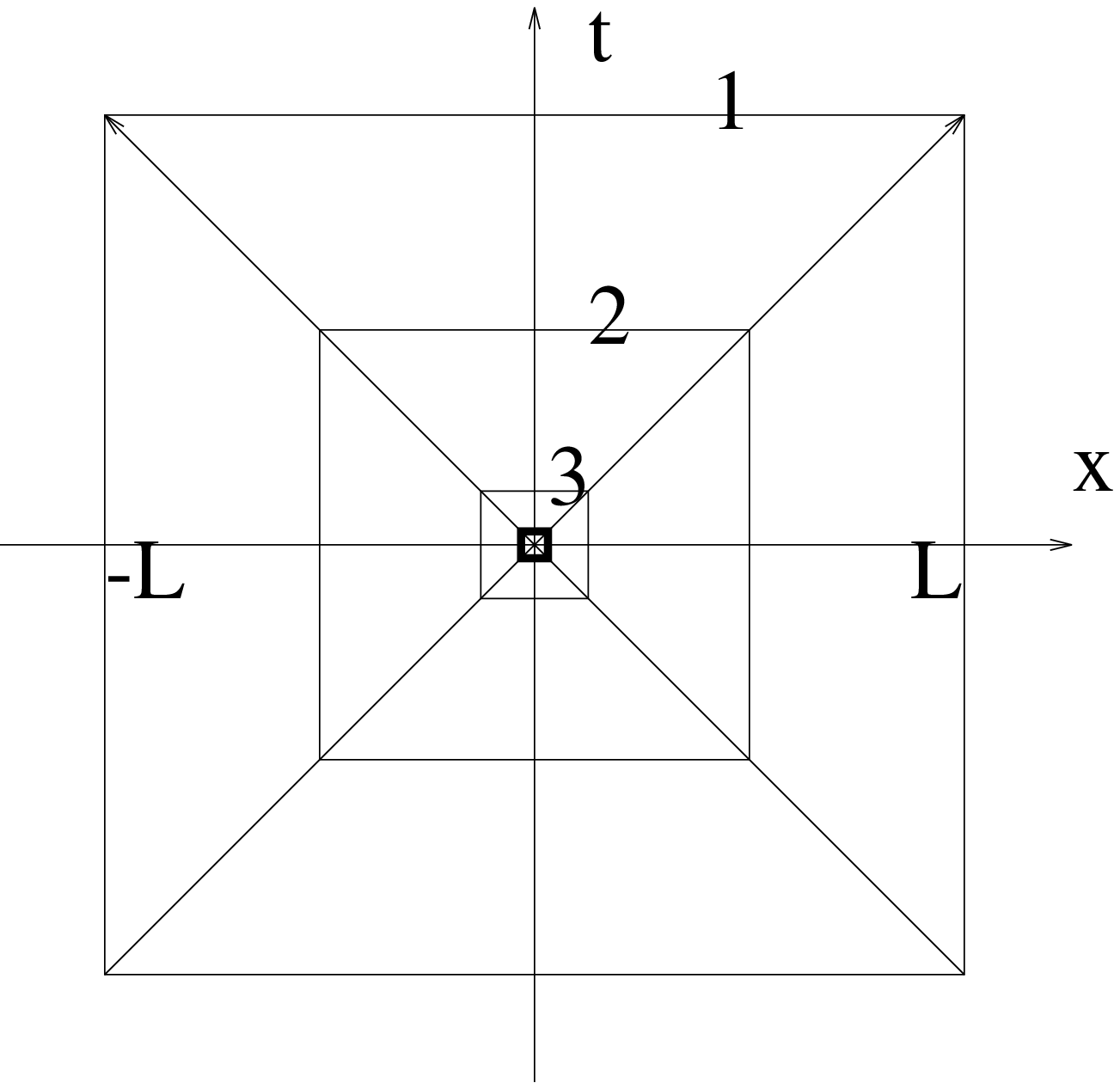,width=\linewidth,angle=-90}  
  \caption{The shrinking of the causality region ${\cal C}$ in 
a fixed frame}\label{CausalityRegion1}
\end{minipage}\qquad  
%
%
%
\begin{minipage}[tr]{0.47\linewidth}
\centering
 \psfig{clip=,figure=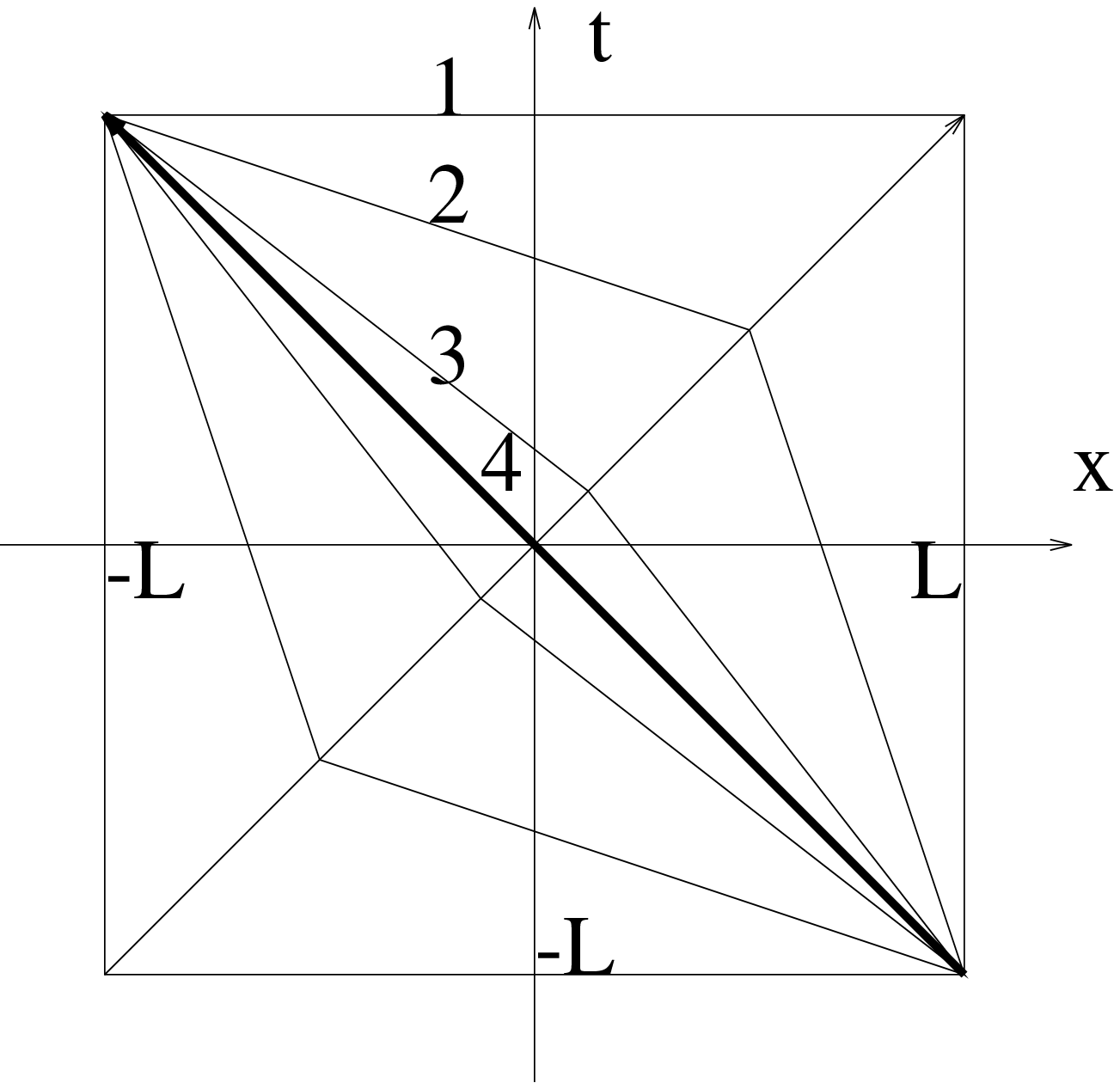,width=\linewidth,angle=-90}  
  \caption{The shrinking of the causality region ${\cal C}_\varepsilon$
in $\varepsilon$ co-ordinates}\label{CausalityRegion2}
\end{minipage}   
\end{figure}
There are problems even worse than those encountered before:
{\it Light-like BCs are 
incompatible with microcausality}. For the sake of a comparison
with ~\Cite{HeinzlWerner94} we do this part in $1+1$ dimensions.
We may restrict ourselves, as~\Cite{HeinzlWerner94}, to a free
field theory since the violation of microcausality arises already for free fields.  
It should be evident that the major result will persist in $3+1$ dimensions.
The breakdown of microcausality is intimately connected to the fact
that light-like boundary conditions, contrary
to space-like boundary conditions, are {\em not} invariant 
under parity or time-reversal. A boundary vector ${\frak{L}\,}$ 
with ${\frak{L}\,}^+=0$ and ${\frak{L}\,}^-\neq 0$ is transformed into ${\frak{L}\,}'$ with
${\frak{L}\,}^{'-}=0$ and ${\frak{L}\,}^{'+}\neq 0$.
This symmetry can not be restored (except for the non-relativistic limit)
even if ${\frak{L}\,}^-\rightarrow \infty$.

Relativistic equations of motion have always causal and acausal 
solutions. 
Hence it does not suffice to find any solutions to the relativistic
equations of motion at all. 
{\bf A Hamiltonian has to generate causal solutions only.}
A necessary condition for relativistic field theories to be causal
ist that
any commutator of local operators must
\be
[{\cal O}_1(x),{\cal O}_2(y)]
\ee
vanish for space-like $(x-y)^2<0$ distances $x-y$. For our purposes
it will suffice to take ${\cal O}_1,{\cal O}_2=\varphi$.
Non-interacting scalar operators $\varphi(x)$ 
quantised in the IF fulfill this requirement since the
Klein-Gordon propagator
\be
i\Delta(x-y)=[\varphi(x),\varphi(y)]
=\int {dp^3\over  2\pi\omega(\vec p) } \sin(-px)
\ee
vanishes for space-like
distances. This fact is not changed if we impose {\em space-like} periodic
BCs $\varphi(t,-L)=\varphi(t,+L)$ 
as long as
we restrict ourselves to the finite region 
\index{C@${\cal C}$}
\be  \label{CausalityRegion}
{\cal C}
\dn
\{x|
-{\frak{L}\,}^3<x^3<{\frak{L}\,}^3 ;\text{   }  -{\frak{L}\,}^3<x^0<{\frak{L}\,}^3
\}
\ee
which we shall refer to as the \Index{causality region}
illustrated in~\figRef{CausalityRegion1}. 
For times $|x^0|<{\frak{L}\,}^3$, the periodic propagator coincides exactly
with the propagator in an infinite volume since the effects of the
boundary conditions cannot propagate faster than the speed of light. 
This can easily be seen by 
inspection of the periodic propagator
\be
\Delta(x-y;{\frak{L}\,}^3)= 
\sum_{p^3}  {1\over \omega(\vec p) (2{\frak{L}\,}^3) } \sin(-px)
=
\sum_{ n} \Delta(x^0-y^0,x^3- y^3 + 2{\frak{L}\,}^3 n) \qquad.
\ee
Contrary to this, the FF propagator with light-like periodicity ${\frak{L}}$
\be
\Delta_{\text{FF}}(x,{\frak{L}\,}^-)=
\sum_{p_-}  {1\over 2p_- {\frak{L}\,}^- } \sin(-px)
\ee 
does {\em not} converge 
against $\Delta(x)$ in the region $-{\frak{L}\,}^-<x^-<{\frak{L}\,}^-$,
$-{\frak{L}\,}^-<x^+<{\frak{L}\,}^-$. How can this be? 
Firstly, the projection
\be
\sum_{ n} \Delta(x^+-y^+,x^- - y^- + 2{\frak{L}\,}^- n) 
\ee
of $\Delta$ onto its periodic part is divergent unless 
we {\em explicitly} subtract
the $p_-=0$ mode responsible for this divergence
\be
-{1\over 2p_- {\frak{L}\,}^+} \sin({m^2\over 4p_-})
\ee
from the propagator\footnote{It is precisely the FFConstraint~\Ref{ZeroModesFF.2}
which avoids such divergences since this constraint reads $\varphi(x)=0$ for $g_0=m_0=0$. 
Cf. the discussion of 
zero-modes on page \pageref{ZeroModesFF.2}.}.
Secondly, let us assume that one can approximate an integral of the type
$\int_{-\infty} ^\infty dk f(k)$ 
by the Riemann sum $\sum_n \triangle k f(n\triangle k)$ such that
this series converges against the integral. Then there is no guarantee that
the same integral 
\be
\int_{g^{-1}(-\infty)} ^{g^{-1}(+\infty)}
   g'(l)dl f(g(l))
\ee 
can be approximated by a (equidistant) 
sum if the co-ordinate transformation
$g:k=g(l)$ 
contains a singularity.
Now, the transformation $g : k^+=g(k^3)=\omega(k^3)+k^3$ 
which transforms 
$k^3$ into $k^+$ is indeed singular since $dk^3={dk^+\over k^+}\omega(k^3)$
diverges at $k^+=0$. Furthermore, 
the integrand $\sin\left[{m^2x^+\over 2k^+}\right]$ oscillates violently
in terms of $k^+$, 
close to $k^+=0$ which is incompatible with Riemann integrability.
An explicit calculation confirms this: 
Since the FF propagator transforms as
\be
\Delta_{\text{FF}}({x^+\over\kappa},x^- \kappa;{\frak{L}\,}^-\kappa)
=
\Delta_{\text{FF}}(x^+,x^-;{\frak{L}\,}^-)
\ee
under \Index{co-ordinate boosts} it does not 
make sense to say ${\frak{L}\,}^-$ is large or small. We can make it
arbitrarily large or small by means of a co-ordinate boost. 
Therefore we can always assume that it is arbitrarily small
such that   $\triangle k^- \gg m$. 
Thus in the neighbourhood of the 
quantisation surface (i.e. $x^+$ sufficiently small)
\be
\Delta_{\text{FF}}(x^+,x^-;{\frak{L}\,}^-,m)
\approx
\Delta_{\text{FF}}(x^+,x^-;{\frak{L}\,}^-,0)
= -{1\over 4}\text{sgn}(x^-)+{x^-\over 4{\frak{L}\,}^-}
\qquad. 
\ee
This means
a full-scale violation of microcausality since $\Delta_r$ does
not disappear in the space-like region $x^+<0,x^->0$; it disappears
in the time-like region instead. 
The fact that $\Delta_r$ does
not vanish for arbitrarily small $x^+$ 
in the space-like region close to light-front $x^+=0$ cannot be changed
by boosts, and therefore microcausality is violated for
large ${\frak{L}\,}^-/ \kappa$ also. 
This may be reexpressed in a somewhat sloppy way:
Since there is no preferred frame in the FF, we can
make every particle behave as if it were massless to any accuracy
by choosing a frame with almost infinite momentum spacing $\triangle k_-$
where particle-masses can be neglected. Now, 
massless left-movers cannot be quantised in the FF-1 without a 
full-scale violation of microcausality. Consequently, the FF violates
causality irrespective of particle masses. 
 
\subsection{The Instant Form Perspective} 

Our FF construction on a space-like surface (in section~\secRef{ConstraintsAndBoudary}) 
is able to illustrate these arguments.
Inside the causality region, the massive
propagator $\Delta(x,m)\approx \Delta(x,0)$
coincides with the massless propagator up to order ${\cal O}(L^2)$
corrections. 
We divide the IF field $\varphi(x^0,x^3)$ 
into the right-moving part $\varphi_r$ and
the left-moving part $\varphi_l(x^0,x^3)$ which contains all operators which
create or annihilate left-movers and 
$\varphi_0(x^0)$ which contains the IF zero-modes.
Since $\varphi_r$, $\varphi_l$ and $\varphi_0$ commute mutually,
it follows that
\be
[\varphi(x),\varphi(y)]
=
[\varphi_r(x),\varphi_r(y)]+
[\varphi_l(x),\varphi_l(y)]+
[\varphi_0(x),\varphi_0(y)]
\ee
and, therefore, the causal propagator 
\begin{equation}
\begin{split}
i\Delta(x-y;{\frak{L}\,}^3,m)
&=
i\Delta_l(x-y;{\frak{L}\,}^3)
+
i\Delta_0(x^0-y^0;{\frak{L}\,}^3)
+
i\Delta_r(x-y;{\frak{L}\,}^3) \\
&\approx 
-{1\over 4} 
\left[
\text{sgn}(x^+)+\text{sgn}(x^-) 
\right]
+{\cal O}(L^2)
\end{split}
\end{equation}
is the sum of three acausal parts
where 
\be
\Delta_l(x^0,x^3)
=
\Delta_r(x^0,-x^3)
\approx 
\Delta_{\text{FF}}(x^+,x^-)
\ee
 is
just the parity-reversed massless propagator $\Delta_r$
and
\begin{equation}
\begin{split}
\Delta_0(x^0)
={1\over 2m{\frak{L}\,}^3} \sin(-mx^0)\approx -{x^0\over 2{\frak{L}\,}^3}
=
-{1\over 4{\frak{L}\,}^3}(x^++x^-)
\end{split}
\qquad.
\end{equation}
The last part only holds for $|t|<{\frak{L}\,}^3$.
This nicely illustrates how a balanced, parity-invariant combination
of left-movers and right-movers ensures microcausality even though
microcausality is strongly violated by 
both $\Delta_l$ and $\Delta_r$.  
The mathematical reason for why $\Delta(x)$ vanishes
in the space-like region 
is that  there is a parity reversed partner $-k^3$ 
for every $k^3$: both contributions cancel. Such a cancellation is
rendered impossible in the FF-1 since neither light-like BCs nor light-like quantisation
surfaces can be parity-invariant.

\proPub{
Elimination of left-movers on the classical level 
(cf. Sec.~\secRef{secTheIMF})
yields the same unphysical propagator as FF quantisation.  
The restriction of microcausality to the causality region ${\cal C}$ 
has salient consequences of the \Index{strong IMF} as well. In the 
limit $L\rightarrow 0$ the causality region of the IMF collapses
into a point,
as illustrated in~\figRef{CausalityRegion1}. This means, that the strong IMF is only capable of
describing space-time events that take place in an infinitely small
space-time region. In particular, this means that only instantaneous
events may be described in this frame. 
In an arbitrarily small region, a massive propagator looks like
a massless propagator. The propagation of free particles is almost massless. 
The causality region of the strong IMF defined in $\varepsilon$
co-ordinates, however, does not collapse into a space-time point. It collapses
into a space-time line instead as illustrated in~\figRef{CausalityRegion2}.
Inside the causality region,
these results, obtained for free fields, are likely to hold
for interacting fields, too, as long as $g/\triangle k^3$ is 
order ${\cal O}(L)$ or, equivalently, if $g/m$ is order  ${\cal O}(1)$.

These results show why the IMF works so well, perturbatively,
in order to formulate
Feynman's parton model. In the na{\"\i}ve parton model,
the time it takes for the electron 
in order to scatter off a quark is arbitrarily short and the
entire process may take place in an arbitrarily short volume. 
This legitimises the use of the strong IMF even though the strong IMF
is only capable of describing events which take place arbitrarily
fast and inside an arbitrarily small volume.
If vacuum effects are explicitly excluded, the IMF may even
be used in order to derive the leading order DGLAP equations: these
equations do not contain loops, i.e. they are classical equations
except for the usage of the running coupling constant. 
In the light of our results 
on the IMF, it is our conviction that a successful {\em non-perturbative}
computation of QCD distribution functions in the strong IMF is 
not possible. }
We would also draw attention to the fact that the propagator $\Delta_r$ 
propagates free fields $\varphi$ faster than the speed of light;
a fact which is, of course, intimately related to the breakdown
of microcausality. So is the unphysical possibility of time-travel
which we have discussed in the last section.  
\proPub{
\subsection{Light-Like Boundary Conditions in the Instant Form}

The periodic free field $\varphi_r(x)$ (or $\varphi_l(x)$) Eq. \Ref{phi:r}
is compatible with
light-like periodic BCs ${\frak{L}}'$ up to order ${\cal O}(L)$ 
\be
\varphi(x+{\frak{L}}')-\varphi(x-{\frak{L}}')
\approx
0
\ee
inside the causality region. 
Here, ${\frak{L}}'=(L/2,-L/2)$. 
The complete field $\varphi(x)$, however, does not fulfill this requirement. 
Excluding $\varphi_l$ and $\varphi_0$ is, therefore,
equivalent to the introduction of light-like BCs if a free theory is 
concerned. 
Inside the causality region, this connection can be expected to remain
valid even 
for an interacting theory, as long as the coupling constant $g$ 
is small when compared to the scale $\triangle k^3$. 

\section{The Limit of Infinite Light-Like Volume}
We have shown that the FF propagator $\Delta_{\text{FF}}(x)$
is completely unphysical if 
the light-like length $\frak{L}^-$ is finite. So are time-ordered
propagators or Feynman propagators. Yet FF perturbation
theory, in the limit $\frak{L}^-\rightarrow \infty$, 
seems to be equivalent to covariant perturbation theory at least
in the case of some simple 
Feynman diagramms~\Cite{Langnau:1993iq,Ligterink:1995tm}. 
What happens in the limit
$\frak{L}^-\rightarrow \infty$?
In this limit,
\be
\lim_{{\frak{L}\,}^- \rightarrow \infty}
\Delta_{\text{FF}}(x,{\frak{L}\,}^-)=
\int {dk_-  \over 2\pi k_-} \sin(-k\cdot x)
\ee 
one is faced with the problem that the quantity 
\index{alpha@$\alpha-$}
\be
1/k_- = {\cal P} [1/k_-] + \delta(k_-)\;\alpha(x^+)
\ee
is ambiguous\footnote{
In fact, this expression does not even exhaust the set of all possible
prescriptions. It suffices, however, to make clear that there is 
an infinite number of possible prescriptions.
}. 
There is an infinite number of possible prescriptions
---parametrised by the arbitrary function $\alpha(x^+)$---
in order to properly define what $1/k_-$ means. Setting $\alpha(x^+)=0$,
for instance, means choosing the \Index{principal value prescription}
$ {\cal P}[1/k_-] $.\index{P@${\cal P}$}
This means that the expression $\Delta_{\text{FF}}$ may be anything: The 
propagator depends crucially
on the choice of \Index{prescription}. In other words: the FF theory has an
infinite number of free parameters. Such a theory
is not sensible. The freedom of choosing physically inequivalent prescriptions
for $\frak{L}^-=\infty$ signals a defect of the FF reflecting the fact that the limit
\be
\lim_{\frak{L}^-\rightarrow\infty} \Delta_{\text{FF}}(x;\frak{L})
\ee
does not exist except for the \Index{FF causality region}\footnote{
Cf.~\figRef{CausalityRegion2}.} 
$x^+=0$
where all prescriptions yield the same propagator. 
Of course, this ambiguity is
intimately related to the fact that two light-like planes are needed
to define a relativistic inital value problem. 

One might argue that the prescription is not ambiguous but
determined as the prescription which yields the correct physical
result. Therefore, it seems that one need not worry about
the prescription as long as the correct propagator is obtained.
There is a flaw in this argument (a {\em petitio principii}), however. 
{\ttfamily There is reason to worry}. 
By chosing the correct propagator, we
feed the theory with the data we would like to obtain. 
{\em The correct value of $\alpha$ does not follow from the FF quantisation
procedure}. 
IF quantisation, in contrast, unambiguously defines
the propagator without additional {\em ad hoc} assumptions.
The fact that a prescription may be found which reproduces
the correct IF propagator\footnote{This may no longer be true
in the presence of interactions} 
does not mean that the FF and the IF are equivalent for $\frak{L}^-=\infty$.
If they were then 
the FF would also be "equivalent" ---in the same sense--- to an arbitrary 
number of completely unphysical theories. 
 
The choice of prescription is often referred to as
{\em regularisation} in the literature.
We do not agree with this misleading parlance.  
In the IF, the propagator
$\Delta(x)$ is defined {\em without} regularisation in a {\em free} QFT
contrary to other quantities such as the energy $H$ which has to be regularised
even if a free QFT is concerned. 
Furthermore, in the IF, a regularisation does {\em not} introduce ambiguities into
quantities which need to be regularised as long as the correlation
length is sufficently large\footnote{
More precisely, ambiguities are present but related to irrelevant
operators. The influence of these operators is strongly suppressed
in a system close to the critical point.} 
and as long as one is dealing with a renormalisable
theory. 
Similarly,  
in the FF the problem is {\em not} that the expression $1/k_-$ would require 
regularisation ---it does not. The problem is
that $\Delta_{\text{FF}}$ is {\em ambiguous} yet {\em finite}. 
Calling the choice of $\alpha$ a regularisation is a misnomor.

It is interesting to establish a connection between the $\alpha$-ambiguity
and $\varepsilon$ co-ordinates. We have seen, that the causality
region of $\varepsilon$ co-ordinates collapses into the line $x^+=0$. 
Cf.~\figRef{CausalityRegion2}. 
A boost $B(v=1)$ even collapses the universe onto the subspace $x^+=0$. 
The net result is the same, 
even though a boost $B(v)$ with $v=1$ is singular and $T(\varepsilon=0)$ is not:
Quantisation
with respect to the transformed co-ordinates unambiguously describes the collapsed
subspace $x^+=0$ only. Outside this region, neither of both approaches
is able to provide (unambiguous) predictions. 

\section{Situations Where the FF May Be a Good Approximation}

 
\subsection{The Non-relativistic Limit}

In the non-relativistic limit $c\rightarrow\infty$, 
all space-like and light-like hyper-planes
\be
x^0-{v\over c^2} x^3=0
\ee
collaps onto the unique, non-relativistic quantisation surface $x^0=0$. 
The FF and the IF coincide in this limit. 
The question arises, if the FF remains accurate close to the relativistic
limit. The {\em zero-hypothesis} is that the FF-corrections to the non-relativistic
limit are not accurate.

\subsection{Perturbation Theory With External Input} 

Only free propagators go into perturbation theory.
The ambiguity of free propagators in the FF
allows to specify the parameter of ambiguity $\alpha$ such
that the free time-ordered propagators 
coincide with the corresponding IF propagators.
Therefore, it is possible to correctly describe some elementary diagrams
of perturbation theory but {\em only if} external input is used: 
One has to choose amongst an infinity of possible prescriptions in order
to regularise $1/k_-$ singularites, i.e. one has to fix an infinity of
free parameters $\alpha$. FF quantisation itself does not specify the
prescription $\alpha$. 

\subsection{Theories Without Vertices Which Connect at Least Four Bosons} 

We shall see in Chapter~\secRef{CNum}
that fermionic self-energies need not {\em necessarily} be inaccurately
described by the FF
because they do not contain left-movers.
The FF may or may not come close to the correct solution to a QFT which does
not contain vertices
which connect at least four bosons (except for the ultra-relativistic limit).
QCD, however, contains a four-gluon vertex. 

\subsection{Accidental Cases}
There is a case where the FF {\em accidentaly} reproduces the
exact mass spectrum of a QFT. 
The massless \Index{Schwinger model}, treated in~\secRef{CSchwinger}, is equivalent
to the \Index{chiral Schwinger model} in the 
unphysical limit $L\rightarrow 0$. The chiral Schwinger model does not
contain left-movers. We shall see in~\secRef{CSchwinger} how these features
lead to the fact that the mass spectrum of the massless Schwinger model 
is exactly described in the IF, the IMF and the FF. 

\subsection{Phenomenology}

The FF may be used as a bookkeeping device in phenomenology. The 
zero-hypothesis is that
every phenomenolgical problem which can be easily treated in the FF,
may equally well be treated in the IF if the volume is taken to be sufficiently 
small. We do not know of any example where the FF is useful but the IF 
is not. 


\section{Other Relativistic Forms}

Every relativistic form represented in~\tabRef{StabilityGroups}
is a valid quantisation surface except for the FF. 
May these forms be useful in order to quantise QFTs? There is reason to 
doubt this. 
Firstly, the classification given in~\Cite{Leutwyler:1978vy} 
does not take discrete symmetry into account. Yet time reversal
and parity symmetry are as important as the continous part of the
Lorentz group. There is only one form whose stability group comprises time-reversal:
the instant form. The quantisation surface
$x^0=0$ is the only quantisation surface from~\tabRef{StabilityGroups}
which is invariant under time reversal. 
Important theorems such as the PCT theorom naturally apply in the IF and only
in the IF. In every other form, time reversal invariance can be realised
dynamically at best. The FF is worst in this respect since 
even parity invariance is destroyed. 

Another aspect, which is neglected in the literature is the aspect of implicit
time dependence. 
For instance, even though the boosts ${\sf J}_{03}$ and ${\sf J}_{\perp,-}$
are in the FF stability group, they do not commute with the FF time evolution
operator
\be
U_+(x^+)\dn 
\exp(-i\hat P_+ x^+)
\qquad.
\ee 
This means that these operators have an {\em explicit} time dependence: 
${\sf J}_{03}={\sf J}_{03}(x^+)$. Such a symmetry is not very helpful. 
In paricular, the diagonalisation of the Hamiltonian is not facilitated. 
We therefore introduce the notion of the \Index{stability group proper}
which consists of operators which do not explicitly depend on time
and the \Index{ephemeral stability group} which consists of operators which
do not commute with the time evolution operator. 
Geometrically, this difference of usefulness may be visualised as follows. 
In the IF, a valid quantisation surface $x^0=0$ is transported into another
valid quantisation surface $x^0=T^0$ by means of the operator $U_0(T^0)$
of temporal evolution. Any quantisation surface obtained by means of a 
temporal evolution is again symmetrical with respect to the full IF
stability group. 
The IF is the only form which boasts this feature. 
The light-front, for instance, is transported onto a quantisation
surface which is no longer symmetrical under the original symmetry group.
It is easy to realise that the elements of the original symmetry group 
which cease to be elements of a time-evolved quantisation surface
do not commute with the time evolution operator of a given relativistic form. 
Hence all forms, except for the IF, single out an arbitrary instant of time. 
\begin{table}[t]\begin{center}

\begin{tabular}{|c|c|c|c|}
\hline
Relativistic form & Quantisation-surface 
& 
\begin{tabular}{c} Algebra of the \\stability group proper \end{tabular}\\
 \hline
Front form   & $x^+=e^{(+)}\cdot x=0$ 
& $\vec P_\perp,{\mathsf J}_{12};P_-$\\  \hline
Instant form & $x^0=e^{(0)}\cdot x=0$ 
& $\vec P,{\mathsf J}_{mn}$ \\  \hline 
Point form   & $x\cdot x=\kappa^2$, $x^0>0$ 
& ${\mathsf J}_{mn}$\\  \hline
Line form   & 
$\begin{matrix}
 (x^0)^2-(\vec x^\perp)^2=\kappa^2,\\
  x^0>0
 \end{matrix}$ 
& $P_3,{\mathsf J}_{12}$\\  \hline
\begin{tabular}{c} Extended front form  \end{tabular}
& $x^+ x^-=\kappa^2$, $x^0>0$ 
& $\vec P_\perp,{\mathsf J}_{12}$\\  \hline
\end{tabular}

\caption[The Useful Stability Group]{Relativistic forms under the aspect
of usefulness.  
The parameter $\kappa$ is a real number.}\label{StabilityGroupsII}
\end{center}\end{table}
Another implication is this: in the IF, symmetries are stable. The "time"
evolution operator does not transform an irreducible representation
of the stability group into a different
irreducible representation. 
This does not hold true in any other form. 
Temporal evolution of irreducible representations of a time-dependent
stability group are mixed by the evolution operator. These irreducible
representations should, of course, become irreducible representations
of the stablility group of the {\em evolved} quantisation surface.
The relation between the original representation and the evolved representation
is, however, dynamical, i.e. it depends on the interaction. 
A realistic comparison of different relativistic forms based on the
stability group proper is given in table~\tabRef{StabilityGroupsII}.

} 

%% file: CSchwinger.i

\section*{Introduction}

The \Index{Schwinger model}~\Cite{Schwinger:1962tn,Schwinger:1962tp}
is quantum electrodynamics in two space-time dimensions
\index{QED@QED$(1+1)$}
with massless fermions. This model can be 
solved {\em analytically} 
by \Index{bosonisation}~\Cite{Schwinger:1962tp}:
the Schwinger model with coupling constant $g$ 
is equivalent to a model 
of free bosons with mass $M_B={g\over\sqrt{\pi}}$.
\index{MB@$M_B
$}
The original solution to the Schwinger model was
given by J.~Schwinger~\Cite{Schwinger:1962tp}
using the action formalism; in this chapter we shall be using
a solution in the
Hamiltonian formalism given by N.S. Manton ~\Cite{Manton:1985jm}.
We are interested in the numerical treatment of the
\Index{massive Schwinger model}, (i.e. QED$(1+1)$) with massive fermions $m=0$)
because 
we shall {\em numerically}
approximate the solutions to the \Index{massive
Schwinger model}
in Chapter~\secRef{CNum}.
This numerical solution will be based on various
assumptions and approximations that require justification:
In particular, we shall assume that
(1) left-movers may be neglected in small volumes, i.e.
the number of virtual fermions moving contrary to the motion 
of physical boson states is negligible as long as the
lattice size $2L$ is sufficiently small,  
(2) axial gauge ---which is unphysical in a large volume---
preserves the masses of the physical bosons as long
as long as these move sufficiently fast,
(3) the vacuum has a simple, almost trivial structure 
on a small lattice,
(4) left-movers are negligible in fast-moving states on a small lattice. 

While the good description of this
model which we shall obtain in~\secRef{CNum},
can already be regarded as an {\it a posteriori} justification of the
approximations made, it is preferable, nontheless,
to seek a more thorough understanding
of why these approximations preserve the physical character of QED$(1+1)$. 
Thence, 
the exact solution to the Schwinger model in the {\em Hamiltonian}
frame-work~\Cite{Manton:1985jm}
constitutes an ideal laboratory to study the impact of various approximations 
in more detail. 

Furthermore, we would like 
%
to investigate to what extent physics is affected if
%

(a)  a na{\"\i}ve cut-off regularisation
is employed instead of a proper gauge-invariant regularisation or

(b) to illustrate 
that in the infinite volume limit, the usual distinction of valence- and
sea partons has to be supplemented by \Index{vacuum} partons
which are related to the
left-movers. These vacuum-partons do not contribute to the structure functions
in the rest-frame. 

(c) to calculate the proper distribution functions showing that
they {\it never} coincide with the raw distribution
functions on a large lattice, not even in the infinite momentum frame

(d) to demonstrate that non-triviality of the vacuum in general and
of the chiral condensate in particular
is inseparably linked to the presence of left-movers. 



%
We argue that topological effects contribute primarily to 
the rest-frame sector $\vec P=0$ where the De Broglie wave-length
$\lambda={2\pi\over |\vec P|}=\infty$ is larger than any finite
lattice size $2L$.
\index{L@$L$} 
Some introductory comments are in order.

\begin{enumerate}
\item
We tried to keep our notation as close as 
possible to both the notation used for DIS as well as to Manton's
notation. 
However, both notations do not coincide and hence
some compromise had too be made.
For example, we shall refer to particles with positive chirality
as right-handed which corresponds to the usual definition 
of this word whereas Manton refers to negative chirality particles
as right-handed. Left-handed fermions in ~\Cite{Manton:1985jm} seem to
move right and right-handed fermions seem to move left, contrary to the
usual convention. The reason for the different terminologies is a
subtle one:
Manton implicitly
abridges $x=x^1$ and $p=p_x=p_1=-p^1$ as the reader 
may easily verify 
(We are using the 3-direction instead of the 1-direction and
therefore Manton's $p$ corresponds to $-p^3$ in our notation).  
This may be a good idea from the point of view of Chapter~\secRef{CFF}
as $\hat P_3$ rather than $\hat P^3$ is a generator of translations. 
Such a convention implies, however, that 
the state $a^{\dag} _{1,r}\cket{0}$ has momentum $k^3=-k_3=-r$ 
rather than $+r$ ($r\in \Bbb{N}$) since the argument of Manton's
operators are assumed to be {\em co}-variant rather than contra-variant
operators. A state with $k_3=r>0$ moves \Index{left} rather than \Index{right}, 
i.e. it moves against the $x^3$ axis.  
\item
It cannot be the scope of this thesis to repeat the article 
Ref. ~\Cite{Manton:1985jm} in every detail; herein, we shall
take most results for granted. The reader will {\bf NOT} be able
to understand this section unless he has studied Ref.~\Cite{Manton:1985jm}.
 
\item 
A finite periodic box (circle) with length $2L=2\pi$ is used
as in Ref.  ~\Cite{Manton:1985jm}. It will be shown that without this
volume-regularisation, the mean number of fermions in physical states
 ---even in the vacuum---
would be infinite. 

\item 
The raw distribution functions we calculate can not
be related to the structure functions since $F_1$ does not exist
in two dimensions and the remaining longitudinal structure-function
$F_L$ vanishes due to the Callan-Gross relation which is not spoiled
by perturbative corrections because quarks cannot radiate gluons
with non-zero momentum. 
 
\item
In one space dimension, the symbols $\vec k$ and $k^3$ may be used
interchangeably. 
 
\end{enumerate}

\section{The Exact Solution to the Schwinger Model:\\
A Brief Review}

\subsection{Terminology and Notation}

We start with a brief review of the solution to the
Schwinger model given by  N.~S.~Manton~\Cite{Manton:1985jm}. 

The Lagrangian of the Schwinger model in standard notation is
\begin{eqnarray}
{\cal L} 
=
\bar \psi(i\not\!\partial-g\not\! A)\psi
-
{1\over 4} F^{\mu\nu} F_{\mu\nu}
\end{eqnarray}
with $\psi=(\psi_R,\psi_L)$ the fermionic field and $F^{\mu\nu}$ 
the field strength tensor. 
We require the 
fermionic fields to be
$2L$-periodic, i.e. 
the Schwinger model is defined 
on a circle with circumference $2L$.

In order to be compatible with the usual three-dimensional notation
(where the three-direction is the longitudinal direction)
we use the co-ordinates $x=(x^0,x^3)$ rather than $x=(x^0,x^1)$.
\index{x@$x$}
The metric $g$ is the Minkowsky metric with signature $(+1,-1)$. 
\index{g@$g$}
Absorbing the coupling $g$  into the connection (vector potential) $A^\mu$
\begin{eqnarray}
{\mathsf A}\dn g A
\end{eqnarray}
\index{A@${\mathsf A}$}
\index{A@$A$}
and expressing the field strength $F^{\mu\nu}$ in terms of 
${\mathsf A}$ yields Manton's form
\begin{eqnarray}
{\cal L} 
=
\bar \psi(i\not\!\partial-\not\!{\mathsf A})\psi
+
{1\over 2g^2}(\partial_0 {\mathsf A}_3 - \partial_3{\mathsf A}_0)^2
\end{eqnarray}
of the Lagrangian 
\index{L@${\cal L}$: ! Lagrangian density, Schwinger model}
except that he uses the co-ordinates $(x^0,x^1)$
instead of $(x^0,x^3)$. Please 
note that the field ${\mathsf A}$ which includes the coupling is written in
{\it sans serif} style to facilitate the distinction. 
Since we are working in two dimensions, the generators $\gamma^\mu$ 
\index{gamma@$\gamma$ : Dirac matrices}
\index{sigma@$\sigma$ : Pauli matrices}
of the spinor group SPIN(2) $\equiv$ SU(2) 
are represented by the familiar $\sigma$ matrices.
Here, the chiral representation 
\begin{xalignat}{2}
\gamma^0&=+\sigma^1 \; ,&
\gamma^3&=-i \sigma^2  \;,\\
\gamma^5&=\sigma^3=\gamma^0 \gamma^3 
\end{xalignat}
from Ref.  ~\Cite{Manton:1985jm} will be used. 
The Lagrangian is invariant under the \Index{gauge transformation} $\iota(x)$
\index{i@$\iota$}
\begin{align}\label{GaugeTransformation}
\psi(x) 
&\overset{\iota(x)}{\longmapsto}
   e^{-i\iota(x)}\psi(x) \\
{\mathsf A}_\mu(x) 
&\overset{\iota(x)}{\longmapsto}
  {\mathsf A}_\mu(x)+\partial_\mu \iota(x)
\qquad.
\end{align}
The gauge parameter $\iota(x)$ is a function subjected to the constraint
\index{nW@$n_W$}
\be\label{GaugeConstraint}
{1\over 2\pi}
\left[
\iota(x^0,x^3+L)-\iota(x^0,x^3-L)
\right]
=
n_W
\in \Bbb{Z}
\ee
but otherwise arbitrary. 
This constraint has to be imposed in order to
protect the periodicity
of the fermionic field $\psi(x)$. Cf. Eq.~\Ref{GaugeTransformation}. 
The number $n_W\in\Bbb{Z}$ is called \Index{winding number}. 
Gauge transformations
with non-vanishing winding number ---i.e. non-periodic gauge transformations $\iota(x)$--- are called \Index{topologically
non-trivial gauge transformations} or ---shorter--- \Index{large gauge
transformations}.
The constraint~\Ref{GaugeConstraint} prevents the 
\Index{zero-mode of the gauge field} ${\mathsf A}_3(x)$
\be
{\mathsf A}_3
\dn
\int_{-L}^L {dx^3\over 2L}{\mathsf A}_3(x)
\ee 
from being gauged away. Following Manton's minor abuse of notation, the symbol 
${\mathsf A}_3$ now stands simultaneously  for the gauge field {\em and} 
its zero mode. 
The zero-mode ${\mathsf A}_3$ is
reduced to the domain\footnote{The so-called \Index{fundamental 
modular domain}.} $0\le{\mathsf A}_3<\pi/L$
\be
{\mathsf A}_3
\overset{\iota(x)}{\longmapsto} 
{\mathsf A}_3
+
\int_{-L}^L {dx^3\over 2L} \partial_3\iota 
=
{\mathsf A}_3
+
n_W {\pi/L}
\ee 
by a large gauge transformation. This reflects the fact that
the \Index{global Wilson loop} 
\be
\exp\left[i\int_{-L}^L dx^3{\mathsf A}_3(x) \right]
=
\exp\left[i\int_{-L}^L dx^3{\mathsf A}_3(x) +2\pi i \Bbb{Z}\right]
\ee
is the only gauge degree of freedom on a circle. The requirement
that the global Wilson loop be invariant under 
gauge transformations would have been an alternative way of deducing
the constraint~\Ref{GaugeConstraint}. 
Axial gauge ${\mathsf A}_3=0$ would remove the zero-mode. Therefore,
axial gauge is incompatible with periodic BCs. 
\Index{Coulomb gauge}, in contrast, 
 \begin{eqnarray}
\label{CoulombGauge}
\partial_3 {{\mathsf A}_3}(x) =0
 \end{eqnarray}
%
%
%
is weaker than axial gauge. While leaving large gauge transformations
intact, Coulomb gauge completely eliminates every
{\em local} gauge
transformation generated by Gauss law\footnote{Gauss' law
is a \Index{first class constraint} in the terminology of the Dirac-Bergmann
quantisation procedure~\Cite{dirac:generalized,SundermeyerBuch} and as such it
may be considered the generator of {\em local}
gauge transformations.}, i.e. it eliminates
the gauge transformations with trivial winding number $n_W=0$.  
Thus \Index{Gau{ss}' law} 
 \begin{eqnarray}\label{Gauss}
-\partial_3^2{\mathsf A}_0 = g^2\psi^{\dag} \psi
\end{eqnarray}
can be solved for $A_0$ such that the (un-regularised) \Index{Hamiltonian}
\index{HNO@$H_{\text{N.O.}}$}
\begin{equation}
\begin{split}
&H_{\text{raw}}
=
\int_{-L} ^{L} dx^3
 \psi^{\dag}(i\partial^3 - {\mathsf A^3})\gamma^0\gamma^3\psi
+
{g^2\over 2} \int _{-L} ^{L} dx^3
(\psi^{\dag} \psi){1\over -\partial_3 ^2}(\psi^{\dag} \psi)
+
{L\over g^2} {\mathsf E}^2 \\
&=
\int_{-L} ^{L} dx^3
 \psi^{\dag}i\partial^3\gamma^5\psi+{\mathsf A_3}{\cal Q}_A
+
{g^2\over 2} \int _{-L} ^{L} dx^3
(\psi^{\dag} \psi){1\over -\partial_3 ^2}(\psi^{\dag} \psi)
+
{L\over g^2} {\mathsf E}^2
\end{split}
\end{equation}
contains only one dynamical boson field degree of freedom: 
the gauge field zero mode ${\mathsf A}_3$. The electric field ${\mathsf E}={\mathsf F}^{03}=g F^{03}$ 

plays the r{\^o}le of  the canonical momentum
\index{E@${\mathsf E}$}
\be
[{\mathsf E},{\mathsf A}_3]= -i{g^2 \over 2L}
\ee
conjugate to the zero mode ${\mathsf A}_3$ of the gauge field.

\subsection{Operators in Momentum Space}

Particles are most conveniently described in momentum space. 
There are several equivalent ways of expanding 
the fermionic field $\psi(x)$ 
\index{psi@ $\psi(x)$ : fermionic field }
in terms of creation and annihilation operators. 
Usually, one would choose an expansion in terms of particles and
anti-particles with positive kinetic energies. 
It turns out, however, that the description of the Schwinger mode
is drastically simplified 
in the  
\Index{particle-hole representation}\Cite{Kiskis:1978tb,Callan:1978gz,Ambjorn:1983hp,Manton:1985jm}
wherein anti-particles are realised as holes in the Dirac sea. 
In this representation, the axial anomaly has an intuitive interpretation
as the effect of a fluctuating Fermi level. 
We therefore expand the fermionic field in terms of particle annihilation
operators
\begin{equation}
\begin{split}
\psi(0,{{x^3}})
&= 
 \sum_{k^3} {1 \over \sqrt{2\pi}} 
(
w_R a   _{{{R}} k^3} e^{ik^3\cdot x^3}
+
w_L a   _{{{L}} k^3} e^{+ik^3\cdot x^3}
)
\qquad;
\end{split}
\end{equation}
with \Index{chiral spinors} 
\index{wR@$w_R$} 
\index{wL@$w_L$} 
\begin{xalignat}{2}
w_R
=
 \left(
\begin{array}{rr}
1\\
0\\
\end{array}
\right) 
  & \qquad
w_L
=
\left(
\begin{array}{rr}
0\\
1\\
\end{array}
\right)   
\end{xalignat}
in a way {\em similar}\footnote{
Please note that $a_{R,n}^{\dag}$ creates a {\em right}-handed particle
with momentum $k^3=n$ whereas in Ref.~\Cite{Manton:1985jm},
the similar operator $a_{1,n}^{\dag}$,
creates
a {\em left}-handed particle with momentum $k^3=-k_3=-n$.
}
to Ref.~\Cite{Manton:1985jm}. 
The momenta $k^3$ assume multiple values of $\triangle k$, i.e. 
$k^3/\triangle k\in \Bbb{Z}$.
The creation operator $a_{R,\vec k}^{\dag}$ creates a {\em right}-handed particle
with momentum $\vec k$ and
$a_{L,\vec k}^{\dag}$ creates a {\em left}-handed particle. 
The creation operators obey the usual
anti-commutation relations
\begin{eqnarray}
\{a_{\alpha,{{k^3}}},a^{\dag} _{\beta,{{k^3}}'}\}
=
\delta_{\alpha,\beta}\;\delta_{{{k^3}},{{k^3}}'}
\end{eqnarray}
where $\alpha,\beta=R,L$.  
\subsection{The Hamiltonian in Manton's basis}

Now we are able to express the Hamiltonian $H$ 
\index{H@$H$: Hamiltonian! of the Schwinger model} 
in terms
of the annihilation and creation operators defined above 
\index{HNO@$H_{\text{raw}}$}
\begin{equation}
\begin{split}
H_{\text{raw}}
&=
\sum_{p_3} a^{\dag} _{{R},{{p^3}}} a_{{R},{{p^3}}}\;(p^3+{\mathsf A}_3)
+
\sum_{p_3} a^{\dag} _{{L},{{p^3}}} a_{{L},{{p^3}}}\;(-p^3-{{\mathsf A}_3})\\
&+
{g^2\over 4L} 
 \sum_{p^3\neq 0} \tilde {{j}}^0 _{{{p^3}}}{1\over ({p^3})^2} \tilde {{j}}^0_{-{{p^3}}}
+
{L\over g^2}{\mathsf E}^2 
\qquad.
\end{split}
\end{equation}
This Hamiltonian is 
{\em formally} equivalent to the expression given in ~\Cite{Manton:1985jm}.
Note, however, the subtle yet important difference that we are using
contra-variant momenta $p^3$ in lieu of co-variant momenta. 
The symbol $\tilde j^\mu$ stands for the spatial Fourier transform
\be
\tilde {{j}}^0(P^3)
\dn
\int dx^3\; e^{iP^3 x^3} j^0(x)
=
\rho_R(P^3)+\rho_L(P^3)
\ee
of the \Index{vector current} $j^\mu\dn \bar\psi\gamma^\mu\psi$
where
\index{j@$j^\mu$}
\index{rhoR@$\rho_R$}
\index{rhoL@$\rho_L$}
\index{j@$\tilde{{j}}^\mu$}
\be
\rho_\alpha(\vec P)
\dn \int dx^3\; e^{ik^3 x^3} \psi_\alpha^{\dag}\psi_\alpha
=
\sum a_{\alpha,\vec P+\vec k}^{\dag}a_{\alpha,\vec k}
\ee
with $\alpha=R,L$.
Classically, both the \Index{vector current}
\be\label{VectorCurrent}
j^\mu(x)\dn \bar\psi(x)\gamma^\mu\psi(x)
\ee
and the \Index{axial current}
\be\label{AxialCurrent}
j^\mu_A(x)\dn \bar\psi(x)\gamma^\mu\gamma^5\psi(x)
\ee
are conserved quantities because the Lagrangian
is invariant under the vectorial $U(1)$ transformation
\be
\psi(x)\longmapsto e^{-i{\eins}\theta_V }\psi(x)
\ee
and the axial $U(1)$ transformation
\be
\psi(x)\longmapsto e^{-i\gamma^5\theta_A }\psi(x)
\qquad.
\ee
In the quantised theory, however, the Hamiltonian $H$, 
the \Index{vector charge} ${\cal Q}$ and the \Index{axial charge} ${\cal Q}_A$
\be
{\cal Q}
\dn
\int_{-L} ^L dx^3\; j^0
=
{\cal Q}_R+{\cal Q}_L
\quad,\quad
{\cal Q}_A
\dn
\int_{-L} ^L dx^3\;
j^0_A 
=
{\cal Q}_R-{\cal Q}_L
\ee
with
$
{\cal Q}_\alpha 
=\rho_\alpha(0)
$
have to be \Index{renormalised} by subtracting their respective vacuum expectation
values.
The vacuum energy is the only part of the Hamiltonian that contains ultra-violet divergences. 
The renormalisation constants are unity. 
This property is related to the fact 
that the Schwinger-model is \Index{super-renormalisable} (i.e. 
the coupling constant $g$ has the dimension of a mass). 

Manton performs the renormalisation in the limit $g\rightarrow 0$ 
at first. 
The divergent vacuum energy of $H_{\text{raw}}$ is
contained in the first part of $H_{\text{raw}}$  
\index{Traw@$T_{\text{raw}}$}
\begin{equation}
\begin{split}
T_{\text{raw}}
&=
\sum_{p_3} a^{\dag} _{{R},{{p^3}}} a_{{R},{{p^3}}}\;(p^3+{\mathsf A}_3)
+
\sum_{p_3} a^{\dag} _{{L},{{p^3}}} a_{{L},{{p^3}}}\;(-p^3-{{\mathsf A}_3})
\quad.\\
 \end{split}
\end{equation}
Eigenvectors of $T_{\text{raw}}$
can be constructed as follows:
Let us define the \Index{a-vacuum}
$\cket{\Omega_a:{\mathsf A}_3}$
as an eigenstate $\cket{\Omega_a:{\mathsf A}_3}$ of the vector potential
operator
\begin{eqnarray}
\hat {\mathsf A}_3
 \cket{\Omega_a:{\mathsf A}_3}
=
{\mathsf A}_3
\cket{\Omega_a:{\mathsf A}_3} 
\end{eqnarray}
which is annihilated by
the operators $a_{\alpha,{{k^3}}}$ 
\index{Omegaa@$\cket{\Omega_a:{\mathsf A}_3}$}
\index{a-vacuum@$a$-vacuum ($\cket{\Omega_a:0,{\mathsf A}_3}$)    }
\begin{eqnarray}
\forall_{\alpha;k^3}\;
a_{\alpha,{{k^3}}} \cket{\Omega_a:{\mathsf A}_3}=0
\qquad. 
\end{eqnarray}
This state is an eigenstate to $T_{\text{raw}}$ with eigenvalue zero. 
For ${\mathbf A}_3=0$ we may construct the state with lowest 
(na{\"\i}ve) energy
(i.e. the vacuum of $T_{\text{raw}}$) by filling the a-vacuum with
left-moving right-handed particles and with right-moving right-handed 
particles as illustrated in \figRef{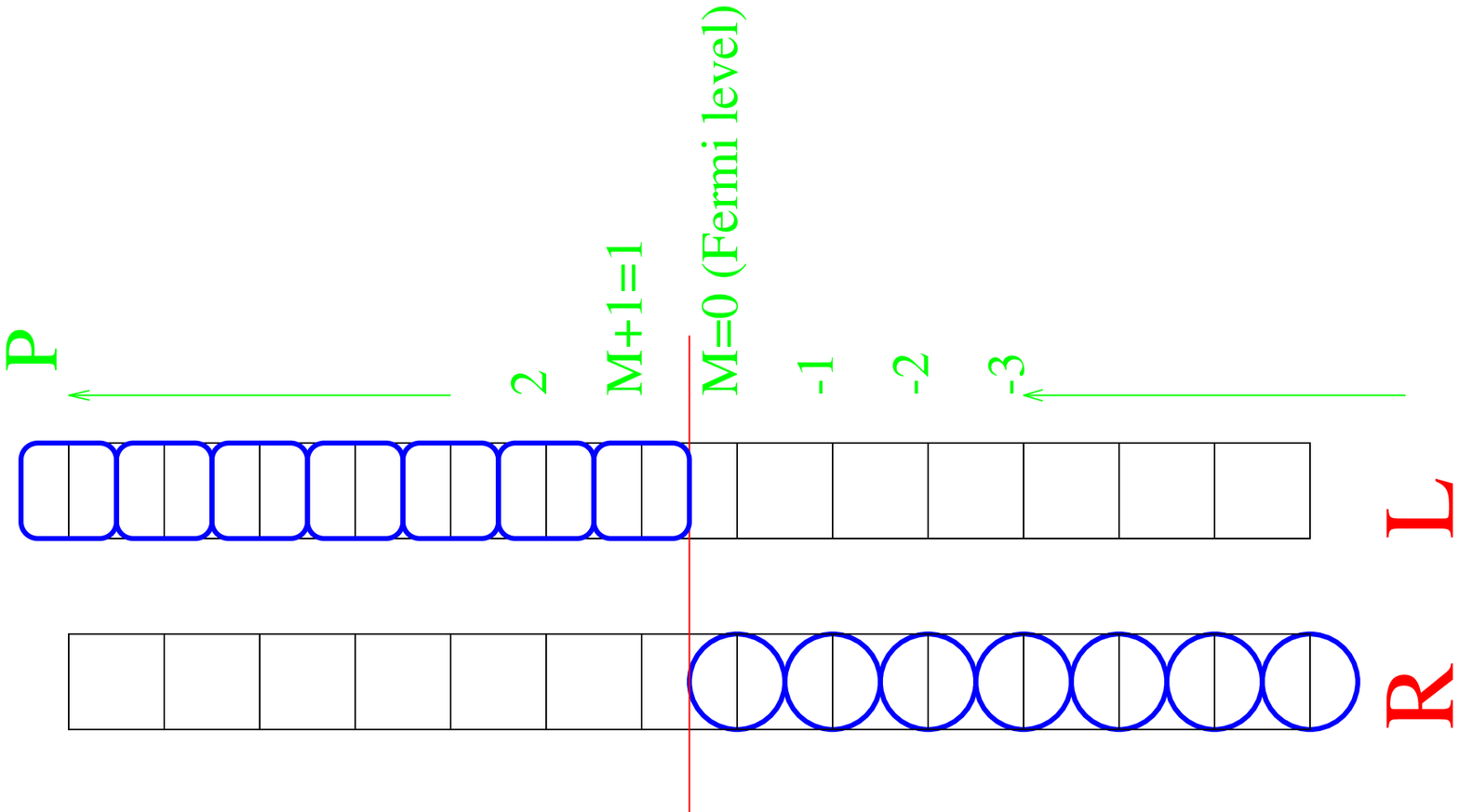}.
Choosing ${\mathsf A}_3\neq 0$, the so-constructed state is no longer 
the state with lowest energy.  
Filling up the $a$-vacuum around a different 
\Index{discrete Fermi level} ${\cal M}$
with right-handed fermions 
$a_{R,k^3};k^3\le {\cal M}$ and
left-handed fermions  $a_{R,k^3};k^3> {\cal M}$ 
we obtain the \Index{axial vacuum}
\be
\cket{\Omega:{\mathsf A}_3,{\cal M} }
\dn
\prod_{k^3> {\cal M}} 
   a^{\dag}_{R,{\cal M}+1-\vec k}
   a^{\dag}_{L,\vec k}
\cket{\Omega_a:{\mathsf A}_3}
\ee
with \Index{na{\"\i}ve axial charge} $:{\cal Q}_A:=2{\cal M}$. This construction
is illustrated in \figRef{MantonVak0.eps}.
Please note that
the vacuum symbols cannot be found in Manton's article. We have introduced 
them for later convenience. 

Now, there is a serious problem with this argument, of course. 
The energies of the states
$\cket{\Omega:{\mathsf A}_3,{\cal M} }$ are, in fact, infinite. 
So is their charge ${\cal Q}$. Little does it make sense, consequently,
to speak of "the state with lowest energy". 
The axial charge, finally, is finite but ill-defined. 
The subtraction of 
infinite quantities has to be rendered well-defined by means
of a controlled regularisation procedure.
The difference of an infinite number of left-movers
and an infinite number of right-movers
is then defined via the \Index{Continuum limit}, i.e. the limit
in which the regularisation is removed. 
It turns out that different regularisation procedures
yield different results in the continuum limit.
A cut-off regularised Hamiltonian, for instance, does not become
gauge-invariant in the continuum limit whereas the vectorial charge and the
axial charge are both conserved for every cut-off parameter $\Lambda$. 
\pstxt
{Symbolic representation of the vacuum 
$\cket{ \Omega:\mathsf{A}_3,{\cal M} }$ with ${\cal M}=0$ 
and $\mathsf{A}_3=0$}
{MantonVak0.eps}
{The circles represent right-handed fermions filling the Fermi
sea from below. The rounded squares
represent left-handed fermions filling the Fermi sea from above.}
Manton
employs the gauge-invariant \Index{heat-kernel regularisation}
in order to regularise the vacuum energy of ${\cal Q}$, ${\cal Q}_A$ and
$H_{\text{raw}}$. Basically, this regularisation can be seen
as the gauge-invariant version of a \Index{smooth cut-off} $\Lambda_h$
\index{Lh@$\Lambda_h$} which consists in
damping the number of right-handed fermions 
$a^{\dag}_{R,\vec k}a_{R\vec k}$
and
left-handed fermions 
$a^{\dag}_{L,\vec k}a_{L\vec k}$
with the factors 
$\exp({k^3+{\mathsf A}_3\over \Lambda_h})$ and 
$\exp(-{k^3+{\mathsf A}_3\over \Lambda_h})$ respectively\footnote{
A conventional/gauge-variant smooth cut-off would employ the factors 
$\exp(\pm{k^3\over \Lambda_h})$ instead.}.
This procedure renders infinite sums such as $\sum_{k^3\le 0} 1$ 
finite. Manton finds that the divergent vacuum expectation values 
\begin{align}
\brac{\Omega:{\mathsf A}_3,{\cal M} }
{\cal Q}_R
\cket{\Omega:{\mathsf A}_3,{\cal M} }
&=
c_\infty
+({\cal M}+{\mathsf A}_3+{1\over 2})  \\
\brac{\Omega:{\mathsf A}_3,{\cal M} }
{\cal Q}_L
\cket{\Omega:{\mathsf A}_3,{\cal M} }
&=
c_\infty
-({\cal M}+{\mathsf A}_3+{1\over 2})  \\
\brac{\Omega:{\mathsf A}_3,{\cal M} }
T_{\text{raw}}
\cket{\Omega:{\mathsf A}_3,{\cal M} } 
&= 
c_\infty^H
+
({\cal M}+{\mathsf A}_3+{1\over 2})^2
\end{align}
may be decomposed in a finite, field-dependent part and field-independent
constants $c_\infty\approx \Lambda_h+{\cal O}(1/\Lambda_h)$, 
$c_\infty^H\approx -2\Lambda_h^2+{\cal O}(1/\Lambda_h)$  
which diverge in the continuum limit 
$\Lambda_h\rightarrow \infty$. 
Subtracting these constants from the regularised operators
${\cal Q}^{\text{reg}}_R   \dn {\cal Q}_R-c_\infty$, 
${\cal Q}^{\text{reg}}_L   \dn {\cal Q}_L-c_\infty$ and
$H_{\text{reg}}          \dn H_{\text{raw}}-c_\infty^H$ 
yields the renormalised operators. 
It is crucial that gauge-invariant regularisation 
produce expressions which depend on the gauge-field. The finite gauge-dependent
part of the {\em vacuum energy}
acts now as an additional interaction term 
of the Hamiltonian. The finite gauge-dependent part of the {\em axial
charge} is not conserved and will be responsible for the anomaly.
The regularised charge
${\cal Q}_{\text{reg}}$ is zero by construction:
a non-vanishing vectorial charge would be
incompatible with Gau{ss}' law~\Ref{Gauss}.
The regularised axial charge may be written as
${\cal Q}_A^{\text{reg}}=2\hat {\cal M}+2{\mathsf A}+1$,
where $\hat {\cal M}$ is the discrete Fermi level. As a result, the quantity
${\cal Q}_A^{\text{reg}}/2$ may now be regarded as a continuous 
Fermi label. This means that we end up with the ---somewhat paradoxical---
result that the difference between the (infinite) number of right-movers
and the infinite number of left-movers is not an integer. This result becomes
less paradoxical, of course, if one realises that heat-kernel regularisation
consists in giving fermions a fractional, ${\mathsf A}_3$-dependent
particle number. Hence we should
not be overly surprised that the axial charge remains fractional and
${\mathsf A}_3$-dependent in the continuum
limit. The same comment applies to alternative gauge-invariant regularisation
techniques.

The heat-kernel regularised charges are invariant under \Index{large gauge transformations}
$U_G$
\index{UG@$U_G$}
\be
U_G\cket{\Omega:{\mathsf A}_3,{\cal M} }
=
\cket{\Omega:{\mathsf A}_3-1,{\cal M}+1}
\ee
which simultaneously change ${\cal M}$ and ${\mathsf A}_3$. 
Gauge-invariant states are irreducible representations
of the group of large gauge transformations, generated
by $U_G$ with group characters $e^{-i\theta}$. 
They are called \Index{$\theta$-states}.

Na{\"\i}ve renormalisation
based on cut-off regularisation
partially destroys gauge invariance. Conversely, if the renormalisation
procedure chosen 
is gauge-invariant, then the {\em renormalised} axial charge ceases
to be conserved. This phenomenon is called \Index{axial anomaly}.
The technical details involved in order to describe the axial
anomaly vary remarkably, depending on the way quantisation is 
performed:
(1) In \Index{perturbation theory},
the anomaly arises due to ambiguities
in divergent loop integrals of a certain class of 
Feynman diagrams,
the easiest of which is the so-called 
\Index{triangle diagram}. For details, the reader is referred to Ref.~\Cite{Adler:1969av,Bell:1969re}.
(2) In \Index{path-integral quantisation}~\Cite{Fujikawa:1979ay}, the
renormalisation ambiguity can be traced back to the path integral
measure which has to be regularised.
All of these manifestations have in common that the anomaly 
stems from the necessity of choosing 
amongst different regularisation procedures 
one of which is gauge invariant. 
If one uses an infinite volume or fixed BCs, then the anomlay
manifests itself in yet another way. In this case axial gauge
may be chosen and ${\mathsf A}_3$ disappears. The $\theta$ angle
is then due to the fact that adding an {\em unperiodic} term
$\theta x^3$ to a solution ${\mathsf A}_0$ to Gau{ss}' law~\Ref{Gauss}
transforms it into another solution~\Cite{Coleman:1975pw,Coleman:1976uz}.
This term may then be interpreted as a constant background field~\Cite{Coleman:1975pw,Coleman:1976uz}. 
We note in passing that one may regularise the axial charge such that
it is conserved and the vector charge is not. 
In the Hamiltonian approach, however, the non-conservation
of (vector) charge is incompatible with Gau{ss}' law.

In order to diagonalise the heat-kernel regularised 
Hamiltonian, Manton 
expresses the Hamiltonian
in terms
of the bosonic fields $\Pi$ and $\Phi$
\begin{align}
\Pi(\vec p)
&=
+{ \rho_{{R}}(\vec p)-\rho_{{L}}(\vec p) \over \sqrt{2} }
=
+{ 1  \over \sqrt{2}}
\tilde {{j}}^3_{\vec p}     \\
%
\Phi(\vec p)
&=
-{\rho_{{R}}(\vec p)+\rho_{{L}}(\vec p) \over \sqrt{2} i p^3}
=
-{\tilde {{j}}^0_{{{p^3}}} \over \sqrt{2} i p^3}    \\
%
%
%
\Pi(0)
&= 
\sqrt{2} (\tilde {\mathsf A} +{1\over 2} )         \\
\Phi(0)
&= {1\over g^2\sqrt{2}} {\mathsf E} 
\end{align}
\begin{eqnarray}\label{BozoCommut}
[\Pi(-{{p^3}}),\Phi({{p^3}}')]\cket{\text{phy}}
=[\Pi^{\dag}_{{{p^3}}},\Phi({{p^3}}')]\cket{\text{phy}}
=
-i\delta(p,p')\cket{\text{phy}}
\qquad.
\end{eqnarray}
The bosonic commutation relations~\Ref{BozoCommut} hold on a \Index{physical
subspace} of the total Fock space only. Physical states $\cket{\text{phy}}$
are states
with finite regularised energy, i.e. $\brac{\text{phy}}H_{\text{reg}}\cket{\text{phy}}<\infty$. 
With these operators, the Hamiltonian assumes a form 
\begin{eqnarray}
H_{\text{\text{reg}}}
=
{1\over 2} \sum_{p^3} 
(\Pi^{\dag}(p^3)\Pi(p^3) + \Omega^2_{{{p^3}}} \Phi^{\dag}(p^3)\Phi(p^3))
\end{eqnarray}
which is
manifestly equivalent to a free theory of bosons with mass $M_B\dn g/\sqrt\pi$
and 
energy 
is
\index{Omega@$\Omega({{p^3}})$}
\begin{eqnarray}
 \Omega({{p^3}})
\dn
\sqrt{({p^3})^2+{g^2\over \pi}}
\qquad.
\end{eqnarray}
These bosons are called \Index{Schwinger bosons}. 
Let us finally look, however briefly, at the nature 
of the axial anomaly exhibited by the Schwinger model. 
The axial symmetry of the Lagrangian is not realised in the mass spectrum. 
This might easily lead us to the conclusion that the heat kernel
regularisation breaks the axial symmetry in the sense that 
$[H_{\text{reg}},:{\cal Q}_A:]=0$ 
no longer holds. This is wrong, $[H_{\text{reg}},:{\cal Q}_A:]=0$  
{\em does} hold. Axial symmetry and gauge symmetry are incompatible on the
quantum mechanical level since both symmetries are 
now merged into a larger, non-commutative group: $[U_G,:Q_A:]\neq 0$. In other
words, $U_G$ and
$:{\cal Q}_A:$ form a non-commutative algebra, i.e. they cannot be measured
simultaneously.  
The axial symmetry is not realised in the spectrum because
states with different axial charges are related via $U_G$: therefore,
states with different axial charges must have the same energy. 

\Index{Axial gauge} breaks 
the symmetry generated by $U_G$ and, consequently,
destroys the degeneracy of the mass spectrum: spurious states appear.

\section{Translation into the Particle-Antiparticle Picture}
In the framework of the parton model, fermion fields are expressed 
\begin{eqnarray}
\psi(0,{{x^3}})
= 
 \sum_{k^3} {1 \over \sqrt{2\pi}} 
(
{\mathsf u}({{k^3}}) b       _{{{k^3}}} e^{+i{{k^3}}\cdot{{x^3}}}
+
{\mathsf v}({{k^3}}) d^{\dag}_{{{k^3}}} e^{-i{{k^3}}\cdot {{x^3}}}
)
\end{eqnarray}
in terms of particle creation operators and anti-particle
creation operators rather than in terms of particles and holes.  
The usual spinors (for massless fermions)
\begin{align}
{\mathsf u}({{k^3}})
&=
\left\{
\begin{array}{ll}
+w_R &  \text{if } k^3 > 0 \\
+w_L &  \text{if } k^3 \leq 0 \\ 
\end{array}
\right\}     \\
{\mathsf v}({{k^3}})
&=
\left\{
\begin{array}{ll}
+w_R &  \text{if } k^3 > 0 \\
-w_L &  \text{if } k^3 \leq 0 \\ 
\end{array}
\right\}
\end{align}
are used. We have absorbed the term ${1/\sqrt{\omega({{k^3}})}}$ usually appearing
in the expansion into the spinors ${\mathsf u}$ and ${\mathsf v}$.
\index{u@${\mathsf u}$} 
\index{v@${\mathsf v}$} 
The choice for $k^3=0$ is arbitrary, reflecting the
fact that there are two degenerate vacua of the free, massless
fermionic theory. 
Naturally, the operators defined above fulfill the standard
anti-commutation relations
\index{b@$b$}
\index{d@$d$}
\begin{xalignat}{2}
\{b_{{{k^3}}},b^{\dag} _{{{k^3}}'}\}=\delta_{{{k^3}},{{k^3}}'}
 & \qquad
\{d_{{{k^3}}},d^{\dag} _{{{k^3}}'}\}=\delta_{{{k^3}},{{k^3}}'}
\qquad.
\end{xalignat}
In what follows
we shall set $\triangle k=1$ (i.e. $L=\pi$).
In order to transform the exact solution to the Schwinger model
into the partonic picture,
we relate the different creation and annihilation
operators. This can be done
by a simple comparison of these two expansions 
\index{a@$a$}\index{b@$b$}
\be
\displaystyle{
\begin{pmatrix}
a       _{R, {{k^3}}}\\
a       _{L, {{k^3}}}
\end{pmatrix}
}
=
\begin{cases}
\begin{pmatrix}
 b          _{{{k^3}}}\\
+d^{\dag}  _{-{{k^3}}}
\end{pmatrix}  & \text{if $k^3 > 0$}\\
\quad\\
\begin{pmatrix}
-d^{\dag}  _{-{{k^3}}}\\
 b       _{{{k^3}}}
\end{pmatrix}  & \text{if $k^3\le 0$}\\
\end{cases}
\ee
We call the vacuum 
$\cket{\Omega:{\mathsf A}_3,0}$
which is annihilated by $b_{\vec k}$ and $d_{\vec k}$
\index{bd-vacuum}
\index{OmegaA@$\cket{\Omega:{\mathsf A}_3,0}$}
\begin{xalignat}{2}
\forall_{{{k^3}}}\;
b_{{{k^3}}}\cket{\Omega:{\mathsf A}_3,0 }
=0
& \qquad
\forall_{{{k^3}}}\;
d_{{{k^3}}}\cket{\Omega:{\mathsf A}_3,0 }
=0
\end{xalignat}
the bd-\noIndex{vacuum}. 
This vacuum is identical to the vacuum
$\cket{\Omega:{\mathsf A}_3,{\cal M}}$ constructed in the last subsection
if we set ${\cal M}=0$.  
In particular, $\cket{0}\dn \cket{\Omega:{\mathsf A}_3=0,0}$ is the 
\Index{perturbative vacuum} of the theory in axial gauge ${\mathsf A}_3=0$. 

A change of the discrete Fermi level ${\cal M}$ has to be seen as
a creation or annihilation of particles if the bd-basis is used
rather than the a-basis. For example,
$\cket{\Omega:{\mathsf A}_3,{-1} }$ is obtained from 
$\cket{\Omega:{\mathsf A}_3,0}$ by lowering the Fermi-level with
the operator $a_{R,0}a^{\dag}_{L,0}$
and therefore
\be
\cket{\Omega:{\mathsf A}_3,{-1} }
=
a^{\dag}_{L,0}a_{R,0}   \cket{\Omega:{\mathsf A}_3,0}
=
b^{\dag}_0 d^{\dag}_0 \cket{\Omega:{\mathsf A}_3,0}
\qquad.
\ee
Similarly, the vacuum for ${\cal M}=-2$ is obtained as
\be
\cket{\Omega:{\mathsf A}_3,{-2} }
=
a^{\dag}_{L,0} a_{R,0}  \cket{\Omega:{\mathsf A}_3,-1}
=
 b^{\dag}_{-1} d^{\dag}_{+1}  b^{\dag}_0 d^{\dag}_0 \cket{\Omega:{\mathsf A}_3,0}
\ee
and so on.

Now we have to regularise the Hamiltonian. 
Heat-kernel regularisation works 
only on an infinite lattice. 
Numerical methods, however, can handle a finite number of 
(effective) degrees of freedom only. 
Therefore, we shall not repeat this procedure in spite of its elegance.
Instead, we suggest a {\em heuristic} procedure 
which will turn out to be equivalent
to heat-kernel regularisation. 
%
%
We start with a na{\"\i}ve cut-off regularisation and
regularise the divergent vacuum energy by normal-ordering.
As this destroys the invariance of the
Hamiltonian under \Index{large gauge transformations}
(i.e. the topologically non-trivial gauge transformations that remain
in spite of Coulomb gauge ~\Cite{Manton:1985jm})
we have to restore gauge invariance {\em by hand} in order
to obtain physically meaningful results. 
We shall show that gauge-invarince can be repaired simply 
by adding a term to the Lagrangian which disappears in the  
infinite volume limit. It will turn out that the 
addition of this term exactly reproduces the same Hamiltonian 
as heat-kernel regularisation if the cut-off is removed.
The un-regularised Hamiltonian in the $bd$ basis reads 
\index{Hraw@$H_{\text{\text{raw}}}$}
\begin{equation}
\begin{split}
H_{\text{raw}}=
T
+
{\mathsf A}_3 {\cal Q}_A
-
\Lambda(\Lambda+1)
+
{L\over g^2}{\mathsf E}^2
+
{g^2\over 4L} 
 \sum_{p^3\neq 0} \tilde {{j}}^0_{{{p^3}}}{1\over ({p^3})^2} \tilde {{j}}^0_{-{{p^3}}}
\end{split}
\end{equation}
where $T$ is the normal-ordered \Index{kinetic energy}
\index{T@$T$}
\begin{eqnarray}
T=
\sum_{p^3} 
(b^{\dag} _{{{p^3}}} b_{{{p^3}}} 
+
 d^{\dag} _{{{p^3}}} d_{{{p^3}}})
|p^3|
\end{eqnarray}
and
\index{M@${\cal M}$}
$\Lambda(\Lambda+1)$ is the constant which 
arises when the kinetic energy is normal ordered. 
\index{QA@${\cal Q}_A$}
This constant has to be regularised since it 
contains the short-distance singularity $\Lambda(\Lambda+1)$.
Normal ordering the Hamiltonian with respect to the $bd$ operators 
corresponds to removing this term. 
Subtracting the kinetic energy of the state $\cket{\Omega:{\mathsf A}_3,{\cal M}}$
\begin{eqnarray}
\hat{\cal M}(\hat{\cal M}+1) 
\end{eqnarray}
from the kinetic energy $T$,
we define the \Index{modified kinetic energy}
\index{modified kinetic energy}
\begin{eqnarray}
{\frak T}\dn
\sum_{p^3} 
(b^{\dag} _{{{p^3}}} b_{{{p^3}}} 
+
 d^{\dag} _{{{p^3}}} d_{{{p^3}}})
|p^3|
-
\hat{\cal M}(\hat{\cal M}+1) 
\end{eqnarray}
\index{T@${\frak T}$}
which is insensitive to the discrete Fermi-level ${\cal M}$.
While the states 
\begin{equation}
b^{\dag} _{{{k^3}}}
d^{\dag} _{{{p^3}}}
\cket{\Omega:{\mathsf A}_3,0}
\text{  and  }
b^{\dag} _{{{k^3}}}
d^{\dag} _{{{p^3}}}
\cket{\Omega:{\mathsf A}_3,{\cal M} }
\end{equation}
have different kinetic energies $T$, for instance, 
(i.e. $|k^3|+|p^3|$ resp. $|k^3|+|p^3|+{\cal M}({\cal M}+1)$)
their modified energies $|k^3|+|p^3|$ coincide. 
The same holds for mor complicated states. 
We have introduced ${\frak T}$ because it is invariant 
under topologically non-trivial gauge transformations $U_G$ which 
change the axial charge whereas the conventional kinetic energy is {\it not}.  
Both operators have the same eigenstates but their eigenvalues
coincide solely in the zero axial charge ${\cal Q}_A=0$ sector. 
The distinction between $T$ and ${\frak T}$ (not explicitly introduced in Ref. ~\Cite{Manton:1985jm})
is very important.  
The bosonized kinetic energy in Ref. ~\Cite{Manton:1985jm} 
corresponds in fact to the {\it modified} kinetic energy ${\frak T}$
rather to the usual kinetic energy $T$.
Now we insert the new quantity ${\frak T}$ into the 
normal-ordered (with respect to $bd$) Hamiltonian 
\begin{eqnarray}
H_{\text{N.O.}}
=
{\frak T}+
{\cal M}({\cal M}+1)
+
2{\cal M} {{\mathsf A}_3}
+ 
{{\mathsf A}_3} 
+
{L\over g^2}{\mathsf E}^2
+
{g^2\over 4L} 
 \sum_{p^3\neq 0} \tilde {{j}}^0_{{{p^3}}}{1\over ({p^3})^2} \tilde {{j}}^0_{-{{p^3}}}
\end{eqnarray}
because we want to isolate the terms which are not invariant 
under \Index{large gauge transformations} $U_G$
\index{UG@$U_G$}
\begin{eqnarray}
\hat{\cal M}\rightarrow U_G\hat{\cal M}U^{-1}_G=\hat{\cal M} +1 
\end{eqnarray}
\begin{eqnarray}
\hat{{\mathsf A}_3}\rightarrow  U_G\hat{{\mathsf A}_3}U^{-1}_G\hat{{\mathsf A}_3}-1.
\end{eqnarray}
After completion of the square 
\begin{eqnarray}
H_{\text{N.O.}}
=
{\frak T}+
(\hat{\cal M}+{{\mathsf A}_3} + {1\over 2})^2 - {\mathsf A}^2 _3 -{1\over 4}
+
{L\over g^2}{\mathsf E}^2
+
{g^2\over 4L} 
 \sum_{p^3\neq 0} \tilde {{j}}^0_{{{p^3}}}{1\over ({p^3})^2} \tilde {{j}}^0_{-{{p^3}}}
\end{eqnarray}
we see that the quadratic term $-{\mathsf A}_3 ^2$ is the
only term which varies under large gauge transformations.
Subtracting  ${\mathsf A}_3 ^2$ from the Lagrangian
or, equivalently, adding it to the Hamiltonian
\begin{eqnarray}
H_{reg}=H_{\text{N.O.}}+{\mathsf A}_3 ^2+{1\over 4}
\qquad,
\end{eqnarray}
the Hamiltonian becomes identical to the one obtained
by \Index{heat-kernel regularisation}. 
{\bf
This means that the quantisation of the {\em gauge-variant} Lagrangian
${\cal L}'\dn {\cal L}-{L\over \pi} {\mathsf A}_3 ^2$ 
with a na{\"\i}ve cut-off regularisation, is equivalent to
quantisation of the {\em gauge-invariant} Lagrangian
${\cal L}$ with a heat-kernel regularised 
Hamiltonian (in the limit $\Lambda\rightarrow \infty$). 
}
The addition of one single field variable ${\mathsf A}_3^2$
might seem small a change in the infinite volume limit
where an infinite number of fields is present: 
the term ${L\over \pi}({\mathsf A}_3) ^2$ (units restored) even vanishes 
in the infinite volume limit $L\rightarrow \infty$ because ${\mathsf A}_3$
can always be gauged to be smaller than $\triangle k={\pi\over L}$. 
Yet this step has a profound influence on the spectrum
and the condensate 
(in the rest frame and only in the rest frame).
It is remarkable that Hamiltonians which have the same 
classical infinite volume limits should have different quantum mechanical 
infinite volume limits.  
While the heuristic procedure proposed here is not necessary
in the Schwinger model where the only divergence in the Hamiltonian
is the vacuum energy, a similar procedure may turn out to 
be helpful for theories where the Hamiltonian itself diverges (e.g. 
QCD($3+1$)). 
The Hamiltonian $H_{\text{reg}}$ is invariant under {\em two}\footnote{
Note that the anomaly does {\em not} mean that the axial symmetry is
broken: the point is that the generator of the axial symmetry
does not commute with the generator of the large gauge symmetry. As
physical states have to be gauge invariant, the conserved quantity
$:{\cal Q}_A:$ cannot be observed whereas the
non-conserved quantity ${\cal Q}_A^{\text{reg}}$ can be: the axial symmetry
is invisible.

}
symmetries: \Index{axial transformations}
 generated by $:{\cal Q}_A:$ and
large
\Index{gauge-transformations} $U_G$. This distinguishes $H_{\text{reg}}$ from a large
class of Hamiltonians which differ by quantity that vanishes in the 
limit $l\rightarrow \infty$. 

\section{Solutions to the Schwinger Model and the Influence of Approximations}

In order to study the influence of several approximations it 
is convenient to introduce 
some Hamiltonians
constructed using these approximations: 
The \Index{Hamiltonian in axial gauge}
\begin{eqnarray}
H_{\text{axial}}=
H_{\text{N.O.}}-{L\over g^2}{\mathsf E}^2 -(2\hat{\cal M}_A+1){\mathsf A}_3 
\end{eqnarray}
would have been obtained through the choice of the 
{\it unphysical} (for large volumes, see~\Cite{Manton:1985jm}) axial gauge
plus normal-ordering. This Hamiltonian is not invariant under large gauge
transformations.
If we
subtract the term ${\cal M}({\cal M}+1)$, however, the resulting
\Index{artificial Hamiltonian}
\begin{align}
H_{\text{arti}}
&=
H_{\text{axial}}-\hat {\cal M}(\hat {\cal M}+1)
=
H_{reg}- {L\over g^2}{\mathsf E}^2 - (\tilde {\mathsf A}+{1\over 2})^2 \\
&=
{\frak{T}}
+
{g^2\over 4L} 
 \sum_{p^3\neq 0} \tilde {{j}}^0_{{{p^3}}}{1\over ({p^3})^2} \tilde {{j}}^0_{-{{p^3}}} 
\end{align}
%
%
%
with $\tilde {\mathsf A}\dn {\cal M}+{\mathsf A}_3$
\index{A@$\tilde {\mathsf A}$}
is again invariant under large gauge transformations as it differs 
from the heat-kernel Hamiltonian by a gauge-invariant quantity. 

Now, we would like to find a complete solution to all of these
Hamiltonians and interpret them in terms of quark- and gluon- degrees 
of freedom (in the literature, the fermions of the Schwinger
model are often suggestively 
referred to as \Index{quarks}, the gauge boson field as \Index{gluons}. 
We follow this convention).

In addition to Ref.~\Cite{Manton:1985jm}
it is useful to introduce the operators 
%
\index{A@${\cal A}$}
\begin{equation}
\begin{split}
   {\cal A}^{\dag} _{{{k^3}}}        
&\dn
{1\over \sqrt 2}( \sqrt{\Omega({{k^3}})} \Phi^{\dag}(-{{k^3}}) - 
  {1\over \sqrt{\Omega({{k^3}})}}\Pi({{k^3}})i)   \\
 &=
-i{ \rho_{{R}}({{k^3}})({\Omega({{k^3}})\over k^3}+1) + \rho_{{L}}({{k^3}}) ({\Omega\over k^3}-1) 
 \over 2\sqrt{\Omega} }
\end{split}
\end{equation}
\begin{equation}
\begin{split}
   {\cal A}_{\vec 0}           
& =
{\sqrt{\Omega}\over \sqrt 2}( \Phi(\vec 0) + 
  {1\over {\Omega}}\Pi^{\dag}(\vec 0)i) \\
& =
{\sqrt{M_B}\over \sqrt 2}( {i\over\sqrt{2}} {d\over d\tilde {\mathsf A}} + 
  {1\over {M_B}} \sqrt{2} (\tilde {\mathsf A} +{1\over 2}) )i)     \\
& =
-{1\over \sqrt{M_B}}({\mathsf E}/M_B - (\tilde {\mathsf A}+{1\over 2}i)
\end{split}
\end{equation}
which create and absorb Schwinger bosons.
The artificial Hamiltonian
\begin{eqnarray}
H_{\text{arti}}
=
\sum_{p^3\neq 0}
{\cal A}^{\dag}_{{{p^3}}} {\cal A}_{{{p^3}}}
\;
\Omega({{p^3}})
\end{eqnarray}
written in this form is then equivalent to the
heat-kernel regularised Hamiltonian except that $H_{\text{arti}}$
lacks the pure gauge sector
\be
H_{\text{gauge}}
\dn
H_{\text{reg}}-
H_{\text{arti}}
=
{L\over g^2}{\mathsf E}^2 + (\tilde {\mathsf A}+{1\over 2})^2
=
{\cal A}^{\dag}_{\vec 0} {\cal A}_{\vec 0}
\;
\Omega(\vec 0)
\ee
describing the rest frame. 
Consequently, the axial Hamiltonian
\begin{eqnarray}
H_{\text{axial}}
=
\sum_{p^3\neq 0}
{\cal A}^{\dag}_{{{p^3}}} {\cal A}_{{{p^3}}}
\;
\Omega({{p^3}})
+\hat{\cal M}(\hat{\cal M}+1)
=
H_{\text{arti}}+\hat{\cal M}(\hat{\cal M}+1)
\end{eqnarray}
is not Lorentz co-variant except for states with ${\cal M}=0,-1$:
all other states are \Index{spurious states} whose
masses depend strongly on $\vec P$
because their four-momenta
\be
P_{\cal M}
\dn
\begin{pmatrix}
\Omega(\vec P)\\
\vec P
\end{pmatrix}
+
\begin{pmatrix}
{\cal M}({\cal M}+1)\\
0
\end{pmatrix}
\ee
do not lie on the mass-shell, i.e. $P_{\cal M}^2\neq M_B^2$. 
In summary, axial gauge destroys both Lorentz covariance and
the axial anomaly. 

It is interesting though that these illnesses of axial gauge 
are swept under the carpet 
in the (unphysical) limit $L\rightarrow 0$.
The unphysical term ${\cal M}({\cal M}+1)\triangle k$ diverges
if physical units are restored (i.e. $L\neq \pi$) because
$\triangle k$ diverges. Consequently, spurious states with 
${\cal M}({\cal M}+1)\neq 0$ are removed 
from the low-energy mass spectrum. If only the mass-spectrum is concerned,
the peculiar limit $L\rightarrow 0$ mimicks the anomaly.
This phenomenon is a {\em fake-realisation} of the anomaly
since two wrongs ---i.e. zero lattice size and axial gauge--- make a right
and this is not acceptable.

\section{The Schwinger Model on an Infinitesimal Lattice}
In this section, we 
point out that the mass spectrum of the Schwinger model
may be reproduced in the \Index{constituent quark model} 
{\em without phenomenological input} if (1)
the lattice size $L$ is sufficiently small
and (2) no $\theta$ states are constructed from axial states.  
We demonstrate that the
vacuum of the Schwinger model
is trivial and that the Schwinger bosons are pure two-fermion
states if the limit of an infinitesimally small
lattice size is performed.
This holds except for the rest-frame
where the Schwinger boson may be interpreted as a topological glue-ball.
Remarkably, the limit $L\rightarrow 0$ does not affect
the mass-spectrum. 
In the ensuing section, we shall then demonstrate
that the large volume solutions to the Schwinger model
may be obtained by a simple, unitary transformation. 
Herein, the \Index{zero volume basis}
\index{A@$\Bbb{A}$}
\be                                  
 \Bbb{A}_{\,{{P^3}}}
\dn
\begin{cases}
 {i\over \sqrt{|P^3|} } 
     \rho_{{R}} (-{{P^3}})
&
\text{for $P^3>0$} \\
%
%
%
-{i\over \sqrt{|P^3|} }
 \rho_{{L}} (-{{P^3}})
&
\text{for $P^3<0$} \\
\end{cases}
\ee
will be used.   
This basis appears in the limit 
\be
M_BL\rightarrow 0 \Leftrightarrow g/\triangle k\rightarrow 0
\ee
which may be interpreted either as the zero-volume limit $2L\rightarrow 0$
(if the physical mass $M_B$ is constant) or 
as the limit of vanishing Schwinger boson mass $M_B$ (if the volume
is kept constant). 
\begin{eqnarray}
\lim_{p^3/M_B\rightarrow \infty} 
{\cal A}^{\dag}_{{{p^3}}} =
-i{ \rho_{{R}}({{p^3}}) 
 \over 2 \sqrt{p^3} }
=
\Bbb{A}_{\,p^3} 
\end{eqnarray}
where $\Omega(\vec p)\rightarrow |\vec p|$. 
The zero volume basis has the interesting property that 
these operators annihilate the perturbative
vacuum for $\vec P\neq 0$. The same holds for every axial vacuum
$
\cket{\Omega:{\mathsf A}_3,{\cal M}}
$
---a fact that is immediately clear in Manton's $a$-basis: the operators
$\rho_{{R}}$ cannot move fermions onto sites with negative momenta
as they are already occupied. The same way, $\rho_{{L}}$ is unable to
shift fermions deeper into the occupied Fermi sea.
This proves
\begin{Theorem}[Triviality of the Vacuum]\label{VakTh}
The axial vacua $\cket{\Omega:{\mathsf A}_3,{\cal M}}$
and
$\cket{\Omega:0,{\cal M}}$
are eigenstates of the Hamiltonians $H_{\text{axial}}$ and
$H_{\text{arti}}$ if $g\ll \triangle k$. 
The axial vacua $\cket{\Omega:{\mathsf A}_3,{\cal M}}$ are vacua
(i.e. lowest eigenstates) of $H_{\text{arti}}$. $H_{\text{axial}}$,
in contrast, has only two vacua $\cket{\Omega:0,0}$
and $\cket{\Omega:0,-1}$.
\end{Theorem}
The reason for why the axial Hamiltonian $H_{\text{axial}}$
has only two vacua 
$\cket{\Omega:0,0}$ and $\cket{\Omega:0,-1}$ 
is that the energies ${\cal M}({\cal M}+1)$
of the other states $\cket{\Omega:0,{\cal M}}$
are larger than zero.
Theorem~\Ref{VakTh} implies that the \Index{perturbative vacuum} 
$\cket{0}\dn\cket{\Omega:0,0}$ is a vacuum of both Hamiltonians.
In other words, both Hamiltonians have 
one \Index{trivial} vacuum and the other vacua are almost trivial. 
We may construct $\theta$ vacua as 
projectively gauge-invariant linear combinations of the trivial vacua
\index{OmegaAtheta@$\cket{\Omega:\tilde {\mathsf A};\theta }$}
\begin{equation}\label{AThetaVacua}
\begin{split}
\cket{\Omega:\tilde {\mathsf A};\theta}
=
\sum_{{\cal M}}
\cket{\Omega:\tilde {\mathsf A}-{\cal M},{\cal M} } {1 \over \sqrt{2\pi} } e^{-i \theta {\cal M}}
\qquad.
\end{split}
\end{equation}
These states remain eigenstates of $H_{\text{arti}}$ but not of $H_{\text{axial}}$. 

The states $\cket{\Omega:{\mathsf A}_3,{\cal M}}$ are not 
vacua of $H_{\text{reg}}$ because they are not eigenvectors of
the sector $M_B {\cal A}_0 ^{\dag} {\cal A}_0$ which consists
of gauge fields only.  
Using the eigenfunctions ${\mathsf H}_n({\mathsf A}/M_B)$ of an harmonic oscillator
\index{Hn@${\mathsf H}_n$}
in the spatial representation, however,
we may build harmonic oscillator states 
\index{Omegan@$\cket{\Omega_n:{\cal M}}$}
\be   
  \cket{\Omega_n:{\cal M}}
\dn
\int d{\mathsf A}  
\;
 {\mathsf H}_n({{\mathsf A}\over {M_B}}) 
\cket{\Omega:{\mathsf A}_3,{\cal M} }
\ee 
that are eigenstates of $H_{\text{reg}}$.
These, in turn, may be super-imposed to $\theta$ vacua  
\begin{equation}\label{nThetaVacua}  
\begin{split}    
  \cket{\Omega_n;\theta }  
&=
\left[\Bbb{A}^{\dag}_{0}\right]^n
\cket{\Omega_0;\theta }
=
\int d\tilde {\mathsf A}  
\;
 {\mathsf H}_n({\tilde {\mathsf A}\over {M_B}}) 
\cket{\Omega:\tilde {\mathsf A},\theta }   \\
&=
\int d\tilde {\mathsf A}  
\;
 {\mathsf H}_n({\tilde {\mathsf A}\over {M_B}}) 
\sum_{{\cal M}}
\cket{\Omega:\tilde {\mathsf A}-{\cal M},{\cal M} } {1 \over \sqrt{2\pi} } e^{-i \theta {\cal M}}
\end{split}
\end{equation}
\index{Omegantheta@$\cket{\Omega_n;\theta}$}
of $H_{\text{reg}}$.
The index $n$ in $\cket{\Omega_n;\theta}$ or $\cket{\Omega_n,\cal M}$
indicates the number of
Schwinger bosons with 
momentum $\vec P=0$. Consequently, the state 
$\cket{\Omega_0;\theta }$ is the physical vacuum. 
The states labeled with ${\cal M}$ are eigenstates of $H_{\text{reg}}$ and
the na{\"\i}vely axial charge $:{\cal Q}_A:=2{\cal M}$. 
States labeled with ${\theta}$ are eigenstates of $H_{\text{reg}}$ and 
the heat-kernel regularised axial charge 
${\cal Q}_A^{\text{reg}}=2{\cal M}+2{\mathsf A}_3+1$.
\index{QAreg@${\cal Q}_A^{\text{reg}}$}
\index{QA@$:{\cal Q}_A:$}
Only the latter states are physical because the former states are
not gauge-invariant\footnote{One might question that $\theta$ states
are more physical than axial states. This question is interesting
but hypothetical
since only for massless fermions
does this question arise.  We do not observe massless particles which bear colour.}.  
All eigenstates of the Hamiltonians under consideration
can now simply be obtained by applying the creation operators 
$\Bbb{A}^{\dag}_{\,\vec P}$
onto the respective vacua. 
A corrolary of theorem~\Ref{VakTh} reads as follows
\begin{Theorem}[Costituent Picture]
The eigenstate 
$\Bbb{A}_{\;\vec P}^{\dag} \cket{\Omega_n:0}$ of $H_{\text{reg}}$ 
contains either right-movers and zero modes
only ($P^3>0$), left-movers and zeromodes only ($P^3<0$) or
pure gauge configurations only ($P^3=0$).
\end{Theorem}
In other words: the state of a Schwinger boson is a 
\Index{topological glueball} in the
rest-frame and a {\it pure two-fermion state} in all other frames 
as long as $M_B\ll \triangle k$ and as long as this state is an eigenstate
to $\hat {\cal M}$ rather than to ${\cal Q}_A^{\text{reg}}$.
Proof: the state
\begin{equation}
\begin{split}
\Bbb{A}_{\;\vec P}^{\dag} \cket{\Omega_n:0}
&=
{1\over\sqrt{P^3}}
\sum_{1\le k^3 \le P^3} a_{R,k^3}^{\dag}a_{R,k^3-P^3}\cket{\Omega_n:0} \\
&=
{1\over\sqrt{P^3}}  
\sum_{1\le k^3 \le P^3} b_{k^3} ^{\dag}d_{P^3-k^3} ^{\dag}\cket{\Omega_n:0}
\end{split}
\end{equation}
contains only right-movers for $P^3>0$ as is easily seen
in Manton's a-basis. This is illustrated in \figRef{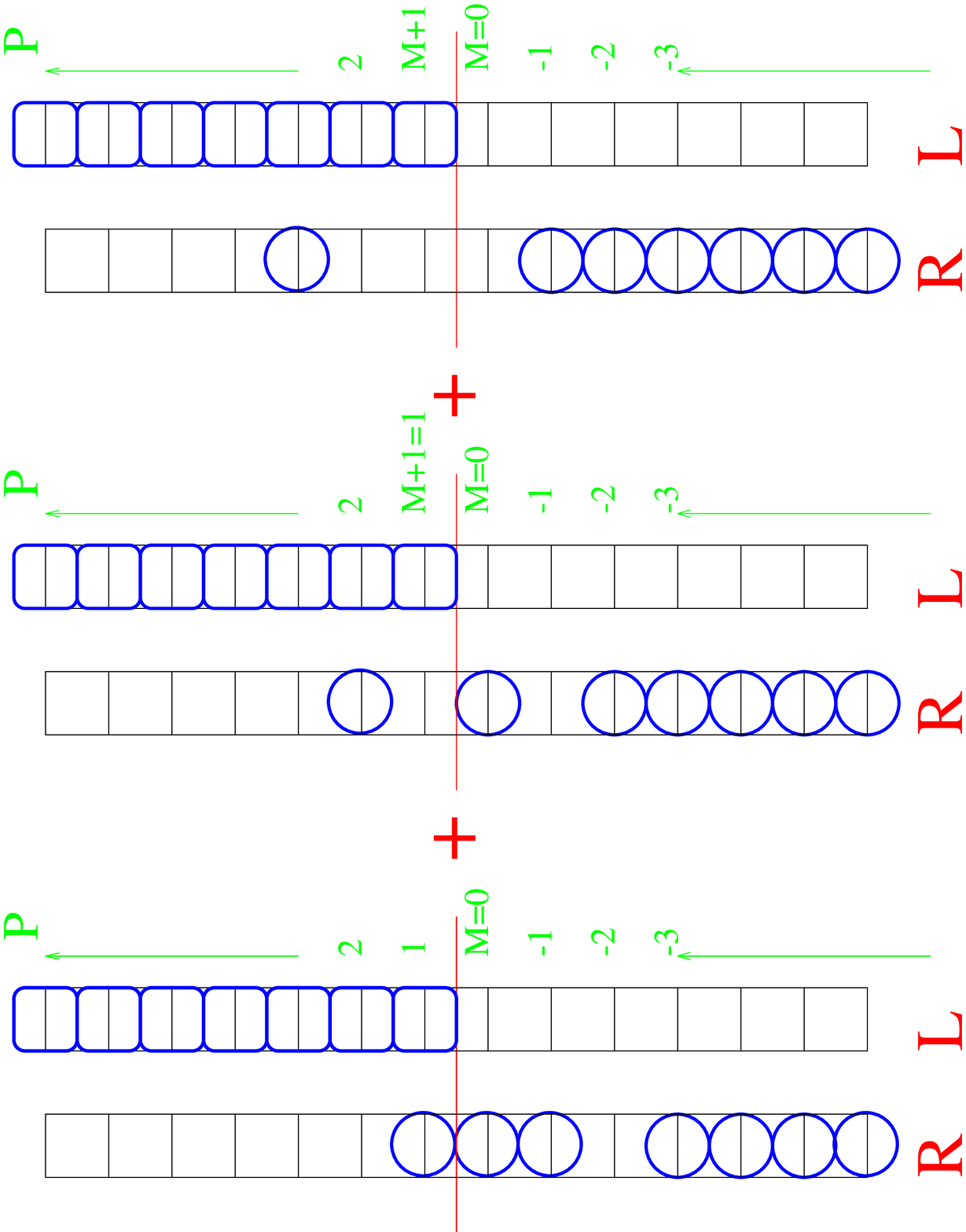}
for the case $P^3=3$. Analogously for $P^3<0$. 
The vacuum $\cket{\Omega_0:0}$ does not contain fermions. Neither does
a boson at rest
$\Bbb{A}_{\,0} ^{\dag}\cket{\Omega_0:0}=\cket{\Omega_1:0}$.
The vacuum state and bosons at rest are purely gluonic except that 
the presence of fermions is responsible ---via the anomaly---
for the potential energy
$(\tilde A+1/2)^2$ of the boson. 
The  potential energy
$(\tilde A+1/2)^2$may thus be considered
as an effective interaction due to the presence of fermions. 
These features bear some resemblance with zero-mode dominance of QCD glueballs 
described by Van Baal et
al. ~\Cite{Koller:1987yk,vandenHeuvel:1994ah}.

The ${\cal M}=0$ vacuum and the bosonic states are even simpler for the axial
or the artificial Hamiltonian: Their ${\cal M}=0$ vacuum is the perturbative
vacuum $\cket{0}$ and their Schwinger boson state
\begin{equation}\label{SchwingerBosonState}
\begin{split}
\Bbb{A}_{\;\vec P}^{\dag} \cket{0}
=
{1\over\sqrt{P^3}}  
\sum_{1\le k^3 \le P^3} b_{k^3} ^{\dag}d_{P^3-k^3} ^{\dag}\cket{0}
\end{split}
\end{equation}
is built on this vacuum without gluonic structure.

\section{Theta Vacua and Cut-Off Regularisation}

 \begin{table}[tr]

\begin{center}
\begin{tabular}{|l|l|}
\hline\hline
perturbative vacua $\cket{\Omega\dots}$, $L\rightarrow 0$ & 
Symbol \\ \hline\hline
perturbative axial vacuum &
$\cket{\Omega:{\mathsf A}_3,{\cal M}}$ \\
perturbative $\theta$-vacuum &
$\cket{\Omega:\tilde{\mathsf A};\theta}$ \\ \hline\hline
$n$ boson vacua, $L\rightarrow 0$ & Symbol \\ \hline\hline
perturbative axial vacuum &
$\cket{\Omega_n:{\cal M}}$ \\
perturbative $\theta$-vacuum &
$\cket{\Omega_n;\theta}$ \\ \hline\hline
artificial vacua $\cket{\Psi\dots}$, $L\rightarrow \infty$ & Symbol \\ \hline\hline
artificial axial vacuum &
$\cket{\Psi:{\mathsf A}_3,{\cal M}}$ \\
artificial $\theta$-vacuum &
$\cket{\Psi:\tilde{\mathsf A};\theta}$ \\ \hline\hline
$n$ boson vacua, $L\rightarrow \infty$ & Symbol \\ \hline\hline
full axial vacuum &
$\cket{\Psi_n:{\cal M}}$ \\
full $\theta$-vacuum &
$\cket{\Psi_n;\theta}$ \\ \hline\hline
\end{tabular}
\caption{Vacua: a systematic collection}\label{VacuumCollection}

\end{center}

 \end{table}

\pstxt
{Symbolic representation of a Schwinger boson}
{MantonExc0.eps}
{The circles represent right-handed fermions filling the Fermi
sea from below. The rounded squares
represent left-handed fermions filling the Fermi sea from above.}

Since the vacua $\cket{\Omega:{\mathsf A}_3,{\cal M}}$ of $H_{\text{reg}}$ are degenerate,
every linear combination of these vacua is again a vacuum. For $\Lambda=\infty$,
a special case of this statement are the $\theta$ vacua defined
in~\Ref{nThetaVacua}. 
States like
\begin{equation}\label{ThetaVacuaN}
\begin{split}
\cket{\Omega:\tilde {\mathsf A};\theta,N_\theta }
&=
\sum_{{\cal M}=-N_\theta}^{N_\theta}
\cket{\Omega:\tilde {\mathsf A}-{\cal M},{\cal M} } {1 \over \sqrt{2\pi} } e^{-i \theta {\cal M}}
\end{split}
\end{equation}
\index{OmegaAtheta@$\cket{\Omega:\tilde {\mathsf A};\theta }$}
are an other special case. These states are no irreducible
representations of the gauge group; for they are not eigenstates
of the gauge-transformation operator $U_G$ unless $N_\theta=\infty$. 
They may, however, be considered as an approximation of $\theta$ vacua 
in the sense that the expectation value of $U_G$ 
with respect to these states is approximately $e^{i\theta}$.
We introduce  
a \Index{cyclic group} ${\frak C}_{2N_\theta+1}$
with generator $U_G^{(N_\theta)}$
\index{UGN@$U_G^{(N_\theta)}$}
\index{C}
defined as 
\be
U_G^{(N_\theta)}\cket{\Omega:{\mathsf A_3},{\cal M}}
\dn
\begin{cases}
\cket{\Omega:\frak{m}({\mathsf A_3}-1),
             \frak{m}({\cal M}+1)} 
& \text{if $|{\cal M}|\le N_\theta$} \\
\cket{\Omega:{\mathsf A_3},{\cal M}}, & \text{else}
\end{cases}
\ee
where 
$
\frak{m}(a)\dn 
 [(a+N_\theta) \mod (2N_\theta+1)]-N_\theta
$
denotes the \Index{modulo division} which imposes the periodicity 
$2N_\theta+1$ upon the argument $a$. 
The states $\cket{\Omega:\tilde {\mathsf A};\theta,N_\theta }$
which we shall call \Index{cyclic vacua} 
form irreducible representations of this group 
if 
\be\label{thetas}
\theta=
\theta_n
\dn
{\pi n\over N_\theta+1}
\ee
with $n\in\Bbb{Z}$. 
Therefore,
${\frak C}_{2N_\theta+1}$ may be considered as a finite approximation
to the infinite gauge group. 
We shall use the irreducible representations 
of $U_G^{(N_\theta)}$ in~\secRef{CNum} for the special case $N_\theta=0$. 
{\bf
i.e. we shall consider right-movers and zero-modes only. 
According to~\Ref{thetas}, this implies that 
we shall be able to describe two $\theta$ angles: $\theta=0$ and $\theta=\pi$.}
If $N_\theta\ll\Lambda$ and if $H$ is restricted to a subspace
of n-boson states with $n|\vec P|+N_\theta<\Lambda$,
then $U^{(N_\theta)}_G$ is a genuine symmetry 
of the restricted Hamiltonian. This allows 
the infinite gauge group to be reduced in a way
suitable for numerical computations. 
\section{Conclusions for Numerical Computations}

{\bf
When diagonalising the {\em massless} 
Schwinger model, we have the choice between 
gauge-invariant $\theta$ eigenstates and gauge-variant axial eigenstates. 
The eigenstates of the {\em massive} Schwinger model, however, are necessarily
$\theta$ states. This renders numerical computations (in a partonic framework)
very difficult since $\theta$ states are states with an infinite parton
content. Replacing Coulomb gauge by axial gauge seems to solve this numerical
problem. Alas, axial gauge is unphysical and introduces spurious eigenstates. 
Herin we propose a simple numerical technique which allows us to 
identify and repair the 
damage inflicted on the mass spectrum
by axial gauge. In Chapter~\secRef{CNum}, we shall demonstrate
that and how this technique works. 
}

We have shown that the spectrum of the artificial Hamiltonian and
the heat-kernel regularised Hamiltonian are identical for all momenta
$\vec P$ except for $\vec P=0$. This fact has important consequences for 
numerical calculations in the unphysical axial gauge. 
Axial gauge destroys the invariance of the Hamiltonian under large
gauge transformations (gauge invariant regularisation does not make
sense in this gauge). Since the degeneracy of ground states of the Hamiltonian
is a result of the invariance of the Hamiltonian under large gauge transformations, axial gauge destroys the degeneracy of the axial
vacua except for ${\cal M}=0,-1$. 
States with ${\cal M}\neq 0,-1$ 
manifest themselves as spurious states
polluting the mass spectrum. 
The spectrum of states with ${\cal M}\neq 0,-1$, however,
does not contain spurious states. Hence 
it is correctly described by $H_{\text{axial}}$ (except for the rest frame
spectrum ($\vec P=0$) the only frame
affected by axial gauge). 

The discussion above suggests that the problem of spurious states
can be solved ---even for ${\cal M}\neq 0,-1$---
by the following {\em prescription} (1) Diagonalise the axial Hamiltonian
(2) subtract the energy ${\cal M}({\cal M}+1)$ of the spurious
state in the 
$:{\cal Q}_A:=2{\cal M}$ sector from the energies $E_n'$ of all other
states in this sector $E_n=E'_n-{\cal M}({\cal M}+1)$. 
This procedure
is equivalent to a computation of the mass spectrum of the artificial 
Hamiltonian and, therefore, restores the degeneracy of the mass spectrum
(except for the rest-frame $\vec P=0$). 
Furthermore, it is easy to show that the degeneracy of $2N_\theta$ states
is restored as well if the energy of the vacua~\Ref{ThetaVacuaN}
is subtracted
from the energies of states built on these vacua.
In Chapter~\secRef{CNum}, finally, we shall demonstrate that this
technique works even for the {\em massive} Schwinger model (if the fermion
masses are small) where only the $\theta$ vacua are degenerate
whereas the boson-masses acquire a $\theta$ dependence: it will be shown
that the $\theta$ dependence
of the vector boson and the scalar boson 
is correctly reproduced by the above-described technique.

Needless to say that the anomaly is irreparably destroyed by the
axial gauge if we choose the rest-frame.


\section{The Schwinger Model on a Large Lattice}

We have seen, that a \Index{constituent picture} arises
for $L\approx 0$. 
For large volumes $2L$, however, the perturbative
vacua (labeled with $\Omega$) are no longer eigenstates of the Hamiltonian. 
But it is not difficult to construct vacua belonging to large volumes
in terms of the already constructed zero-volume vacua.
The operators ${\cal A}_{\vec P}$ ($P^3>0$) can be expressed 
\begin{align}\label{AArelations}                            
{\cal A}_{\vec P}  
&=
{ \Bbb{A}_{\,\vec P} 
\left(|\vec P|+\Omega(\vec P)\right) +
  \Bbb{A}^{\dag}_{\,-\vec P} 
\left(|\vec P|-\Omega(\vec P)\right) 
\over 2\sqrt{\Omega(\vec P)|\vec P|} }    \\
{\cal A}^{\dag}_{-\vec P}  
&=
{ \Bbb{A}_{\,\vec P} 
\left(\Omega(\vec P)-|\vec P|\right) + 
  \Bbb{A}^{\dag}_{\,-\vec P} 
\left(|\vec P|+\Omega(\vec P)\right) 
\over 2\sqrt{\Omega(\vec P) |\vec P|} }
\end{align}                                                                                                                        %
in terms
of the $L\rightarrow 0$ basis and
therefore, a finite-volume vacuum $\cket{\Psi\dots}$ 
(all these vacuum states are represented in \tabRef{VacuumCollection},
 e.g. $\cket{\Psi\dots}=\cket{\Psi_n:\theta}$)
corresponding to a large volume
Hamiltonian $H_{\text{reg}}$ is obtained as the solution to the equations
\index{Psintheta@$\cket{\Psi_n;\theta}$}
\be\label{PsiDef}
{\cal A}_{\vec P}\cket{\Psi\dots}\equiv 0
\ee
Since this equation allows to construct  $\cket{\Psi_0;\theta}$
in terms of  $\cket{\Omega_0;\theta}$ and the $\Bbb{A}$ operators 
and since we know how the $\Bbb{A}$ operators are built from fermions,
we are able to construct $\cket{\Psi_0;\theta}$
in terms of fermions. 
In fact we are able to construct a large-volume state (labeled with $\Psi$)
from any zero-volume state (labeled with $\Omega$)
by means of this procedure.
The large-volume state
\index{UB@$U_B(L)$}
\index{Psi@$\cket{\Psi\dots}$}
\index{Omega@$\cket{\Omega\dots}$}
\be
\cket{\Psi\dots}=U_B(L)\cket{\Omega\dots}
\ee
may formally be obtained by applying
the unitary operator $U_B(L)$
on a small-volume state $M_B L\rightarrow 0$.
$U_B(L)$ can be explicitly constructed
in terms of $\Bbb{A}_{\vec P}$ operators. But for our purposes,
we do not need the explicit
expression;
what we are interested in are fermion distribution functions.
For an explicit construction, the reader is referred to 
Ref. ~\Cite{Bogalyubov,BogalyubovR}. 
The reason for this simple relation between zero-volume
and large-volume states is the fact that the
spectrum of the Schwinger model is the spectrum of a free theory. 
Note that for $L=\infty$ neither the ${\cal A}$ basis nor 
the $U_B(L)$ operator are well defined
in terms of fermionic degrees of freedom. This is an illustration
of Haag's theorem~\Cite{HaagsTheorem}.

\section{The Vacuum Distribution Function:\\ Axial Vacua}

Let  $\cket{\Psi\dots}$ be one of the {\em axial} vacua 
listed in \tabRef{VacuumCollection}. 
It is now elementary to calculate the distribution of $\Bbb{A}$ bosons
inside the vacuum $\cket{\Psi\dots}$
\begin{eqnarray} \label{BosonNumberInVacuum}
h(\vec k)\dn                                   
 \brac{ \Psi\dots } \Bbb{A}^{\dag} _{\,\vec k} \Bbb{A}_{\,\vec k} 
\cket{ \Psi\dots }
=
{(|\vec k|-\Omega(\vec k))^2 \over 4|\vec k|\Omega(\vec k)}
\end{eqnarray}  
or inside the large-volume Schwinger boson 
${\cal A}^{\dag} _{\,\vec P}  \cket{ \Psi\dots }$
\pstxt{The meson cloud}{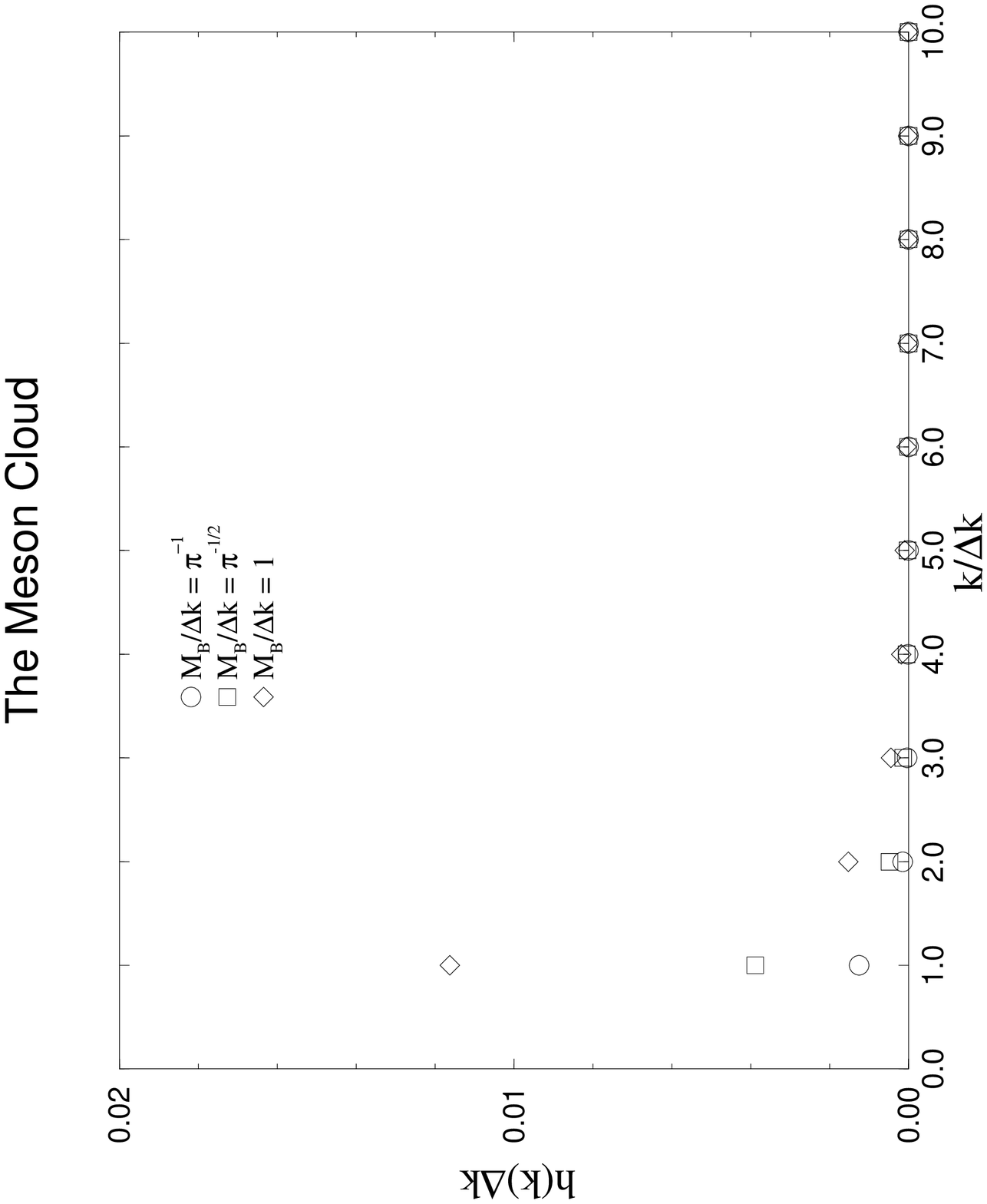}
  {The probability of finding a $\Bbb{A}$ boson inside the vacuum for 
   lattices with small but finite size.}
\index{h@$h$}       
\begin{eqnarray} 
h(\vec k;\vec P)\dn                                   
 \brac{ \Psi\dots } {\cal A}_{\,\vec P}
\;
\Bbb{A}^{\dag} _{\,\vec k} \Bbb{A}_{\,\vec k} 
\;
{\cal A}^{\dag} _{\,\vec P}
\cket{ \Psi\dots }
=
\delta(\vec k-\vec P)+h(\vec k) 
\qquad.
\end{eqnarray} 
%
%
The \Index{parton distribution} of quarks and anti-quarks
inside the zero-volume state $\Bbb{A}_{P^3}\cket{\Omega_n:{\cal M}}$, in turn, 
can be calculated 
\begin{equation} 
\begin{split}
{\mathsf f}(\vec k,P^3;\Omega:{\cal M})\triangle k^3
&\dn
\brac{\Omega_n:{\cal M}}
\left[
 b^{\dag}_{\vec k}b_{\vec k}
-d^{\dag}_{-\vec k}d_{-\vec k}
\right]
\cket{\Omega_n:{\cal M}}  \\
&=
\brac{\Omega_n:{\cal M}}
\left[
 a^{\dag}_{R,\vec k}a_{R,\vec k}
+a^{\dag}_{L,\vec k}a_{L,\vec k}
-1
\right] 
\cket{\Omega_n:{\cal M}} \\
&=
\theta({\cal M}+1\leq k \leq {\cal M}+P^3)/P^3 \\
&-
\theta({\cal M}\leq k \leq {\cal M}+P^3-1)/P^3
\end{split}
\end{equation}   
using Mantons solution (graphically
represented in \figRef{MantonExc0.eps}).
We also need the \Index{raw distribution functions} for ${\cal M}=0$
\begin{equation}\label{SchwingerRawDist}
\begin{split}
f(\vec k,P^3;\Omega:0)\triangle k^3
\dn
\brac{\Omega_n:0}b^{\dag}_{\vec k}b_{\vec k}\cket{\Omega_n:0}
&=
\theta(1\leq k \leq P^3)/P^3 \\
\end{split}
\end{equation}   
and the corresponding raw distribution of an anti-quark
\begin{equation} 
\begin{split}
\bar f(\vec k,P^3;\Omega:0)\triangle k^3
\dn
\brac{\Omega_n:0}d^{\dag}_{\vec k}d_{\vec k}\cket{\Omega_n:0} 
&=
\theta(0\leq k \leq P^3-1)/P^3 \\
\end{split}
\end{equation}   
which follow from Eq~\Ref{SchwingerBosonState}.
Here, $\theta(x)$ denotes the Heaviside step function. 
\index{thetax@$\theta(x)$} 
The quark distribution inside the {\em large volume} vacuum 
$\cket{\Psi_0,0}$ is approximately the convolution
\be\label{FuerCNum}
f(\vec k;\Psi:0)
\approx
\sum_{\vec P}\triangle k^3\;
h(\vec P) f(\vec k;P^3:\Omega:0)
\ee
of these
two functions. 
If $h(\vec P)$ is small so is the convolution. We use
this fact in order to demonstrate that the approximations
utilised in Chapter~\secRef{CNum} are accurate, i.e. we can easily show now
that the vacuum is almost trivial for $g=\triangle k$
a value of g that we shall use in Chapter~\secRef{CNum}.
The graph of $h$ is represented in \figRef{vacPi.ps} for $g=\triangle k$.
We conclude that smallness of $h$ justifies the assumption
of Chapter~\secRef{CNum} that the vacuum is almost
trivial. 
Moreover, we conclude from Eq~\Ref{BosonNumberInVacuum} that
the total number of particles in the vacuum
${\cal N}_{\text{vac}}\dn \sum_{\vec k} h(\vec k)$
diverges in the limit $L\rightarrow \infty$ 
(i.e. $M/\triangle k\rightarrow \infty$): another illustration
of Haag's theorem. It is useless 
to start with $L=\infty$ and ---strictly speaking--- even wrong.

\section{The Vacuum Distribution Function:\\ Gauge Invariant Vacua}

So far we have been calculating distribution functions
in axial states. Physical states, however, are gauge-invariant. 
In a {\em gauge-invariant} $\theta$ state,
the expectation value of the particle number operator 
$
a^{\dag}_{\alpha,\vec k}a_{\alpha,\vec k}
$
is no longer well-defined 
because the number of partons is no gauge-invariant quantity. 
In order for distribution functions of $\theta$ vacua to 
be well defined, they have to be defined relative to the fermion level,
i.e. via the gauge invariant particle number operators
$
a^{\dag}_{\alpha,\vec k+{\cal M}}a_{\alpha,\vec k+{\cal M}}
$
or
$
a^{\dag}_{\alpha,\vec k-{\mathsf A}_3}a_{\alpha,\vec k-{\mathsf A}_3}
=
a^{\dag}_{\alpha,\vec k+{\cal M}-\tilde {\mathsf A}}
       a_{\alpha,\vec k+{\cal M}-\tilde{\mathsf A}}
$.
In the first case, the distribution functions of $\theta$ states
are identical to the distribution functions in states with ${\cal M}=0$:
the calculation of gauge-invariant distribution functions
in a $\theta$ state can thus be replaced by the simpler
calculation of a gauge-dependent distribution function
in a axial state. 
In the second case, the distribution functions of $\theta$ states
are smeared through the oscillation of $\tilde A$ with an oscillation
amplitude of roughly $\triangle \tilde{\mathsf A}\approx M_B$. 

In fact, structure functions are related
to distribution functions of the second type
(cf. the remark on p.\pageref{dR}) if axial gauge cannot be chosen:
Therefore, gauge-invariant distribution functions of theta states
are obtained 
by "smearing out" gauge-variant distribution functions of axial states.  
These considerations illustrate nicely why the inclusion of a gauge string $U_A$ 
into the correlation function $\tilde\mathsf{q}_\mu(y|PS)$ is necessary.

\proPub
{
\section{Epilogue: The Front Form and the Schwinger Model}

If the Schwinger model is quantised in the FF, only right-movers are
present. We have already demonstrated that this quantisation
is not equivalent to $QED$ but rather to chiral $QED$
in zero volume. In this theory, $:{\cal Q}_A=:{\cal Q}:=0$. Therefore,
$\theta$
vacua are impossible to implement in this approach\footnote{I.e. they
cannot be implemented using {\em periodic BCs}},
as they constitute
superpositions of states with different axial charge ${\cal Q}_A$.  
McCartor and Robertson~\Cite{McCartor:1997pe} treated the Schwinger model by
quantising the fermionic fields on two light-like quantisation
surfaces. They quantised right-movers on the quantisation surface
$x^+=0$ and left-movers on the quantisation surface $x^-=0$. 
This allowed them to include $\theta$ vacua. They found, however,
that the condensate becomes zero in the limit of infinite light-like 
periodicity. 
We are now able to explain this puzzle. 
Consider the quantisation surface used by McCartor and Robertson
depicted in~\figRef{AdmissibleBCs}. Also depicted is an equivalent
quantisation surface which we shall prefer for reasons that will
be come clear soon. 
The proper way 
of quantising on a 'bent' quantisation surface is using bent co-ordinates
$\check x^\mu$
defined as
\be
\check x^3\dn x^3
\qquad
\text{and}
\qquad
\check x^0=
\begin{cases}
x^+ &\text{if } x^3<0\\
x^- &\text{if } x^3\ge 0
\end{cases}
\ee
in terms of which the bent quantisation surface reads $\check x^0=0$. 
\begin{center}
 \begin{figure}[t]
\hbox{
\psfig{clip=,figure=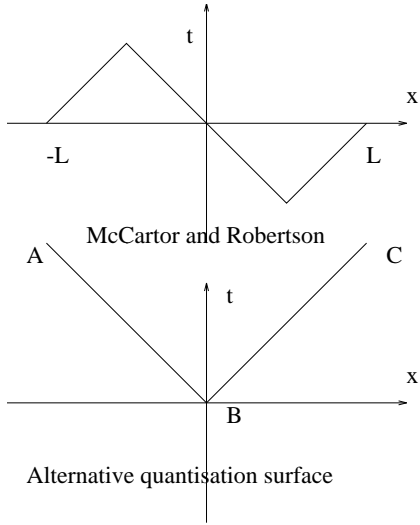,width=0.4\linewidth,angle=-90}  
     }
\caption{Admissible BCs for the FF}\label{AdmissibleBCs}
 \end{figure}
\end{center}
Now we have to impose boundary conditions upon the fields $\varphi$. 
If we require periodicity in $\check x^3$ then this periodicity 
is {\em space}-like because the four-vector $2{\frak{L}}$
which joins $A$ and $B$ is space-like.
As such, ${\frak{L}}$ defines a finite, non-zero renormalisation
scale. Quantisation
in a space with space-like periodicity, however, is most conveniently
performed in the IF formalism; quantisation in bent co-ordinates
is much more cumbersome to put it mildly. 

One might as well ---following McCartor and Robertson--- introduce TWO
{\em light}-like BCs on the right-moving and the left-moving component
respectively joining the point $O$ with the points $A$ and $B$ respectively. 
This defines two {\em light}-like boundary four-vectors. Consequently, 
one is working in an infinitesimally small box. 
Such a theory is not equivalent to the IF as such. 
Such a theory is equivalent to the \Index{strong IMF} instead, i.e.  
to the IF on an infinitesimally small lattice $L=0$. 

In the Schwinger model, the kinetic energy 
does not necessarily render left-movers infinitely heavy in the IMF. The reason
for this is that the kinetic energy $T$ is replaced by the modified
kinetic energy 
${\frak{T}}$
which is zero for every axial vacuum even though axial vacua may
contain an arbitrary number of left-movers.
Consequently, the FF Schwinger model is not an effective theory 
for the IMF Schwinger model because
an effective theory 
for the IMF Schwinger model must not exclude the left-movers which
are responsible for the $\theta$ vacuum. 
What is the effective Hamiltonian then? Even though we cannot eliminate
left-movers, we are able to eliminate every operator which
{\em mixes} left-movers and right-movers. 
The justification for this is 
the fact that axial vacua $\cket{\Omega_0:{\cal M}}$ 
are exact solutions to the zero volume Schwinger model,
as we have already shown. This means that operators which enforce
mixing of left-movers and right-movers are completely suppressed in the limit 
$L\rightarrow 0$:
From these considerations we conclude that
the effective IMF Hamiltonian for the Schwinger model is the IMF Hamiltonian
with left-right-mixing operators removed. Once these terms are removed, 
the plus-component $P^+(e^{(0)})$ of the IF quantised Hamiltonian
consists merely of right-movers and the minus-component $P^+(e^{(0)})$ consists
merely of left-movers. 
This effective Hamiltonian has now exactly the same structure as a theory constructed 
on two light-fronts~\Cite{McCartor:1997pe}.
This also explains why this theory is unable to describe the 
axial condensate of the Schwinger model. The condensate is due to
the fluctuating particle number in the vacuum
generated by the Bogolubov transformation
$U_B(L)$: it can only be
correctly described on a large volume $L\rightarrow \infty$. 
The volume of the strong IMF, however, is infinitesimally small and so
is the volume of the two-light-front theory~\Cite{McCartor:1997pe}. 
We shall elaborate these arguments in a later publication. 
We summarise: {\bf
The two-light-front theory is equivalent to the zero volume
Schwinger model whereas the FF is equivalent to the chiral zero volume
Schwinger model.} 

{\tiny\ttfamily
\begin{verse}
Bornons ici cette carri\`ere.\\
Les longs ouvrages me font peur.\\
Loin d'\'epuiser une mati\`ere\\
On n'en doit prendre que la fleur.\\
Il s'en va temps que je reprenne\\
Un peu de forces et d'haleine\\
Pour fournir \`a d'autres projets.\\
\end{verse}
[La Fontaine: Livre VI]
}

}

%% file: CNum.i
 

In this chapter we diagonalise the Hamiltonian of the massive Schwinger model
in a frame where the \Index{vector boson} (i.e. the lightest physical particle)
with four-momentum 
\be
P=(E,0,0,P^3)
\ee
\index{P@$P$}
moves at a velocity $v^3=P^3/E$ close to the
speed of light. This frame will be referred to as 
\Index{large momentum frame} (LMF).
\index{LMF}
\index{v@$v$}
We compare our results with the computations of Hamer et al.~\Cite{Hamer:1997dx}
who used the Kogut-Susskind Hamiltonian 
\footnote{For the first computations with Wilson fermions 
see~\Cite{Fang:1992bi,Luo:1990cb,Luo:1990fc,Chen:1990ej}.
Since we are working in momentum space, we do not have to deal 
with fermion doubling. 
}.
The LMF has several advantages. 
{\em Firstly}, the \Index{radius} $R$ 
\index{R@$R$}
of the lightest physical particle is Lorentz-contracted
to
\be
R'=\sqrt{1-v_3^2} \;R
\ee
which is smaller than the \Index{correlation length} $a\xi ={1\over M}$
\index{xi@$\xi$}
if the velocity $v^3$ is sufficiently high
(The actual size of $R$ is not important for our argument). 
$M=M_V$ is the mass of the vector boson.
\index{M@$M$}\index{MV@$M_V$}
In the rest-frame, in contrast, 
the correlation length of a particle is usually smaller than its extension. 
Therefore, the physical requirement that {\em both} the correlation length 
{\em and} the physical size of the boson be smaller 
than the size $L$ of the lattice
 \be
  a\xi ,\;R < 2L
 \ee
may be replaced by the much
weaker condition
\be
a\xi < 2L
\ee
which allows us to use a \Index{lattice size} $L$ 
\index{L@$L$}
much smaller than if we had
chosen the rest-frame--- albeit not an infinitely small one. 
This advantage is crucial since pair-creation  
leads to a finite density of \Index{virtual particles} in any 
interacting field theory. The number of virtual particles is thus roughly
proportional to the volume
and, consequently, the total number of particles is drastically reduced
if the volume $2L$ is small. 
A {\em second} advantage is the fact that most constituents 
inside a fast-moving object can be expected to move in the same 
direction as the fast moving object itself {\em as long as}
the number of  \Index{virtual particles associated with the vacuum} 
can be neglected when compared to the number of virtual particles
associated with the moving object. 
Under such circumstances, the neglection of \Index{left-movers}, i.e.
particles moving in the direction of the negative 3-axis, may thus
be a very good approximation. Having no left-movers,
in turn,  
constitutes a {\em third advantage} since it renders the number
of states in the Fock space finite (except for bosonic zero-modes) as 
has been explained
in Ref.~\Cite{Kroger:1997hj,Kroger:1997jm}. 
We are going to show now that this approximation
results indeed in an excellent description
of the Schwinger-boson mass as well as of the condensate of the massless 
Schwinger model.  

The defining Lagrangian density of the massive Schwinger model is
\be
{\cal L}
=
\bar \psi(i\not\!\!D+m)\psi
-
{1\over 4g^2}{\sf F}^{\mu\nu} {\sf F}_{\mu\nu} 
\ee
where $m$, $g$ and $D=\partial+i{\sf A}$ denote the mass of the fermions, the coupling constant
and the covariant derivative, respectively.
\index{m@$m$: ! mass of a virtual fermion in the Schwinger model}
\index{g@$g$: coupling constant in the Schwinger model}
\index{D@$D$: covariant derivative}
 ${\sf A}^\mu$ is the gauge-field
connection and ${\sf F}$ the corresponding field-strength. We have 
absorbed 
\be
{\sf A}^\mu=g A^\mu
\ee
the coupling constant $g$ in the photon fields $A^\mu$. 
Canonical quantisation in axial gauge $A^3=0$ yields the Hamiltonian
\begin{eqnarray}
H
=
\int_{-L} ^{L}dx^3 (
\bar\psi \gamma^3 i\partial_3 \psi
+
m\bar\psi \psi
)
+
{g^2\over 2} \int _{-L} ^{ L} dx^3
(\psi^{\dag} \psi){1\over -\partial_3 ^2}(\psi^{\dag} \psi)
\end{eqnarray}
where we have introduced a finite lattice size $2L$. Periodic boundary
conditions are assumed from now on. 
Strictly speaking, axial gauge is incompatible
with periodic boundary conditions
since the global Wilson loop
\be
\exp(i\int_{-L} ^{L} dx^3{\sf A}^3)
\ee
is a physical observable. 
It will turn out, however, that removing this physical degree
of freedom does not hamper the computation of the mass-spectrum and
the vacuum condensate if the lattice size is sufficiently small. 
The deeper reason for this is the fact
that topological effects are less important when working in tiny volumes
as we have already demonstrated in Chapter\secRef{CSchwinger} 

We expand the fermionic fields
\begin{equation}
\psi(t=0,\vec x )
=
\sum_{\vec k}
{1\over 2\sqrt{L \omega(\vec k)} }
(
u_{\vec k} b_{\vec k}\; 
 e^{+i\vec k\cdot \vec x}
+
v_{\vec k} d^{\dag} _{\vec k}\; 
 e^{-i\vec k\cdot \vec x}
)
\end{equation}
in the usual way in terms of annihilation and creation operators
where 
\be
\omega(\vec k)
=
\sqrt{m^2+\vec k^2}
\ee
spinors appearing in this equation are defined as solutions of the free, massive 
Dirac equation with the normalisation: 
\be
\bar u_k u_k =2m
\ee
\be
\bar v_k v_k =-2m                \quad.
\ee
We diagonalise the Hamiltonian in the $\vec P=0$ sub-sector of the
total Fock space and obtain
the vacuum energy $E'_0$. This sub-sector is spanned by a finite
number of Fock-states since we have
\begin{enumerate}
\item
truncated the total Fock space to states that do not contain 
any left-moving particles with $k^3<0$ 
\item 
at most $N=P^3/\triangle k$ particles with $k^3>0$
are able to share the total momentum $P^3$
\item
there is only a finite number (i.e. two) of fermionic zero-modes due
to the Pauli exclusion principle 
\item
we have removed the bosonic zero-mode by our (unphysical) choice of axial gauge
(contrary to fermions, there may be an arbitrary numbers of bosons at $\vec k=0$).
\end{enumerate}
In fact, there are two vacua corresponding to theta angles
$\theta=0$ and $\theta=\pi$.
\index{theta@$\theta$}
If we had restricted the fermion momenta to the region
\be
-n\triangle k \le k^3\le P^3+n\triangle k
\ee
we would have had 
$2n$ $\theta$-vacua (or rather approximants thereof, cf. ~\Ref{ThetaVacuaN}). No 
restriction on the momenta at all  
yields an infinity of them (the number of 
these vacua is, of course, equal to the number of distinct values that the na{\"\i}ve axial charge
can take on). 
Here, we shall  primarily be concerned with  
the case $\theta=0$ as our approximation is able to
describe the sector $\theta=\pi$ 
for small fermion masses only.

Numerical diagonalisation of the Hamiltonian for the momentum $\vec P$
yields the energy spectrum $E' _n$ (relative to the energy of the
perturbative vacuum).  
The physical energies are now 
\be
E_n(\theta)=E' _n(\theta) - E'_0(\theta)
\ee
where $E'_0(\theta)$ is the energy of a $\theta$-vacuum. 
The physical masses, in turn, 
are obtained from the relativistic dispersion relation
\be
M_n=\sqrt{E^2 _n-\vec P^2}
\ee
Let $M=M_1={1\over a\xi}$ be the mass of the \Index{vector boson}, i.e. 
the boson with smallest mass. 
$\Lambda={\pi\over a}$ is the cut-off and $a\rightarrow 0$ is the 
lattice-spacing. 
As in the case of the massless Schwinger model, we are allowed to
the continuum-limit $a\rightarrow 0$. The
only renormalisation necessary in the {\em super-renormalisable} Schwinger model
is the subtraction of the vacuum energy $E'_0$. 

\section{Scaling Window and Region of Validity}

On a finite lattice, not every pair of parameters $m,g$ is physically 
meaningful. 
It is well-known that 
the computed mass spectrum is physically meaningful~\Cite{MontvayMunsterBuch} only
if the correlation length
$a\xi$ is smaller than the lattice size $2L$ and larger than the lattice spacing
$a$. Therefore, the parameters $m$ and $g$ have to be tuned such that 
the computed correlation length lies inside the scaling window
\be\label{ScalingWindow}
a < \xi a < 2L  \qquad.
\ee
We have set $a=0$ and 
therefore the scaling window seems to be open 
\be
\xi a < 2L
\ee
in the ultra-violet region 
\footnote{(it is remarkable that --- in contrast to
the $\varphi^4$ model--- the cut-off $\Lambda=\pi/a$ cannot be made smaller
than $2P^3$ without a massive deterioration of the numerical results)}. 
The physical region in the parameter space, however, is smaller than
the scaling window~\Ref{ScalingWindow} since we have used the 
{\em approximation}
that left-moving particles are not important--- an assumption
which can only be justified if the velocity $v^3=P^3/E$ of the boson is sufficiently close to the velocity
of light, 
i.e. if $Ma=\xi^{-1}\ll aP^3$.
%
%
Therefore, our results can only be trusted inside the {\it accuracy window} 
\be\label{Accuracy}
(1/P^3) \ll \xi a < 2L \quad.
\ee
This constraint is still not strong enough, however, since it only ensures
the absence of left-moving  
virtual particles associated with the moving object. Since the vacuum
in interacting 
relativistic field theories is not empty, there are always 
\Index{virtual particles associated with the vacuum} (rather than with the moving object itself). In particular, the parity invariance of the vacuum
implies
that it contains an equal number of 
right-movers and left-movers. The mere 
presence of a fast-moving physical particle 
does not suffice to modify the properties of the vacuum,
let alone render it trivial.
This is why we choose the length $2L$ of the system as small 
as possible such that the number of particles associated with 
the vacuum is negligible or such that at least most particles associated with
the vacuum are concentrated at $\vec k=0$. 
The choice $L=\infty$ right from the onset is 
impossible~\Cite{HaagsTheorem} and un-necessary. 
In the context of lattice gauge theory one
works in a volume $2L$ which is
large enough to comfortably contain the physical object of diameter $2R$ and
yet small enough for practical computations. Our
choice of the moving frame allows us to boost the size $2R'$ of the 
physical object to arbitrarily small values thus enabling us to further reduce
the number of vacuum particles dramatically when compared to what is
possible in the rest-frame.   

Naively we could even reduce the number of particles in the vacuum to zero
 (and all extensive quantities with it)  by the extreme choice of $L\approx 0$
which seems to be legitimate in the 
\Index{infinite momentum frame}
(IMF) $v^3\rightarrow 1$ where the Lorentz-contracted \Index{size} $R\sqrt{1-v_3^2}\rightarrow 0$ of 
any physical object vanishes(cf. Chapter\secRef{CFF}).
This, however, would be 
in blatant contradiction with the scaling window~\Ref{ScalingWindow} which requires that the size of the lattice be larger than $1/M_1$. 
To our knowledge,
this deficiency
of the IMF has nowhere
been discussed in the literature. 
We shall 
present our results for a coupling of $g=\triangle k=\pi/L=1$ corresponding
to a Schwinger-boson mass of 
\be
M_B={g\over \sqrt{\pi} }\approx 0.56
\ee
\index{MB@$M_B$}
which minimises the volume and yet is compatible with the scaling window: 
The volume
$2L=2\pi$ is then small enough to drastically reduce the number of particles
in the vacuum and yet sufficiently large for the correlation length 
$1/M_B\approx 1.77<2L=2\pi$ to lie comfortably 
inside the scaling window ~\Ref{ScalingWindow}: 
We have already demonstrated in chapter\secRef{CFF} that the vacuum of the exact solution to the massless Schwinger model
is indeed almost trivial
for $\triangle k=\pi/L \approx g$. 
Since the number of virtual particles in the vacuum
must decrease when they become heavier, 
it follows that the vacuum is also trivial for $m/g>0$ ($g\approx\triangle k$).
For $m/g>0$ the vector boson mass $M_1(m,g)/\triangle k$ increases and we could
further reduce the volume in physical units
$2LM_1=2L/(a\xi)$ (i.e. in units of $a\xi$)  
by choosing an even smaller
value of $g$ without leaving the window.
Indeed, fixing $M_1(g)$ at the
value of $M(g,m)\approx M_1(g,0)=\triangle k/\sqrt{\pi}$ accelerates the convergence
of physical quantities for larger fermion masses. At given
$m/g\approx 0.25$, for instance,
the convergence is approximately $2.3$ times better for $g=1/\sqrt\pi$
than for $g=1$
\be
{M^{(N=384)} _1 (g=1.0) - M^{(N=6)} _1 (g=1.0)
\over
M^{(N=384)} _1 (g={1\over\sqrt{\pi}}) - M^{(N=6)} _1 (g={1\over\sqrt{\pi}})
}
\approx 2.3
\ee
and the distance between $M^{(N)}(m/g=0.25,g=1/\sqrt{\pi})$ and
the corresponding mass of chiral perturbation theory
is divided by a factor of roughly three when compared to 
$M^{(N)}(m/g=0.25,g=1)$.
Inverting the function
$M_1(g)$, however, is necessary for this purpose and  
requires several self-consistent diagonalisations for each value
of $m/g$ instead of only one.
As our results are already excellent for fixed $g\approx 1$,
we prefer the straight-forward choice $g=1$ and do not seek to further increase 
precision by fixing $M_1(g)\equiv \triangle k/\sqrt{\pi}=\text{const}$. 
We observed that for masses as small as $m/g=0.5$ already, 
the dependence on the scale $g/\triangle k$ is almost negligible.

\section{Convergence and Covariance}

Physical particle masses are Lorentz scalars. Therefore, an important test
of a relativistic model is the approximate independence of the 
computed physical
masses of the momentum. Diagonalising the mass squared operator
in sectors of different momenta must {\em approximately} 
yield the same result. Ideally, this should hold for every momentum
$-\Lambda\ll P^3 \ll \Lambda$. In praxis, however, the 
physically relevant region of $\vec P$ may be further restricted
by additional approximations as, for instance, the assumption that only right-movers are important. 
In what follows, we shall use the requirement that physical masses be scalars
in order to localise the momenta 
for which our approximation {\it does not} hold: For if the computed masses
depend strongly 
on $P^3$ in some region of momentum space then this indicates that our approximation
cannot be trusted in this region. This is the case for the region of small 
total momenta $P^3$ (i.e. momenta that are small  
compared to the boson mass $M_n$) because left-movers do become important 
if the physical particle moves slowly.

A few comments are in order here as
care must be taken to make a clear distinction between
the \Index{continuum limit} $\Lambda\rightarrow \infty$
and what we shall call
the \Index{covariance limit} N=$P^3/\triangle k\rightarrow \infty$. 
%
In the Schwinger model, the only ultraviolet regularisation necessary
is the subtraction of the vacuum energy. Contrary to renormalisable
theories, we are able
to perform the continuum limit $\Lambda/M_1\rightarrow \infty$ without
rendering the Hamiltonian singular. This
means that the actual number of lattice points $2\Lambda/\triangle k$
we are working with is
infinite. 
If, however, only right-moving particles $\vec k\ge 0$ are important, this
implies that particles with momenta $\vec k$ larger than the total momentum $\vec P$ cannot play a r{\^o}le, either. Hence only particles lying on the
\Index{effective lattice} $0\le k^3 \le P^3$ have to be taken into account
for the {\em practical} numerical calculation. 
Consequently, the quantity which determines the {\em minimal size}
of the Hamiltonian matrix to be diagonalised (which
determines the actual numerical effort)
is the number $N+1$
\be
N\dn {P^3\over \triangle k}
\ee
of lattice points between $\vec k=0$ and $\vec k=\vec P$ whereas the
cut-off $\Lambda$ primarily influences the {\em form} of the Hamiltonian. 
For this reason we shall call the number $N+1$ of numerically
important lattice sites the \Index{effective lattice size}
\index{effective lattice size}---
a number which must not be confused with
the real lattice size 
$2\Lambda/\triangle k=\infty$. 
One should be careful not
to confuse the limit $N\rightarrow \infty$ with the \Index{continuum limit}
$\Lambda/M_1\rightarrow \infty$ or with the thermodynamic limit 
$M_1 L\rightarrow
\infty$
\footnote{The last point makes clear that our results can,
at best,
approximate
the correct results since we never perform thermodynamic limit(!)}
.

\section{Numerical Results}

We have computed the mass spectrum and the parton distributions
of QED$(1+1)$
in terms of $m/g$ in the \Index{ultra-relativistic region} $m/g\lessapprox 0.3$. 
For the sake of comparison with the results of Hamer et al.~\Cite{Hamer:1997dx} we have used
the logarithmically spaced points $m_i/g=1/8,1/4,\dots,16,32$.
We choose $g=\triangle k=\pi/L=1$ for all numerical calculations
presented below for the reasons explained in the last chapter.
The results for the FF that are used here were published first in~\Cite{Eller:1987nt}.
The generality of our computer programme allowed us to recalculate these
results in order to compare them with~\Cite{Eller:1987nt}
(thus providing us with an additional control of our software) 
and to create the figures
presented in this chapter.


%
\pstxt{Convergence of distribution functions
in the chiral limit $m/g=1/8$}{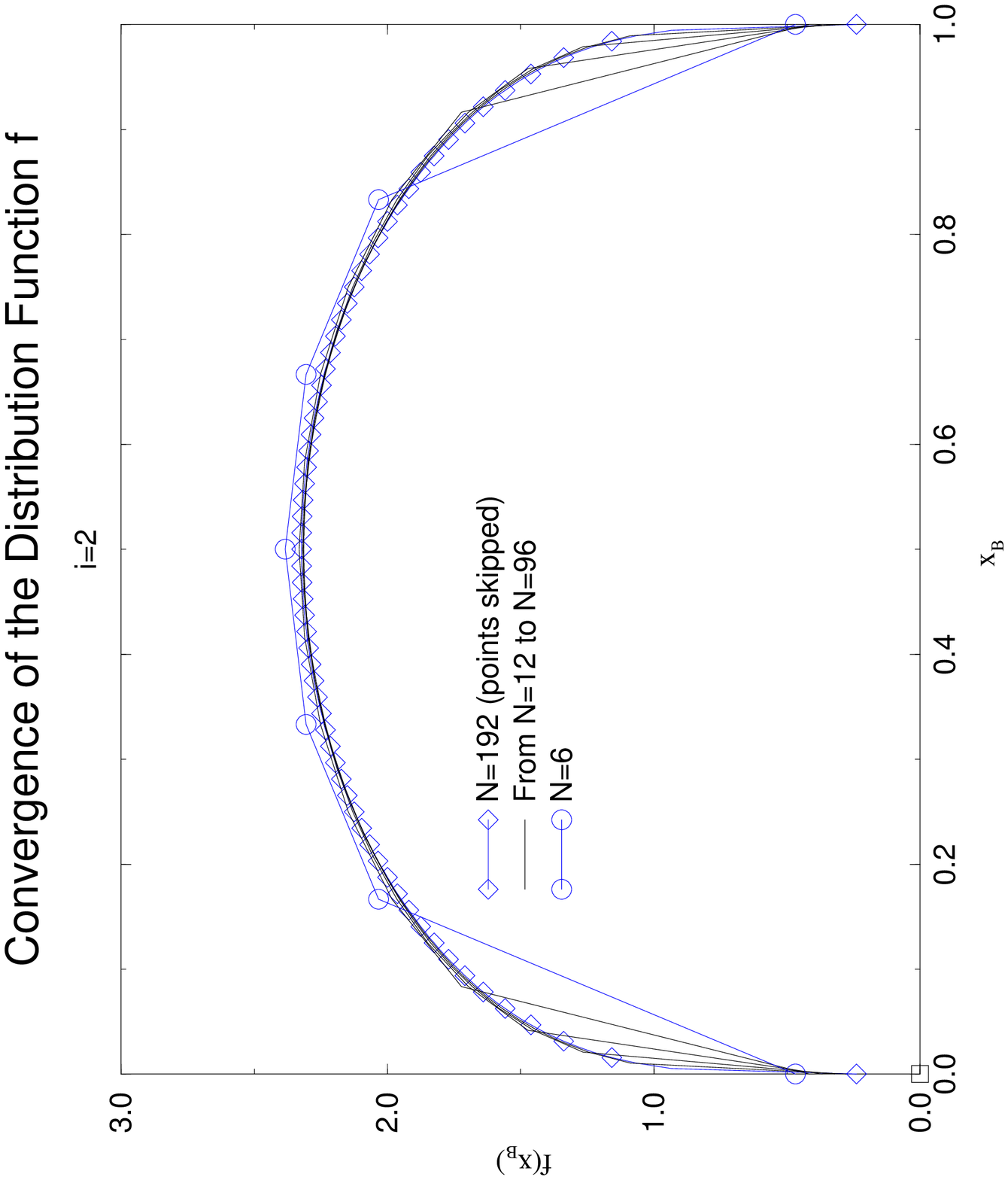}
{The LMF distribution function ${\frak{f}}(x_B,P^3)$, 
defined on the $N+1$ points of the effective 
lattice, is represented for $P^3=N=6,12,24,..,192$. 
In the \Index{covariance limit} $N\rightarrow\infty$, 
the variable $x_B$ becomes
continuous.}
\pstxt{Convergence of the distribution functions in the FF, $m/g=1/8$}{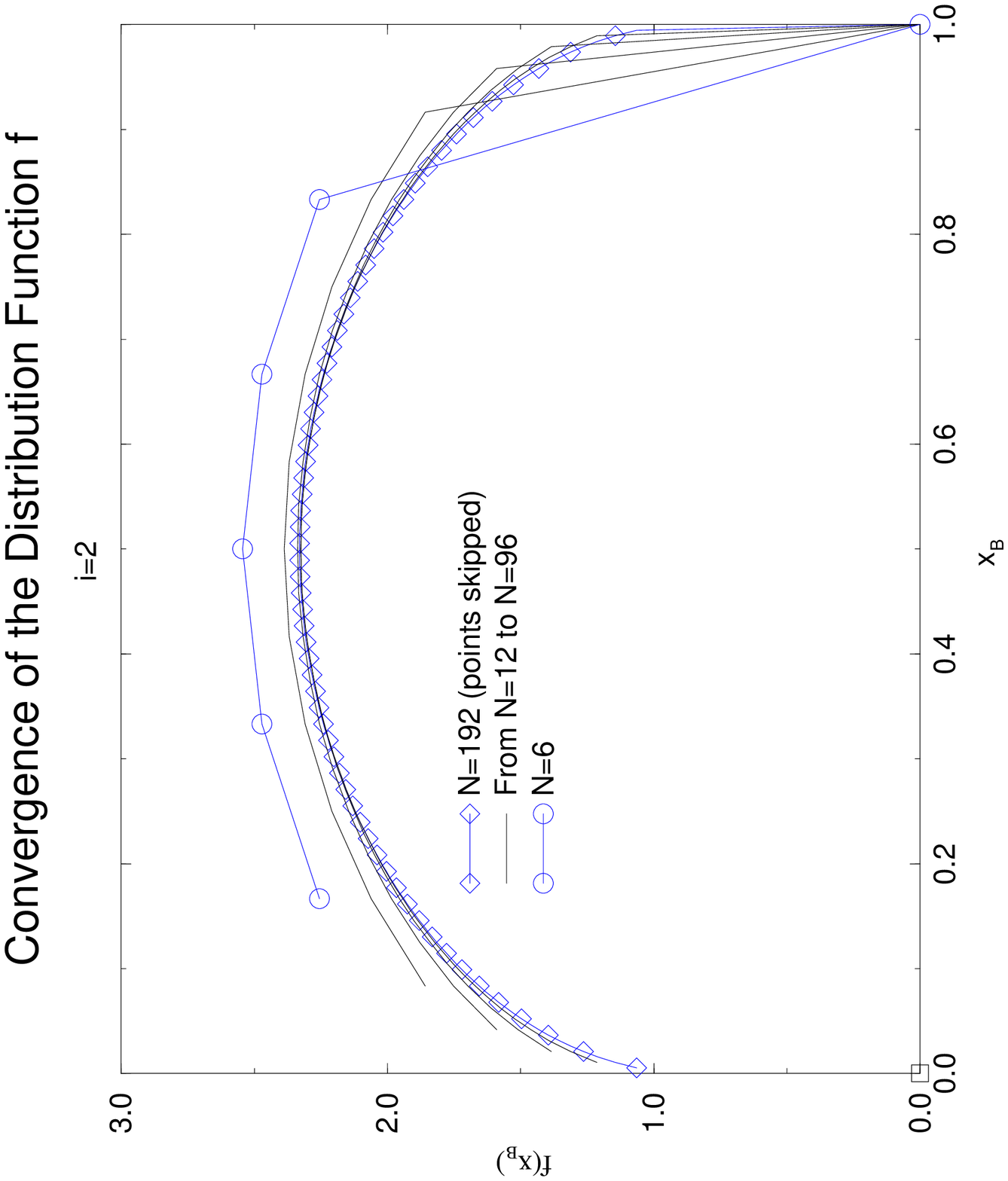}
{The FF distribution function ${\frak{f}}(x_B,P^3)$, 
defined on the $N+1$ points of the effective 
lattice, is represented for $P^3=N=6,12,24,..,192$.
In the \Index{covariance limit} $N\rightarrow\infty$, 
the variable $x_B$ becomes
continuous.}
%
%
\pstxt{The number of fermions in a boson}
{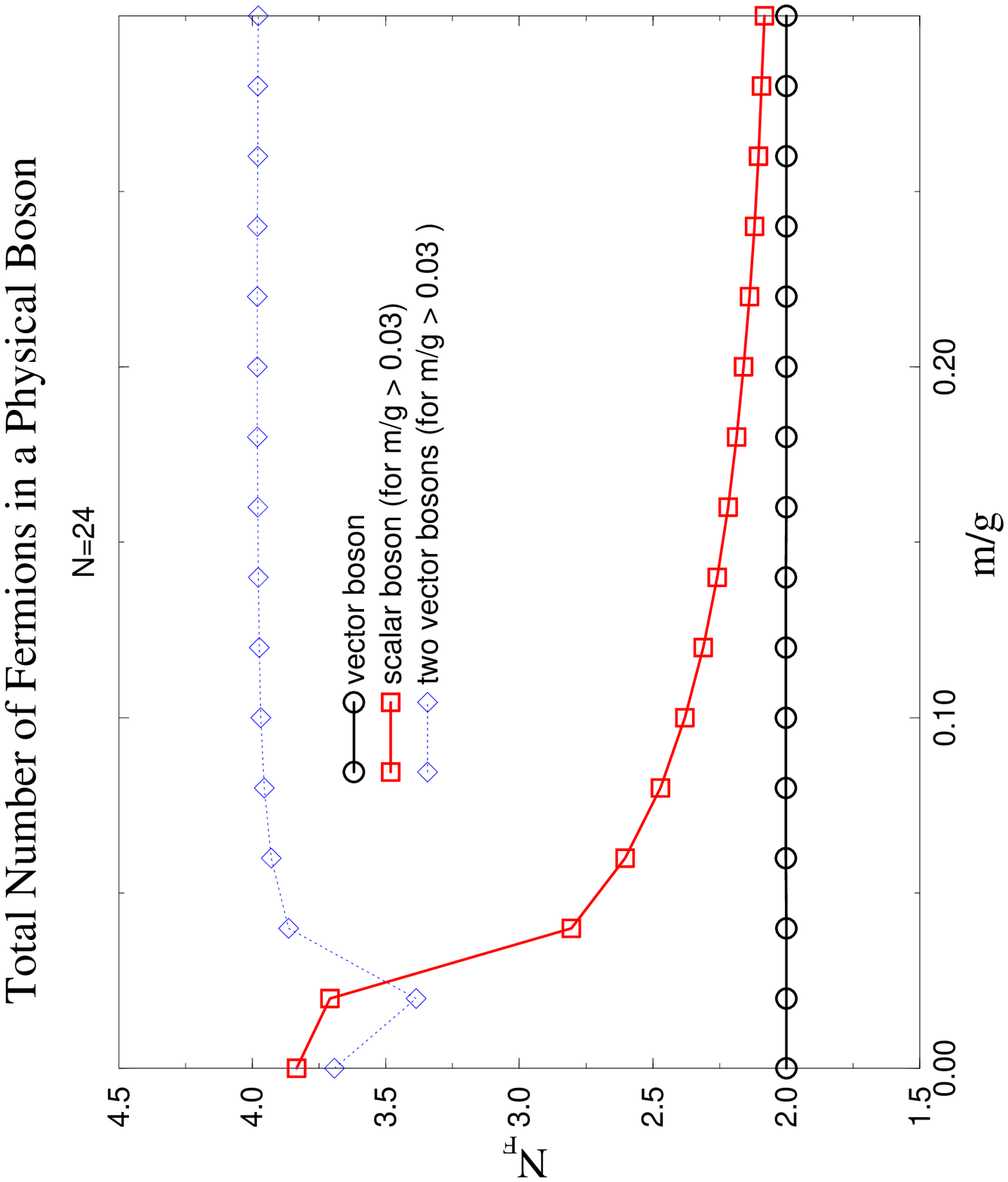}
{The number of fermions in the vector boson, the 
scalar boson and in a two-boson bound-state respectively,
plotted against the fermion mass $m/g$.}
\pstxt{The number of fermions in a boson (FF)}
{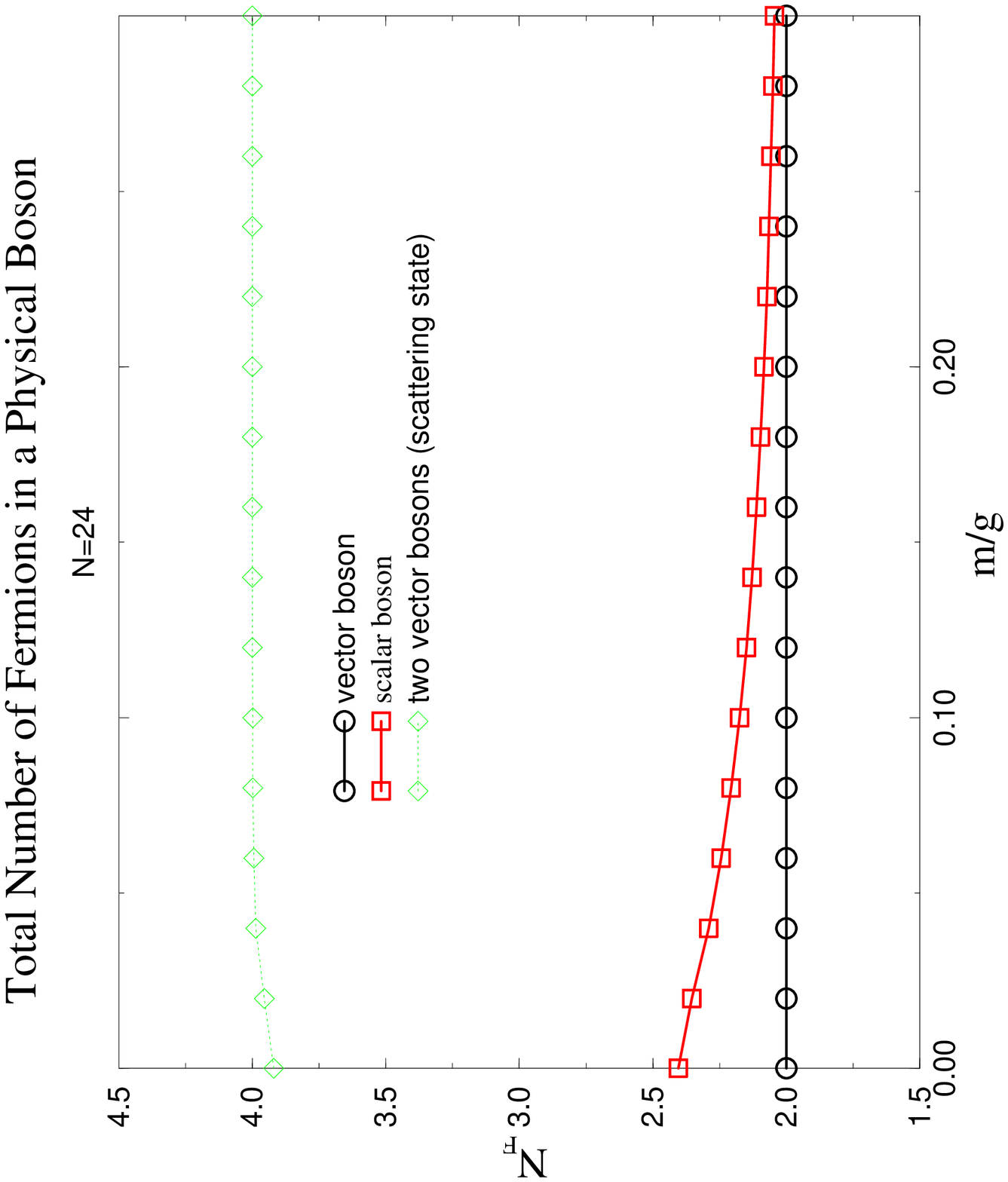}
{The number of fermions in the vector boson, the 
scalar boson and in a two-boson bound-state respectively,
plotted against the fermion mass $m/g$.}
%
%
%
\pstxt{Convergence of distribution functions for $m/g=32$}{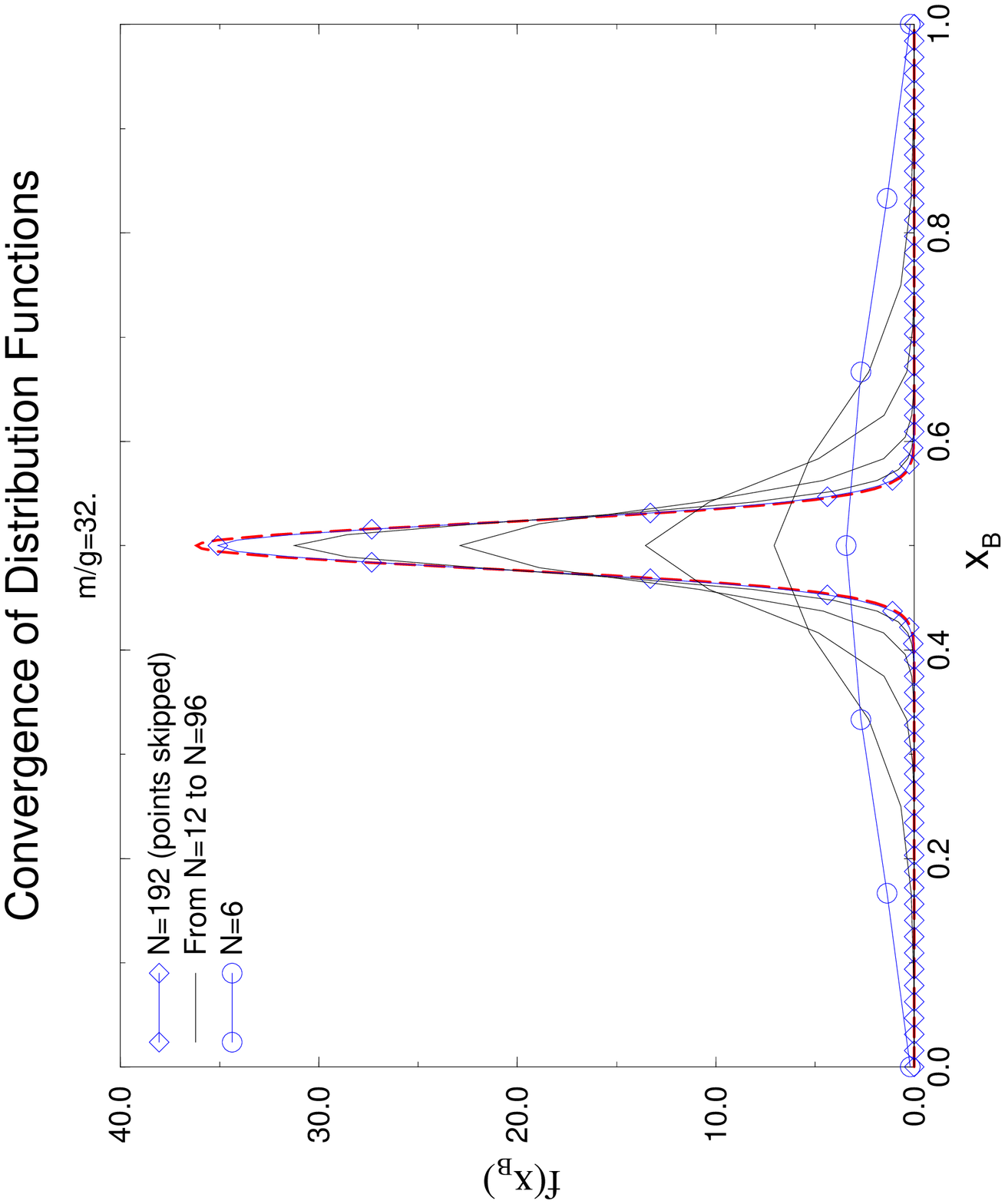}
{The distribution functions ${\frak{f}}(x_B,N)$ for a large fermion
mass ($m/g=32$) ar plotted against the Bjorken scaling variable $x_B$.}
\pstxt{Overview of distribution functions on the entire range 
of fermion masses}{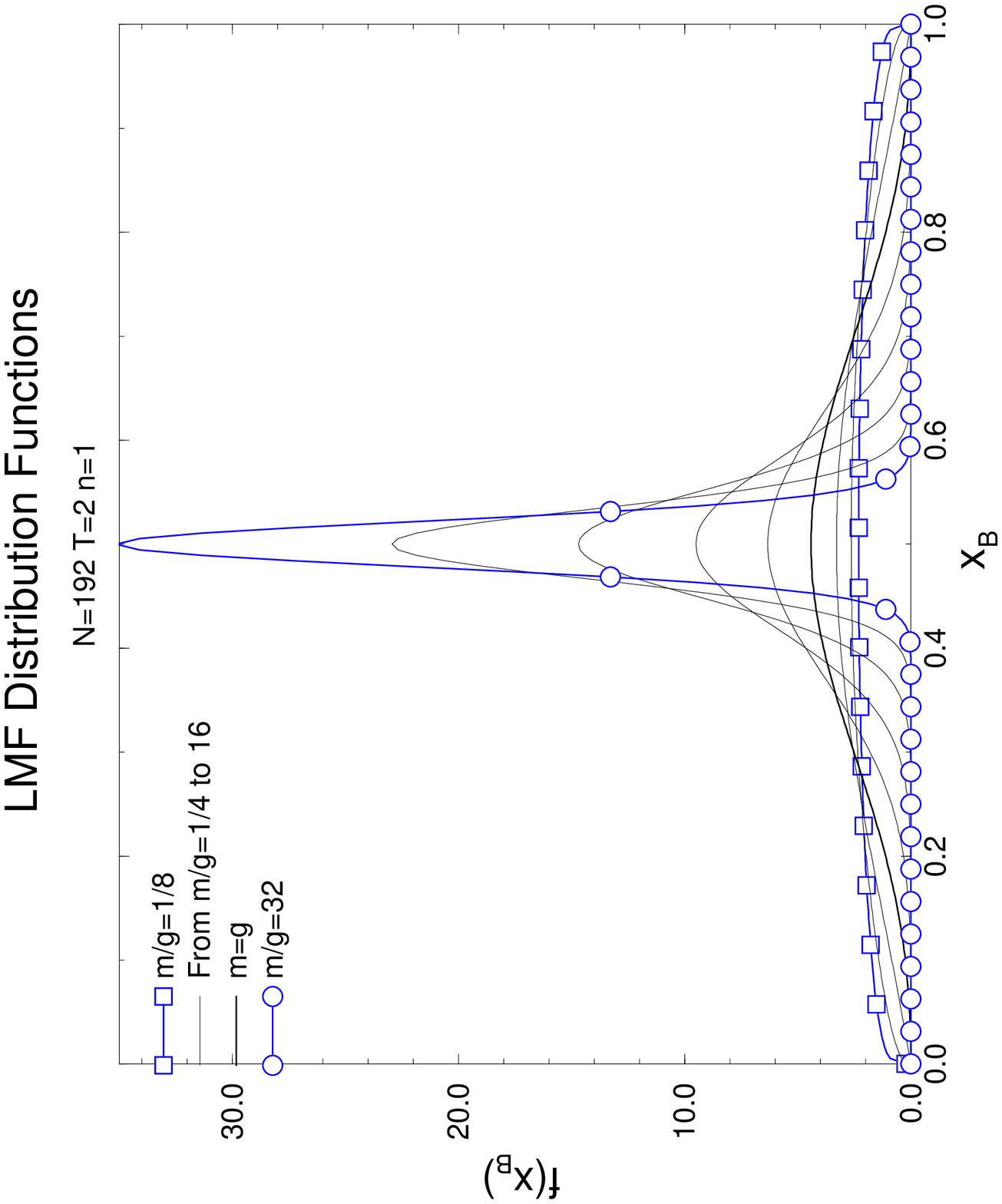}
{The distribution functions ${\frak f}(x_B,P^3)$ for 
different fermion masses $m/g=1/8,1/4,..,32$
are plotted against the Bjorken variable $x_B$.}
\pstxt{Distribution functions (LMF) in the chiral limit $m/g\rightarrow 0$}
{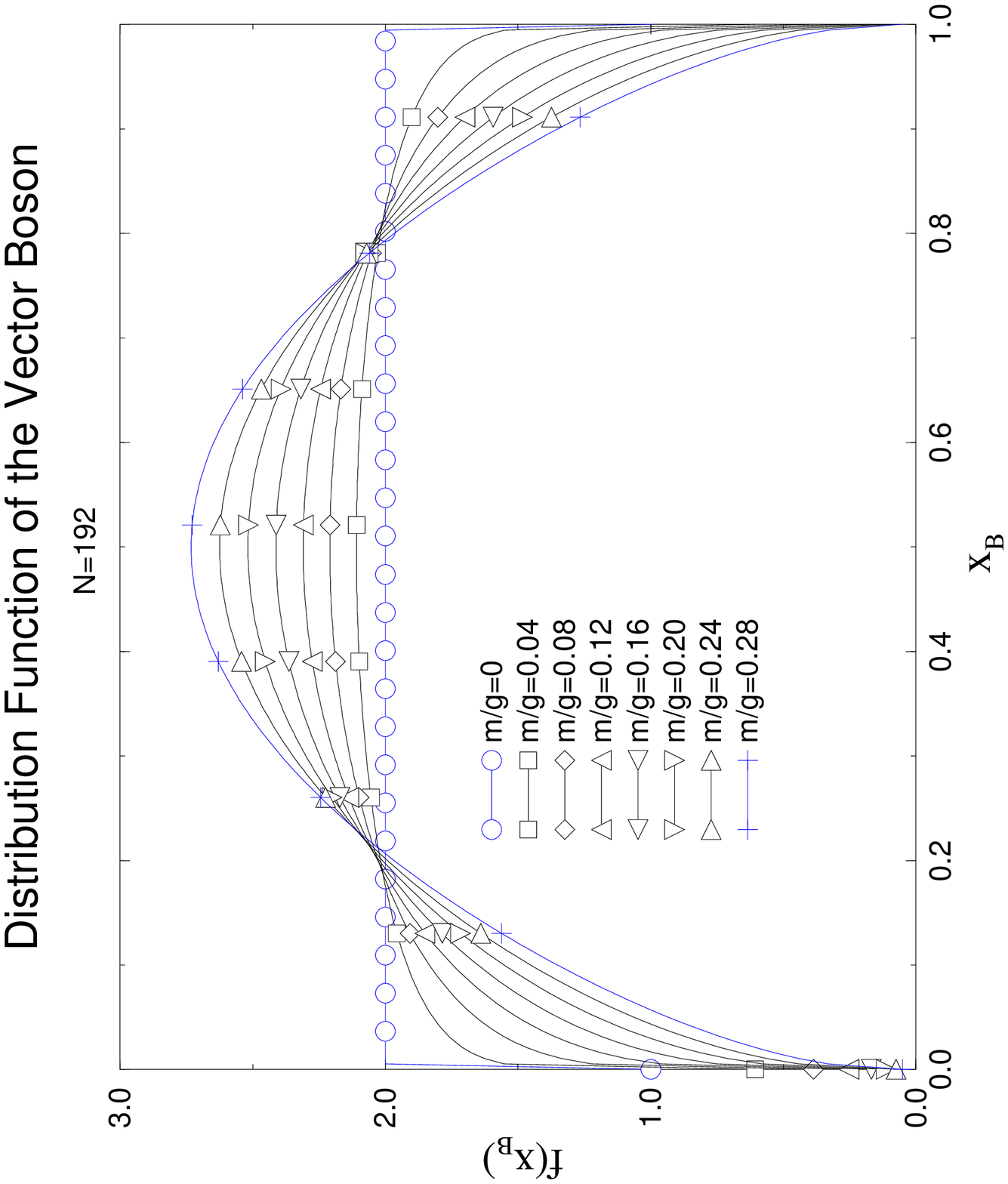}
{The distribution function ${\frak f}(x_B;N=192)$ is plotted against
the Bjorken variable $x_B$ for small masses $m/g\lessapprox 0.3$.}
\pstxt{Distribution functions (FF) in the ultra-relativistic parameter region}{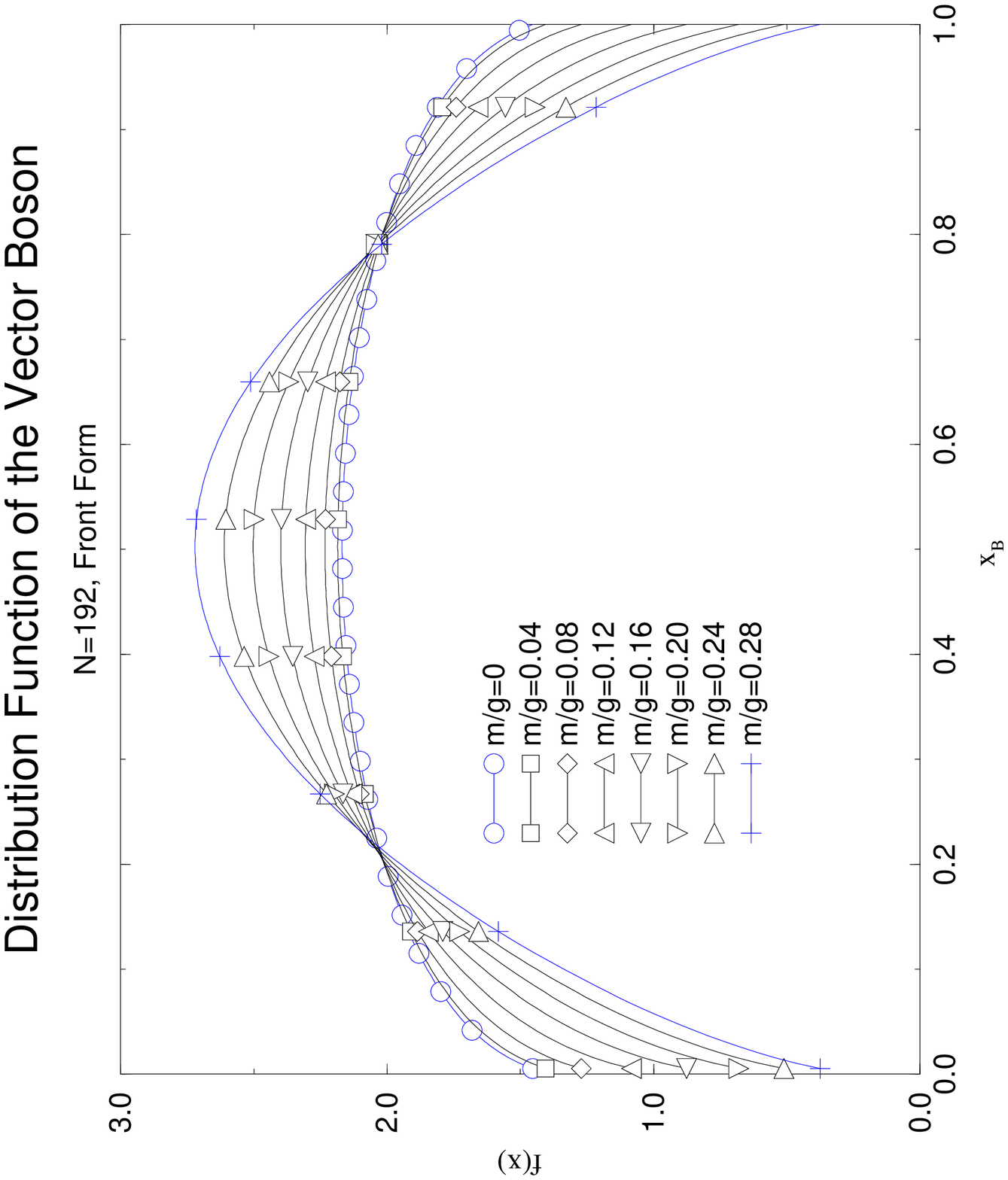}
{The FF distribution function ${\frak f}(x_B;N=192)$ is plotted against
the Bjorken variable $x_B$ for small masses $m/g\lessapprox 0.3$.}


\section{Distribution Functions}

We define the fermion distribution function
\be\label{distribution.function1D}
{\frak f}(x_B;P^3)\triangle x_B
=
{\frak f}_{x_B}(P^3)
\dn
\brac{P}
( b^{\dag} _{x_B \vec P} b _{x_B \vec P}
+ d^{\dag} _{x_B \vec P} d _{x_B \vec P})\cket{P}
\ee
as the mean number of
virtual particles or antiparticles at momentum $\vec k = x_B \vec P$
inside the boson state $\cket{P} $. 
~\figRef{B2NStru.ps} shows rapid convergence
of this function for small fermion masses ($m/g=1/8$).
This indicates that the computed
distribution functions become approximately boost invariant for 
{\em large} momenta which is all the more remarkable
as distribution functions cannot be boost invariant for {\em small} momenta $P^3$(to see this, set $P^3=0$ 
in Def.~\Ref{distribution.function1D}).
The corresponding FF result~\figRef{B2NStruLM.ps} also shows approximate
boost-invariance. 
It is remarkable that the distribution functions computed in the 
large momentum frame(LMF)
are boost-invariant to a much higher degree than the ones computed
using the FF. This illustrates our comment of 
 chapter\secRef{CFF} that a finite lattice spacing $\triangle k$ destroys boost-invariance in all relativistic
forms including the FF. 

The (discrete) integral over the (vector boson)
distribution function is $2.0$ for every
value $m/g$ of the fermion mass as can be inferred
from~\figRef{xMoment.ps}. 
The vector boson state is practically a pure two fermion
state in our LMF approximation (the same thing happens in the FF). 
~\figRef{xMomentLM.ps} 
shows that the same holds true in the FF (as already remarked
in~\Cite{Eller:1987nt}).
Indeed, the vector boson mass remains
unchanged if
we truncate to the Fock space such that states with more than two bosons
are excluded. 
The fact that the number of fermions does not fluctuate 
implies that the \Index{probability} ${\cal P}_{x_B}$
\index{P@${\cal P}_{x_B}$}
of finding a fermion with momentum fraction $x_B$ is proportional 
\be
{\cal P}_{x_B}
=
{1\over 2} {\frak f}_{x_B}
\ee
to the 
distribution function. This is so because the number operator becomes
the projector on the two-particle state
$\cket{x_B}\dn b^{\dag}_{x_B \vec P}d^{\dag}_{x_B \vec P}\cket{0}$
\be
b^{\dag} _{x_B \vec P} b _{x_B \vec P}=\cket{x_B}\brac{x_B}
\ee
if the particle number is fixed. 
In~\figRef{B10aNStru.ps} we represent convergence 
of ${\frak f}(x_B)={\frak f}(x_B,N\triangle k)$ for an almost 
non-relativistic value $m/g=32$ of the fermion mass. 
On the one hand, 
a finer (effective) lattice is required
in non-relativistic domain in order to resolve the narrow peak of the distribution
function. 
On the other hand, however, we observe that ${\frak f}(x_B)$ is zero for
$x_B>0.6$ or for $x_B<0.4$ which means that we could
have considerably reduced the numerical effort
by choosing an {\em effective} lattice between $x_B=0.4$ and $x_B=0.6$
which is five times smaller than the original one.
~\figRef{BStru.ps} depicts the dependence of the distribution
functions on the fermion mass $m$ from $m/g=1/8$ to $m/g=32$,
~\figRef{BCStru.ps} shows the corresponding results
from $m/g=0$ to $m/g=0.3$. 
Comparing the LMF distribution functions~\figRef{BCStru.ps}
with the same functions~\figRef{BCStruLM.ps} obtained
in the FF reveals a substantial disparity in the region where 
$x_B$ or $1-x_B$ is small. 
In particular, our LMF result coincides exactly with the 
exact solution
to the Schwinger model~\Ref{FuerCNum} on a small lattice (see Chapter\secRef{CSchwinger})
whereas the FF result differs considerably.

\section{The Mass of the Vector Boson}

%
%
\pstxt{The vector boson mass $M_V$. A comparison with chiral perturbation theory}
{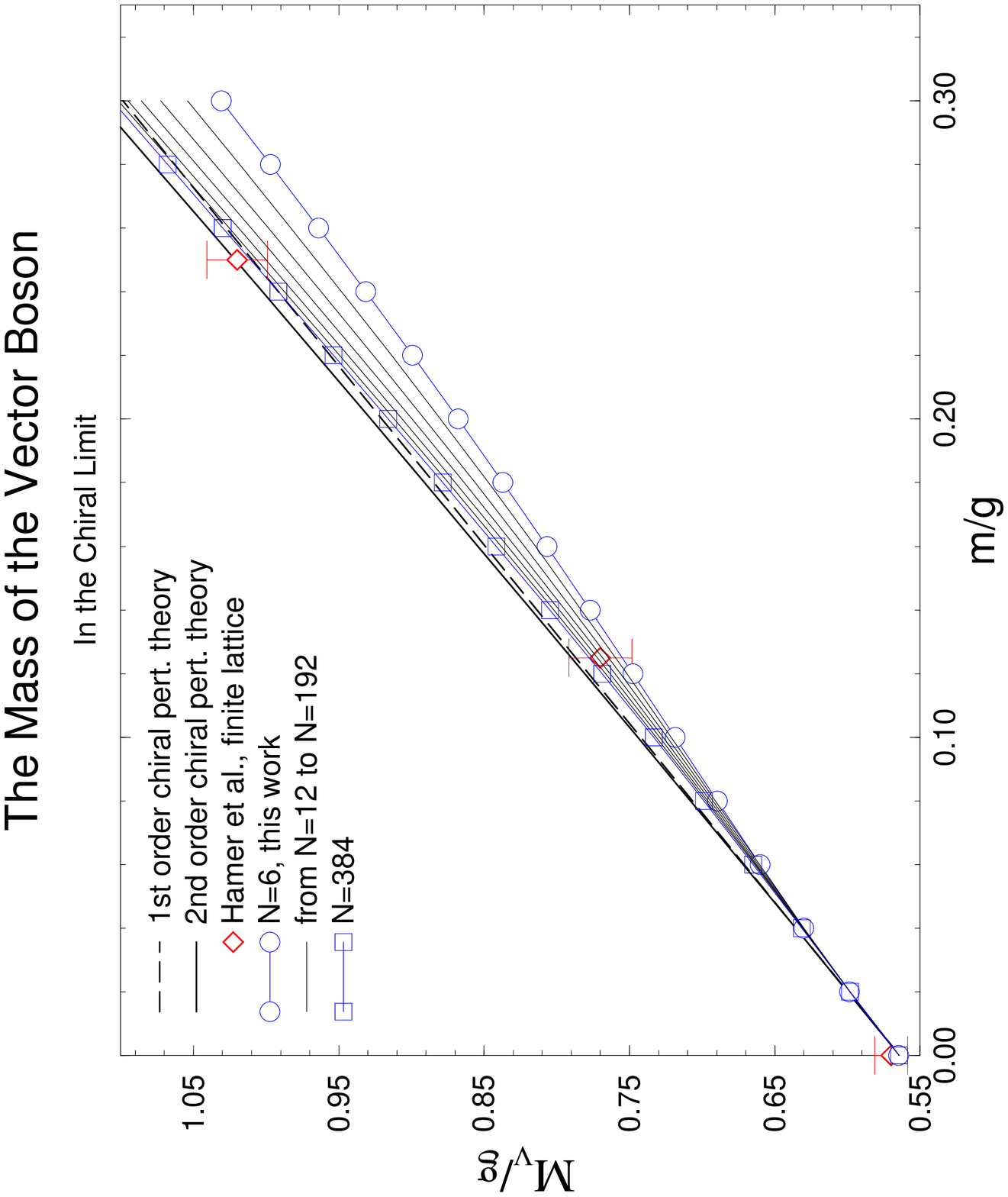}
{The mass of the vector boson ({\bf LMF}) 
is plotted against the fermion mass
$m/g\lessapprox 0.3$ and compared to chiral perturbation theory and
the results of Hamer et al. The dimensionless momentum $N$ varies 
logarithmically according to $3\cdot 2^i$.   }
\pstxt{The vector boson mass in the FF. A comparison with chiral perturbation theory}
{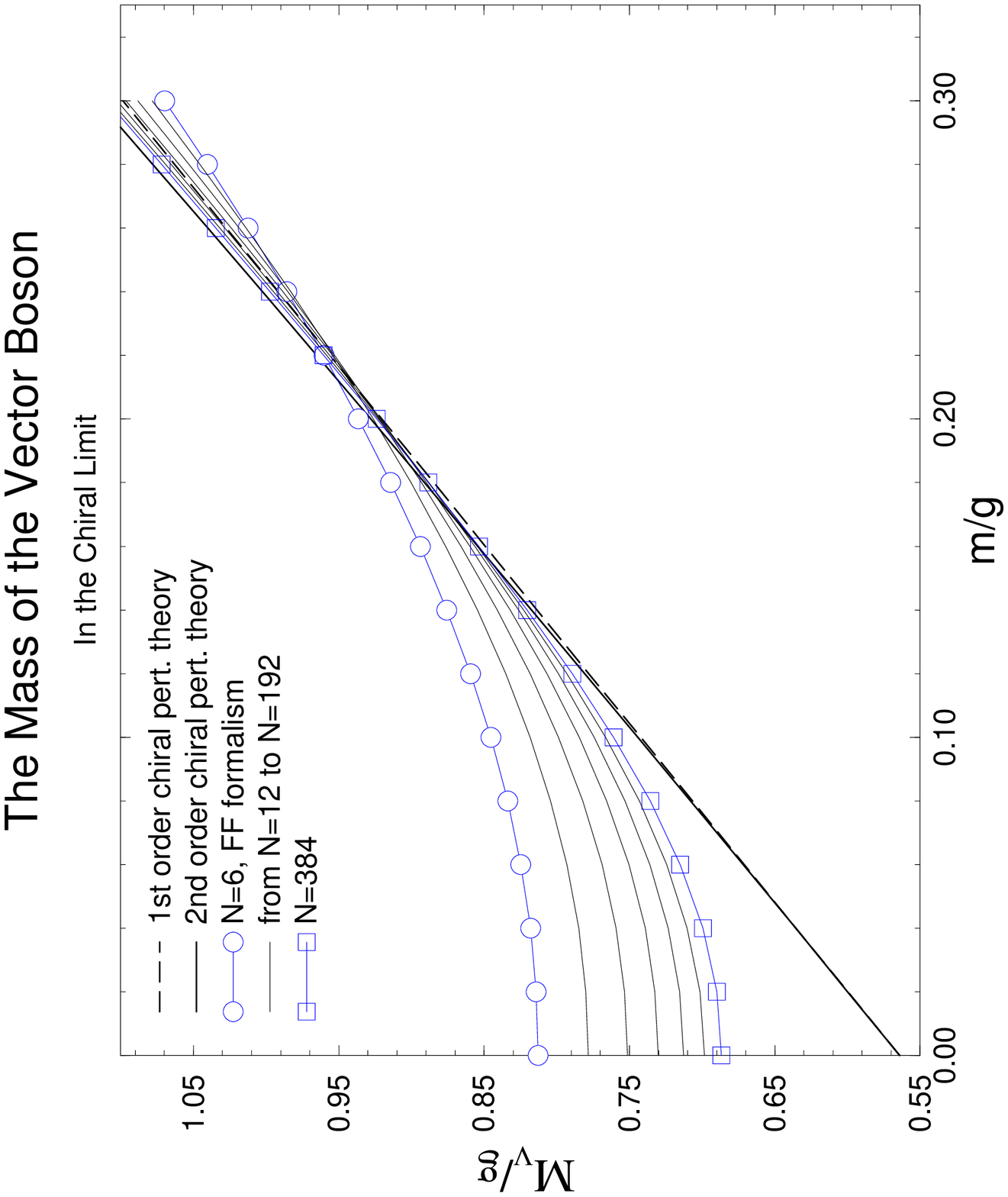}
{The mass of the vector boson computed in the {\bf FF formalism} 
is plotted against the fermion mass
$m/g\lessapprox 0.3$ and compared to chiral perturbation theory.
The dimensionless momentum $N$ varies 
logarithmically according to $3\cdot 2^i$.   }
\pstxt{The binding energy of the vector boson}
{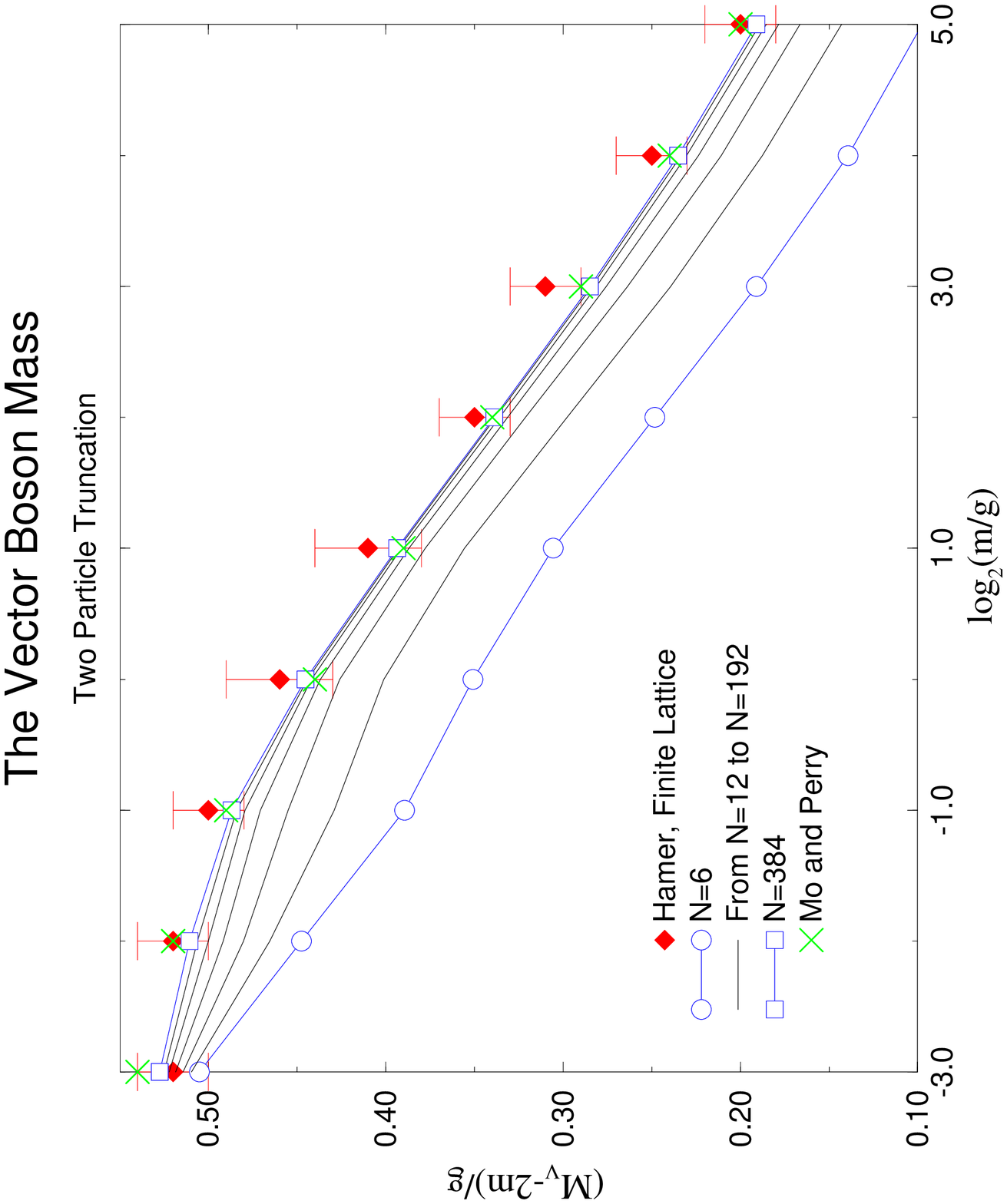}
{The binding energy of the vector boson $(M_V-2m)/g$ is
plotted against $\log_2(m/g)$. 
The Fock-space has been truncated to the two fermion subspace.}
\pstxt{The binding energy of the vector boson in the FF}
{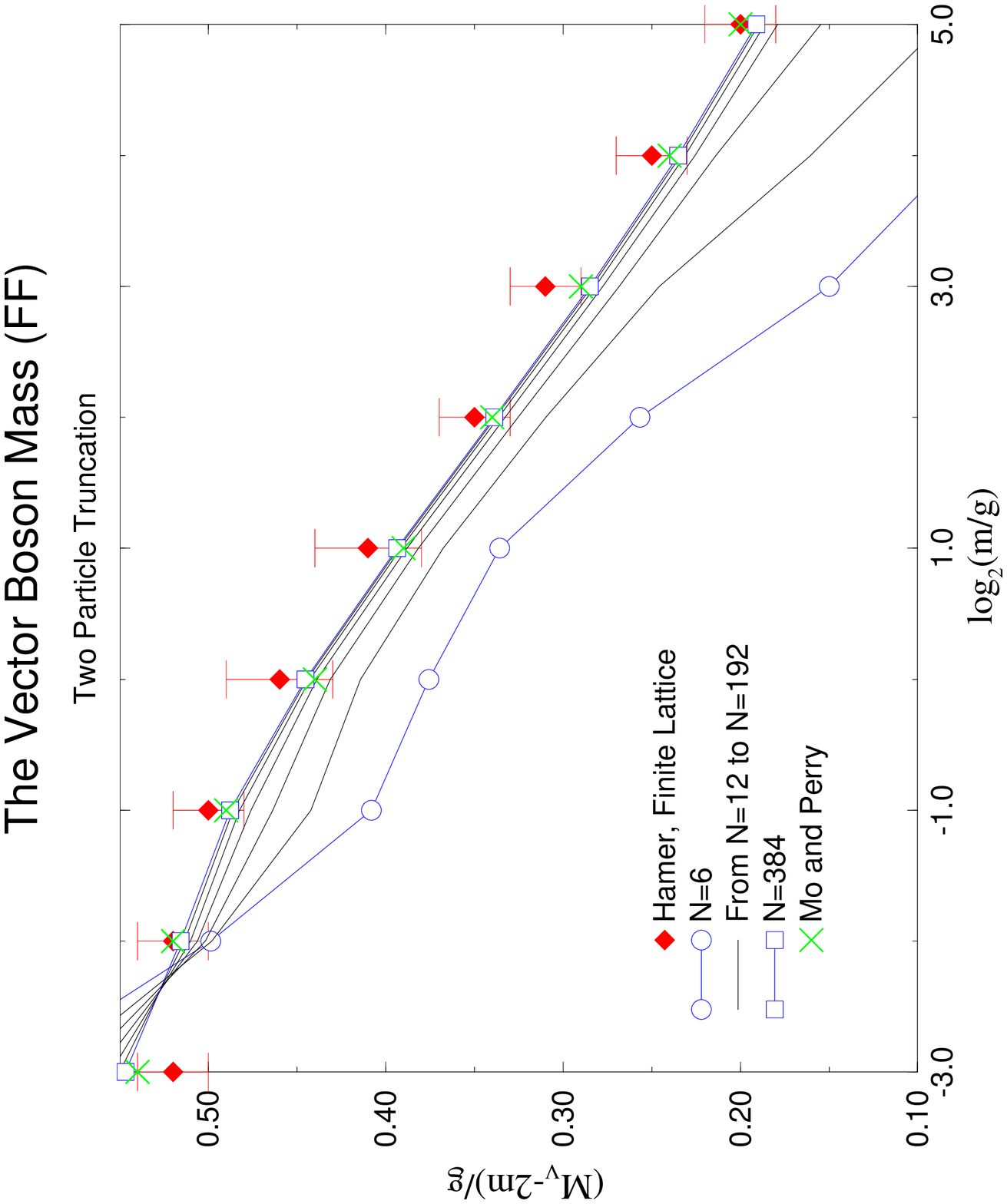}
{The binding energy of the vector boson $(M_V-2m)/g$, computed
in the FF, is
plotted against $\log_2(m/g)$. 
The Fock-space has been truncated to the two fermion subspace.}

%


~\figRef{BCole.ps} shows that vector boson mass (mass of the 
lightest physical particle) as a function of  $m/g \lessapprox 0.3$. 
Chiral perturbation theory predicts
\be
{M_1(g)\over g}
\approx
0.5642+1.781 {m\over g} + 0.1907 ({m\over g})^2
\ee
for the vector meson mass
~\Cite{Coleman:1975pw,Coleman:1976uz,Vary:1996uc,Vary:1994vd,Hamer:1997dx} 
for small values of $m/g$. Agreement with 
second order chiral perturbation theory as well as with the
points computed in Ref.~\Cite{Hamer:1997dx} is observed. 
The excellent 
convergence in terms of $N$ indicates 
that the physical
mass becomes approximately independent 
of the momentum even for very moderate values of $N$. 
The Schwinger limit $m\rightarrow 0$ is special in that 
the physical mass $M$ is entirely 
independent of the momentum for every $N>0$. Moreover, the mass-spectrum
is correctly described even if the correlation length $1/M_1$ is
much larger than the volume $2L$. These peculiarities are
due to the fact that the (massless) Schwinger model is equivalent
to a theory of point-like free bosons.  
These results have to be contrasted with the results obtained in the
FF approach depicted in~\figRef{BColeLM.ps} where the convergence is so
slow that it is hardly possible to accurately 
estimate the exact boson mass between $m/g=0$ and $m/g=0.2$ even on
huge lattices.  
The instant form computation, in contrast, 
~\figRef{BCole.ps} reproduces the linear
dependence of $M$ on the fermion mass even on effective
lattices as small as $N=6$.

We see that the physical mass computed in 
our IF  approximation is approximately 
boost-invariant already at very small momenta whereas 
the mass computed in the FF is far from being invariant even 
for the largest momenta $N$. This is most remarkable since
one of the alleged advantages of the FF over the IF  is the manifest
boost-invariance of the former. We see that in a non-perturbative
calculation on {\em any finite} lattice, neither the LMF {\em nor} the
FF are boost-covariant. The IF , however, is almost boost-covariant even
on small lattices
whereas a gigantic effective lattice is needed in the FF 
in order to
obtain covariance to the same degree as 
in the IF.

Non-perturbatively, the IF  is boost-invariant to a much higher
accuracy on a finite lattice than is the FF which means that it is
exceedingly preferable to use the IF in order to optimise covariance.

It has been shown~\Cite{Mo:1992sv} 
that the FF reproduces approximately the correct vector boson
mass
on an effective lattice which is infinitely large 
(i.e. if the limit $N\rightarrow\infty$ is performed). 
The massive Schwinger model is special in the sense that
the FF, the IMF and the LMF coincide
on an infinite lattice $N=\infty$. This is why
the results of ~\Cite{Mo:1992sv} may be regarded as the results of
the FF, the IMF and the LMF for an infinite effective lattice. 
It is easy
to show that the absence of (a) four-boson couplings in QED$(1+1)$ and 
(b) massless
mesons is 
responsible for this result.
We shall show later on that this does not hold for the two-flavour massive
Schwinger
model. 
{\bf
Thence one has to realise that the successful computations
in~\Cite{Mo:1992sv} are successful only because of this quirky coincidence
of three theories (FF, IMF, LMF) which do not coincide for most QFTs}. 
A further comment is in order. Even though the results of~\Cite{Mo:1992sv}
are quite successful, they are not exact nontheless. The reason for this
is that one has never performed the infinite volume limit because 
in the infinite
volume limit, the vecor boson state consists of an arbitrary numer of
quarks and anti-quarks. Cf. Chapter~\secRef{CSchwinger}. It is impossible
to see this in the FF since the FF always implies an infinitely small
volume and hence precludes the infinite volume limit.

\figRef{xVecBindung.ps} and~\figRef{xVecBindungLM.ps} depict
the \Index{binding energy} $(M_V-2m)/g$ of the vector boson with mass $M_V$. 
One observes that
the manifest inequality between the IF{ }  
and the FF at finite momenta $N$ 
changes drastically in the non-relativistic
limit $m/g\rightarrow \infty$ as well as in the moderately relativistic region
$m/g\approx 1$ where the FF and the IF{ }  are practically identical.
The reason for this is that matrix elements involving only right-movers
$k^3\ge 0$ are practically identical in the 
LMF, the IMF  and the FF. Therefore, the numerical results
are almost identical for larger fermion masses $m/g\lessapprox 0.25$ since
virtual
particles with zero momenta $\vec k=0$ are not important in this case. 
A glimpse at the particle distribution function ${\frak f}(x_B;P^3)$ makes this
point clear: 
In~\figRef{BCStru.ps}, the distribution function in the LMF is depicted
for different values $m/g=0,0.05,\cdots,0.3$ of the fermion mass. 
For $m=0$, our results coincide with 
the exact distribution function~\Ref{SchwingerRawDist}
\begin{equation}
\begin{split}
\label{SchwingerRawDist2}
\frak{f}(x_B)&=f(x_B P^3,P^3;\Omega:0)+
         \bar f(x_B P^3,P^3;\Omega:0)\\
&=
\Theta(1\le x_B \le P^3)
+
\Theta(0\le x_B \le P^3-1)
\end{split}
\end{equation}
computed in Chapter~\secRef{CSchwinger}.
~\figRef{BCStruLM.ps} represents the same function as computed in the FF. 
It is clear that the contribution of the zero-momentum sector $\vec k=0$
cannot be neglected in the ultrarelativistic region and this
is exactly the region where the FF and the LMF differ substantially.
At $m=0$ the probability ${\cal P}_0={1\over 2}{\frak f}_0=0.5/N$ 
to find a particle at $\vec k=0$ is 
as important as any other mode $0<k^3<P^3$
and can not be discarded. The FF distribution
function for $m=0$ differs considerably from
the exact result~\Ref{SchwingerRawDist2}. 
Only if the limit $N\rightarrow \infty$
is performed can this mode be neglected; although the probability for
finding a particle at $\vec p=0$ converges quickly $\propto 1/N$, 
the influence it exerts on the physical
spectrum remains extremely strong even for momenta as large as $N=384$.
A comparison
of~\figRef{BCole.ps} and~\figRef{BColeLM.ps} immediately
makes this clear. 
This explains nicely the following conundrum:
while
Mo's and Perry's "continuum" version~\Cite{Mo:1992sv} ($N=\infty$ and therefore $\triangle x_B=\triangle k/P^3=0$) of the FF is able to reproduce 
the linear behaviour 
\be
M_1(m)-M_1(0) \propto m + {\cal O}(m^2)
\ee
of the vector boson mass in terms of the fermion mass.
But
there is always a quadratic behaviour
\be
M_1(m)-M_1(0) \propto m^2 + {\cal O}(m^3)
\ee
around $m=0$ for {\em any} finite $N$. 
If $N=\infty$, the probability ${\cal P}_0$ of finding a zero-mode fermion
is zero, which allows for the
possibility that the FF and the LMF are equivalent at $N=\infty$.
We note in passing that this result is unlikely to hold in four-dimensional
QCD
since the experimentally measured quark distribution functions diverge
at $x_B=0$. 
It should also be stressed that the possible equivalence of the LMF and
the FF in the limit $N=\infty$ does by no means imply the equivalence
of the FF and the IF: We have already stressed that keeping a small
volume prevents us from performing the continuum limit. Therefore, 
our model is an approximation of the
complete IF dynamics. An excellent one--- but an approximation it is.

\section{The Mass of the Scalar Boson}

%
\pstxt{The binding energy of the scalar boson}
{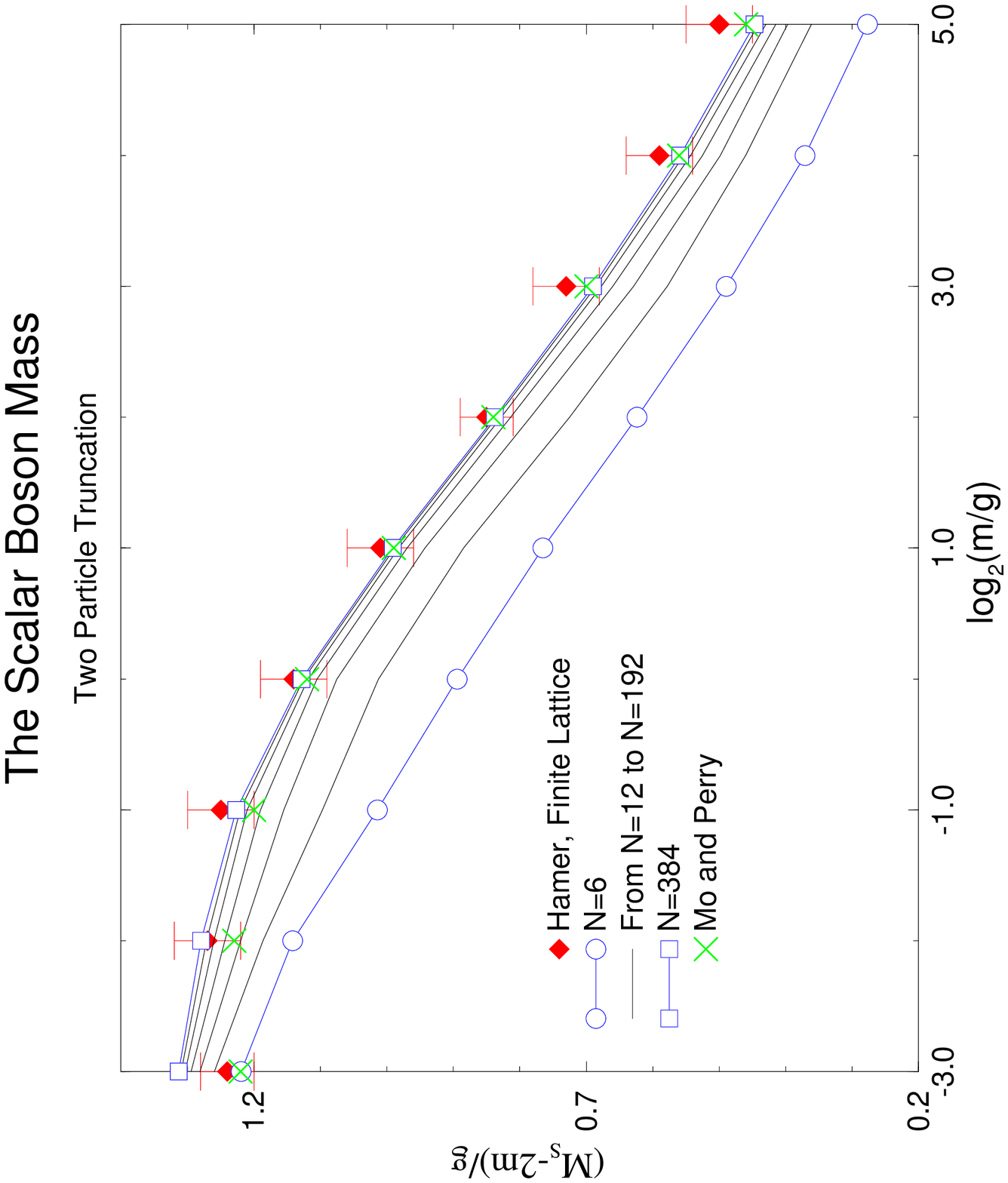}
{The binding energy of the scalar boson $(M_S-2m)/g$ is
plotted against $\log_2(m/g)$. 
The Fock-space has been truncated to the two fermion subspace.}

\pstxt{The binding energy of the scalar boson in the FF}
{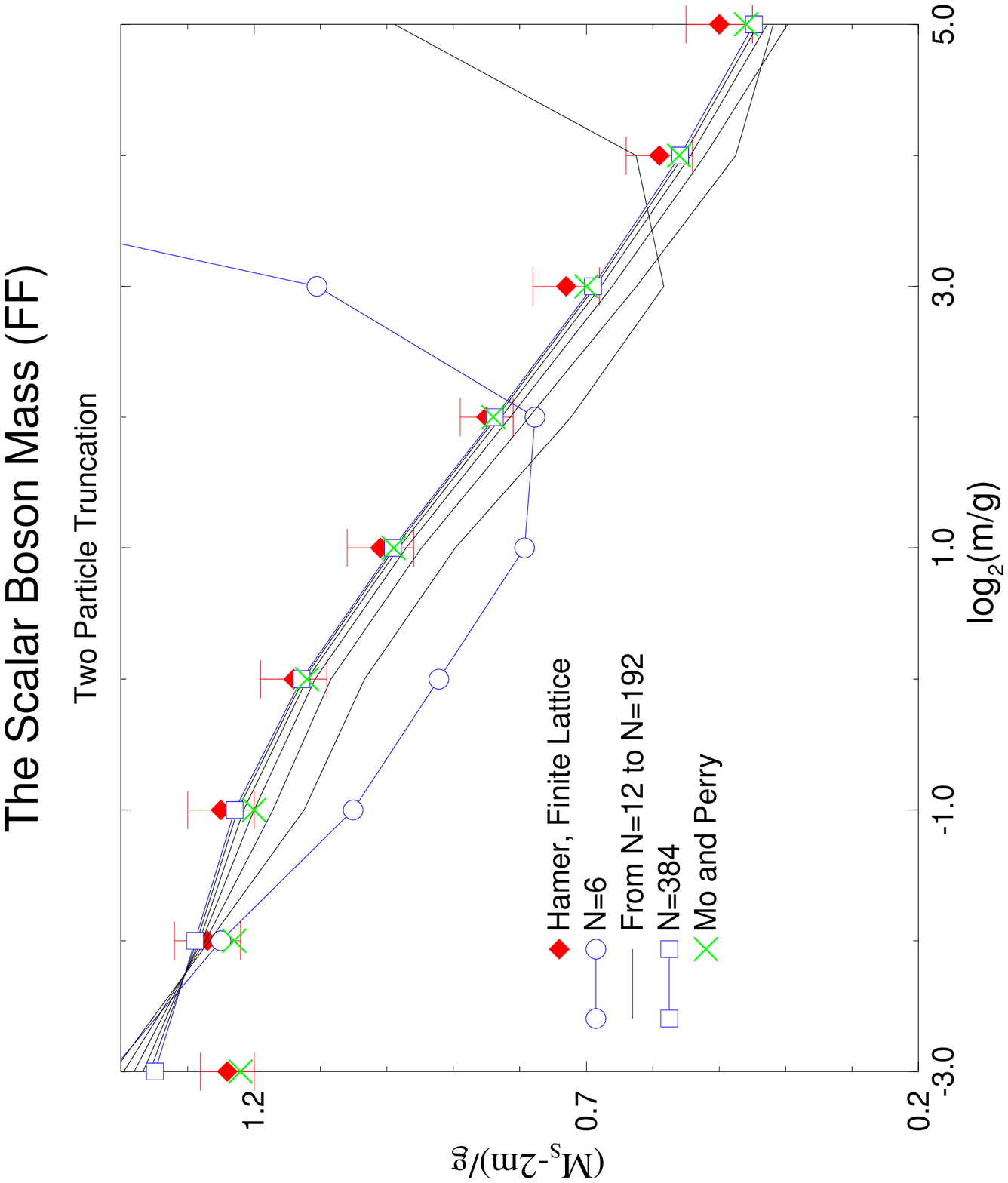}
{The binding energy of the scalar boson $(M_S-2m)/g$,
computed in the FF, is
plotted against $\log_2(m/g)$. 
The Fock-space has been truncated to the two fermion subspace.}
%


The \Index{binding energy} $(M_S-2m)/g)$ of the scalar boson mass $M_S$ 
\index{MS@$M_S$}is 
shown in~\figRef{xScaBindung.ps}
with a logarithmic
abscissa for the points used in ~\Cite{Hamer:1997dx}
including the ultrarelativistic region $m\ll g$ and
the non-relativistic region. The Fock-space has been truncated to the two fermion subspace. Excellent convergence in terms
of the effective lattice size $N+1$ is observed except for
very small fermion masses $m\lessapprox 1/4$ where the two-particle truncation
is no longer valid as shown in~\figRef{xMoment.ps}.
The binding energy of the scalar boson mass $M_S$ in the FF
is presented in~\figRef{xScaBindungLM.ps}. 
The Fock-space has been truncated to the two fermion subspace. Excellent convergence in terms
of the effective lattice size $N+1$ is observed except for
very small fermion masses $m/g\lessapprox 1/4$ where the two particle truncation
is no longer valid as demonstrated elsewhere.

\section{The Modified Front Form}
%
%
%
\pstxt{The vector boson mass in the modified FF. A comparison with chiral perturbation theory}
{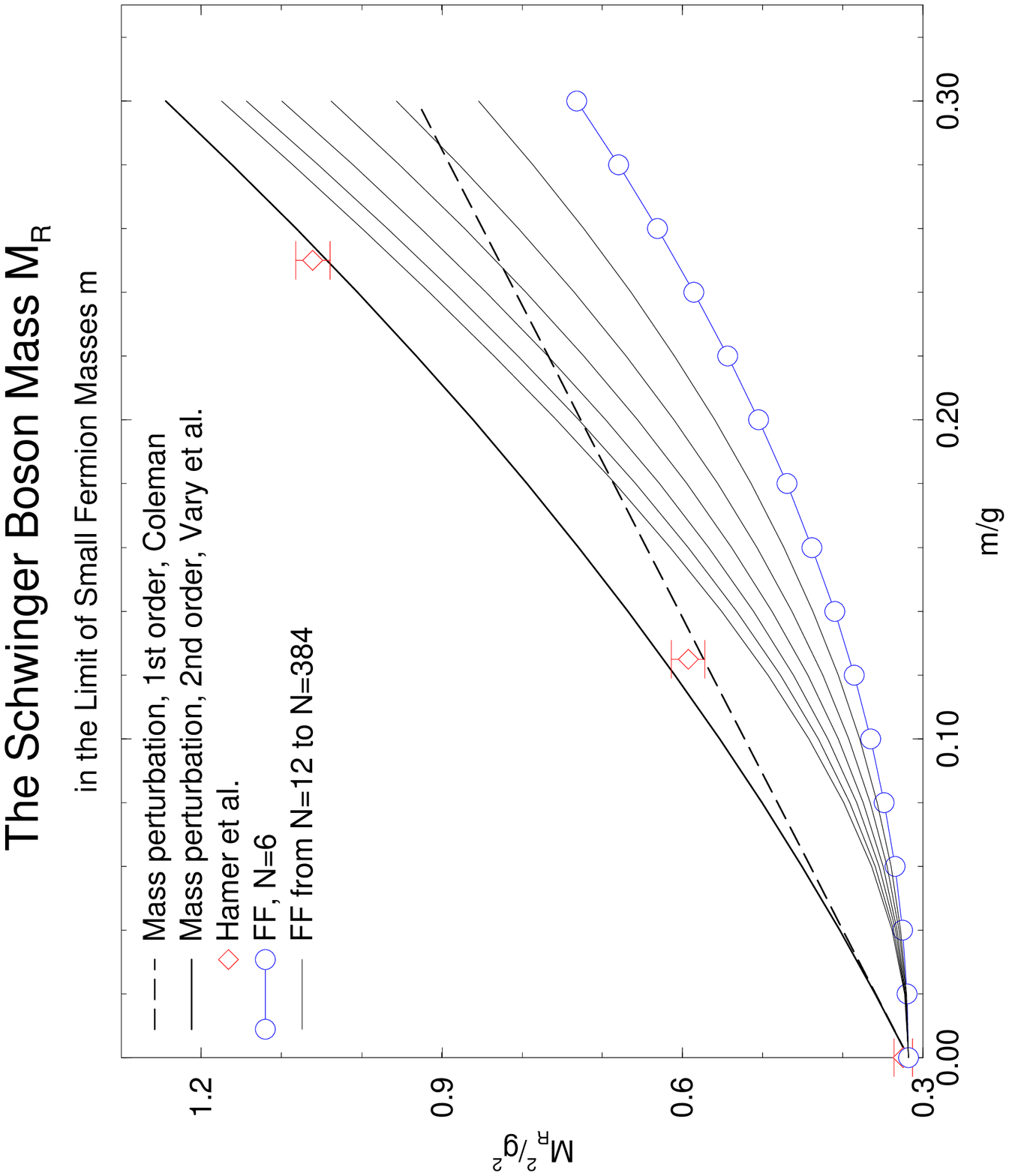}
{The mass of the vector boson in the {\bf modified FF} 
is plotted against the fermion mass
$m/g\lessapprox 0.3$ and compared to chiral perturbation theory and
the results of Hamer et al. The dimensionless momentum $N$ varies 
logarithmically according to $3\cdot 2^i$.   }

{}
\pstxt{The number of fermions in a boson (modified FF)}
{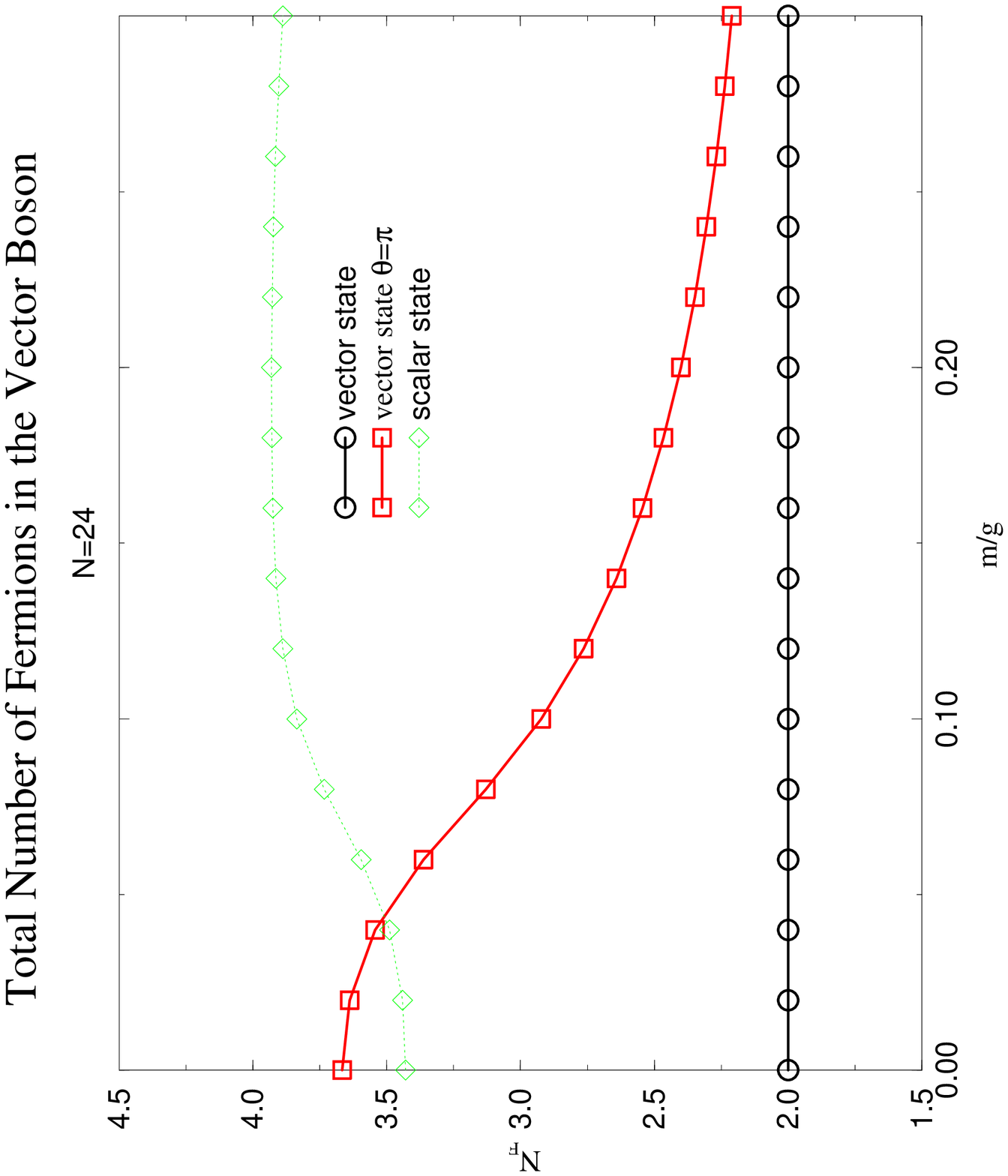}
{The number of fermions in the vector boson, the 
scalar boson and in a two-boson bound-state respectively,
plotted against the fermion mass $m/g$.} 
\pstxt{The binding energy of the vector boson in the modified FF}
{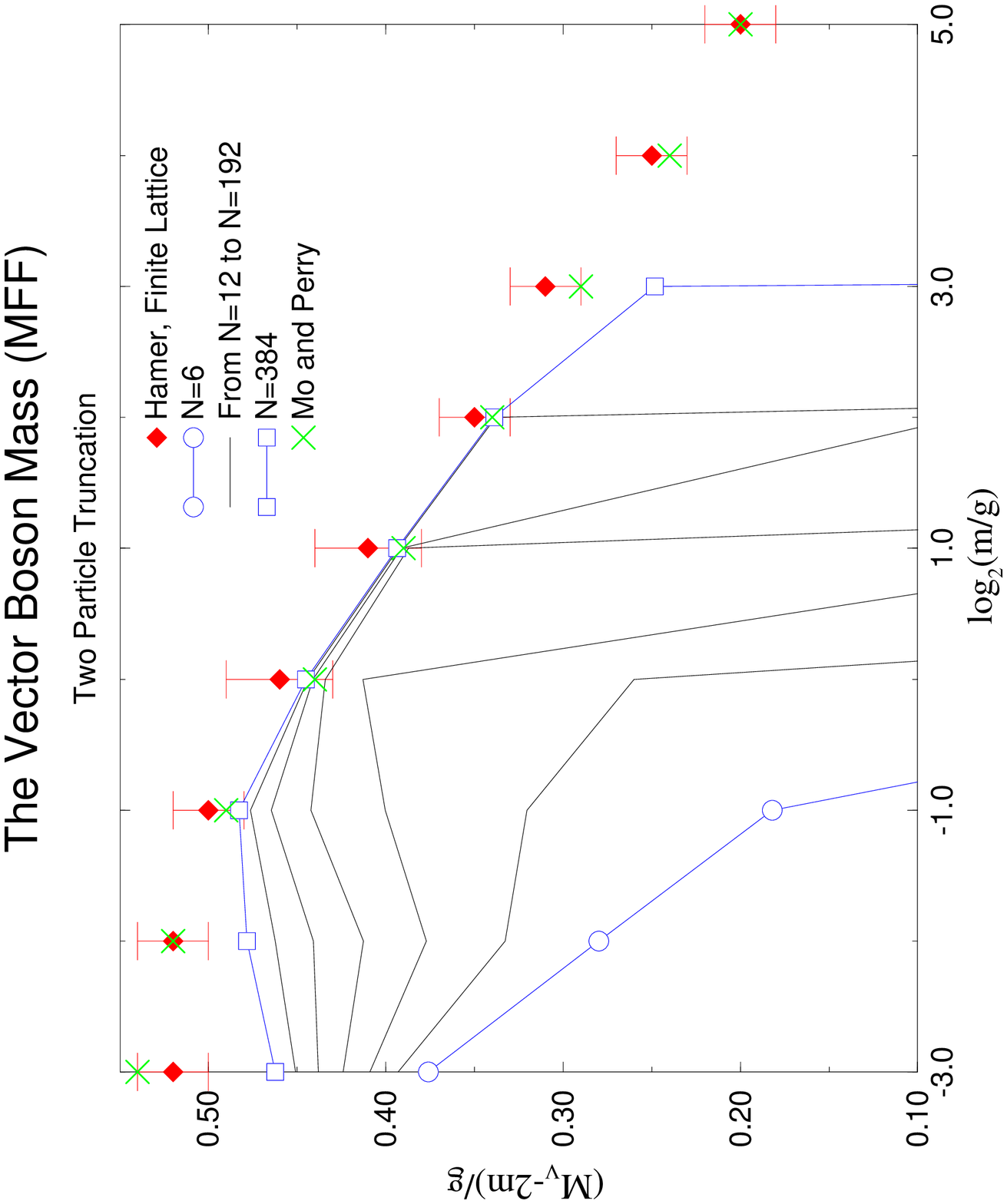}
{The binding energy of the vector boson $(M_V-2m)/g$,
computed in the modified FF, is
plotted against $\log_2(m/g)$. 
The Fock-space has been truncated to the two fermion subspace.}
{
The mass of the vector boson computed using the {\bf modified
FF formalism} as compared to chiral perturbation theory and
the results of Hamer et al. The dimensionless momentum $N$ varies 
logarithmically according to $3\cdot 2^i$. No convergence is observed in the
entire ultra-relativistic region. The {\em only} point of convergence
is the massless limit.}
\pstxt{Distribution functions in the chiral limit $m/g\rightarrow 0$
(modified FF)}
{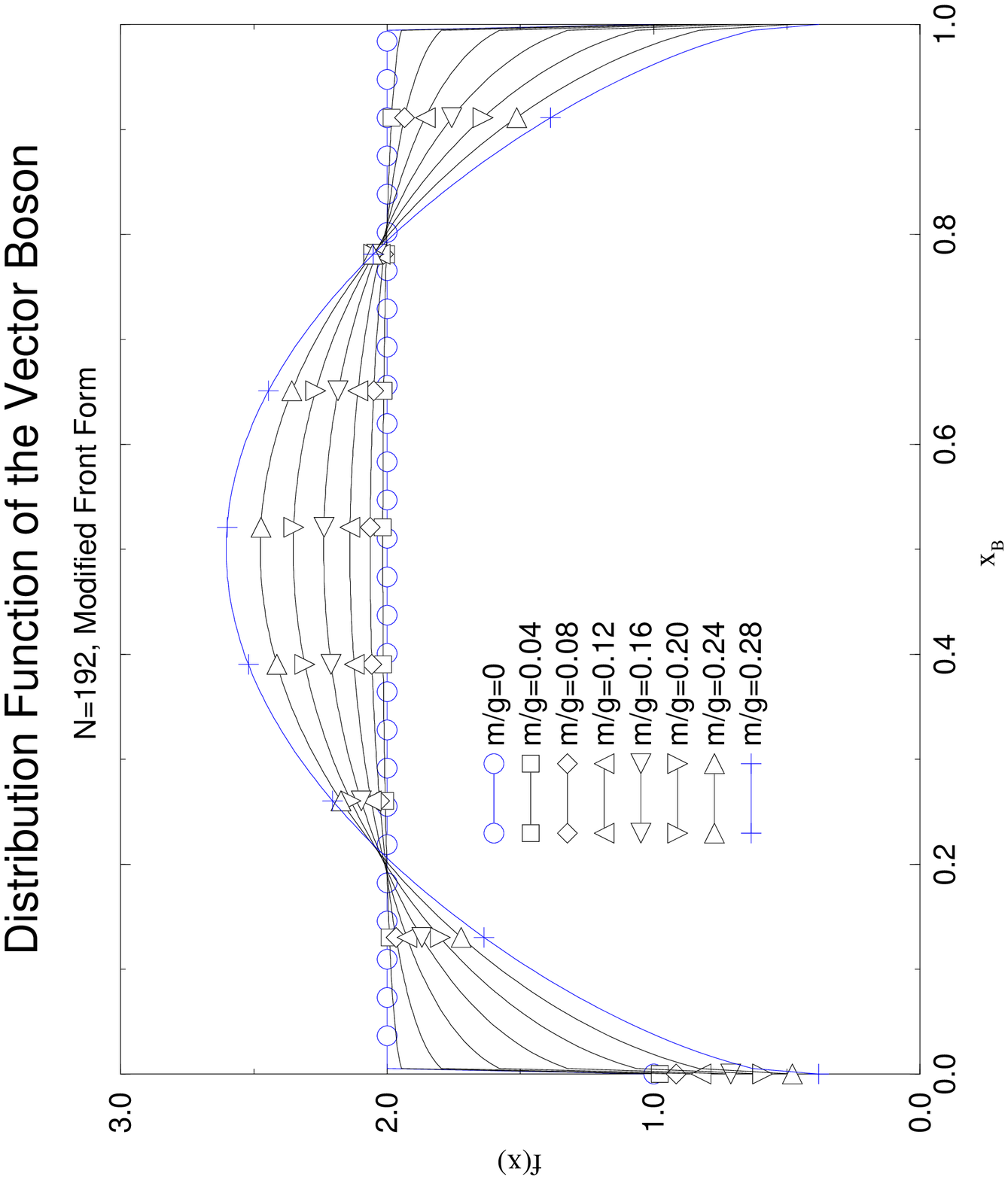}
{The distribution function ${\frak f}(x_B;N=192)$ is plotted against
the Bjorken variable $x_B$ for small masses $m/g\lessapprox 0.3$.}

\pstxt{The binding energy of the scalar boson in the modified FF}
{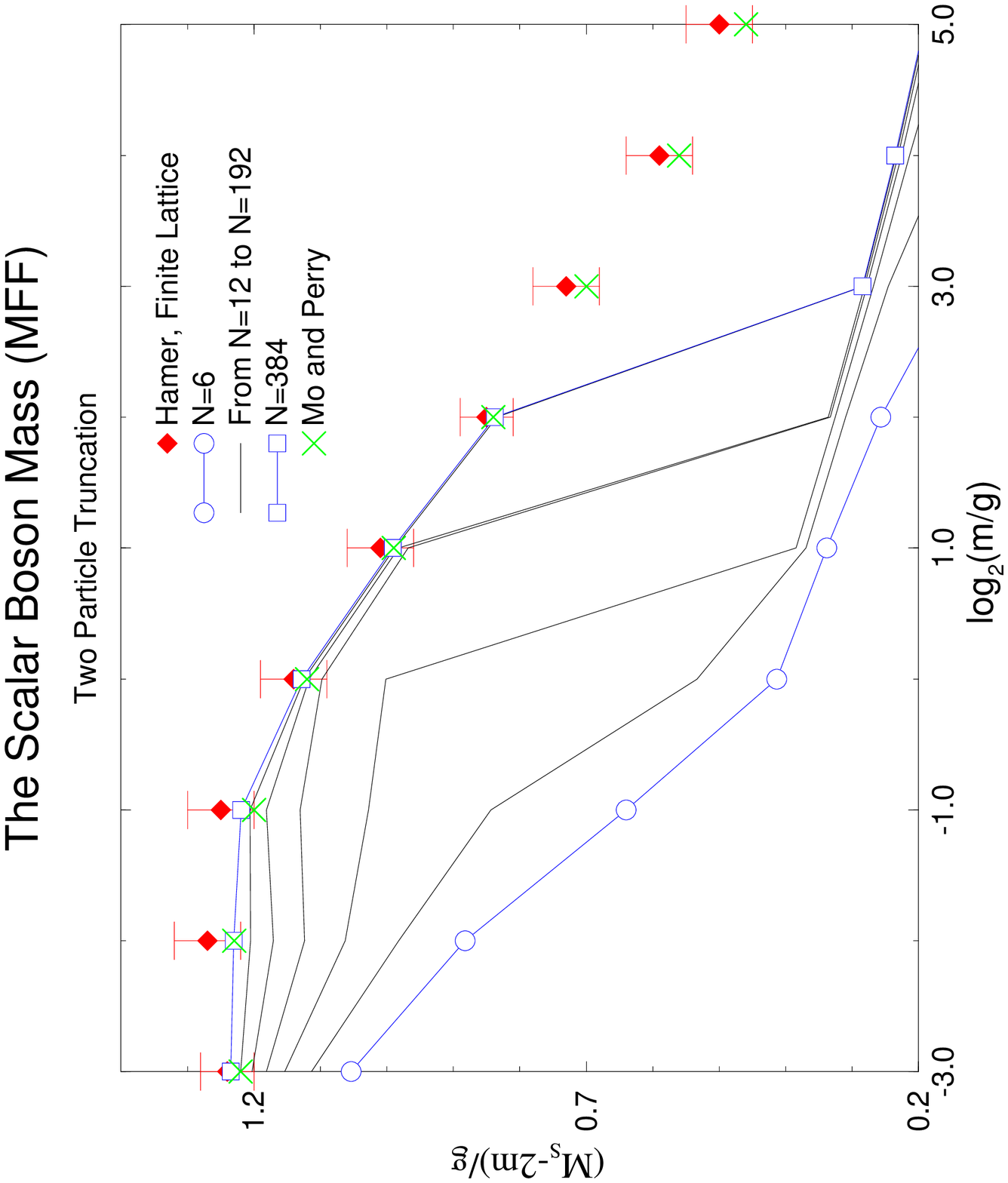}
{The binding energy of the scalar boson $(M_S-2m)/g$,
computed in the modified FF, is
plotted against $\log_2(m/g)$. 
The Fock-space has been truncated to the two fermion subspace.}
%

%
%

It is possible to expand the FF Hamiltonian in terms of massless
fields rather than in terms of massive fields. Details are explained
in ref.~\Cite{Eller:1987nt} where this approach was first introduced. 
This expansion has the advantage that
zero-modes are not suppressed as in the massive expansion and 
the correct Schwinger-boson mass is reproduced in the limit $m=0$. 
We shall refer to the massless expansion as 
\Index{modified front form}(MFF)
\index{MFF}
since this expansion
cannot be obtained by means of (Dirac-Bergmann) quantisation of the 
massive Schwinger
model~\Cite{Kalloniatis:1994fk,Mustaki:1990uq} 
on a light-like quantisation plane
(except for the special case $m=0$).
A comparison of the vector boson mass as obtained in the MFF
~\figRef{BColeL0.ps} to the corresponding LMF result~\figRef{BCole.ps}
shows that the modified
FF is unable to reproduce chiral perturbation
theory even though the MFF is able to
reproduce the correct Schwinger-boson mass at $m=0$.
The reason for this is that the FF for $m=0$ is equivalent to 
\Index{chiral QED} (QED with right-handed fermions only) if
(a) the axial gauge is chosen and 
(b) the $L\rightarrow 0$ limit is performed and
(c) the physical states are restricted to the ${\cal M}=0$ sector.   
The spectrum of this {\em unphysical} theory 
\footnote{
unphysical since it
does not permit large gauge transformations~\Cite{Manton:1985jm})
}
is identical to the spectrum of QED except that it contains right-moving
bosons only as the reader may easily verify using the chiral Hamiltonian
\begin{eqnarray}
H_{\text{chiral}}
=
\sum_{P^3>0}
{\Bbb A}^{\dag}_{\,{{P^3}}} {\Bbb A}_{\,{{P^3}}}
\;
\Omega({{P^3}})
+\hat{\cal M}(\hat{\cal M}+1)
\end{eqnarray}
obtained by removing the left-movers
from $H_{\text{axial}}(L\rightarrow 0)$
(of course, this is a strong violation of Lorentz invariance). 
The fact that the modified FF cannot be obtained by Dirac-Bergmann
quantisation is further illustrated by a comparison
of~\figRef{xMomentL0.ps},~\figRef{xVecBindungL0.ps},~\figRef{BCStruL0.ps}
and
~\figRef{BColeL0.ps}
with the corresponding figures 
~\figRef{xMoment.ps}
,
~\figRef{xVecBindung.ps}
,
~\figRef{BCStru.ps}
and
~\figRef{BCole.ps}
.
In the relativistic limit as well as in the non-relativistic
limit this method fails to reproduce the correct results except for 
$m=0$ and an intermediate region. The breakdown of the modified FF
in the non-relativistic region is due to the fact
that the zero-modes are massless in the modified FF. Consequently
their kinetic energy is zero which makes them ''easily excited'':
If $N$ is too small to resolve the peak at $x_B=1/2$, the wave-function
spreads towards $x_B=0$ and $x_B=1$, zero-modes are excited and the spectrum
breaks down as can be seen 
in~\figRef{BCStruL0.ps},~\figRef{xVecBindungL0.ps} 
and~\figRef{xScaBindungL0.ps}.
The phenomenon that neither the FF nor the modified FF are able
to reproduce chiral perturbation
theory was first described (but not explained)
in Ref.~\Cite{StephanElsersArbeit} by means of a numerical computation.

\section{The Infinite Momentum Frame }
%
%
%
\pstxt{The vector boson mass in the IMF. A comparison with chiral perturbation theory}
{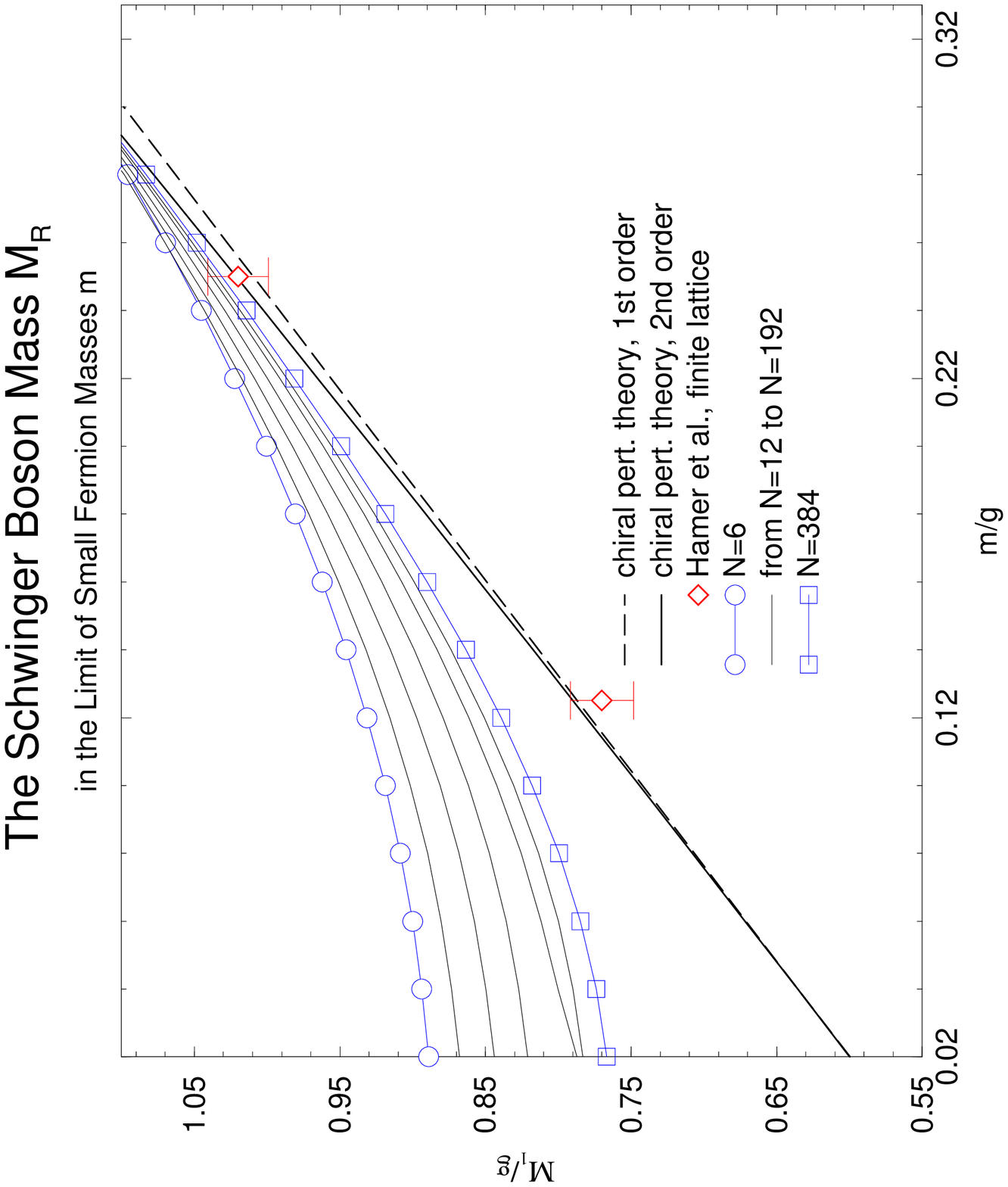}
{The mass of the vector boson in the {\bf IMF} 
is plotted against the fermion mass
$m/g\lessapprox 0.3$ and compared to chiral perturbation theory and
the results of Hamer et al. The dimensionless momentum $N$ varies 
logarithmically according to $3\cdot 2^i$.   }
\pstxt{The binding energy of the vector boson in the IMF}
{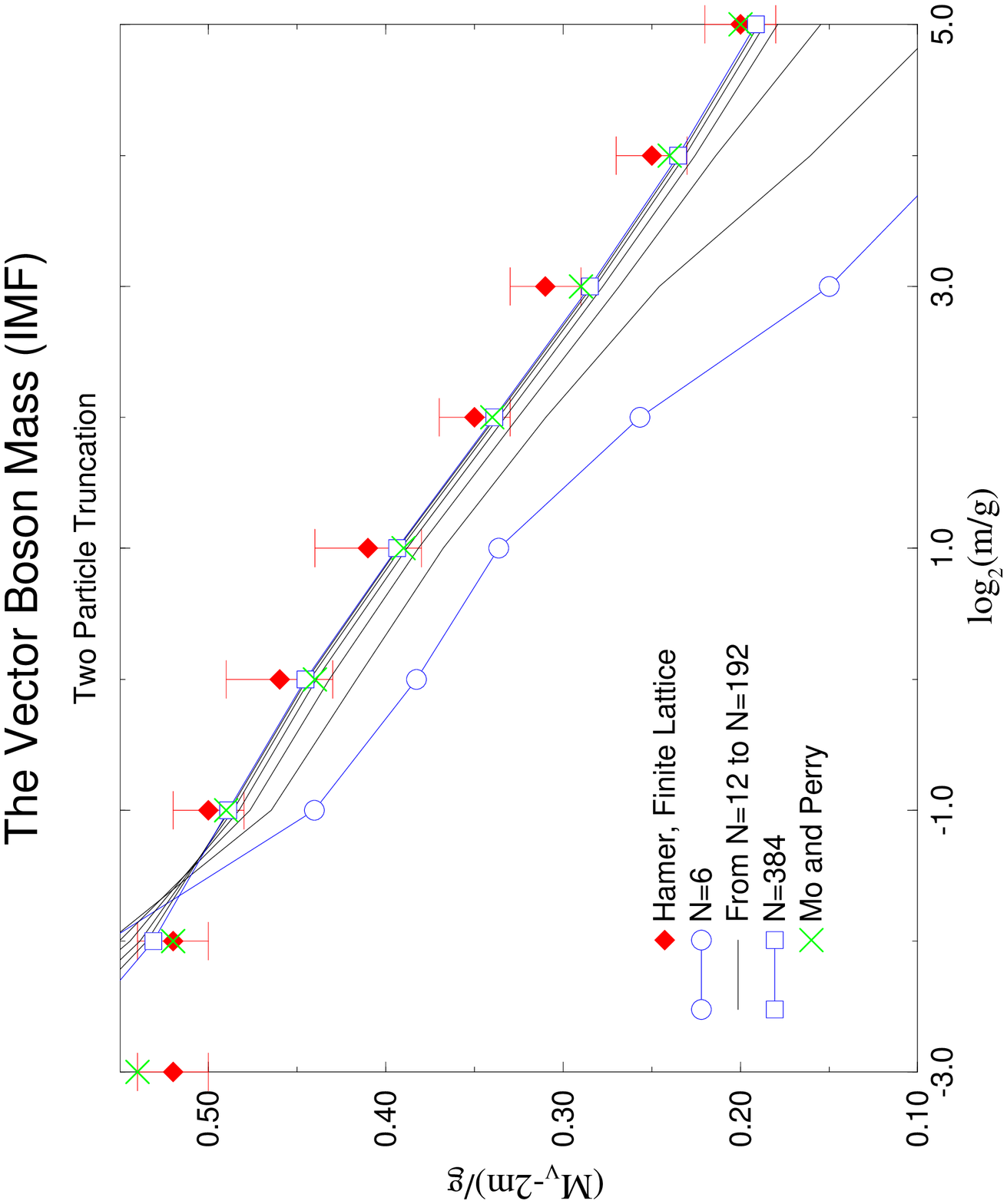}
{The binding energy of the vector boson $(M_V-2m)/g$,
computed in the IMF, is
plotted against $\log_2(m/g)$. 
The Fock-space has been truncated to the two fermion subspace.}
\pstxt{The binding energy of the scalar boson in the IMF.}
{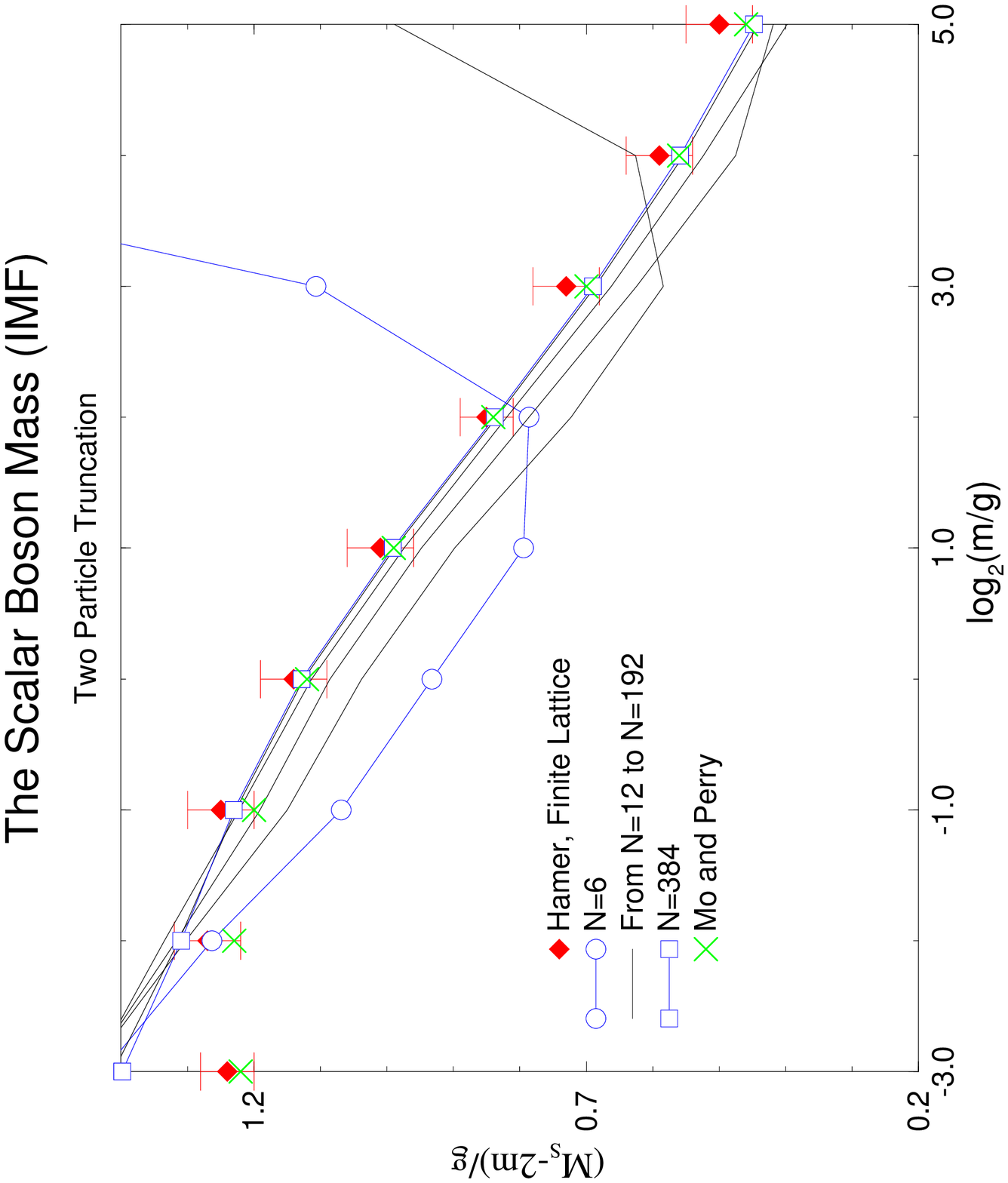}
{The binding energy of the scalar boson $(M_S-2m)/g$,
computed in the IMF, is
plotted against $\log_2(m/g)$. 
The Fock-space has been truncated to the two fermion subspace.}
%

%

Are the LMF, the IMF  and the FF equivalent? In general not, as we 
have
proved in Chapter\secRef{CFF}. The massive Schwinger model in the 
unphysical axial gauge 
exhibits a remarkable property which distinguishes it from
most other field theories: the equivalence
between the IMF  and the FF as well as an almost exact equivalence between
the IMF /FF and the LMF 
if the effective lattice is infinite ($N=\infty$).
If $N$ is finite, these three approximations are clearly 
inequivalent as we have already seen in the case of the LMF/FF.

In the previous subsections we have chosen $g\approx 1$ in order
to comply with the constraint on $M_1/\triangle k$
imposed by the scaling window~\Ref{ScalingWindow}. 
In this subsection we shall investigate what happens if a very small value
of $g/\triangle g$ is chosen such that $M_1(g)$ lies outside the scaling window.
This corresponds to choosing the IMF
as $\triangle k$ is very large compared to the vector boson
mass $M_1$ and the fermion mass $m$ such that the approximation
\be
\omega(\vec k)
\approx k^3 
+ 
{m^2\over 2k^3} 
\ee
for the kinetic energy of a given virtual particle
becomes exact in the $m\triangle k\rightarrow 0$ limit for every momentum
$k^3=n \triangle k>0$.
The kinetic energy at $k^3=0$, in contrast, remains $\omega(0)=m$. 
A given physical mass 
\be
M_1^2=\sqrt{H^2-P^2_3}\approx 2P^3(H-P^3)
\ee
is the sum of the kinetic squared mass
\be
M_{1,kin}^2   
=
2P^3(m(b^{\dag}_0 b_0+
  d^{\dag}_0 d_0)
+\sum_{0<n<N} {m^2\over 2n\triangle k}(b^{\dag}_n b_n +d_n ^{\dag}d_n) )
\ee
and the contribution of the interaction $H_{pot}$.
It is immediately clear that the contribution of a zero-mode
to a physical mass $M$ is infinitely larger than the contribution
of a right-mover since the ratio
\be
{m\over m^2/(2n\triangle k)}
=
2n {\triangle k\over m}
\ee
goes to infinity in the limit $\triangle k\rightarrow \infty$. 
Consequently, zero-modes are kinetically repressed in the IMF . 
As long as the kinetic contribution is not cancelled by a similar
contribution from the interaction term, the diverging ratio 
\be
{\omega(0)\over 
\omega(\vec k^3)}>0
\ee
has the same effect for $m>0$ as the explicit constraint
that removes zero-modes in the FF.

The contribution $m/g^2$ of the zero-mode to the total kinetic mass squared
$M^2/g^2$ 
is finite for any finite fermion mass $m/g$. This contribution
even diverges in the IMF  where $g\rightarrow 0/$, $m/g=const$ which
means that the zero-mode can not be excited in the IMF  limit. Therefore, 
only the FF contribution to the total mass squared is important. Hence the
IMF  and the FF are approximately equivalent for the Schwinger model
(this does not hold for most other models, in particular renormalisable
ones) for $m\neq 0$. For $m=0$, however,
the kinetic energy of the zero-mode cannot diverge and the zero-mode
is free, even in the IMF  limit. 
~\figRef{BCole.ps} illustrates this. In particular, we see
that the IMF  is equivalent to the LMF for $m=0$ whereas for $m\neq 0$
the IMF  and the FF are almost equivalent. In~\figRef{xColeI.ps}
we chose a small but finite coupling $g/ \triangle k=0.001$ in order
to simulate the IMF  where $a\xi\gg L$. 
This scale independence (exact conformal symmetry) of the LMF in the 
limit $m=0$ is responsible for the observed exact covariance of the spectrum
for $P^3\ge \triangle k$. Even on a grotesquely small effective lattice
with $N+1=2$ only two lattice points
does one obtain the exact Schwinger-boson mass.

For larger fermion masses $m$ where dynamics becomes similar to
non-relativistic dynamics, the IMF is numerically indistinguishable
from the FF as can be inferred by comparing the IMF mass spectrum
~\figRef{xVecBindungI.ps}
,
~\figRef{xScaBindungI.ps}
with the FF mass spectrum 
~\figRef{xVecBindungLM.ps}
,
~\figRef{xScaBindungLM.ps}
. 
\section{Why the Schwinger Model is Special}
This might suggest that the FF (or the IMF) are in general equivalent
to the IF as long as the continuum version $N=\infty$ of the FF or
the IMF is used. It is easy to show that this is assumption is
wrong in general. The simplest counter-example is the two-flavour
Schwinger model with small fermion mass
where theories defined on an infinitesimal volume $2L$
as the IMF and the FF must fail. 
In ~\Cite{Hetrick:1995wq,Hosotani:1995zg} it is shown that 
the mass $M_\pi$ of the lightest boson of this theory is 
proportional to $m^{2/3}$ for
$mL\sqrt{M_v L}\gg 1$
whereas it is proportional to $m$ for 
$mL\sqrt{M_v L}\ll 1$ with $M_v=\sqrt{g^2\over 2\pi}$ the mass 
of the vector boson of this theory (we set $\theta=0$). 
For small volumes (i.e. $L\ll 1/M_v$ and $L\ll 1/m$), however,
the lightest boson is always proportional to $m$; therefore,
the FF and the strong IMF cannot reproduce the physical result
of the infinite volume limit $M_\pi\propto m^{2/3}$.
The weak IMF for $N=\infty$ 
(being merely the $N\rightarrow \infty$ limit
of the LMF) can reproduce this result in principle
but it does so with much more numerical effort necessary than that needed in the
LMF on a finite lattice.  
Note that this counter-example cannot be blamed on 
the well-known
difficulties of describing chiral symmetry breaking in the FF:
The problem is more fundamental, for there is no 
spontaneous symmetry breaking in two space-time
dimensions according to Coleman's theorem~\Cite{ItzyksonBuch}.

Things are likely to get worse for the FF for three-dimensional or 
renormalisable theories: The dominance of the FF contribution
to the mass squared operator is a {\em very particular} feature of the
Schwinger model since, amongst others, it 
(a) 
does not contain dynamic bosons (except for the  zero-mode) that would 
(b)
fermionic self-energies
\begin{equation}\label{SchwingTadpoles}
\begin{split}
\triangle \omega_b(\vec k)
&\dn 
\brac{0} b_{\vec k} :H: b^{\dag}_{\vec k }\cket{0} \\
&=
\sum_{l^3\neq \vec k}
\left(
{\sf u}^{\dag}_{\vec k} {\sf u}_{\vec l} {1\over (\vec k-\vec l)^2} 
{\sf u}^{\dag}_{\vec l} {\sf u}_{\vec k}
-
{\sf v}^{\dag}_{\vec l} {\sf u}_{\vec k} {1\over (\vec k+\vec l)^2}
{\sf u}^{\dag}_{\vec k} {\sf v}_{\vec l}
\right) \\
&\approx
\sum_{\vec k\neq l^3\ge 0}
\left(
{\sf u}^{\dag}_{\vec k} {\sf u}_{\vec l} {1\over (\vec k-\vec l)^2} 
{\sf u}^{\dag}_{\vec l} {\sf u}_{\vec k}
-
{\sf v}^{\dag}_{\vec l} {\sf u}_{\vec k} {1\over (\vec k+\vec l)^2}
{\sf u}^{\dag}_{\vec k} {\sf v}_{\vec l}
\right) \\
\end{split}
\end{equation}
associated with right-movers ($\triangle k^3>0$) 
vanish for fast-moving left-movers and vice versa
(c)
it is super-renormalisable.
In general, we are not able to perform such a limit in a meaningful way,
especially in renormalisable theories where $P^3<\Lambda$ has to be {\em finite}. 
In general, the importance of zero-modes does not disappear in the 
limit $N\rightarrow \infty$ especially in the presence of condensates 
 
\section{The Influence of the $\theta$ Angle}
%
\pstxt{The lowest-lying mass spectrum (for very small fermion masses)}{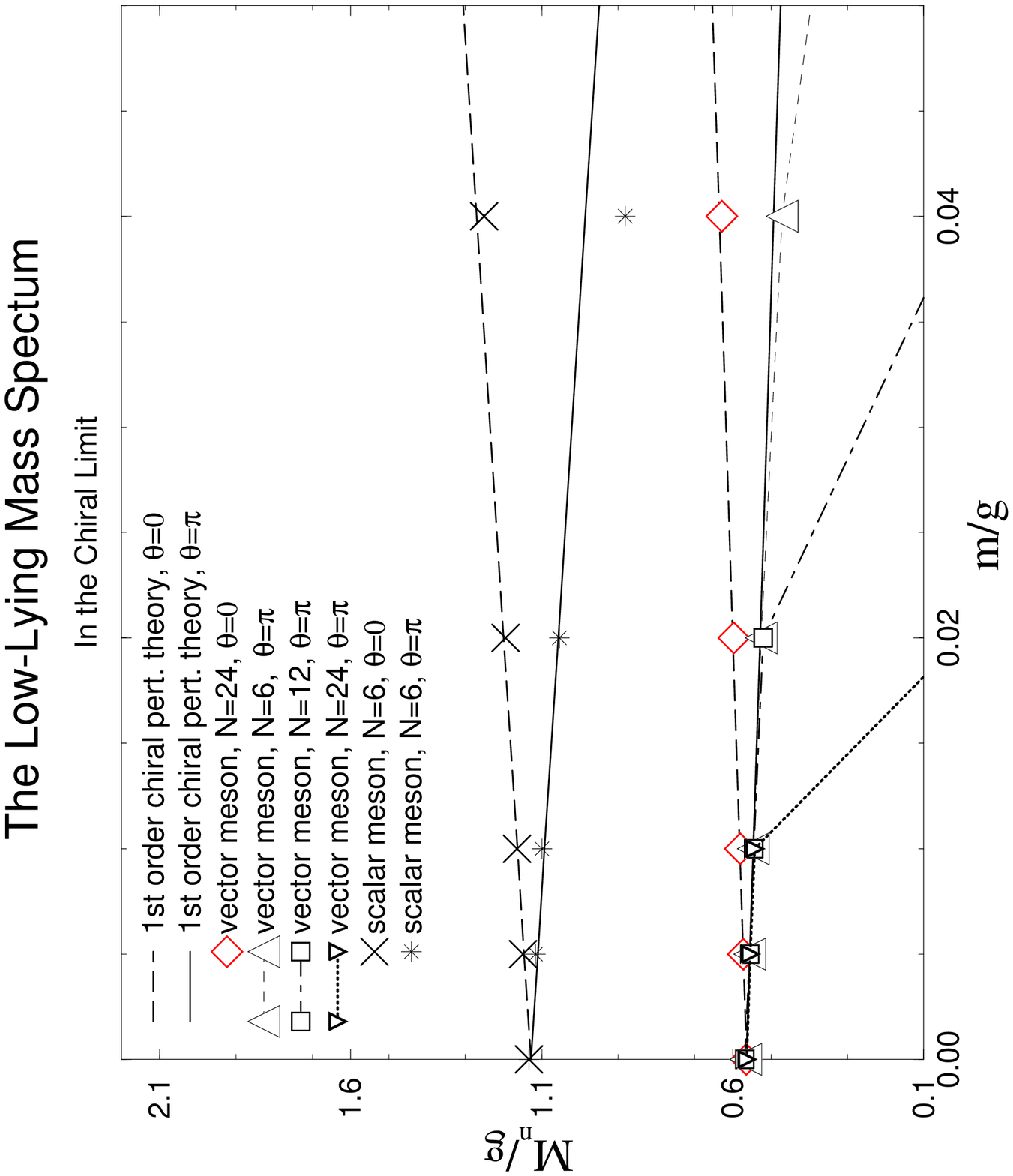}
{The mass of the vector meson ($\theta=0$ and $\theta=\pi$) and
the scalar meson ($\theta=0$ and $\theta=\pi$) 
are plotted against $m/g$ and compared to
the respective linear approximations of chiral perturbation theory. 
The bifurcations of the vector mass towards $M=0$ indicate the
breakdown of our approximation at a level-crossing.}
\pstxt{The na{\"\i}ve mass spectrum (for very small fermion masses)}{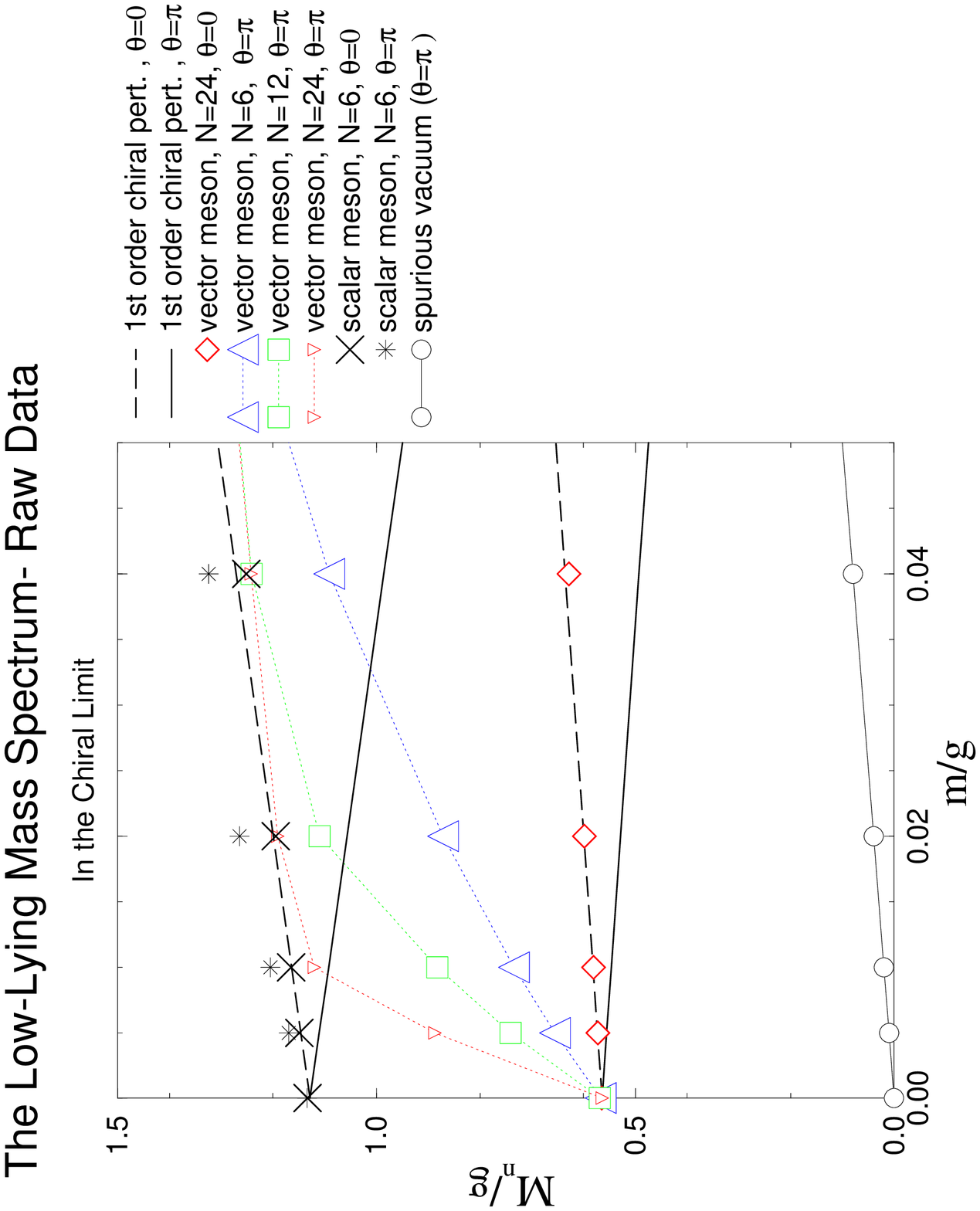}
{The mass of the vector mesons ($\theta=0$ and $\theta=\pi$) and
the scalar mesons ($\theta=0$ and $\theta=\pi$) are plotted
against $m/g$ and compared to
the respective linear approximations of chiral perturbation theory. 
In this figure, the energy of the $\theta=\pi$ vacuum is {\em not}
subtracted from the energies of the $\theta=\pi$ states
in order to study the level-crossings.}
%

The $\theta$ angle in the massless Schwinger model 
is a super-position of all axial vacua weighted with the factor 
$
e^{i{\cal M} \theta}
$
\index{M@${\cal M}$}.
If left-movers are excluded from the Fock space, however,
only two axial sectors ${\cal M}=0$ and ${\cal M}=-1$ can be described
because states with ${\cal M}<-1$ contain
at least ${\cal M}+1$ left-movers. 
Following Chapter\secRef{CFF}, 
a rudimentary version of the $\theta$ vacua may be constructed,
\be
\cket{\theta=0}
=
{1\over\sqrt 2}
(\cket{\Omega:0,0}+\cket{\Omega:0,-1})
\ee
and
\begin{equation}
\begin{split}
\cket{\theta=\pi}
&=
{1\over\sqrt 2}
(\cket{\Omega:0,0}-\cket{\Omega:0,-1})   \\
&=
{1\over\sqrt 2}
(\cket{\Omega:0,0}+e^{-i\pi}\cket{\Omega:0,-1})
\end{split}
\end{equation}
corresponding to the theta angles $\theta=0$ and $\theta=\pi$ which
both have a non-vanishing condensate. The state $\cket{\theta=\pi}$
corresponds to the states 
$\cket{\Omega:\tilde {\sf A};\theta,N_\theta }$ 
introduced in Chapter~\secRef{CFF}
with $N_\theta=0$ except that ---contrary to Chapter~\secRef{CFF}---
we have chosen axial gauge for numerical computations. 
Axial gauge destroys covariance and
the degeneracy of $\theta$ states as well. Therefore,
the bare vacuum energy $E_0'(\theta)$ depends on the the $\theta$ angle. 
In axial gauge, states corresponding to $\theta\neq 0$ appear
as spurious states. In our case, the energy $E'_0(-1)$ 
of the (approximate)  
vacuum with $\theta=-1$ is slightly larger than the energy $E'(0)$ of 
the $\theta=0$ vacuum. This means that the state $\cket{\theta=-1}$
ceases to be a vacuum since a vacuum ---in common parlance---
is defined as the state with lowest energy. 
If we define the renormalised energy spectrum as 
$\tilde E_n(\theta)\dn E_n'(\theta)-E'_n(0)$ then the 
"vacuum" with $\theta=\pi$ appears as a non-covariant spurious state with
a finite, momentum-dependent mass. 

In this section, we demonstrate that the harm inflicted upon the mass
spectrum by axial gauge can be repaired ---to a remarkable degree---
by treating even degenerate "vacua" as vacua: this means that we define
the renormalised energy $E_n(\theta)\dn E_n'(\theta)-E_0'(\theta)$
by subtracting the vacuum energy $E_0'(\theta)$ from the bare energy
$E_n'(\theta)$
rather than subtracting the energy $E_0'(0)$ of the vacuum proper. 
This procedure renders the renormalised
energies 
$E_0(\theta)\dn E_0'(\theta)-E_0'(\theta)=0$
of
every $\theta$ vacuum degenerate --- by construction. 
This renormalisation prescription also allows a correct
description of the mass spectrum at very small fermion mass $m$ as
shown in~\figRef{xSpezial.ps}: one observes that 
our approximation allows us to reproduce the linear 
approximation (see Ref.~\Cite{Coleman:1975pw,Coleman:1976uz}) of the scalar boson mass
\be
M_V
\approx
M_B + 2\pi {C(\infty)\over M_B} m
\approx M_B+1.78 \cos(\theta) m
\ee 
and vector boson mass
\be
M_S\approx 2M_V
\ee  
quite nicely in both $\theta$-sectors.
Here, 
\index{C@$C(L)$}
\be
C(L)\dn 
\brac{\Omega_0;\theta}
\bar\psi\psi
\cket{\Omega_0;\theta}
= 
{e^{\gamma_E} \over 2\pi} M_B \cos(\theta)
\ee
is the fermionic \Index{condensate}, $\gamma_E$\index{gamma@$\gamma_E$}
is the \Index{Euler number}.

We are faced with one practical problem, however. In order to 
perform the correct renormalisation of the bosonic
energies, we have to find out
which value of $\theta$ a physical state corresponds to. 
We can do this either by inspection of the wave function for 
{\em every} fermion mass or 
by inspection of the wave function for one fermion mass and then 
using the continuity of $M_n$ in terms of the fermion mass. 
We chose the second possibility which is much simpler from the
numerical point of view because it can be easily automated. 
The first possibility is more powerful though. We could not implement 
it before the end of the thesis. 
A disadvantage of using the continuity of $M_n(m)$ is that
this scheme does only work before the bare energy $E_n'(\theta)$
crosses the level $E_{n+1}'(\theta)$ of a state with a different
$\theta$ angle. Then this scheme breaks down as can be seen 
in~\figRef{xSpezial.ps} because the wrong vacuum is subtracted
in this case. The spurious mass $M_n$ 
dives into the vacuum and becomes imaginary. The larger the momentum $N$
the smaller the fermion mass where this happens. 

In~\figRef{xuSpezial.ps} we depict the vector and scalar masses
computed under the assumption that there is only one vacuum: the vacuum
corresponding to $\theta=0$. In this case, the second vacuum appears
as a state of its own in~\figRef{xuSpezial.ps}.
The $\theta=\pi$ states do not only have the wrong masses 
without appropriate treatment of the vacuum,
they also loose their covariance (i.e. they acquire $N$-dependence)
in a drastic way. Already at very small fermion masses does
the $\theta=\pi$
vector state crosse the $\theta=0$ scalar state and a level-crossing
occurs. It is this level crossing which limits the fermion mass
region in which we are able to describe the $\theta=\pi$ sector correctly
as can be seen through a comparison of~\figRef{xSpezial.ps} and
~\figRef{xuSpezial.ps}. 
The subtraction of the wrong vacuum energy $E'_0(\theta)$ is responsible
for the breakdown of covariance and other unphysical effects in
~\figRef{xuSpezial.ps} and~\figRef{xSpezial.ps}


Parts of this  chapter have been presented on the
IX th International Conference on Recent Progress in Many-Body Theories
in Sidney, Australia~\Cite{Australien}.
A brief summary of these results (and additional figures) is accepted for publication~\Cite{PhysLettSchwing}.

%% file: CPhi4.i
Our results concerning the $\varphi^4$ theory have been published in
\Cite{Kroger:1997jm} and \Cite{Kroger:1997hj}. These papers are enclosed at the end of
this thesis. A further paper (on both the $\varphi^4$ theory
and the Schwinger model) will appear in
\Cite{Australien}.  
The results described in \Cite{Kroger:1997jm,Kroger:1997hj} can be
further improved upon in order to better deal with
the logarithms appearing in the four-dimensional scalar model.
We shall describe how
to do this in yet another publication.

%% file: CConclusion.i
 
{\ttfamily\small\footnotesize
\begin{citation}

Que dites-vous? ... C'est inutile? ... Je le sais!\\
Mais on ne se bat pas dans l'espoir du succ\`es!

\end{citation}

[Edmond Rostand: Cyrano de Bergerac]
}

\section{Relativistic Forms}

 For more than a decade, considerable effort has been invested
in order to solve QCD and other field theories 
using FF techniques.  It was thought that
the FF could accompish what the IF could not: describing hadrons
as a bound-state of few constituents. 
We have shown that the FF is not equivalent to FF. We have also
shown that the FF is unphysical to a high degree. 
Some problems and insufficiencies of the FF  have been known for
a long time. It has been known, for instance, that
the FF is unable to describe 
massless left-movers and that time-ordered propagators depend on an 
arbitrary choice of prescription.  As a reaction to
these previously known problems,
several disparate philosophies had emerged in the literature
 \begin{itemize}
\item[A1]
The {\bf pragmatical} philosophy: some researchers argued that the known 
problems did
not justify abandoning a promising appoach: in particular, the successes of the
FF in two space-time dimensions were impressive, the massive
Schwinger model being a prominent example. It was long claimed 
---without proof--- that these successes
could not be repeated in the conventional IF --- at least not in such
a simple way.
\item[A2] 
{\bf Other researchers}, however, became suspicious 
and conjectured that
the FF and the IF were not equivalent, acknowledging intrinsic shortcomings.
Fascinated by the relative simplictiy of the FF, 
they tried to overcome these shortcomings by treating the FF as an
effective theory which may be further improved upon if additional
effective interactions are introduced. Some of these researchers tried
to turn the badly broken Poincar\'e symmetry of FF quantisation
into an advantage:
the class of additional effective interactions ---so they conjectured--- is
defined as the class of operators which restore these broken symmetries
when added to the Hamiltonian. 
This programme might be called  
FF-bootstrap. 
\item[A3]
{\bf A third class of researchers}, finally, 
tried to trace every problem of the FF to the fact that most researchers
neglect zero modes. 
\end{itemize}
The results of this Ph.D. thesis  
render all of these philosophies questionable or unnecessary. 
\begin{itemize}
\item[R1]
{\bf Firstly}, we have shown that the IF and the FF are not equivalent
and that the FF in its pure form leads to completely unphysical
predictions such as infinite speed of light, time travel and the
violation of both microcausality
and causality. There is one theory where the FF yields an {\em exact}
mass-spectrum: The massless Schwinger model. We have traced this
to the complete scale-invariance of the Schwinger boson mass and to
the fact that the unconsistent chiral Schwinger model becomes consistent
for axial boson states in an infinitesimally small box. Already the massive
Schwinger model can be solved only approximately in the FF
--- albeit with good accuracy. 
\item[R2]
{\bf Secondly}, we have demonstrated that the FF is not, in general, 
a low-energy (or low mass)
effective theory since it does not do what a low-energy
effective theory is
supposed to do: reproducing the low-energy spectrum of the 
full theory. A prominent exception is the FF Hamiltonian of the Schwinger
model on an {\em infinite} effective lattice because 
(a) no left-movers contribute to the
self-energy and (b) the contribution of the zero mode disappears
on an infinite lattice. 
Our findings show that
symmetry considerations do not sufficiently
constrain the number of possible effective interactions which
could improve the FF Hamiltonian. In one spatial dimension, for example,
we may replace the FF Hamiltonian $H_R$ by the IMF Hamiltonian
or by the effective IMF Hamiltonian $H_{\text{eff}}$. In the IMF there is no
need to add additional, effective interactions in order to restore, say,
parity symmetry, since the IMF Hamiltonian does not violate this symmetry. 
Yet the strong IMF is unphysical as we have seen. Thus, parity is not helpful
in order to distinguish the unphysical IMF Hamiltonian from 
a physical Hamiltonian. 
One might argue that in three spatial dimensions, rotation symmetry
is explicitely broken in the IMF, too so that rotation symmetry
may serve in order 
to distinguish the strong IF from the true
effective Hamiltonian. Yet if the symmetry argument
does not work in one dimension, this FF bootstrap is questionable in three dimensions
as well.  
\item[R3]
{\bf Thirdly}, the problems of the FF cannot be blamed on the
zero-modes as we have shown. On the contrary, they even
aggravate these problems in some circumstances.
For example, a correct treatment of zero-modes
renders the violation of microcausality worse not 
better. 
Constraints on zero modes arise due to unphysical BCs
or ---equivalently--- due to the attempt to construct an
effective theory on the classical level. Constraints
on zero modes are additional symptons of unphysical assumptions rather
than the solution of problems which arise due to exactly these assumptions. 
Our finding that the FF is the {\em classicaly} effective theory of the 
strong IMF explains why the zero modes of the scalar $\varphi^4$ model
obey the {\em classical} constraint of the Landau mean field theory 
if other modes are discarded. This explains why the FF is unable
to go beyond mean-field results in the scalar model. 
Thus, the problems of the FF do not go away if
the zero modes are properly taken into account via the 
Dirac-Bergmann algorithm. 
\end{itemize}

A quantisation method that is incapable of describing relativisitic propagators can hardly be justified. One might argue that some FF successes 
in one-dimensional theories provide such a justification. Not so. 
We have shown for QED($1+1$) that these successes may
easily be reproduced within the traditional IF formalism. The 
FF is not needed. 
What is more, we have laid bare the mechanism which enables the FF and the IMF to be applied upon the massive
Schwinger model: The LMF treatment of this model is almost
{\em scale independent} on a small {\em infinite} effective lattice. The massless
Schwinger model is {\em exactly scale-independent} even for finite lattices. 
We have also shown that this scale-independence --- which allows
for the lattice scaling window to be broken --- is peculiar to the
Schwinger model: already the two-flavour Schwinger model does 
not exhibit this feature. 

The FF only works (a) if it coincides with a viable approximation
in the IF formalism (e.g. the special form of fermionic contractions)
or (b) if an infinity of parameters are left free which can
be adjusted so as to fit into reality (e.g. perturbation theory)
or (c) in the non-relativistic limit.

More or less byzantine methods may be devised in order to salvage
FF perturbation theory (or the FF description of the axial anomaly) 
using an infinity
of parameters as input. 
One cannot help but noticing, however, that such
efforts are
but a clumsy, awkward and questionable way of repeating what can be
done with relative ease in the IF or with covariant methods (beyond
simple Feynman diagramms).
Even some successful applications of the FF on mass spectra of 
two-dimensional theories
become bleak when compared to analogous computations in the IF.
Therefore, it is necessary to ask the question: Is there any
advantage of the FF over the IF/LMF?
What do we
avail by using the FF if everything that the FF does can be done
equally well in the IF and usually with much more ease (but not vice versa)?

\section{$\varepsilon$ Co-ordinates}

We have shown that $\varepsilon$ co-ordinates are completely equivalent
to the IF quantisation for $\varepsilon\neq 0$ and equivalent to the 
FF quantisation for $\varepsilon=0$. 
The advantages of $\varepsilon$ co-ordinates quantisation are by no means due to
$\varepsilon$-quantisation itself. We have traced these advantages to
the {\bf implicit} choice of small lattice sizes $2L(\varepsilon)$ in the limit
$\varepsilon\rightarrow 0$. In this limit, $\varepsilon$ quantisation 
becomes identical to the IMF which violates the lattice scaling window
even though it is more physical than the FF and does not destroy microcausality. 
Only if $\bar \frak{L}^3(\varepsilon)$ is increased 
in the limit $\varepsilon\rightarrow 0$ does the lattice size not decrease.
In this case, there is no advantage in using these co-ordinates.

\section{The Massive Schwinger Model}

These results seem to be rather negative since they seem to eliminate
the hope of describing QFTs in the Hamiltonian approach. 
This would be true if the conventional IF were much too complicated in order
to describe QFTs as is often claimed in the FF literature. For instance,
it is often claimed
that the relatively accurate description of the Schwinger model
cannot be repeated in the IF, at least not with the ease in which this
is done in the FF.  Fortunately, this claim has no foundation: we have demonstrated, in Chapter~\secRef{CNum}, 
that the opposite is true. We have shown that
the mechanism which renders the FF (or $\varepsilon$ co-ordinates
or the IMF) simple is {\em not} a specific quantisation surface: 
the relevant mechanism is rather the {\em implicite choice of an arbitrarily small volume}.
{\em This} is why
the vacuum appears to be almost trivial (i.e. trivial except for zero-modes). 
In the literature, the (almost) triviality of the FF vacuum
is is usually claimed to be a result of fact that (a) the vacuum 
is annihilated by $P_-$ and (b) no constituent particle
is allowed to have a negative $p_-$ momentum. 
This, however, is only a {\em sufficient} reason for the triviality
of the vacuum (if zero-modes are discarded) but by no means
{\em necessary}. 
The vacuum of the IMF is (almost) trivial even though $P_3$ is
{\em not} non-positive or non-negative.  
Unfortunately, an infinitely small volume is unphysical, as we have seen,
since it is not compatible with the LGT scaling window. 
This is why we propose to use a volume that is sufficiently small so as
to profit from a simplified vacuum yet sufficiently
large so as to fulfill the scaling window of LGT. 

Based on this observation, 
we have put forward another method, 
the LMF, which shares every advantage
of the IMF or the FF while avoiding their respective disadvantages.  
We have tested this method on two models, $QED(1+1)$ (Chapter\secRef{CNum}) and the scalar $\varphi^4$ model (Chapter\secRef{CPhi4})
in four space-time dimensions. 
Excellent agreement with other methods is found.   
In the scalar model, our results are compatible with the semi-analytical
results obtained by L{\"u}scher and Weisz in the domain of applicablilty 
of our method\footnote{Our application of the LMF method on 
the scalar model needs some improvement, though (in order to better
describe logarithmic corrections). 
This will be the
subject of another publication}.
For the Schwinger model our results agree well with the computations
of Hamer et al. 
Contrary to the FF, we were able to verify predictions of chiral perturbation
theory on a {\em finite} lattice. 

Chapter\secRef{CSchwinger} uses the exactly known solution of massles $QED(1+1)$
in order to justify the approximations that went into our computations. 
Furthermore we are able to rigorously demonstrate {\em that} and 
{\em why} a moving frame
is advantageous when compared to the rest frame.
We pointed out that topological effects, the chiral
condensate and the axial anomaly
are only essential in the lattice rest-frame. In the LMF, in contrast,
discarding topological effects, destroing the chiral
condensate or treating
the anomaly inadequately does not hamper the correct description
of the mass-spectrum in the LMF. 
We caution that the massive Schwinger model is a rather peculiar theory.
Consequently, it is not very likely that all of these results hold
in the case of QCD$(3+1)$ as well. 
Our method works best if the correlation length $a\xi=1/M$ is sufficiently
small when compared to the diameter $2R$ of the physical particle in 
question. For pions however, the product $2MR$ is very close to one. 
We may therefore expect that the assumption of a trivial vacuum
might not allow us to treat pions correctly. 
This problem can be dealt with by allowing for a restricted number
of lattice sites with negative momenta to be occupied. 
In addition to that, it will be useful to introduce a {\em spatially}
anisotropic lattice for reasons that will be detailed in another publication.

\proPub{

\section{The Massless Schwinger Model}

We have found that the mass spectrum of the Schwinger model is remarkably robust
against a number of severe approximations:
\begin{itemize}
\item 
The mass spectrum does not depend on the size $L$ of the lattice. Consequently,
even the strong IMF is able to reproduce the correct spectrum. 
\item
Axial gauge does not modify the mass spectrum for $L\rightarrow 0$ except
for the spectrum in the lattice rest frame. 
\item
The diagonalisation of the Hamiltonian may be performed using a basis
with definite axial charge instead of using gauge-invariant $\theta$ states.
\item
For $L\rightarrow 0$ the forcible exclusion of fermionic left-movers has no
effect whatsoever on the mass spectrum, even if left-movers are removed
on the classical level. We have traced the latter feature to the fact
that {\em fermionic} left-movers do not contribute to 
the self-energy of right-movers. 
\item
The effective size $N+1$ of the lattice may be chosen to be ridiculously small:
One obtains the {\em exact} Schwinger boson mass even on a effective 
lattice with two lattice sites. 
\end{itemize}
This was elaborated in Chapter~\secRef{CSchwinger}. This robustness
towards severe approximations is the reason for why the FF and the IMF are
able to reproduce the boson masses of the Schwinger model. 
If all of the approximations listed above are performed, the Schwinger
model is formally equivalent to the unphysical 
{\em chiral} Schwinger model\footnote{It is remarkable that
the chiral Schwinger model is even inconsistent without these approximations.}
or to the FF. See Chapter~\secRef{CSchwinger} and~\secRef{CNum}.

We found that the volume dependence of the
massless Schwinger model differs drastically from the massive
Schwinger model: the mass spectrum of 
massive Schwinger model is sensitive to 
correct implementation of the scaling window $2L>a\xi$ whereas the mass spectrum of the 
massless Schwinger model comes out correctly even if the 
scaling window is not fulfilled: Schwinger boson masses are independent
of the lattice size $2L$ --- even in the rest frame. 
{\em If and only if} an infinite effective lattice $N=\infty$ is used, the massive
Schwinger model, too, becomes insensitive to the lattice size $2L$ and the 
LMF Schr{\"o}dinger
equation for the vector state coincides with the FF Schr{\"o}dinger equation.
Therefore, the results obtained in Ref.\Cite{Mo:1992sv} for the
vector state may be considered as LMF results
in the limit $N=\infty$. We have traced this coincidence 
to the fact that the Schwinger model does not have contain a four-boson interactions. 

Topological effects contribute primarily to 
the rest-frame sector $\vec P=0$ where the De Broglie wave-length
$\lambda={2\pi\over |\vec P|}=\infty$ is larger than any finite
lattice size $2L$.
\index{L@$L$} 
This implies, so we argue, that the relations between the mass-spectrum
and the chiral condensate obtained in chiral perturbation theory
are fundamentally different for $\vec P=0$ and for $|\vec P|\neq 0$
respectively. Since the mass spectrum of QED$(1+1)$ is almost insensible 
to the lattice size $2L>1/M$ in the LMF, masses computed in the LMF
are related to the fermionic
condensate $C(\infty)$ obtained at
infinite volume rather than to the condensate 
$C(L)=\brac{P}\bar\psi\psi\cket{P}$ at finite volume\footnote{
Intuitively, one might conjecture 
that the chiral condensate extracted from the
$m$-dependence of the physical mass spectrum is equal to
$C(L/\sqrt{1-v^2})$ where $v$ is the velocity of the physical particle. 
For $v\approx 1$ we have $C(L/\sqrt{1-v^2})\approx C(\infty)$.}. 
For this to be
true, it is {\em de rigeur} that no approximation
destroy Lorentz covariance. 
The spectrum at $\vec P\approx 0$, however, is related to the condensate 
$C(L)$ at finite volume.

} 

\section{Structure Functions}
We have demonstrated in Chapter\secRef{CStructure} 
that the impulse approximation which relates structure functions to 
a convolution of distribution
functions and parton cross sections is {\em frame dependent} due
to the non-triviality of the vacuum. 
For frames similar to the Breit-frame, however, this frame dependence is
negligible and the IA is a good approximation. 
It is well known that the rest-frame is not suited in order to interpret
structure functions in terms of quarks and gluons. We have shown, additionaly,
that the IMF is not suited either,
since close to both frames, the structure
functions calculated using the IA show a strong dependence on the 
momentum signaling the breakdown of the IA. 
This frame dependence stems from the fact that the vacuum is 
probed in both extreme cases. The fact that this frame dependence
is so strong stems from the fact that the vacuum
quark distribution (which is not a physical observable) diverges for small
momenta $\vec k$. 
 
Structure functions are related to {\em space-like} correlation functions. They
are {\em not} related to {\em light-like} correlation functions. 
Space-like correlation functions attain their maximal simplicity in \Index{axial gauge}
and {\em not} in the notoriously difficult~\Cite{McCartor:1994mu} 
\Index{axial gauge}. 
The Taylor series of space-like correlation functions with respect to $x^3$
corresponds
to the \Index{operator product expansion} (OPE) in the sense that the
n-th Taylor coefficient is proportional to the n-th moment\footnote{
It is well-known that the same statement holds {\em formally} 
for the Taylor series of light-like
correlation-functions in terms of $x^-$. 
}
of the structure
function $F_2^{\text{IA}}$. This means that the OPE is a {\em short}-distance
expansion (at least in the IA) and {\em not} a problematic light-cone expansion. 
This is conceptually advantageous because the OPE does ---strictly speaking---
only make sense as a short-distance expansion.